\documentclass[12pt,lot,lof]{puthesis}
\newcommand{\proquestmode}{}
\usepackage{amsmath,amssymb}
\usepackage{braket}
\usepackage{hyperref}
\usepackage{graphicx}
\usepackage{color}
\usepackage{verbatim}
\usepackage{multirow}
\usepackage{longtable}
\usepackage{booktabs}
\usepackage{url}
\usepackage[caption=false]{subfig}
\usepackage{epstopdf}
\DeclareMathOperator{\Tr}{Tr}
\renewcommand{\Im}{\text{Im}}
\renewcommand{\Re}{\text{Re}}
\renewcommand{\vec}[1]{\mathbf{#1}}
\newcommand{\RN}[1]{%
  \textup{\uppercase\expandafter{\romannumeral#1}}%
}
    \newcommand{\Eq}[1]		{(\ref{#1})}			%just parenthesis
    \newcommand{\Fig}[1]	{Fig.~\ref{#1}}			%refer to figure
\title{Heisenberg-Langevin formalism for Open circuit-QED systems}

\submitted{November 2017}  % degree conferral date (January, April, June, September, or November)
\copyrightyear{2017}  % year in which the copyright is secured by publication of the dissertation.
\author{Moein Malekakhlagh}
\adviser{Hakan E. T\"ureci}  %replace with the full name of your adviser
\department{Electrical Engineering}
\ifdefined\printmode
\else
\ifdefined\proquestmode
\makeatletter
\hypersetup{pdftitle=\@title,pdfauthor=\@author}
\makeatother

\else
\usepackage{hyperref}
\hypersetup{colorlinks,bookmarksnumbered}

\makeatletter
\hypersetup{pdftitle=\@title,pdfauthor=\@author}
\makeatother

\fi % proquest or online formatting
\fi % printed or online formatting

\ifodd 0
\renewcommand{\maketitlepage}{}
\renewcommand*{\makecopyrightpage}{}
\renewcommand*{\makeabstract}{}

\else

\abstract{We present a Heisenberg-Langevin formalism to study the effective dynamics of a superconducting qubit coupled to an open multimode resonator, without resorting to the rotating wave, two level, Born or Markov approximations. Our effective equations are derived by eliminating resonator degrees of freedom while encoding their effect in the Green's function of the electromagnetic background. We account for the openness of the resonator exactly by employing a spectral representation for the Green's function in terms of a set of non-Hermitian modes. A well-behaved time domain perturbation theory is derived to systematically account for the nonlinearity of weakly nonlinear qubits like transmon. We apply this method to the problem of spontaneous emission, capturing accurately the non-Markovian features of the qubit dynamics, valid for any qubit-resonator coupling strength. Any discrete-level quantum system coupled to the electromagnetic continuum is subject to radiative decay and renormalization of its energy levels. When inside a cavity, these quantities can be strongly modified with respect to vacuum. Generally, this modification can be captured by including only the closest resonant cavity mode. In circuit QED architecture, with substantial coupling strengths, it is however found that such rates are strongly influenced by far off-resonant modes. A multimode calculation over the infinite set of cavity modes leads to divergences unless an artificial cutoff is imposed. Previous studies have not pointed out what the source of this divergence is. Quite interestingly, the renormalization of spectrum is mutual, i.e. the electromagnetic modal structure of the cavity is also modified due to scattering by the atom. In cavity QED, this phenomenon is manifested as a diamagnetic term, known as the $A^2$ contribution. We show that unless the effect of $A^2$ is accounted for up to all orders exactly, any multimode calculations of circuit QED quantities is bound to diverge. Subsequently, we present the calculation of finite radiative corrections to qubit properties that is free of an artificially introduced high frequency cut-off.
}
\acknowledgements{I would like to thank my advisor, Hakan E. T\"ureci, whose guidance and support has always inspired me throughout my studies. I am also very grateful to my dissertation committee: Alejandro Rodriguez, Jeff Thompson, Claire F. Gmachl and Gerard Wysocki. I have learned and benefited a lot from the members of my research group: Manas Kalkurni, Li Ge, Kostas Makris, Camille Aron, Alexandru Petrescu, Anja Metelmann, Mykola Bordyuh, Hassan Shapourian, Omer Malik, Saeed Khan and Gerry Angelatos. My special thanks belongs to my wife, Melissa, my parents, Esmaeil and Nahid, and my brother, Mojtaba, for their endless support.
}
\dedication{To my family.}
\fi % disable frontmatter
\begin{document}
\makefrontmatter
%%%%%%%%%%%%%%% Publications %%%%%%%%%%%%%%
\chapter*{Publications}
\label{App:Publications}
This dissertation is based in part on the following peer reviewed articles \cite{ Sundaresan_Beyond_2015, Malekakhlagh_Origin_2016, Malekakhlagh_NonMarkovian_2016, Malekakhlagh_Cutoff_2017} and conference presentations \cite{Malekakhlagh_Non-Markovian_APS2014, Malekakhlagh_Amplitude_APS2015, Malekakhlagh_Origin_APS2016, Sundaresan_Multimode_APS2016, Malekakhlagh_Non-Markovian_APS2017, Convergent_Petrescu_APS2017}:
\section*{Journal Publications}
\begin{enumerate}
\item Moein Malekakhalgh, Alexandru Petrescu and Hakan E. T\"ureci, "Cut-off free circuit quantum electrodynamics", Phys. Rev. Lett. 119, 073601, 2017
\item Moein Malekakhlagh, Alexandru Petrescu, Hakan E. T\"ureci, "Non-Markovian dynamics of a superconducting qubit in an open multimode resonator", Phys. Rev. A 94, 063848, 2016
\item Moein Malekakhlagh, Hakan E. T\"ureci, "Origin and implications of an $A^2$-like contribution in the quantization of circuit-QED systems", Phys. Rev. A 93, 012120, 2016
\item Neereja M Sundaresan, Yanbing Liu, Darius Sadri, Laszlo J Szocs, Devin L Underwood, Moein Malekakhlagh, Hakan E. T\"ureci, Andrew A Houck, "Beyond strong coupling in a multimode cavity", Phys. Rev. X 5, 021035, 2015
\end{enumerate}
\section*{Conference Presentations}
\begin{enumerate}
\item Moein Malekakhlagh, Alexandru Petrescu and Hakan E. T\"ureci, "Non-Markovian dynamics of superconducting qubit in open multimode resonator", APS March Meeting, 2017

\item Alexandru Petrescu, Moein Malekakhlagh, Hakan E. T\"ureci, "Convergent expressions for Purcell rate and Lamb shift of superconducting qubit in an open multimode resonator", APS March Meeting, 2017

\item Moein Malekakhlagh, Hakan E. T\"ureci, "Origin and Implications of $A^2$-Contribution in the Quantization of Circuit-QED Systems", APS March Meeting, 2016

\item Neereja Sundaresan, Yanbing Liu, Darius Sadri, Laszlo Szocs, Devin Underwood, Moein Malekakhlagh, Hakan E. T\"ureci, Andrew Houck, "Multimode Strong Coupling in Circuit QED", APS March Meeting, 2016

\item Moein Malekakhlagh, Neereja Sundaresan, Yanbing Liu, Darius Sadri, Andrew Houck, Hakan E. T\"ureci, "Amplitude Bistability in the Multimode Regime of Circuit-QED", APS March Meeting, 2015

\item Moein Malekakhlagh, Hakan E. T\"ureci, Dmitry Krimer, Matthias Liertzer, Stefan Rotter, "Non-Markovian Qubit Dynamics in Multimode Superconducting circuit cavities", APS March Meeting, 2014
\end{enumerate}
%%%%%%%%%%%%%%%%% Chapters %%%%%%%%%%%%%%%%
\chapter{Introduction\label{Ch:intro}}

Quantum mechanics, in its early developement,  mainly focused on abstract problems containing only a single particle such as the Hydrogen atom, single particle scattering and quantum tunneling. On the other hand, the real world physical phenomena are governed by interactions between many particles. Besides the theoretical complications of describing a many-body quantum system, it turns out that even a pure numerical simulation of such systems is almost impossible with our current classical computers. The reason lies in the exponential growth of the Hilbert space size of the problem with the number of particles. In 1982, Feynman put forward a famous proposition \cite{Feynman_Simulating_1982} to use quantum computers, a universal set of quantum gates, to directly simulate quantum mechanical phenomena instead. In 1985, Deutsch \cite{Deutsch_Quantum_1985} extended the Church-Turing principle for quantum computers. Later on in 1994, Shor proposed an exponentially faster quantum algorithm \cite{Shor_Algorithms_1994} for integer factorization. Two years later in 1996, Grover \cite{Grover_Fast_1996} introduced a fast quantum search algorithm. These algorithms confirmed for the first time that a quantum computer has great advances in computational power compared to its classical analogue.  

These theoretical propositions motivated the scientific community from a variety of fields to dedicate more attention towards quantum information processing, in which a computational task is performed by manipulating a collection of quantum objects. Various experimental realizations of a quantum computer have been proposed ever since, started with \textit{natural} quantum systems such as trapped ions \cite{Cirac_Quantum_1995}, electron spin \cite{Nowack_Coherent_2007}, and ultracold atoms
in optical lattices \cite{Dickerscheid_Ultracold_2003}. On the other hand, there has been a great progress in experimental control over light-matter interactions at the quantum level with solid-state optical systems in the past decade \cite{Schoelkopf_Wiring_2008, Yamamoto_Semiconductor_2000, Haroche_Exploring_2013}. This opened new doors to employ these \textit{artificial} quantum systems, such as superconducting circuits, as a potential building block for quantum information processing. Besides the improved controllability and tunability with respect to the natural quantum systems, they have an important technological advantage of being fabricated through the same lithography techniques already used in conventional electronics.

The elementary building block of a natural quantum optical systems is the basic Quantum Electrodynamcis (QED) system that consists of a two-level system interacting with a single mode of an electromagnetic field \cite{Scully_Quantum_1997, Gerry_Introductory_2005, Walls_Quantum_2008, Haroche_Exploring_2013, Milonni_Quantum_2013}. On the contrary, the building block of artificial quantum system is a macroscopic  superconducting quantum bit (qubit) that is coupled to a superconducting microwave resonator \cite{Wallraff_Strong_2004, Blais_Cavity_2004}.  The artificial qubit is made of a large number of atoms, which only collectively mimics the behavior of a single atom \cite{Bouchiat_Quantum_1998, Nakamura_Coherent_1999, Friedman_Quantum_2000, Van_Quantum_2000, Koch_Charge_2007, Manucharyan_Fluxonium_2009}. Superconductivity plays a crucial role here by freezing the vast dergees of freedom into a single collective degree of freedom. In both cases, the interaction generates a hybridization between light and matter, which leads to hybrid excitations, called polaritons, with a finite lifetime.  

Besides the applications in quantum information processing, the tunability and scalability of superconducting circuits provides a promising platform to build circuit-QED lattices to study emergent many body quantum phenomena away from equilibrium \cite{Houck_On-chip_2012, Schmidt_Circuit_2013}. In such systems, the frequency of each individual qubit and its coupling to the electromagnetic field can be tuned. Moreover, qubits are able to interact with each other either directly through mutual capacitance \cite{Pashkin_Quantum_2003, Mcdermott_Simultaneous_2005} or inductance \cite{Majer_Spectroscopy_2005}, or indirectly over long distances mediated by a shared photonic mode \cite{Blais_Cavity_2004, Sillanpaa_Coherent_2007, Majer_Coupling_2007}. Since photons in these systems are just circuit excitations, the total particle number is not necessarily conserved. The inevitable photon loss, which makes these systems intrinsically open, is then compensated through continuous external
driving. These experimental advances have brought forth a rich class of driven-dissipative open quantum systems, that exhibit non-equilibrium quantum dynamics that have received considerable attention in recent years. 

Quantum information applications has motivated the majority of theoretical studies on circuit-QED to employ simple phenomenological models like Jaynes-Cummings \cite{Jaynes_Comparison_1963, Shore_Jaynes_1993} and Rabi \cite{Rabi_Space_1937}. Although these models work reasonably for cavity-QED systems, but they fail to explain some experimental results in circuit-QED \cite{Houck_Controlling_2008} due to different regime of parameters that such systems operate in. In particular, stronger light-matter interaction that is achievable in circuit-QED violates rotating-wave approximation, hence using Jaynes-cummings model is not well justified. Moreover, stronger interaction between the qubit and resonator excites far off-resonant resonator modes and requires a multimode theory. However, it has been pointed out that generalizing these phenomenological models to a multimode case leads to results that are divergent in the number of modes \cite{Houck_Controlling_2008, Filipp_Multimode_2011}. Furthermore, in order to reduce the charge noise, most superconducting qubits that are presently employed have weak nonlinearity. Therefore, two-level approximation will break down specially for stronger driving.

The present thesis is an attempt at understanding some of the fundamental aspects of simplest incarnation of these artificial systems, and developing the requisite theoretical and computational techniques for their exploration. In particular, the main focus of this dissertation is to provide a first principles study of the quantum dynamics of an open QED system without adopting the approximations commonly applied to simplify the theory. The dissertation is structured as follows: In chapter~\ref{Ch:Background}, we provide an overview of cavity quantum electrodynamics, summarize the models commonly used to describe these systems and discuss their advantages and limitations. In chapter~\ref{Ch:NonMarkovian}, we introduce our first principles Heisenberg-Langevin formalism, which does not employ rotating wave, two-level, Born or Markov approximations. In chapter~\ref{Ch-OriginOfA2}, we address the long-standing anomaly of multimode divergence in circuit-QED and provide a resolution by emphasizing the role of the diamagnetic $A^2$ contribution. Finally, in chapter~\ref{Ch:SummaryOutlook}, we summarize our main results and discuss potential future directions.
\chapter{Cavity Quantum Electrodynamics\label{Ch:Background}}

Quantum electrodynamics refers to the general theory of interaction between atoms and the electromagnetic field. It borrows and consistently unifies concepts from classical optics, classical electromagnetism and quantum mechanics. The first theoretical study of light dates back to Newton who formulated a geometrical (or ray) description of optical phenomena, where light particles only travel in straight lines. Despite being successful in capturing the physics of lenses and mirrors with curved surfaces, this theory was unable to illustrate the wave-like phenomena such as interference and diffraction. On the other hand, unification of electricity and magnetism, which was achieved by Faraday and Maxwell, revealed that light can be modeled as oscillations of electric and magnetic fields. This led to development of a new branch of optics called physical (or wave) optics. The beginning of twentieth century witnessed two revolutionary experiments, blackbody radiation and photoelectric effect, that in turn challenged the wave nature of light. Planck successfully explained the black body spectrum by assuming that the energy of light was quantized. Later on, Einstein was also able to provide an explanation for the photoelectric effect by treating light as discrete quanta of particles, which he called photon. This wave and particle duality of light remained a mystery for almost two decades, but was eventually resolved by Dirac in a unified theory called Quantum Electrodynamics.

In this chapter, we provide an overview of the field of cavity quantum electrodynamics (CQED) that studies the interaction of atoms and light confined in a cavity, under experimental conditions such that the quantum nature of light becomes crucial. In Sec.~\ref{Sec:Background-cavityQED} we review the quantization of light in a CQED system and provide a brief derivation of light-matter coupling for these systems from Maxwell's equations. Section~\ref{Sec:Background-circuitQED} gives a succinct summary of the field of circuit quantum electrodynamics, which implements similar quantum optical phenomena using superconducting materials. This is the longest section of this chapter, where the discussed material will be used and referred to in the subsequent chapters. Section~\ref{Sec:Background-closedQED} discusses the physics of well-known closed QED models including Jaynes-Cummings and Rabi. The last Sec.~\ref{Sec:Background-openQED} is devoted to the study of open quantum systems and the net effect of a quantum environment on the QED system of interest. Brief discussion of two main treatments of openness, namely Lindblad master equation and input-output formalism, are presented in Secs.~\ref{SubSec:Background-Lindblad} and~\ref{SubSec:Background-In-Out}, respectively.

\section{Cavity quantum electrodynamics}
\label{Sec:Background-cavityQED}
In 1946, Edward M. Purcell published a revolutionary paper showing that the radiative decay of magnetic resonance of nuclear spins could be dramatically modified, with respect to free space, when coupled to a resonant microwave circuitry \cite{Purcell_Resonance_1946}. Motivated by this idea, quantum opticians also realized that by confining atoms in a finite cavity the atomic radiative transition dynamics and in particular the spontaneous emission rate is significantly changed depending on the mode structure and photonic density of states of the cavity. Engineering polarizable media inside high finesse cavities became possible in 1960s, which led to the discovery of laser. By the 1980s, the improvement in experimental control in CQED systems allowed researchers to study single atoms interacting with a single mode of the electromagnetic field. At this time, resonators with higher quality factor and also stronger light matter interaction made it possible to explore new regimes of spontaneous emission, where the atom and the field are able to coherently exchange excitations. As a result, a reversible atomic transition dynamics is observed, where the coherence is eventually lost due to inevitable openness of the system. In the following, we first discuss the quantization of electromagnetic fields inside a cavity in Sec.~\ref{Sec:Background-QuantOfFreeEM}. Next, we briefly discuss how to derive the basic CQED Hamiltonian, the Rabi model, that describes the interaction of a two level system with a single mode cavity mode in Sec.~\ref{Sec:Background-LightMatterInt}.
\subsection{Quantization of electromagnetic field}
\label{Sec:Background-QuantOfFreeEM}
In this section, we review the quantization of free electromagnetic field inside a finite cavity \cite{Scully_Quantum_1997, Gerry_Introductory_2005, Milonni_Quantum_2013}. This simple example is the first step towards building a quantum Hamiltonian for the interaction of atoms and electromagnetic field in a CQED sytem. We start by the classical source-free Maxwell's equations for the electric field $\vec{E}(r,t)$ and magnetic field $\vec{B}(r,t)$ as
\begin{subequations}
\begin{align}
&\nabla\times\vec{E}(\vec{r},t)=-\partial_t\vec{B}(\vec{r},t),
\label{eqn:Background-CQED-CurlOfE}\\
&\nabla\times\vec{B}(\vec{r},t)=\epsilon_0\mu_0\partial_t\vec{E}(\vec{r},t),
\label{eqn:Background-CQED-CurlOfB}\\
&\nabla\cdot\vec{E}(\vec{r},t)=0,
\label{eqn:Background-CQED-DivOfE}\\
&\nabla\cdot\vec{B}(\vec{r},t)=0.
\label{eqn:Background-CQED-DivOfB}
\end{align}
\end{subequations}
The first step towards quantization is to solve for the eigenmodes and eigenfrequencies of the field. Taking the curl of Eq.~(\ref{eqn:Background-CQED-CurlOfE}) and replacing $\nabla\times\vec{B}$ from Eq.~(\ref{eqn:Background-CQED-CurlOfB}), we obtain the equation of motion for $\vec{E}(\vec{r},t)$ as
\begin{align}
\nabla\times[\nabla\times\vec{E}(\vec{r},t)]+\epsilon_0\mu_0\partial_t^2\vec{E}(\vec{r},t)=0.
\label{eqn:Background-CQED-CurlOfCurlOfE}
\end{align}
Since we assume there are no sources, we can simplify Eq.~(\ref{eqn:Background-CQED-CurlOfCurlOfE}) further. Employing the vector calculus identity $\nabla\times(\nabla\times\vec{E})=\nabla(\nabla\cdot\vec{E})-\nabla^2\vec{E}$ and Eq.~(\ref{eqn:Background-CQED-DivOfE}) we find
\begin{align}
\left(\nabla^2-\frac{1}{c^2}\partial_t^2\right)\vec{E}(\vec{r},t)=0,
\label{eqn:Background-CQED-Wave Eq for E}
\end{align}
which is a source free wave equation for the electric field. To solve for the spectrum, we apply the Fourier transform in time
\begin{align}
\tilde{\vec{E}}(\vec{r},\omega)\equiv \int_{-\infty}^{+\infty} dt \vec{E}(\vec{r},t)e^{-i\omega t},
\label{eqn:Background-CQED-Def of tilde(E)(r,om)}
\end{align} 
to obtain a Helmholtz equation
\begin{align}
\left[\nabla^2+\left(\frac{\omega}{c}\right)^2\right]\tilde{\vec{E}}(\vec{r},\omega)=0.
\label{eqn:Background-CQED-Wave Eq for E}
\end{align}
The spectrum will be determined by the geometry and boundary conditions of the cavity.

%%%%%%%%%%%%%% Fig:LC Osc %%%%%%%%%%%%%%
\begin{figure}[t!]
\centering
\includegraphics[scale=0.40]{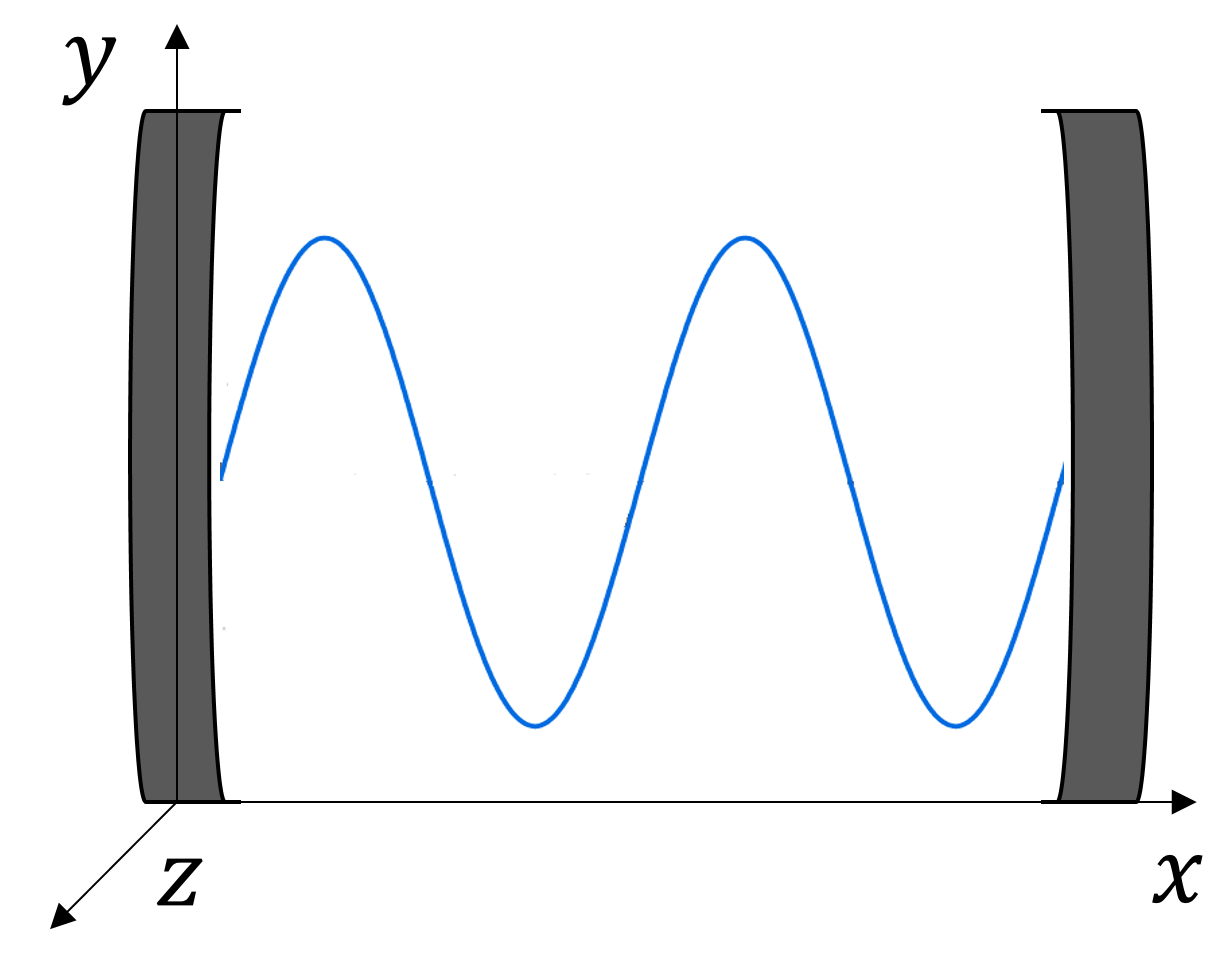}
\caption{Schematic picture of the electromagnetic field inside a one dimensional cavity. The electric ($z$-direction) and magnetic field ($y$-direction) are transverse to the axis of the cavity.}
\label{Fig:Background-CQEDMode}
\end{figure}
%%%%%%%%%%%%%%%%%%%%%%%%%%%%%%%%%%%%%%%%%%
First, we consider a simple one-dimensional geometry shown in Fig.~\ref{Fig:Background-CQEDMode}, where the electric $\vec{E}=E_z(x,t)\vec{u}_z$ and magnetic $\vec{B}=B_y(x,t)\vec{u}_y$ fields only have nonzero transverse components. As a result, Eq.~(\ref{eqn:Background-CQED-Wave Eq for E}) reduces to
\begin{align}
\left[\frac{d^2}{dx^2}+\left(\frac{\omega}{c}\right)^2\right]\tilde{E}_z(x,\omega)=0, 
\label{eqn:Background-CQED-1D Wave Eq for E}
\end{align}
The boundary conditions can be derived from continuity of the transverse electric field as
\begin{align}
\left. \tilde{E}_z(x,\omega)\right|_{x=0,L}=0,
\label{eqn:Background-CQED-1D BC for E}
\end{align}
where we assume that the field outside the cavity is zero. The solution to Eq.~(\ref{eqn:Background-CQED-1D Wave Eq for E}) with BC~\ref{eqn:Background-CQED-1D BC for E} gives the spectrum as 
\begin{align}
\tilde{E}_n(x)=\sqrt{2}\sin(k_n x),\quad k_n\equiv \frac{n\pi}{L},
\label{eqn:Background-CQED-1D Spec for E}
\end{align}
where $k\equiv \omega/c$ is the wave number.
\begin{subequations}

These eigenmodes form a complete basis, and therefore we can use a modal expansion for  for the fields inside the cavity as 
\begin{align}
&E_z(x,t)=\sum\limits_n q_n(t)\sin(k_n x),
\label{eqn:Background-CQED-SpecRep of E}\\
&B_y(x,t)=\sum\limits_n \frac{\epsilon_0\mu_0}{k_n}\dot{q}_n(t)\cos(k_n x),
\label{eqn:Background-CQED-SpecRep of B}
\end{align}
where $q_n(t)$ are a set of time dependent coefficients and modal expansion~(\ref{eqn:Background-CQED-SpecRep of B}) is found from Eq.~(\ref{eqn:Background-CQED-CurlOfB}). Inserting modal representations~\ref{eqn:Background-CQED-SpecRep of E} and~\ref{eqn:Background-CQED-SpecRep of B} into the free classical electromagnetic Hamiltonian
\end{subequations}
\begin{align}
\mathcal{H}_F=\int_V d^3\vec{r} \left[\frac{\epsilon_0}{2} \vec{E}^2(\vec{r},t)+\frac{1}{2\mu_0}\vec{B}^2(\vec{r},t)\right],
\label{eqn:Background-CQED-Free H}
\end{align}
and using the orthogonality conditions 
\begin{subequations}
\begin{align}
&\int_0^{L}dx\sin(k_n x)\sin(k_m x)=\frac{L}{2}\delta_{mn},
\label{eqn:Background-CQED-ortho sinsin}\\
&\int_0^{L}dx\cos(k_n x)\cos(k_m x)=\frac{L}{2}\delta_{mn},
\label{eqn:Background-CQED-ortho coscos}
\end{align}
\end{subequations}
we obtain a modal representation for the Hamiltonian as
\begin{subequations}
\begin{align}
&\mathcal{H}_F=\sum\limits_n \left[\frac{1}{2}m_n\dot{q}_n^2+\frac{1}{2}m_n\omega_n^2q_n^2\right]=\sum\limits_n \left[\frac{1}{2}m_n\dot{q}_n^2+\frac{p_n^2}{2m_n}\right].
\label{eqn:Background-CQED-modal rep of H}
\end{align}
\end{subequations}
In Eq.~(\ref{eqn:Background-CQED-modal rep of H}), we have defined a mass-like quantity $m_n\equiv\frac{\epsilon_0 V}{2\omega_n^2}$ and the conjugate momentum $p_n\equiv m_n \dot{q}_n$ to establish a connection to the Hamiltonian of a free harmonic oscillator.

Next, we quantize each independent normal mode by promoting the conjugate coordinate $q_n$ and momentum $p_n$ into quantum operators that obey the commutation relations
\begin{subequations}
\begin{align}
&[\hat{q}_n,\hat{p}_m]=i\hbar\delta_{mn},\\
&[\hat{q}_n,\hat{q}_m]=[\hat{p}_n,\hat{p}_m]=0.
\end{align}
\end{subequations}
Then, we can write these operators in terms of a set of creation $\hat{a}_n$ and annihilation $\hat{a}_n^{\dag}$ operators such that
\begin{subequations}
\begin{align}
&\hat{q}_n=Q_{n,\text{zpf}}\left[\hat{a}_n(t)+\hat{a}_n^{\dag}(t)\right],
\label{eqn:Background-CQED-hat(q)_n}
\\
&\hat{p}_n=-iP_{n,\text{zpf}}\left[\hat{a}_n(t)-\hat{a}_n^{\dag}(t)\right],
\label{eqn:Background-CQED-hat(p)_n}
\end{align}
\end{subequations}
where the zero-point fluctuation amplitudes can be found in connection with the ones for a simple harmonic oscillator as
\begin{subequations}
\begin{align}
&Q_{n,\text{zpf}}=\left(\frac{\hbar}{2m_n\omega_n}\right)^{1/2}=\left(\frac{\hbar\omega_n}{\epsilon_0 V}\right)^{1/2},
\label{eqn:Background-CQED-Q_(n,zpf)}\\
&P_{n,\text{zpf}}=\left(\frac{\hbar m_n\omega_n}{2}\right)^{1/2}=\left(\frac{\hbar\epsilon_0 V}{4\omega_n}\right)^{1/2}.
\label{eqn:Background-CQED-P_(n,zpf)}
\end{align}
\end{subequations}
Using Eqs.~(\ref{eqn:Background-CQED-hat(q)_n}) and~(\ref{eqn:Background-CQED-hat(p)_n}) we can find the quantum electric and magnetic fields accordingly as
\begin{subequations}
\begin{align}
&\hat{E}_z(x,t)=\sum\limits_n\left(\frac{\hbar\omega_n}{\epsilon_0 V}\right)^{1/2}\left[\hat{a}_n(t)+\hat{a}_n^{\dag}(t)\right]\sin(k_n x),
\label{eqn:Background-CQED-hat(E)_z(x,t)}\\
&\hat{B}_y(x,t)=-i\sum\limits_n\left(\frac{\hbar\omega_n}{\epsilon_0 Vc^2}\right)^{1/2}\left[\hat{a}_n(t)-\hat{a}_n^{\dag}(t)\right]\cos(k_n x).
\label{eqn:Background-CQED-hat(B)_y(x,t)}
\end{align}
\end{subequations}
At last, the quantum Hamiltonian is found as a sum over the Hamiltonians for each normal mode of the system 
\begin{align}
\hat{\mathcal{H}}_F=\sum\limits_n\hbar\omega_n\left(\hat{a}_n^{\dag}\hat{a}_n+\frac{1}{2}\right).
\label{eqn:Background-CQED-hat(H_3d)}
\end{align}

Up to here, we considered a specific one-dimensional cavity. However, these results can be generalized for a cavity with any arbitrary geometry in three-dimensions. The spectral representation of the quantum fields takes the form
\begin{subequations}
\begin{align}
&\hat{\vec{E}}(\vec{r},t)=\sum\limits_{\vec{k},\lambda}\vec{u}_{\vec{k},\lambda}\mathcal{E}_{\vec{k}}\left[\hat{a}_{\vec{k},\lambda}(t)+\hat{a}_{\vec{k},\lambda}^{\dag}(t)\right]\tilde{E}_{\vec{k}}(\vec{r}),
\label{eqn:Background-CQED-hat(E)(x,t)}\\
&\hat{\vec{B}}(\vec{r},t)=-i\sum\limits_{\vec{k},\lambda}\frac{\vec{k}\times \vec{u}_{\vec{k},\lambda}}{\omega_{\vec{k}}}\mathcal{E}_{\vec{k}}\left[\hat{a}_{\vec{k},\lambda}(t)-\hat{a}_{\vec{k},\lambda}^{\dag}(t)\right]\tilde{E}_{\vec{k}}(\vec{r}),
\label{eqn:Background-CQED-hat(B)(x,t)}
\end{align}
\end{subequations}
where $\vec{u}_{\vec{k},\lambda}$ is the unit vector for the mode labeled by wavevector $\vec{k}$ and polarization $\lambda$ and we have defined the electric field amplitude 
\begin{align}
\mathcal{E}_{\vec{k}}\equiv\left(\frac{\hbar\omega_k}{2\epsilon_0 V}\right)^{1/2}
\end{align}
for notation simplicity. The new commutation relations become
\begin{subequations}
\begin{align}
&[\hat{a}_{\vec{k},\lambda},\hat{a}_{\vec{k}',\lambda'}^{\dag}]=\delta_{\vec{k}\vec{k}'}\delta{\lambda\lambda'},\\
&[\hat{a}_{\vec{k},\lambda},\hat{a}_{\vec{k}',\lambda'}]=[\hat{a}_{\vec{k},\lambda}^{\dag},\hat{a}_{\vec{k}',\lambda'}^{\dag}]=0.
\end{align}
\end{subequations}
At last, following the same quantization procedure, the field Hamiltonian again can be written as sum of the Hamiltonians for each normal mode as
\begin{align}
\hat{\mathcal{H}}_F=\sum\limits_{\vec{k},\lambda}\hbar\omega_{\vec{k}}\left(\hat{a}_{\vec{k},\lambda}^{\dag}\hat{a}_{\vec{k},\lambda}+\frac{1}{2}\right).
\label{eqn:Background-CQED-hat(H)}
\end{align}
\subsection{Light-matter interaction}
\label{Sec:Background-LightMatterInt}
In this section, we discuss the interaction of a single-electron atom with the quantized electromagnetic field inside a closed cavity. 
The discussion in this section follows standard textbook treatment of light-matter coupling that starts by writing the total Hamiltonian in the dipole approximation for the system as sum of atomic, field and interaction parts in the dipole gauge \cite{Scully_Quantum_1997, Gerry_Introductory_2005}. We postpone the first principles derivation and discussion of light-matter coupling, where we also include the effect of diamagnetic $\vec{A}^2$ contribution, to chapter~(\ref{Ch-OriginOfA2}) (See also App.~\ref{App:DerivationOfHamCQED}). The full Hamiltonian can be written as 
\begin{align}
\hat{\mathcal{H}}=\hat{\mathcal{H}}_{A}+\hat{\mathcal{H}}_{F}+\hat{\mathcal{H}}_{\text{int}}=\frac{\hat{\vec{p}}_a^2}{2m_a}+U(\hat{\vec{r}}_a)+\int_V d^3\vec{r} \left[\frac{\epsilon_0}{2} \hat{\vec{E}}^2(\vec{r},t)+\frac{\hat{\vec{B}}^2(\vec{r},t)}{2\mu_0}\right]-q_a\hat{\vec{r}}_a\cdot\hat{\vec{E}}(\vec{r}_a,t),
\label{eqn:Background-CQED-Full H}
\end{align}
where $\hat{\vec{r}}_a$ and $\hat{\vec{p}}_a$ are the coordinate and momentum operators of the atomic charged particle and we have considered a dipole interaction of the form $-\hat{\vec{\mathcal{P}}}_a\cdot\hat{\vec{E}}(\vec{r}_a,t)$ with $\hat{\vec{\mathcal{P}}}_a\equiv q_a\hat{\vec{r}}_a$.  

To arrive at a second quantized representation of Hamiltonian~(\ref{eqn:Background-CQED-Full H}), we first have to obtain the spectrum of each subsystem, i.e. atomic and field parts. The field spectrum has been discussed in Sec.~\ref{Sec:Background-QuantOfFreeEM}. For the atomic subsystem, we assume that we know the energy eigenstates and label them as $\ket{n}$ with energy $E_n$ for $n=0, 1, 2,\dots$. Therefore, the atomic Hamiltonian takes the diagonal form
\begin{align}
\hat{\mathcal{H}}_A=\sum\limits_n E_n\hat{P}_{nn},
\label{eqn:Background-CQED-Spec rep of H_A}
\end{align}
where we have define a set of projection operators $\hat{P}_{mn}\equiv\ket{m}\bra{n}$.
Furthermore, these energy eigenstates form a complete basis for the Hilbert space of the atom. Hence, we can represent the coordinate operator $\hat{\vec{r}}_a$ as
\begin{align}
\hat{\vec{r}}_a=\sum\limits_{mn}\bra{m}\hat{\vec{r}}_a\ket{n}\ket{m}\bra{n}=\sum\limits_{mn}\hat{\vec{r}}_{a,\text{mn}}\hat{P}_{mn}
\label{eqn:Background-CQED-Spec rep of hat(r)_a}
\end{align}
Employing the spectral representations~(\ref{eqn:Background-CQED-Spec rep of hat(r)_a}) for the atom and (\ref{eqn:Background-CQED-hat(E)(x,t)}) for the electric field, we are able to reexpress Hamiltonian~(\ref{eqn:Background-CQED-Full H}) in a second quantized form as
\begin{align}
\hat{\mathcal{H}}=\sum\limits_n E_n\hat{P}_{nn}+\sum\limits_{\vec{k}} \hbar\omega_{\vec{k}}\hat{a}_{\vec{k}}^{\dag}\hat{a}_{\vec{k}}+\sum\limits_{mn}\sum\limits_{\vec{k}}\hbar g_{mn,\vec{k}}\hat{P}_{mn}\left(\hat{a}_{\vec{k}}+\hat{a}_{\vec{k}}^{\dag}\right),
\label{eqn:Background-CQED-Full H 2nd quantized}
\end{align}
where we have defined the light matter coupling constants
\begin{align}
g_{mn,\vec{k}}\equiv\frac{-e\bra{m}\hat{\vec{r}}_a\cdot\vec{u}_{\vec{k}}\ket{n}\mathcal{E}_{\vec{k}}\tilde{E}_{\vec{k}}(\vec{r}_a)}{\hbar}.
\label{eqn:Background-CQED-Def of g_k}
\end{align}

If the nonlinearity of the atom is strong, we can model it as a two level system and neglect all higher energy levels except the ground and the first excited state. As a result, the atomic Hamiltonian can be simplified as
\begin{align}
\hat{\mathcal{H}}_a\approx E_0\hat{P}_{00}+E_1\hat{P}_{11}=\frac{E_1-E_0}{2}\left(\hat{P}_{11}-\hat{P}_{00}\right)+\frac{E_1+E_0}{2}.
\end{align}
In terms of Pauli spin-$1/2$ operators 
\begin{align}
\hat{\sigma}^z\equiv \hat{P}_{11}-\hat{P}_{00}, \quad \hat{\sigma}^-\equiv\hat{P}_{01}, \quad \hat{\sigma}^+\equiv\hat{P}_{10},
\end{align}
we can reduce the full Hamiltonian~(\ref{eqn:Background-CQED-Full H 2nd quantized}) to the following multimode Rabi \cite{Rabi_Space_1937} Hamiltonian
\begin{align}
\hat{\mathcal{H}}_{\text{Rabi}}=\frac{\hbar\omega_a}{2}\hat{\sigma}^z+\sum\limits_{\vec{k}} \hbar\omega_{\vec{k}}\hat{a}_{\vec{k}}^{\dag}\hat{a}_{\vec{k}}+\sum\limits_{\vec{k}}\hbar g_{\vec{k}}(\hat{\sigma}^++\hat{\sigma}^-)\left(\hat{a}_{\vec{k}}+\hat{a}_{\vec{k}}^{\dag}\right),
\label{eqn:Background-CQED-H_Rabi}
\end{align}
where we have replaced $g_{\vec{k}}\equiv g_{\vec{k},01}=g_{\vec{k},10}$ for simplicity.

Note that the interaction Hamiltonian consists of four different processes. Terms $\hat{\sigma}^+\hat{a}_{\vec{k}}$ and $\hat{\sigma}^-\hat{a}_{\vec{k}}^{\dag}$ describe number conserving processes where an exchange of excitation occurs between the atom and the field without the total number of excitations being changed. On the other hand, we have number non-conserving processes like $\hat{\sigma}^+\hat{a}_{\vec{k}}^{\dag}$ and $\hat{\sigma}^-\hat{a}_{\vec{k}}$, which are also known as counter rotating terms. These counter-rotating terms become less relevant when the light-matter coupling is weak, i.e. $g_{\vec{k}}\ll \min\{\omega_a, \omega_{\vec{k}}\}$ \cite{Braak_Integrability_2011}. Dropping these non-conserving processes we obtain the multimode Jaynes-Cummings Hamiltonian \cite{Jaynes_Comparison_1963}
\begin{align}
\hat{\mathcal{H}}_{\text{JC}}=\frac{\hbar\omega_a}{2}\hat{\sigma}^z+\sum\limits_{\vec{k}} \hbar\omega_{\vec{k}}\hat{a}_{\vec{k}}^{\dag}\hat{a}_{\vec{k}}+\sum\limits_{\vec{k}}\hbar g_{\vec{k}}
\left(\hat{\sigma}^+\hat{a}_{\vec{k}}+\hat{a}_{\vec{k}}^{\dag}\hat{\sigma}^-\right).
\label{eqn:Background-CQED-H_JC}
\end{align}
The Jaynes-Cummings model is the building block of the field of quantum optics and is the starting point of the majority of analytical calculations. In the following, in sec.~(\ref{Sec:Background-closedQED}), we discuss the physics of both Rabi and Jaynes-Cummings models and also provide numerical results showing the validity of rotating-wave approximation. 
\section{Circuit quantum electrodynamics}
\label{Sec:Background-circuitQED}
There has been a great progress in preparing and controlling the quantum coherence in superconducting microwave circuits in the past two decades \cite{Nakamura_Coherent_1999, Blais_Cavity_2004, Schoelkopf_Wiring_2008, Niemczyk_Circuit_2010}. In analogy to the field of CQED that studies the interaction of \textit{real} atoms with electromagnetic field, the quantum electrodynamical phenomena in these superconducting devices is named circuit quantum electrodynamics (cQED). Although these macroscopic superconducting devices are made out of a large number of cooper pairs, their collective behavior mimics that of real atoms and have a discrete set of energy levels. Despite this generic similarity to conventional CQED systems, cQED provides extreme regimes of light-matter interaction that are not easy to implement with ordinary atoms. In the following, we provide some background information on the theory of cQED that will be used for the rest of the dissertation.
\subsection{Circuit quantization}
\label{SubSec:Background-circuit quantization}

In this section, we review the standard procedure of quantizing electrical circuits \cite{Devoret_Quantum_1995, Bishop_Circuit_2010, Devoret_Quantum_2014}. We treat each ideal circuit component as a lumped element. This approximation works when the size of the element is much less that the wavelength for the frequencies of interest. In microwave engineering, the frequency of interest ranges from $1$ to $300$ Gigahertz, which results in a wavelength range of about $30$ centimeters down to $1$ milimeters.  

Each real circuit is modeled as a network of such ideal lumped elements, which for the purpose of this dissertation, are all two-terminal elements. In conventional microwave engineering, each two-terminal element is described by the voltage-current characteristic across the element. In superconducting circuits, the main nonlinear element is the Josephson junction \cite{Josephson_Possible_1962, Josephson_Discovery_1974}, whose current relation is best described by the generalized flux difference across the junction as
\begin{align}
I(t)=I_c \sin\left(2\pi \frac{\Phi(t)}{\Phi_0}\right),
\end{align} 
where $I_c$ is the critical current, $\Phi_0\equiv h/(2e)$ is the flux quantum and $\Phi_j$ is the flux difference across the junction. Therefore, the convention is to use flux and charge variables to characterize the circuit. These two descriptions can be translated into each other as
\begin{subequations}
\begin{align}
&\Phi_a(t)\equiv \int^{t} dt' V_a(t'),\\
&Q_a(t)\equiv \int^{t} dt' I_a(t'),
\end{align}
\end{subequations}
where $\Phi_a$, $Q_a$, $V_a$ and $I_a$ are the flux, charge, voltage and current across the two-terminal element $a$. Each \textit{real} circuit element may in principle be composed of such \textit{ideal} elements. These ideal elements can be of either capacitive or inductive types.

The general approach towards quantization of electrical circuits can be summarized in the following steps:
\begin{enumerate}
\item Model the circuit as a network of ideal two-terminal circuit elements.
\item Pick a ground (reference) node. The physics of the circuit is independent of the choice for ground node due to gauge symmetry. The remaining nodes are dynamical or active.
\item Pick a loop-free graph $T$ that spans all nodes of the network and is called spanning tree.
\item For each active node of the network, assign a flux coordinate that is the time integral of the voltage across the unique path from the ground as
\begin{align}
\Phi_n(t)\equiv\sum\limits_{a}S_{na}\int^{t}dt'V_a(t')dt',
\end{align}
where $S_{na}$ is $\pm$, depending on the orientation,  if the path on $T$ from ground to $n$ passes through $a$ and otherwise is $0$.
\item Write the Langrangian as the difference between kinetic and potential energy as  
\begin{align}
\mathcal{L}[\Phi_1,\ldots,\Phi_N;\dot{\Phi}_1,\dots, \dot{\Phi}_N]\equiv\mathcal{T}[\dot{\Phi}_1,\dots, \dot{\Phi}_N]-\mathcal{U}[\Phi_1,\ldots,\Phi_N].
\end{align}
\item Find the canonical conjugate momenta $Q_1, \ldots,Q_N$ defined as
\begin{align}
Q_n\equiv\frac{\partial \mathcal{L}}{\partial \dot{\Phi}_n}.
\end{align}
The generalized flux coordinates and the resulting conjugate momenta obey the Poisson bracket algebra
\begin{align}
\{\Phi_m,\Phi_n\}=0, \quad \{Q_m,Q_n\}=0, \quad  \{\Phi_m,Q_n\}=\delta_{mn}.
\end{align}
\item Find the Hamiltonian via a Legendre transformation 
\begin{align}
\mathcal{H}[\Phi_1,\ldots,\Phi_N; Q_1,\dots, Q_N]\equiv\sum\limits_{n} Q_n\dot{\Phi}_n-\mathcal{L}.
\end{align}
\item Promote the classical Poisson bracket algebra into a non-commuting quantum algebra such that
\begin{align}
[\hat{\Phi}_m,\hat{\Phi}_n]=0, \quad [\hat{Q}_m,\hat{Q}_n]=0, \quad  [\hat{\Phi}_m,\hat{Q}_n]=\delta_{mn},
\end{align}
where we have used a hat notation to distinguish between classical variables and quantum operators.
\end{enumerate}

\subsection{Quantum LC oscillator}
\label{SubSec:Background-QuLC circuit}
%%%%%%%%%%%%%% Fig:LC Osc %%%%%%%%%%%%%%
\begin{figure}[t!]
\centering
\includegraphics[scale=0.40]{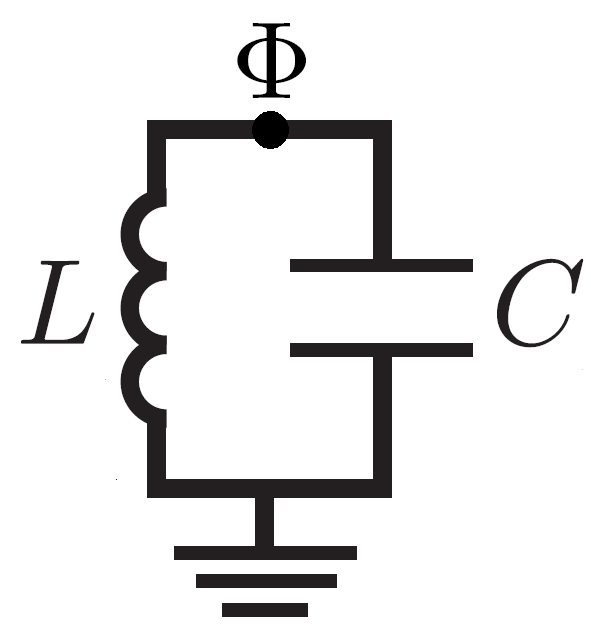}
\caption{Quantum LC oscillator. This circuit has only one active node that is represented by the flux variable $\Phi(t)$}
\label{Fig:Background-LC Oscillator}
\end{figure}
%%%%%%%%%%%%%%%%%%%%%%%%%%%%%%%%%%%%%%%%%%
As a simple example of quantization, consider an LC oscillator shown in Fig.~(\ref{Fig:Background-LC Oscillator}). The only active node of this circuit is labeled by $\Phi(t)$. In terms of the flux coordinate $\Phi$ across the $LC$ circuit, the Langrangian can be written as
\begin{align}
\mathcal{L}=\frac{C\dot{\Phi}^2}{2}-\frac{\Phi^2}{2L}.
\end{align}
The conjugate momentum to the flux is found as
\begin{align}
Q\equiv\frac{\partial \mathcal{L}}{\partial \dot{\Phi}}=C\dot{\Phi},
\end{align}
which is the charge across the capacitor. Therefore, the Hamiltonian is obtained via a Legendre transformation as
\begin{align}
\mathcal{H}\equiv \dot{\Phi}Q-\mathcal{L}=\frac{\Phi^2}{2L}+\frac{Q^2}{2C}.
\end{align}
Next, the coordinate $\Phi$ and its conjugate momentum $Q$ are promoted to quantum operators obeying the canonical commutation relation as
\begin{align}
[\hat{\Phi},\hat{Q}]=i\hbar.
\end{align}
Introducing creation and annihilation operators obeying
\begin{align}
[\hat{a},\hat{a}^{\dag}]=1,
\end{align}
we can rewrite the quantum operators as
\begin{subequations}
\begin{align}
&\hat{\Phi}=\Phi_{\text{zpf}}\left(\hat{a}+\hat{a}^{\dag}\right),
\label{eqn:Background-cQED-Phi interms of a}\\
&\hat{Q}=-iQ_{\text{zpf}}\left(\hat{a}-\hat{a}^{\dag}\right).
\label{eqn:Background-cQED-Q interms of a}
\end{align}
\end{subequations}
where zero-point fluctuation amplitudes $\Phi_{\text{zpf}}$ and $Q_{\text{zpf}}$ are defined
\begin{align}
\Phi_{\text{zpf}}\equiv\sqrt{\frac{\hbar Z}{2}},\\
Q_{\text{zpf}} \equiv\sqrt{\frac{\hbar }{2Z}},
\end{align}
where $Z\equiv\sqrt{L/C}$ is called the characteristic impedance of the oscillator. Plugging Eqs.~(\ref{eqn:Background-cQED-Phi interms of a}-\ref{eqn:Background-cQED-Q interms of a}) into the expression for the Hamiltonian we obtain
\begin{align}
\hat{\mathcal{H}}=\frac{\hbar\omega}{2}\left(\hat{a}^{\dag}\hat{a}+\hat{a}\hat{a}^{\dag}\right)=\hbar\omega\left(\hat{a}^{\dag}\hat{a}+\frac{1}{2}\right),
\end{align}
where $\omega\equiv 1/\sqrt{LC}$ is the oscillation frequency of the circuit.
\subsection{Transmission line resonator}
\label{SubSec:Background-Resonator}
%%%%%%%%%%%%%%% Fig:Resonatro %%%%%%%%%%%%
\begin{figure}[t!]
\subfloat[\label{subfig:Background-Resonator-Schematic}]{%
\includegraphics[scale=0.50]{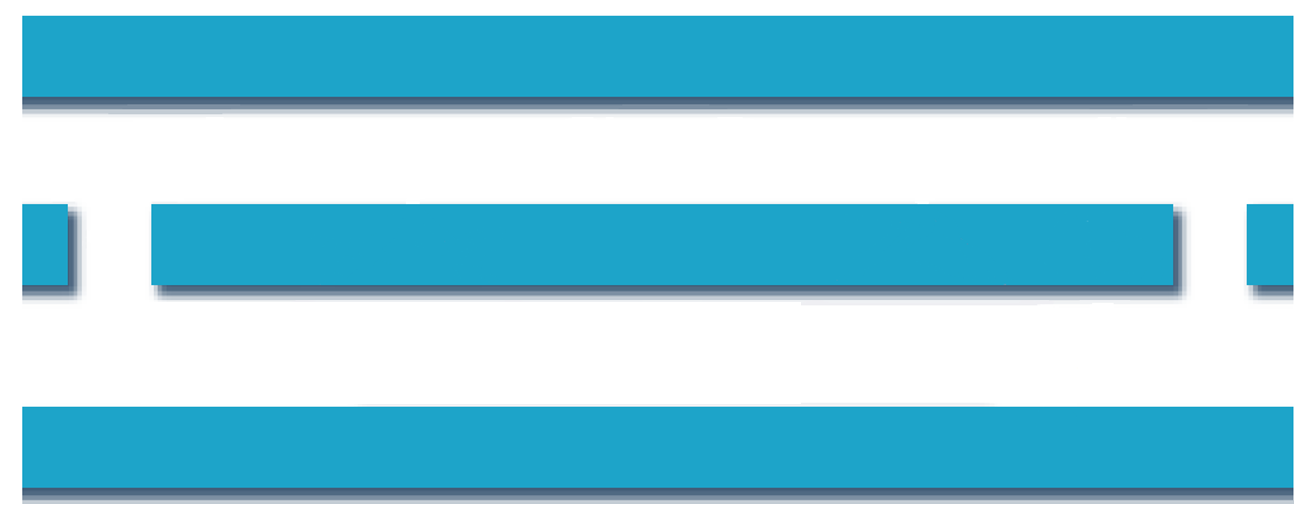}%
}\hfill
\subfloat[\label{subfig:Background-Resonator-EqCircuit}]{%
\includegraphics[scale=0.50]{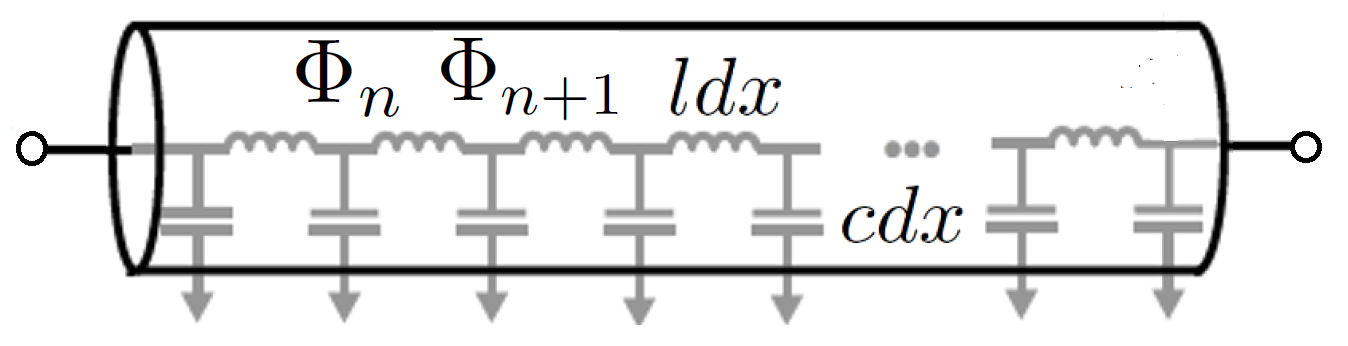}%
}
\caption{Lumped element representation of a transmission line resonator. (a). Device view (b). Equivalent circuit}
\label{Fig:Background-Resonator}
\end{figure}
%%%%%%%%%%%%%%%%%%%%%%%%%%%%%%%%%%%%

In cQED architecture, finite length transmission line resonators play the role of a cavity that exhibit many standing wave resonances, where each independent harmonic oscillator is equivalent to a simple LC oscillator discussed in the previous section. There are various realizations of such resonators, using techniques borrowed from classical microwave engineering, such as coplanar waveguides, coaxial cables and etc. In this section, we review the quantization of the electromagnetic field inside a one dimensional coplanar waveguide, which consists of a superconducting wire fabricated on an insulating substrate and having superconducting ground
wires to each end as shown in Fig.~(\ref{subfig:Background-Resonator-Schematic}). The discrete lumped element equivalent circuit for such a system is depicted in Fig.~\ref{subfig:Background-Resonator-EqCircuit}. For the moment, we assume closed boundary conditions such that the current vanishes at the ends of the resonator.

To obtain the Lagrangian, we begin by modeling the resonator as a finite set of infinitesimal inductors $l \Delta x$ and capacitors $c \Delta x$ (Fig.~\ref{subfig:Background-Resonator-EqCircuit}), where $l$ and $c$ are the inductance and capacitance per unit length and are determined by the material that is used to build the resonator. The discrete Lagrangian is then found as
\begin{align}
\mathcal{L}=\sum\limits_n \left[\frac{c\Delta x}{2}\dot{\Phi}_n^2-\frac{(\Phi_n-\Phi_{n+1})^2}{2l\Delta x}\right]
\label{eqn:Background-cQED-Discrete L}
\end{align}
Next, we go to continuum limit, $\Delta x\to 0$, where now we have a 1D flux field $\Phi(x,t)$. Employing the following identity
\begin{align}
\lim_{\Delta x\to 0}\frac{\Phi_{n+1}(t)-\Phi_{n}(t)}{\Delta x}=\partial_x \Phi(x,t),
\end{align}
we obtain the continuous Lagrangian as
\begin{align}
\mathcal{L}=\int_{0}^{L}dx \left\{\frac{c}{2}[\partial_t\Phi(x,t)]^2-\frac{1}{2l} [\partial_x\Phi(x,t)]^2\right\}
\label{eqn:Background-cQED-Continuous L}
\end{align}
The conjugate momentum to the flux field $\Phi(x,t)$ is obtained as
\begin{align}
\rho(x,t)\equiv\frac{\delta \mathcal{L}}{\delta \dot{\Phi}(x,t)},
\label{eqn:Background-cQED-Def of rho(x,t)}
\end{align}
which is called the charge density field. These conjugate fields obey the classical Poisson bracket algebra 
\begin{align}
\{\Phi(x,t),\Phi(x',t)\}=\{\rho(x,t),\rho(x',t)\}=0,\quad \{\Phi(x,t),\rho(x',t)\}=\delta(x-x').
\label{eqn:Background-cQED-Poisson Bracket}
\end{align} 
By applying Legendre transformation
\begin{align}
\mathcal{H}\equiv \int_{0}^{L}dx \partial_t\Phi(x,t)\rho(x,t)-\mathcal{L},
\end{align}
we find the Hamiltonian in the continuum limit as
\begin{align}
\mathcal{H}=\int_{0}^{L}dx\left\{\frac{\rho^2(x,t)}{2c}+\frac{1}{2l}[\partial_x\Phi(x,t)]^2\right\}.
\label{eqn:Background-cQED-Continuous H}
\end{align}

In order to quantize the classical fields, we need to solve for the eigenmodes and eigenfrequencies of the resonator. The first step is to obtain the Hamilton equations of motion for $\Phi(x,t)$ and $\rho(x,t)$. Using the Poisson bracket relations~(\ref{eqn:Background-cQED-Poisson Bracket}) and $\dot{O}=\{O,\mathcal{H}\}$ we find
\begin{subequations}
\begin{align}
&\partial_t \Phi(x,t)=\frac{\rho(x,t)}{c},
\label{eqn:Background-cQED-dot(Phi)}\\
&\partial_t \rho(x,t)=\frac{1}{l}\partial_x^2\Phi(x,t),
\label{eqn:Background-cQED-dot(rho)}
\end{align}
\end{subequations}
Combining Eqs.~(\ref{eqn:Background-cQED-dot(rho)}-\ref{eqn:Background-cQED-dot(Phi)}) we obtain a 1D wave equation for the flux field as
\begin{align}
\left(\partial_x^2-lc\partial_t^2\right)\Phi(x,t)=0
\label{eqn:Background-cQED-1D Wave Eq}
\end{align}
The boundary conditions for an isolated (closed) resonator can be derived from the fact that no current passes the end nodes. Note that the current field $I(x,t)$ is related to $\Phi(x,t)$ as
\begin{align}
I(x,t)=\lim_{\Delta x\to 0}\frac{\Phi_n(t)-\Phi_{n+1}(t)}{l\Delta x}=-\frac{1}{l}\partial_x \Phi(x,t).
\label{eqn:Background-cQED-I(x,t)}
\end{align}
Therefore the closed boundary conditions for this system is of Neumann type, i.e. $\left.\partial_x \Phi\right|_{x=0,L}=0$. 

Applying Fourier transform in time
\begin{align}
&\Phi(x,t)\equiv \int_{-\infty}^{+\infty}\frac{d\omega}{2\pi}\tilde{\Phi}(x,\omega)e^{-i\omega t},\\
&\tilde{\Phi}(x,\omega)\equiv \int_{-\infty}^{+\infty}dt\tilde{\Phi}(x,t)e^{i\omega t},
\end{align}
the wave Eq.~(\ref{eqn:Background-cQED-1D Wave Eq}) is transformed into the following Helmholtz equation
\begin{align}
\left(\partial_x^2+k^2\right)\tilde{\Phi}(x,k)=0,
\label{eqn:Background-cQED-1D Helm Eq}
\end{align}
where $k\equiv \sqrt{lc}\omega$ is the wave number and $v_p\equiv 1/\sqrt{lc}$ is the phase velocity of the resonator. The solutions to the Helmholtz Eq.~(\ref{eqn:Background-cQED-1D Helm Eq}) with Neumann boundary conditions are only possible at a discrete set of frequencies that read
\begin{subequations}
\begin{align}
&\tilde{\Phi}_n(x)\equiv\sqrt{2}\cos(k_n x),\label{eqn:Background-cQED-Eigmode}\\
&k_n\equiv \frac{n\pi}{L},
\label{eqn:Background-cQED-Eigfreq}
\end{align}
\end{subequations}
where $\tilde{\Phi}_n(x)\equiv\tilde{\Phi}(x,k_n)$ is just a simplified notation. These eigenmodes satisfy two orthogonality condition
\begin{subequations}
\begin{align}
&\int_{0}^{L}dx \tilde{\Phi}_m(x) \tilde{\Phi}_n(x)=L\delta_{mn},
\label{eqn:Background-cQED-Ortho Cond 1}\\
&\int_{0}^{L}dx \partial_x\tilde{\Phi}_m(x) \partial_x\tilde{\Phi}_n(x)=Lk_mk_n\delta_{mn}.
\label{eqn:Background-cQED-Ortho Cond 2}
\end{align}
\end{subequations}

These normal modes form a complete basis for the flux field inside the resonator such that
\begin{align}
\Phi(x,t)=\sum\limits_n \xi_n(t)\tilde{\Phi}_n(x).
\label{eqn:Background-cQED-Spec Rep of Phi(x,t)}
\end{align}
Replacing Eq.~(\ref{eqn:Background-cQED-Spec Rep of Phi(x,t)}) into the classical Lagrangian~(\ref{eqn:Background-cQED-Continuous L}) and employing conditions~(\ref{eqn:Background-cQED-Ortho Cond 1}-\ref{eqn:Background-cQED-Ortho Cond 2}) we find
\begin{align}
\mathcal{L}=\sum\limits_n\frac{Lc}{2}\left[\dot{\xi}_n^2(t)-\omega_n^2\xi_n^2(t)\right],
\label{eqn:Background-cQED-Spec Rep of L}
\end{align}
where by the spectral representation~(\ref{eqn:Background-cQED-Spec Rep of Phi(x,t)}) we have interchanged the field degree of freedom with a discrete but infinite number of simple harmonic oscillators. The Hamiltonian is then found as
\begin{align}
\mathcal{H}=\sum\limits_n\left[\frac{q_n^2(t)}{2Lc}+\frac{Lc}{2}\omega_n^2\xi_n^2(t)\right],
\label{eqn:Background-cQED-Spec Rep of H}
\end{align}
where $q_n(t)\equiv Lc\dot{\xi}_n(t)$ is the conjugate momentum to $\xi_n(t)$. Next, we promote the classical canonical variables $\xi_n$ and $q_n$ into quantum operators such that
\begin{align}
[\hat{\xi}_m,\hat{q}_n]=i\hbar\delta_{mn},
\end{align}
and express them in terms of a set of creation and annihilation operators as
\begin{subequations}
\begin{align}
&\hat{\xi}_n(t)=\left(\frac{\hbar}{2\omega_n cL}\right)^{1/2}\left[\hat{a}_n(t)+\hat{a}_n^{\dag}(t)\right],\\
&\hat{q}_n(t)=-i\left(\frac{\hbar\omega_n cL}{2}\right)^{1/2}\left[\hat{a}_n(t)-\hat{a}_n^{\dag}(t)\right].
\end{align} 
\end{subequations}
After all, we are able to express the quantum fields in the following form
\begin{subequations}
\begin{align}
&\hat{\Phi}(x,t)=\sum\limits_n \left(\frac{\hbar}{2\omega_n cL}\right)^{1/2}\left[\hat{a}_n(t)+\hat{a}_n^{\dag}(t)\right]\tilde{\Phi}_n(x),\\
&\hat{\rho}(x,t)=-i\sum\limits_n \left(\frac{\hbar\omega_n c}{2L}\right)^{1/2}\left[\hat{a}_n(t)-\hat{a}_n^{\dag}(t)\right]\tilde{\Phi}_n(x),
\end{align}
\end{subequations}
in terms of which the Hamiltonian becomes
\begin{align}
\hat{\mathcal{H}}\equiv\sum\limits_n \hbar\omega_n\left(\hat{a}^{\dag}\hat{a}+\frac{1}{2}\right).
\end{align}
\subsection{Superconducting qubits}
\label{SubSec:Background-Super qubits}

We provide a brief background on the developement of superconducting qubits. The term 'qubit' arises from the applications of these circuits in quantum information processing as a quantum two-level system, while none of these systems are two-level in essence. In order to build a qubit, we need to have some anharmonicity to be able to pick a specific pair of energy levels and induce a transition between only those levels. This is not achievable with a quantum harmonic oscillator, whose quantized energy levels are evenly spaced. In cQED, the anharmonicity comes from Josephson junction \cite{Josephson_Possible_1962}\cite{Josephson_Discovery_1974}, that acts as a dissipation free nonlinear inductive element. There are three main categories of superconducting qubits depending on the topology of the circuit and how the nonlinearity is implemented. These are charge \cite{Bouchiat_Quantum_1998, Nakamura_Coherent_1999}, flux \cite{Friedman_Quantum_2000, Van_Quantum_2000} and phase \cite{Martinis_Rabi_2002} qubits. In the following, we only focus on the charge qubit category and discuss the physics of cooper-pair box \cite{Bouchiat_Quantum_1998, Nakamura_Coherent_1999} and transmon qubit \cite{Koch_Charge_2007} in more detail, that will be of interest in the rest of this dissertation.
\subsubsection{Cooper pair box}
\label{SubSec:Background-CPB}
%%%%%%%%%%%%%%% Fig:CPB %%%%%%%%%%%%%%%%%%%%%
\begin{figure}[t!]
\centering
\includegraphics[scale=0.40]{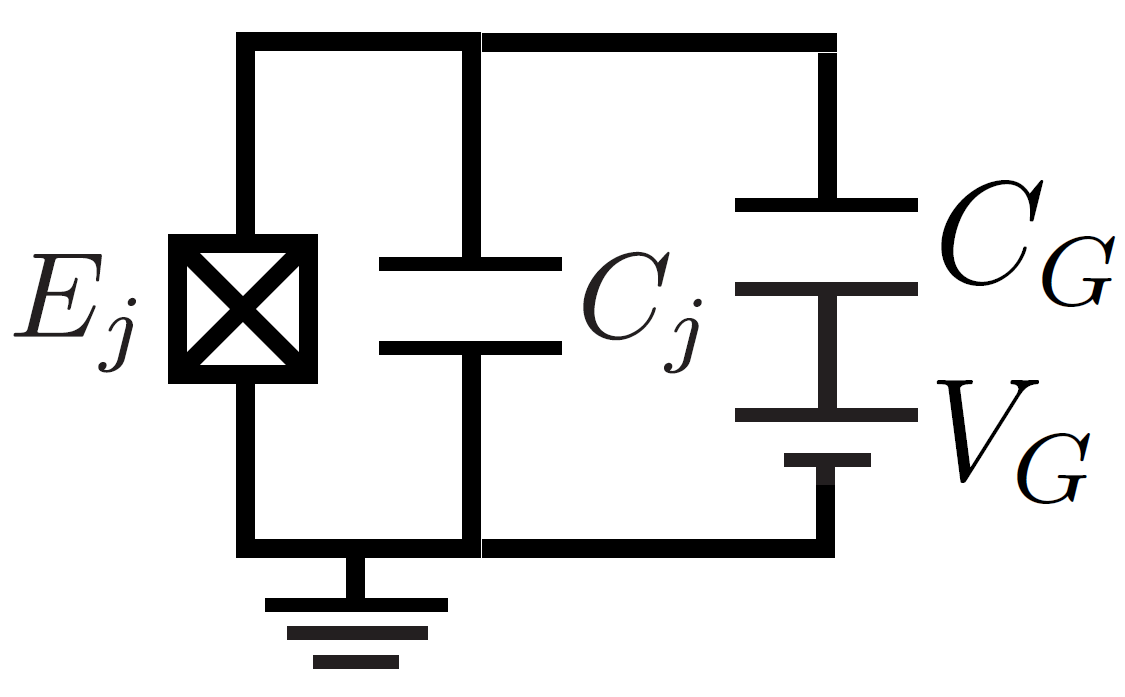}
\caption{Cooper pair box also known as charge qubit. Compared with the quantum LC oscillator, the inductor has been replaced with a nonlinear Josephson junction.}
\label{Fig:Background-CPB}
\end{figure}
%%%%%%%%%%%%%%%%%%%%%%%%%%%%%%%%%%%%%%%%%%%%%
The anharmonicity that is required to build a superconducting qubit is provided by the Josephson junction, that is made out of two superconducting islands separated by a thin insulating layer that allows only a few cooper pairs to tunnel. The low energy Hamiltonian for an ideal Josephson element can be written as 
\begin{align}
\hat{\mathcal{H}}_{j}=-E_j\cos\left(2\pi\frac{\hat{\Phi}_j}{\Phi_0}\right),
\label{eqn:Background-cQED-Josephson Hj 1}
\end{align} 
where $E_j$ is called the Josephson energy and gives the tunneling strength. Up to lowest order, the cosine potential acts like a linear inductor with inductance $L_j\equiv(2\pi/\Phi_0)^2/E_j$, accompanied with a leading order quartic nonlinearity. Therefore, the simplest nonlinear quantum oscillator can be achieved upon replacing the linear inductor in an $LC$ oscillator by a Josephson junction. The quantum Hamiltonian for this system reads 
\begin{align}
\hat{\mathcal{H}}=\frac{\hat{Q}_j^2}{2C_j}-E_j\cos\left(2\pi\frac{\hat{\Phi}_j}{\Phi_0}\right).
\label{eqn:Background-cQED-Simplest qubit 1}
\end{align}
The strength of nonlinearity is determined by the ratio of two energy scales, the Josephson energy and the charging energy of the capacitor. Therefore, the convention is to reexpress Hamiltonian~(\ref{eqn:Background-cQED-Simplest qubit 1}) as
\begin{align}
\hat{\mathcal{H}}=4E_c\hat{n}_j^2-E_j\cos\left(\hat{\varphi}_j\right),
\label{eqn:Background-cQED-Simplest qubit 2}
\end{align}
where $\hat{n}_j$ and $\hat{\varphi}_j$ are untiless phase and number operators defined
\begin{align}
&\hat{n}_j\equiv\frac{\hat{Q}_j}{2e},
\label{eqn:Background-cQED-Def of varphi_j}
\\
&\hat{\varphi}_j\equiv 2\pi\frac{\hat{\Phi}_j}{\Phi_0},
\label{eqn:Background-cQED-Def of n_j}
\end{align}
and $E_c\equiv(2e)^2/C_j$ is called the charging energy. The larger the Josephson energy, the deeper and less nonlinear the cosine potential. Therefore the ratio $E_c/E_j$ is a measure of nonlinearity.

Hamiltonian~(\ref{eqn:Background-cQED-Simplest qubit 2}) is derived for an ideal nonlinear oscillator, while in real experiments there is always some offset charge due to coupling to a gate electrode with a d.c. bias voltage. The dependence of the energy levels of this Hamiltonian on the offset charge is called charge noise. To study this effect, we consider adding a bias voltage and an additional gate capacitor as shown in Fig.~(\ref{Fig:Background-CPB}) to this model, which is now called Cooper pair box or charge qubit \cite{Bouchiat_Quantum_1998, Nakamura_Coherent_1999}. The Hamiltonian for this system is
\begin{align}
\hat{\mathcal{H}}_{\text{CPB}}=4E_c(\hat{n}_j-n_G)^2-E_j\cos\left(\hat{\varphi}_j\right),
\label{eqn:Background-cQED-H_CPB}
\end{align} 
where $n_g\equiv C_G V_G/(C_j+C_G)$ is the charge offset and $E_c=(2e)^2/(C_j+C_G)$. The spectrum of $\hat{\mathcal{H}}_{CPB}$ can be solved exactly in terms of Mathieu functions as \cite{Cottet_Implementation_2002, Koch_Charge_2007, Devoret_Quantum_2014}
\begin{align}
E_n(n_G)=E_ca_{2[n_g+k(m,n_g)]}\left(-\frac{E_j}{2E_c}\right),
\end{align}
where $a_{\nu}(q)$ is Mathieu’s characteristic value, and $k(m,n_g)$ is an integer-valued function that sorts the eigenvalues. 

Another possible treatment that is more suitable for numerical purposes is to represent $\hat{\mathcal{H}}_{\text{CPB}}$ in terms of the number basis. Using the fact that $[\hat{\varphi},\hat{n}]=i$ we can replace\\
\begin{subequations}
\begin{align}
&\hat{n}\equiv\sum\limits_{n=-\infty}^{+\infty} n\ket{n}\bra{n},
\label{eqn:Background-cQED-NumRep of n}\\
&\cos{\hat{\varphi}}=\frac{1}{2}\sum\limits_{n=-\infty}^{+\infty}\left(\ket{n}\bra{n+1}+\ket{n+1}\bra{n}\right),
\label{eqn:Background-cQED-NumRep of cos(varphi)}
\end{align}
\end{subequations}
in Hamiltonian~(\ref{eqn:Background-cQED-H_CPB}) to obtain
\begin{align}
\hat{\mathcal{H}}_{\text{CPB}}=\sum\limits_{n=-\infty}^{+\infty}=\left[4E_c(n-n_G)^2\ket{n}\bra{n}-\frac{E_j}{2}\left(\ket{n}\bra{n+1}+\ket{n+1}\bra{n}\right)\right],
\label{eqn:Background-cQED-H_CPB NumRep}
\end{align}
where $n$ shows the difference in the number of cooper pairs across the junction. Hamiltonian~(\ref{eqn:Background-cQED-H_CPB NumRep}) clearly shows that the energy required for a single cooper pair to tunnel through the junction is $E_j/2$.
\subsubsection{Transmon qubit}
\label{SubSec:Background-transmon qubit}
%%%%%%%%%%% Fig:Charge Noise %%%%%%%%%%%%%%%
\begin{figure}[t!]
\centering
\subfloat[$E_j/E_c=1$\label{subfig:SpEmRateXrXl1Em3Xj5Em2Xg1Em3}]{%
\includegraphics[scale=0.40]{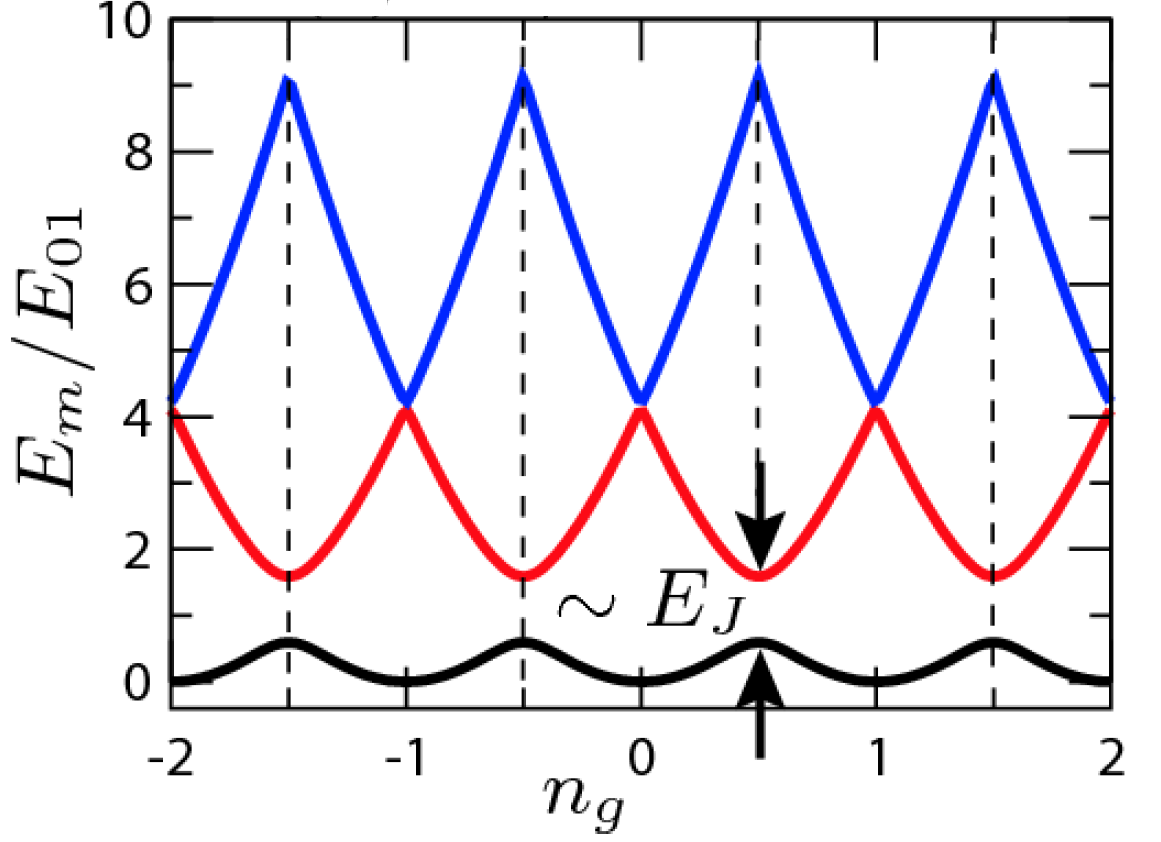}%
}
\subfloat[$E_j/E_c=5$\label{subfig:SpEmRateXrXl1Em3Xj5Em2Xg1Em1}]{%
\includegraphics[scale=0.40]{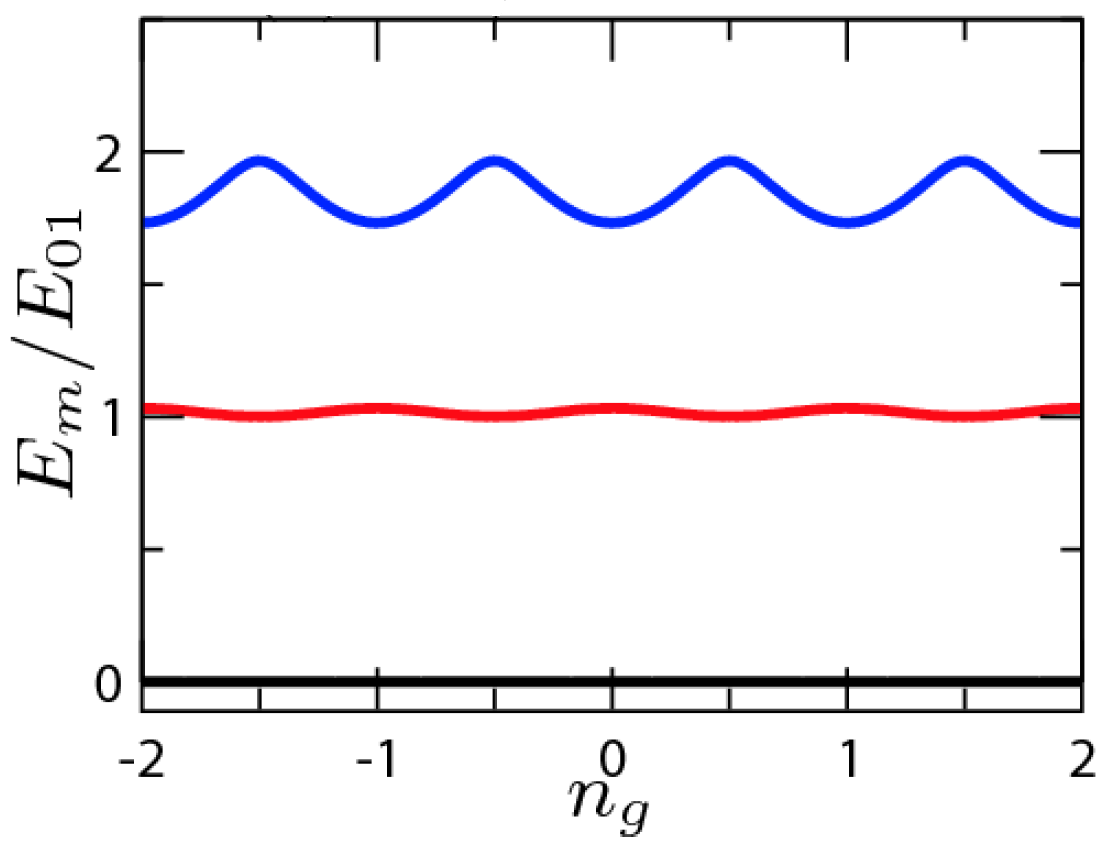}%
}\\
\subfloat[$E_j/E_c=10$\label{subfig:LmbShftXrXl1Em3Xj5Em2Xg1Em3}]{%
\includegraphics[scale=0.40]{Ch-Background/CPBSpectrum3.png}%
}
\subfloat[$E_j/E_c=50$\label{subfig:LmbShftXrXl1Em3Xj5Em2Xg1Em1}]{%
\includegraphics[scale=0.40]{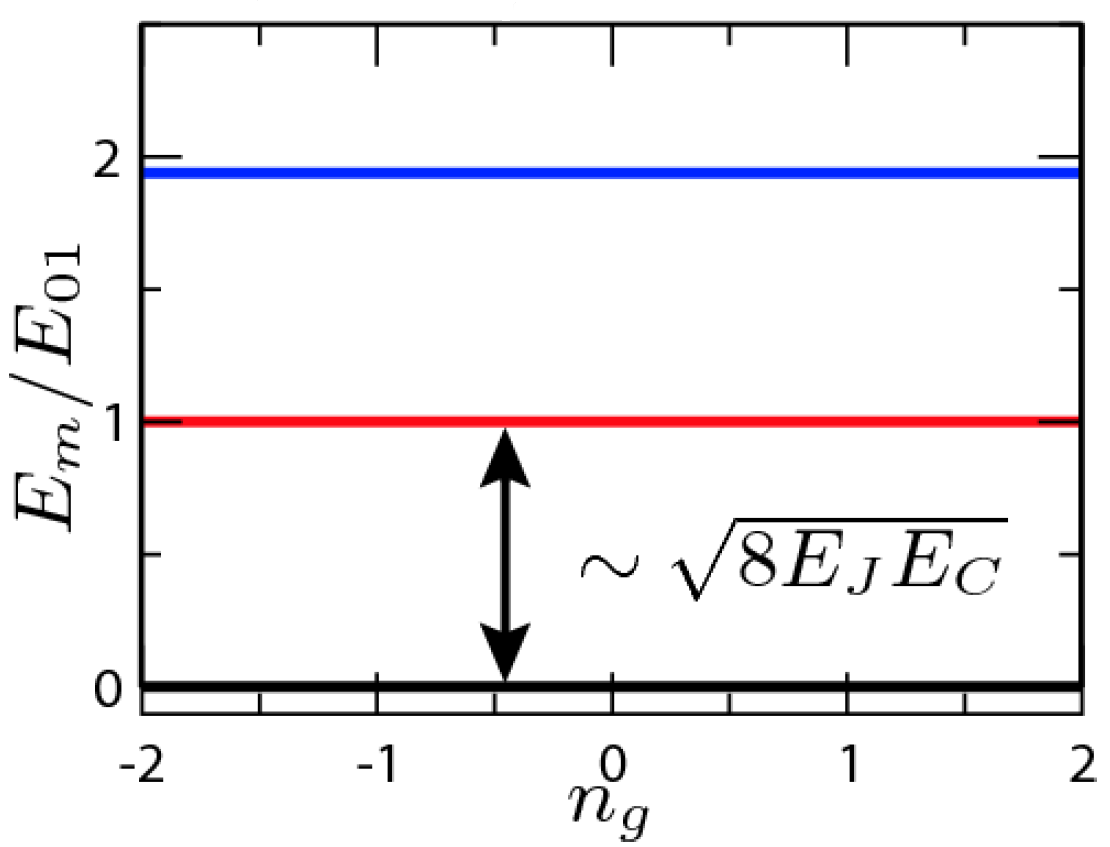}%
}
\caption{First three eigenenergies of $\mathcal{H}_{\text{CPB}}$ as a function effective offset charge. The vertical dashed
line in (a) shows the charge sweet spots at half-integer $n_g$. (This figure has been used with permission from \cite{Koch_Charge_2007}. See Copyright
Permissions.)} 
\label{Fig:ChargeNoise}
\end{figure}
%%%%%%%%%%%%%%%%%%%%%%%%%%%%%%%%%%%%%%%%%%%%%
According to Hamiltonian~(\ref{eqn:Background-cQED-H_CPB}), the spectrum of the charge qubit is dependent on the offset charge $n_G$, which called charge dispersion or charge noise. This effect is experimentally undesirable since it is extremely challenging to engineer the right value for $n_G$ to
avoid unwanted drifts in the transition frequencies between levels.  The original design for the charge qubit \cite{Bouchiat_Quantum_1998, Nakamura_Coherent_1999} operated in highly nonlinear regime $E_c/E_j \approx 1$ as is required to have a well-defined two level system. However, in this regime the energy levels are extremely sensitive to $n_G$. On the other hand, it can be shown that by decreasing the nonlinearity strength $E_c/E_j$ the charge noise gets suppressed \cite{Cottet_Implementation_2002}. In other words, there is an unwanted competition between nonlinearity and charge noise in these systems. A desirable qubit design should exhibit high nonlinearity and low charge noise. However, in order to suppress the noise we need to decrease the nonlinearity.

Figure.~\ref{Fig:ChargeNoise} \cite{Koch_Charge_2007} summarizes this competition by showing the dependence of the first three energy levels on the offset charge for four different values of nonlinearity. According this figure, the highest drift of the energy levels occur at integer and half-integer filling of $n_G$. Therefore, a measure for charge dispersion can be defined as~\cite{Koch_Charge_2007}
\begin{align}
\epsilon_n\equiv E_n(n_G=1/2)-E_n(n_G=0).
\end{align}
Employing asymptotics of the Mathieu functions in the small dispersion limit, i.e. $E_j/E_c\gg 1$, an approximate expression for the charge dispersion can be found
\begin{align}
\epsilon_n=(-1)^n E_c\frac{2^{4n+5}}{n!}\sqrt{\frac{2}{\pi}}\left(\frac{E_j}{2E_c}\right)^{\frac{n}{2}+\frac{3}{4}}e^{-\sqrt{8E_j/E_c}},
\end{align} 
which confirms an exponential suppression of charge dispersion in this limit. 

Based on this idea, the transmon qubit design \cite{Koch_Charge_2007, Schreier_Suppressing_2008} was introduced to resolve the unwanted charge dispersion, which implements an additional large capacitor in order to increase the effective charging energy $E_c$ and hence achieve $E_j/E_c \gg 1$ regime. In this regime, called transmon regime, the nonlinearity is weak and we can treat the cosine potential perturbatively as
\begin{align}
\mathcal{H}_{\text{Transmon}}\approx 4E_c\hat{n}^2+\frac{E_j}{2}\hat{\varphi}^2-\frac{E_j}{24}\hat{\varphi}^4+\mathcal{O}(\hat{n}^6),
\label{eqn:Background-cQED-App H_Trans}
\end{align}
where now the dependence on $n_G$ is removed. Up to leading order, transmon behaves as a quantum Duffing oscillator \cite{Bender_Multiple_1996}. Next, we rewrite the the nonlinear part of the Hamiltonian in terms of the solution for the harmonic part as
\begin{subequations}
\begin{align}
&\hat{\varphi}=\varphi_{\text{zpf}}\left(\hat{a}+\hat{a}^{\dag}\right), \quad \varphi_{\text{zpf}}=\left(\frac{2E_c}{E_j}\right)^{1/4}
\label{eqn:Background-cQED-Def of varphi_zpf}\\
&\hat{n}=-in_{\text{zpf}}\left(\hat{a}-\hat{a}^{\dag}\right), \quad n_{\text{zpf}}=\left(\frac{E_j}{2E_c}\right)^{1/4}
\label{eqn:Background-cQED-Def of n_zpf}
\end{align}
\end{subequations}
where $[\hat{a},\hat{a}^{\dag}]$. Inserting Eqs.~(\ref{eqn:Background-cQED-Def of varphi_zpf}-\ref{eqn:Background-cQED-Def of n_zpf}) into Eq.~(\ref{eqn:Background-cQED-App H_Trans}) we obtain
\begin{align}
\mathcal{H}_{\text{Transmon}}\approx \sqrt{8E_jE_c}\left(\hat{a}^{\dag}\hat{a}+\frac{1}{2}\right)-\frac{E_c}{12}(\hat{a}+\hat{a}^{\dag})^4+\mathcal{O}\left[\left(E_c/E_j\right)^2\right],
\label{eqn:Background-cQED-App H_Trans}
\end{align}
where up to linear approximation transmon frequency is found as $\sqrt{8E_jE_c}$ accompanied by a nonlinear quartic term of strength $E_c$. The correction to energy levels can be found from first order Rayleigh-Schrodinger perturbation theory as
%%%%%%%%%%%%% Fig: Transmon %%%%%%%%%%%%%%%%%
\begin{figure}[t!]
\centering
\includegraphics[scale=0.40]{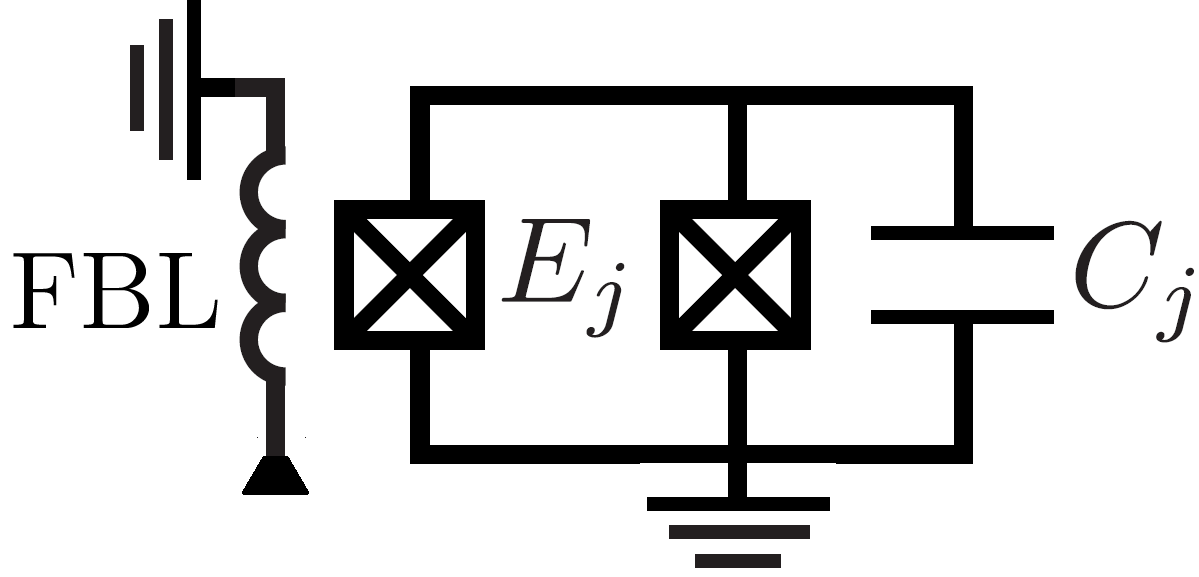}
\caption{Equivalent circuit of a transmon qubit. The effective charging energy has been decreased by a large capacitance $C_j$. Moreover, in contrast to the Cooper pair box, the single Josephson junction has been replaced by a SQUID whose effective Josephson energy can be controlled by an external flux called flux bias line (FBL). As a result, the qubit frequency is tunable.}
\label{Fig:Background-Transmon}
\end{figure}
%%%%%%%%%%%%%%%%%%%%%%%%%%%%%%%%%%%%%%%%%%%
\begin{align}
E_n\approx\sqrt{8E_jE_c}\left(n+\frac{1}{2}\right)-\frac{E_c}{12}\left(6n^2+6n+3\right). 
\label{eqn:Background-cQED-Trans App EigE}
\end{align}
Anharmonicity is defined as the difference between two successive transition frequencies. Based on result~(\ref{eqn:Background-cQED-Trans App EigE}) for transmon, the anharmonicity can be approximated as
\begin{align}
A_n\equiv(E_{n+1}-E_{n})-(E_{n}-E_{n-1})\approx -E_c,
\label{eqn:Background-cQED-Trans Anharm}
\end{align}
which only depends on the charging energy $E_c$.

In addition to solving the problem of charge dispersion, transmon design employs two parallel Josephson junctions (SQUID), instead of a single Josephson junction. Consequently, the effective Josephson energy can be tuned by an external flux called flux bias line (FBL)\cite{Koch_Charge_2007} as 
\begin{align}
E_j=E_{j,max}\cos\left(\pi\frac{\Phi_{\text{ext}}}{\Phi_0}\right),
\end{align}
where $\Phi_{ext}$ is the external control flux. This innovative design provides a qubit whose frequency can be tuned in situ. An important application of this tunability is in better quantum control and measurement of the qubit state, which requires detuning the qubit from all resonator frequencies to avoid unwanted decoherence \cite{Siddiqi_Dispersive_2006, Boissonneault_Dispersive_2009}. The equivalent circuit for a transmon qubit is shown in Fig.~(\ref{Fig:Background-Transmon}). 
\subsection{Transmon qubit coupled to a transmission line resonator}
\label{SubSec:Background-transmon coupled resonator}
%%%%%%%%%%%%%%% Fig:TransCoupledRes %%%%%%%%%%%%
\begin{figure}[t!]
\centering
\subfloat[\label{subfig:Background-TransCoupledResDeviceView}]{%
\includegraphics[scale=0.40]{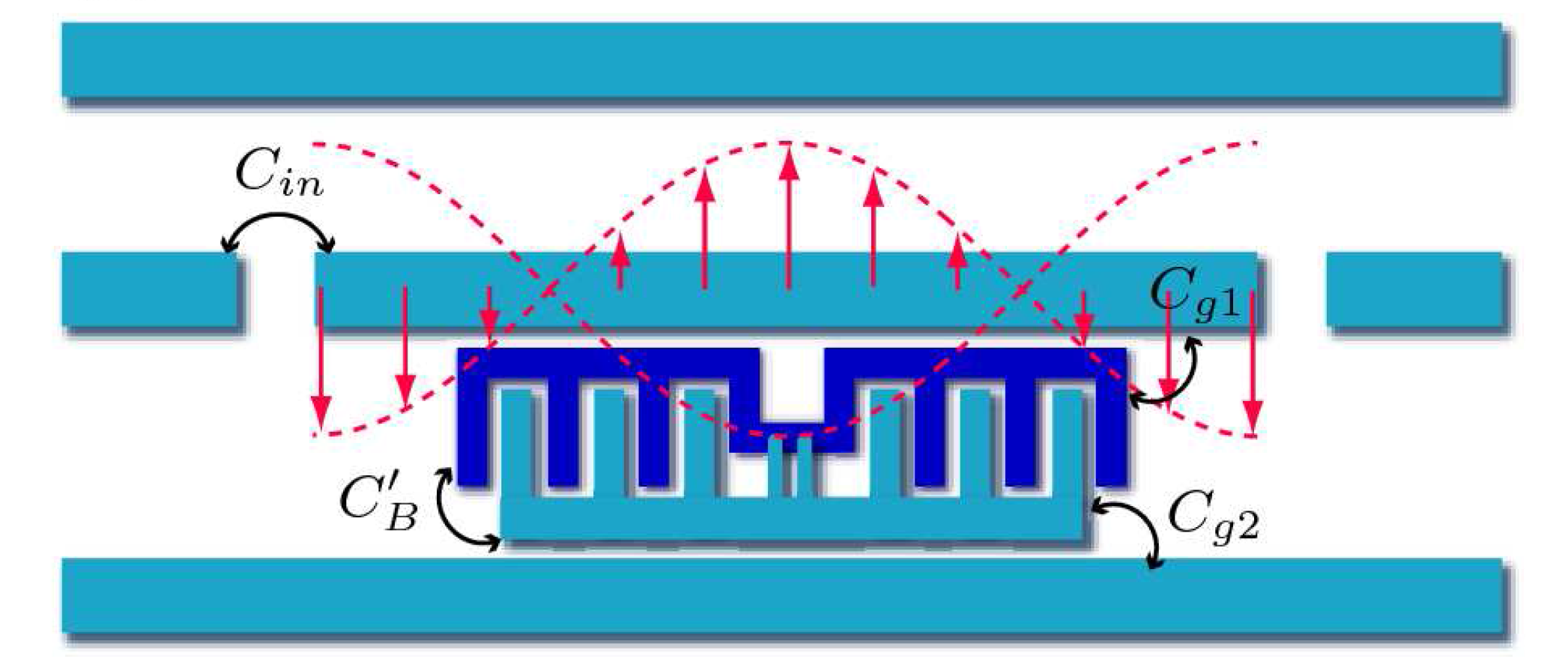}%
}\hfill
\subfloat[\label{subfig:Background-TransCoupledResEqCircuit}]{%
\includegraphics[scale=0.40]{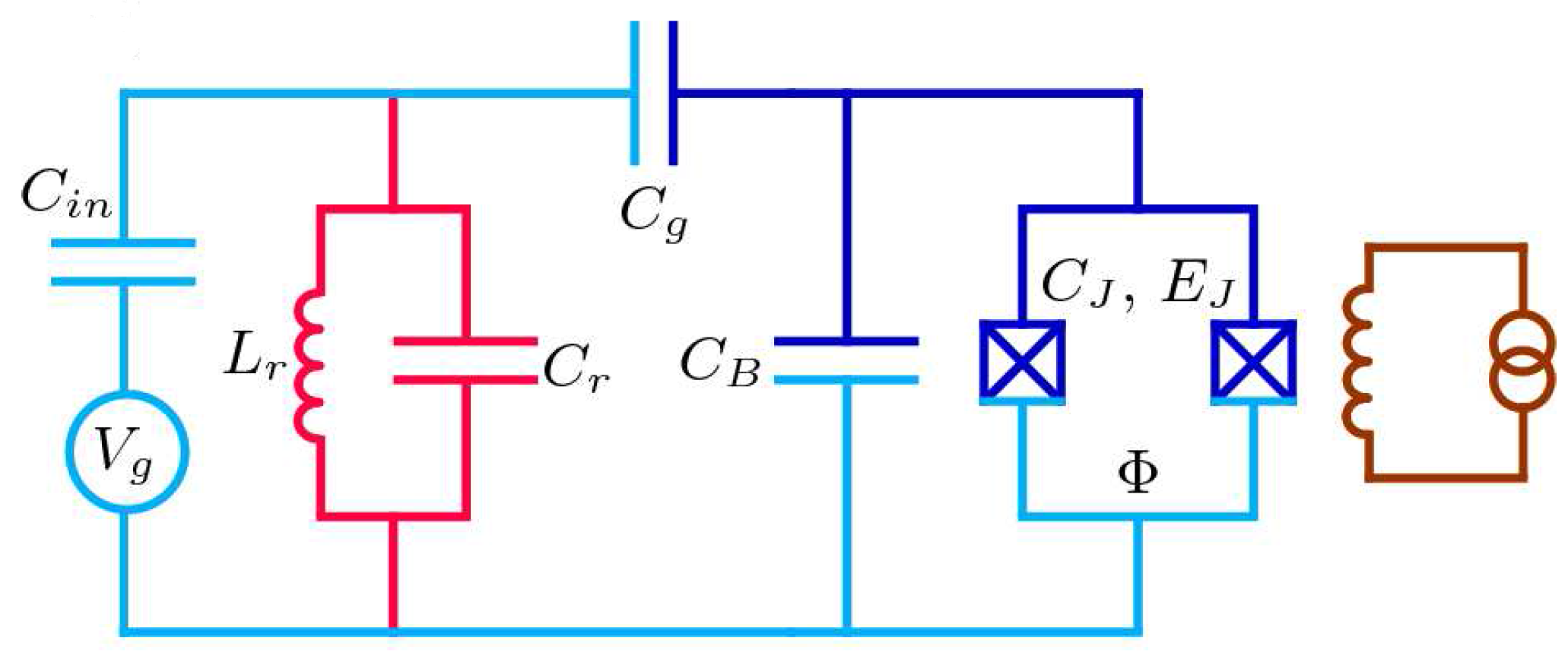}%
}
\caption{A transmon qubit capacitively coupled to a microwave copalanar waveguide.  The SQUID that is made of two Josephson junctions (dark blue and cyan) is characterized with capacitance $C_j$ and Josephson energy $E_j$. Moreover, there is an additional large capacitance $C_B$, which is acheived by the elongated comb-like design in order to decrease the effective charging energy such that $E_j/E_c \gg 1$. The coulomb interaction between trasnmon and the resonator is modeled as capacitive couplings , $C_{g1}$ and $C_{g2}$, for each junction separately. a) Schematic top view of the device (not to scale). b) Equivalent circuit that approximates the short transmission line as a single mode resonator. (This figure has also been used with permission from \cite{Koch_Charge_2007}. See Copyright Permissions.)}
\label{Fig:Background-TransCoupledRes}
\end{figure}
%%%%%%%%%%%%%%%%%%%%%%%%%%%%%%%%%%%%
In Secs.~\ref{SubSec:Background-Resonator} and \ref{SubSec:Background-Super qubits}, we reviewed the basics of the cQED building blocks, i.e. superconducting transmission line resonators and qubits, separately. This prepares us for studying the simplest cQED system, which is achieved by coupling a single superconducting qubit to a coplanar transmission line. This system is able to mimic the same physics as that of a real CQED system discussed in Sec.~(\ref{Sec:Background-cavityQED}). In this section, we follow a phenomenological description of such system and show that the effective Hamiltonian for this system can be mapped to the Rabi Hamiltonian \cite{Koch_Charge_2007}. A first principles study of the Hamiltonian of this system is discussed in detail in chapter~\ref{Ch-OriginOfA2}, where we emphasize on the role of the $A^2$ term in multimode convergence as well \cite{Malekakhlagh_Origin_2016, Malekakhlagh_Cutoff_2017}. 

We consider a transmon qubit that is capacitively coupled to a transmission line resonator as shown in Fig.~\ref{Fig:Background-TransCoupledRes} \cite{Koch_Charge_2007}. A phenomenological Hamiltonian for this system can be written as sum of resonator, transmon and interaction contributions as  
\begin{align}
\hat{\mathcal{H}}=\hat{\mathcal{H}}_j+\hat{\mathcal{H}}_r+\hat{\mathcal{H}}_{\text{int}}=4E_c\hat{n}_j^2-E_j\cos(\hat{\varphi}_j)+\hbar\omega_r\hat{a}_r^{\dag}\hat{a}_r-i\hbar \beta\hat{n}_j\left(\hat{a}_r-\hat{a}_r^{\dag}\right),
\label{eqn:Background-cQED-Phen H_(cQED) 1}
\end{align}
where we have made a single mode approximation of the resonator with frequency $\omega_r$ and assumed a charge-charge coupling of strength $\beta$. The effective parameters of Hamiltonian~(\ref{eqn:Background-cQED-Phen H_(cQED) 1}) can be in principle derived in terms of bare parameters of the circuit network (See App.~A of \cite{Koch_Charge_2007}). Because of strong couplings that is possible in such systems, the basic excitations of this circuit are mixtures of junction and resonator excitations. We refer to this phenomenon as ``hybridization''. This is in contrast to CQED with real atoms where renormalization of bare QED parameters are smaller. 

Equation~(\ref{eqn:Background-cQED-Phen H_(cQED) 1}) can be reexpressed in terms of energy eigenmodes of transmon defined 
\begin{subequations}
\begin{align}
\hat{\mathcal{H}}_j\equiv\sum\limits_{n} E_n\ket{n}\bra{n}.
\label{eqn:Background-cQED-Def of ket(n)}
\end{align}
In this basis, the number operator $\hat{n}_j$ can be written   
\begin{align}
\hat{n}_j=\sum\limits_{mn}\bra{m}\hat{n}_j\ket{n}\ket{m}\bra{n}=-in_{\text{zpf}}\sum\limits_n \sqrt{n+1}\left(\ket{n}\bra{n+1}-\ket{n+1}\bra{n}\right).
\label{eqn:Background-cQED-Spec Rep of n_j}
\end{align}
\end{subequations}
Using Eq.~(\ref{eqn:Background-cQED-Def of ket(n)}) and~(\ref{eqn:Background-cQED-Spec Rep of n_j}), we can rewrite the original Hamiltonian~(\ref{eqn:Background-cQED-Phen H_(cQED) 1})
\begin{align}
\hat{\mathcal{H}}=\sum\limits_n E_n\ket{n}\bra{n}+\hbar\omega_r\hat{a}_r^{\dag}\hat{a}_r-\sum\limits_{n} \hbar g_{n,n+1}\left(\ket{n}\bra{n+1}-\ket{n+1}\bra{n}\right)\left(\hat{a}_r-\hat{a}_r^{\dag}\right),
\label{eqn:Background-cQED-Phen H_(cQED) 2}
\end{align}
where $g_{mn}\equiv \beta \bra{m}\hat{n}_j\ket{n}$. Assuming that the anharmonicity is large enough so that the transmon can be treated as a two-level system, Hamiltonian~(\ref{eqn:Background-cQED-Phen H_(cQED) 2}) is reduced to the Rabi model \cite{Rabi_Space_1937}
\begin{align}
\hat{\mathcal{H}}_{\text{Rabi}}=\frac{\hbar\omega_q}{2}\hat{\sigma}^z+\hbar\omega_r\hat{a}_r^{\dag}\hat{a}_r-g\left(\hat{\sigma}^--\hat{\sigma}^+\right)\left(\hat{a}_r-\hat{a}_r^{\dag}\right),
\label{eqn:Background-cQED-Phen H_(Rabi)}
\end{align}
where $\hbar\omega_q\equiv E_1-E_0$ is the qubit transition frequency and $\sigma$ stands for Pauli operators given as
\begin{align}
\sigma^-\equiv \ket{0}\bra{1}, \quad \sigma^+\equiv \ket{1}\bra{0}, \quad \sigma^z\equiv \ket{1}\ket{1}-\ket{0}\ket{0}.
\label{eqn:Background-cQED-Def of Pauli Ops}
\end{align}
The physics of Rabi model and also its simpler analogue the Jaynes-Cummings model\cite{Jaynes_Comparison_1963, Shore_Jaynes_1993} will be discussed in Sec.~(\ref{Sec:Background-closedQED}).
\section{Closed QED systems}
\label{Sec:Background-closedQED}
Here, we study the spectrum of isolated quantum optical systems without taking into account their coupling to the surrounding environment. In Sec.~\ref{SubSec:Background-JC}, we review the spectrum of the single mode Jaynes-Cummings model, which is an approximation to the more precise Rabi model discussed in Sec.~\ref{Sec:Background-openQED}.    
\subsection{Jaynes-Cummings model}
\label{SubSec:Background-JC}

In 1963, Edwin Jaynes and Fred Cummings proposed a model \cite{Jaynes_Comparison_1963} in order to compare the semiclassical and the quantum theory of radiation in describing the phenomenon of spontaneous emission. The Jaynes-Cummings Hamiltonian is
\begin{align}
\frac{\hat{\mathcal{H}}_{JC}}{\hbar}\equiv\frac{\omega_q}{2}\hat{\sigma}^z+\omega_c\hat{a}^{\dag}\hat{a}+g\left(\hat{a}^{\dag}\hat{\sigma}^{-}+\hat{\sigma}^{+}\hat{a}\right),
\label{eqn:Background-JC-H_JC}
\end{align}
where $\omega_q$, $\omega_c$ and $g$ are the two-level system frequency, cavity frequency and the light-matter coupling strength, respectively. Note that the Rabi model~(\ref{eqn:Background-cQED-Phen H_(Rabi)}) couples the quadratures, i.e. $(\hat{a}+\hat{a}^{\dag})$ and $(\hat{\sigma}^++\hat{\sigma}^-)$, of each subsystem. As a result, it contains both rotating wave $g\left(\hat{a}^{\dag}\hat{\sigma}^{-}+\hat{\sigma}^{+}\hat{a}\right)$ and counter rotating wave contributions $g\left(\hat{a}^{\dag}\hat{\sigma}^{+}+\hat{\sigma}^{-}\hat{a}\right)$ contrary to the JC model that only keeps the rotating wave terms. Neglecting the counter rotating terms is called rotating wave approximation (RWA). In order to understand where this approximation holds, it is helpful to move to the interaction picture, where the rotating wave terms oscillate with aditional phase factor $e^{\pm i(\omega_c-\omega_q)t}$ while counter rotating terms oscillate with $e^{\pm i(\omega_c+\omega_q)t}$. Hence, if the qubit-cavity detunning is small, i.e. $|\omega_q-\omega_c|\ll|\omega_q+\omega_c|$, the counter rotating terms oscillate at a much faster rate. If the coupling strength is sufficiently weak, $g\ll \min\{\omega_c,\omega_q\}$, the fast oscillation of the counter rotating terms can be averaged out since they have a negligible contribution.

Neglecting the counter rotating terms brings a fictitious $\mathbb{U}(1)$ symmetry, which leads to conservation of total number of excitations 
\begin{align}
\hat{N}\equiv\hat{a}^{\dag}\hat{a}+\hat{\sigma}^+\hat{\sigma}^-= \hat{a}^{\dag}\hat{a}+\frac{\hat{1}+\hat{\sigma}^z}{2}.
\label{eqn:Background-JC-H_JC}
\end{align}
Mathematically, this can be formulated as 
\begin{align}
\left[\hat{\mathcal{H}}_{JC},\hat{N}\right]=0,
\label{eqn:Background-JC-[H_JC,N]=0}
\end{align}
which means that $\hat{N}$ is a constant of motion. This makes analytical solutions for the spectrum of JC Hamiltonian feasible since $\hat{N}$ and $\hat{\mathcal{H}}_{JC}$ can be diagonalized in the same basis. Eigenstates of $\hat{N}$ with eigenvalue $n$ are $\ket{g,n}$ and $\ket{e,n-1}$, where $g$ and $e$ stand for ground or excite state of the qubit and the second label shows the number of photons in the cavity. Therefore, it is possible to project $\hat{\mathcal{H}}_{JC}$ into separate $2\times2$ sectors, call it $\hat{\mathcal{H}}_{JC,n}$, spanned by these states. 

For this purpose, we rewrite the original JC Hamiltonian in term of $\hat{N}$ as
\begin{align}
\frac{\hat{\mathcal{H}}_{JC}}{\hbar}=\omega_c\left(\hat{N}-1/2\right)+\frac{\delta_{qc}}{2}\hat{\sigma}^z+g\left(\hat{a}^{\dag}\hat{\sigma}^{-}+\hat{\sigma}^{+}\hat{a}\right),
\label{eqn:Background-JC-H_(JC) in terms of N}
\end{align}
where $\delta_{qc}\equiv \omega_q-\omega_c$ is the qubit-cavity detuning. The JC Hamiltonian~(\ref{eqn:Background-JC-H_(JC) in terms of N}) has a unique ground state $\ket{g,0}$ with no photon in the cavity and qubit in the ground state. The ground state energy is obtained as $E_g/\hbar=\omega_q/2$. The excited energy states can be obtained by $2\times 2$ representation of $\hat{\mathcal{H}}_{JC,n+1}$ over the basis $\left\{\ket{g,n+1},\ket{e,n}\right\}$ as
\begin{align}
\frac{\hat{\mathcal{H}}_{JC,n+1}}{\hbar}=\left(n+\frac{1}{2}\right)\omega_c 
\begin{bmatrix}
1 & 0\\
0 & 1
\end{bmatrix}
+\frac{1}{2}
\begin{bmatrix}
\delta_{qc} & \Omega_n\\
\Omega_n & -\delta_{qc}
\end{bmatrix},
\label{eqn:Background-JC-H_(JC,n+1)}
\end{align}
where $\Omega_n\equiv2g\sqrt{n+1}$ is called the $n$-photon Rabi frequency. The eigenstates of~(\ref{eqn:Background-JC-H_(JC,n+1)}) are found as
\begin{align}
&\ket{n,-}\equiv \cos(\theta_n)\ket{g,n+1}+\sin(\theta_n)\ket{e,n}
\label{eqn:Background-JC-ket(n,-)}\\
&\ket{n,+}\equiv -\sin(\theta_n)\ket{g,n+1}+\cos(\theta_n)\ket{e,n},
\label{eqn:Background-JC-ket(n,+)}
\end{align}
with eigenenergies
\begin{align}
\frac{E_{n,\pm}}{\hbar}=\left(n+\frac{1}{2}\right)\omega_c\pm\frac{1}{2}\sqrt{\delta_{qc}^2+\Omega_n^2},
\label{eqn:Background-JC-E_(n,pm)}
\end{align}
where $\theta_n$ is defined by $\tan(2\theta_n)\equiv\Omega_n/\delta_{qc}$. 
Therefore, the unperturbed eigenstates $\ket{g,n+1}$ and $\ket{e,n}$ are now hybridized due to interaction and their energies are shifted by $\pm\frac{1}{2}\sqrt{\delta_{qc}^2+\Omega_n^2}$. Schematic plots of the JC spectrum for both cases of zero and nonzero detuning are shown in Fig.~\ref{Fig:Background-JC Spectrum}. 
%%%%%%%%%%%%%%% Fig:JC Spectrum %%%%%%%%%%%%
\begin{figure}[t!]
\subfloat[\label{subfig:Background-JC-JCSpectrumNoDetuning}]{%
\includegraphics[scale=0.50]{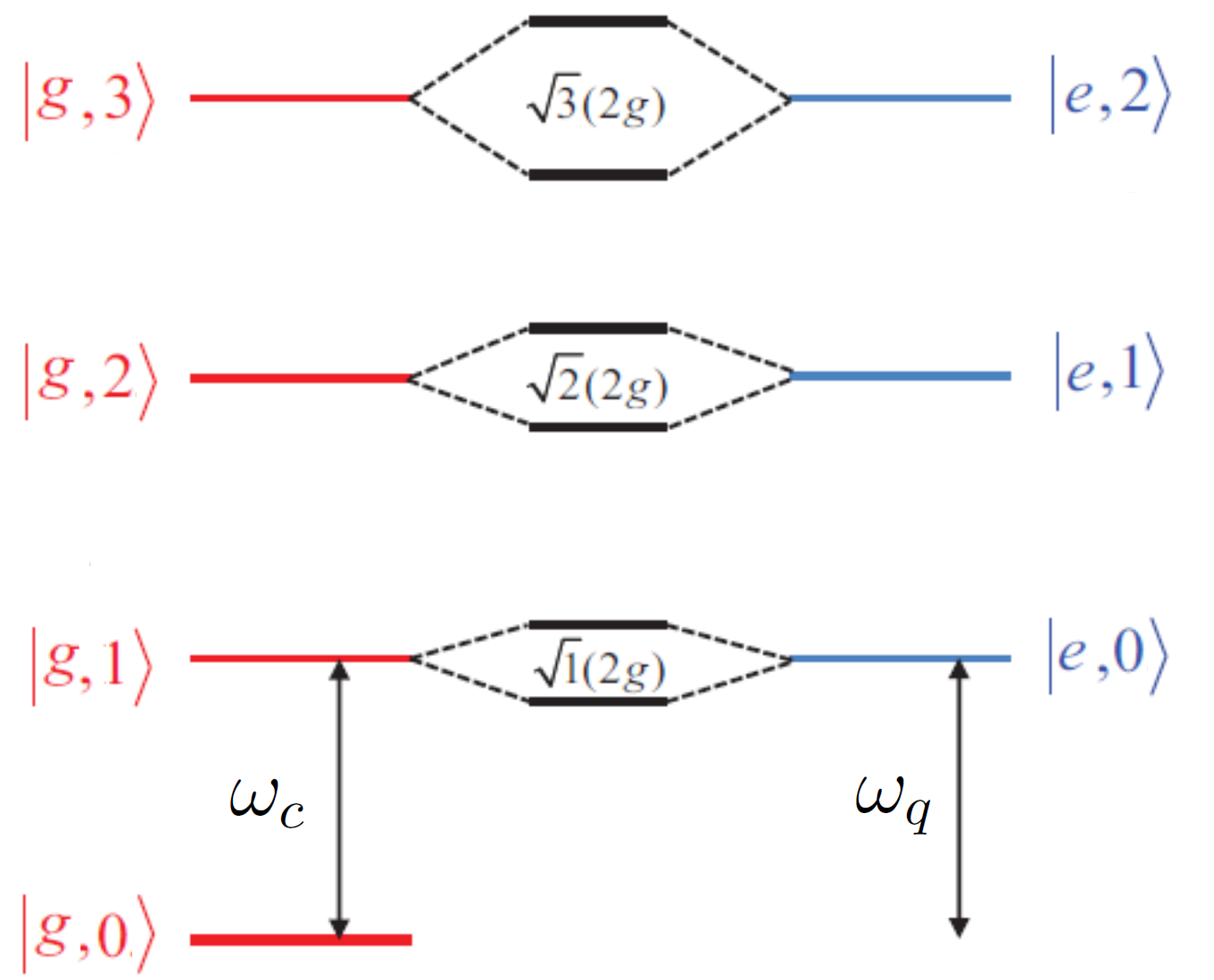}%
}\hfill
\subfloat[\label{subfig:Background-JC-JCSpectrumWithDetuning}]{%
\includegraphics[scale=0.50]{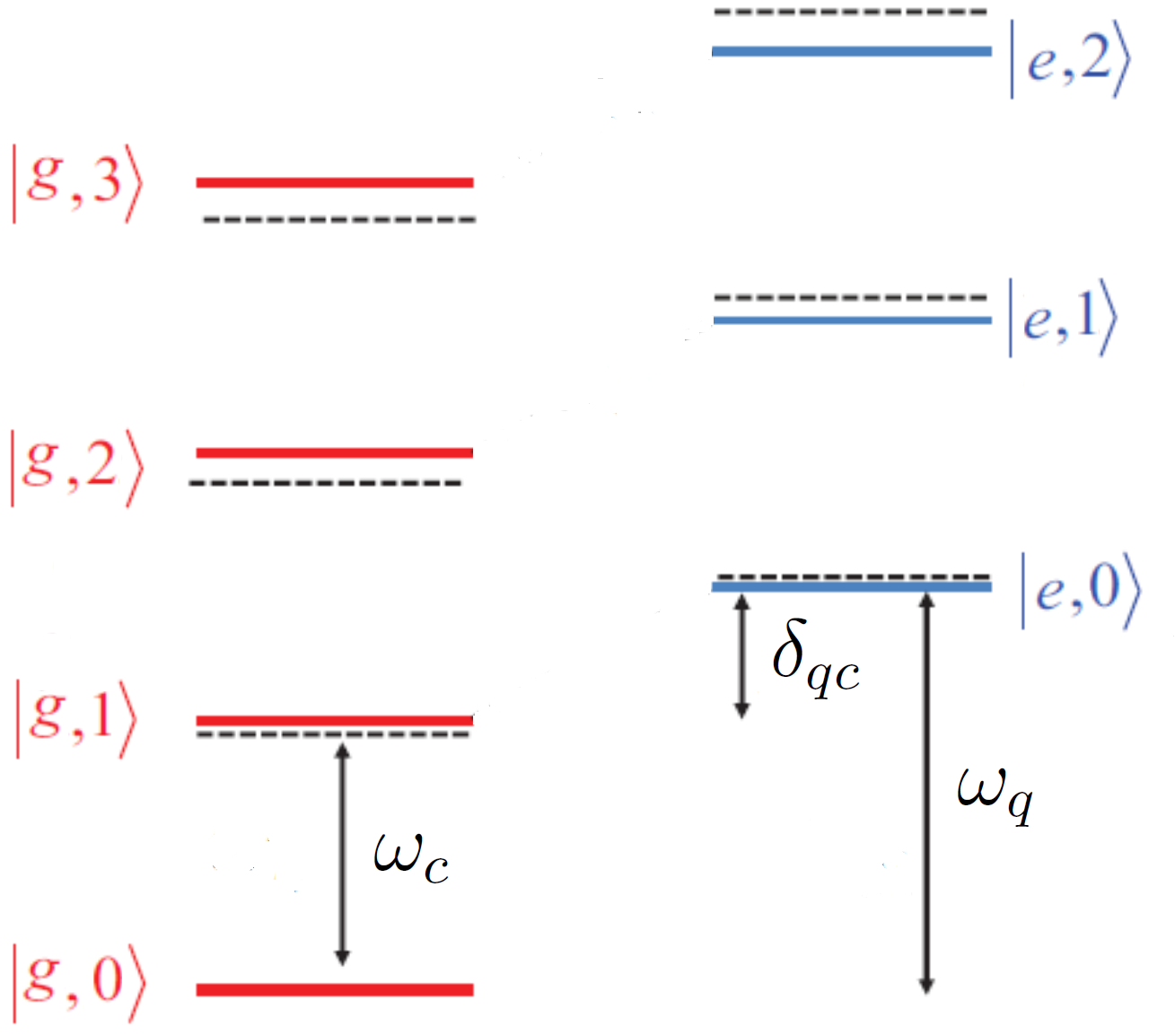}%
}
\caption{Spectrum of the JC Hamiltonian also know as the JC ladder: a) case of no detunig $\delta_{qc}=0$, b) case of non-zero detuning $\delta_{qc}>0$.}
\label{Fig:Background-JC Spectrum}
\end{figure}
%%%%%%%%%%%%%%%%%%%%%%%%%%%%%%%%%%%%
\subsection{Rabi model}
\label{SubSec:Background-Rabi}
%%%%%%%%%%%%%%% Fig:Gfunction %%%%%%%%%%%%%%%
\begin{figure}[t!]
\centering
\includegraphics[scale=0.35]{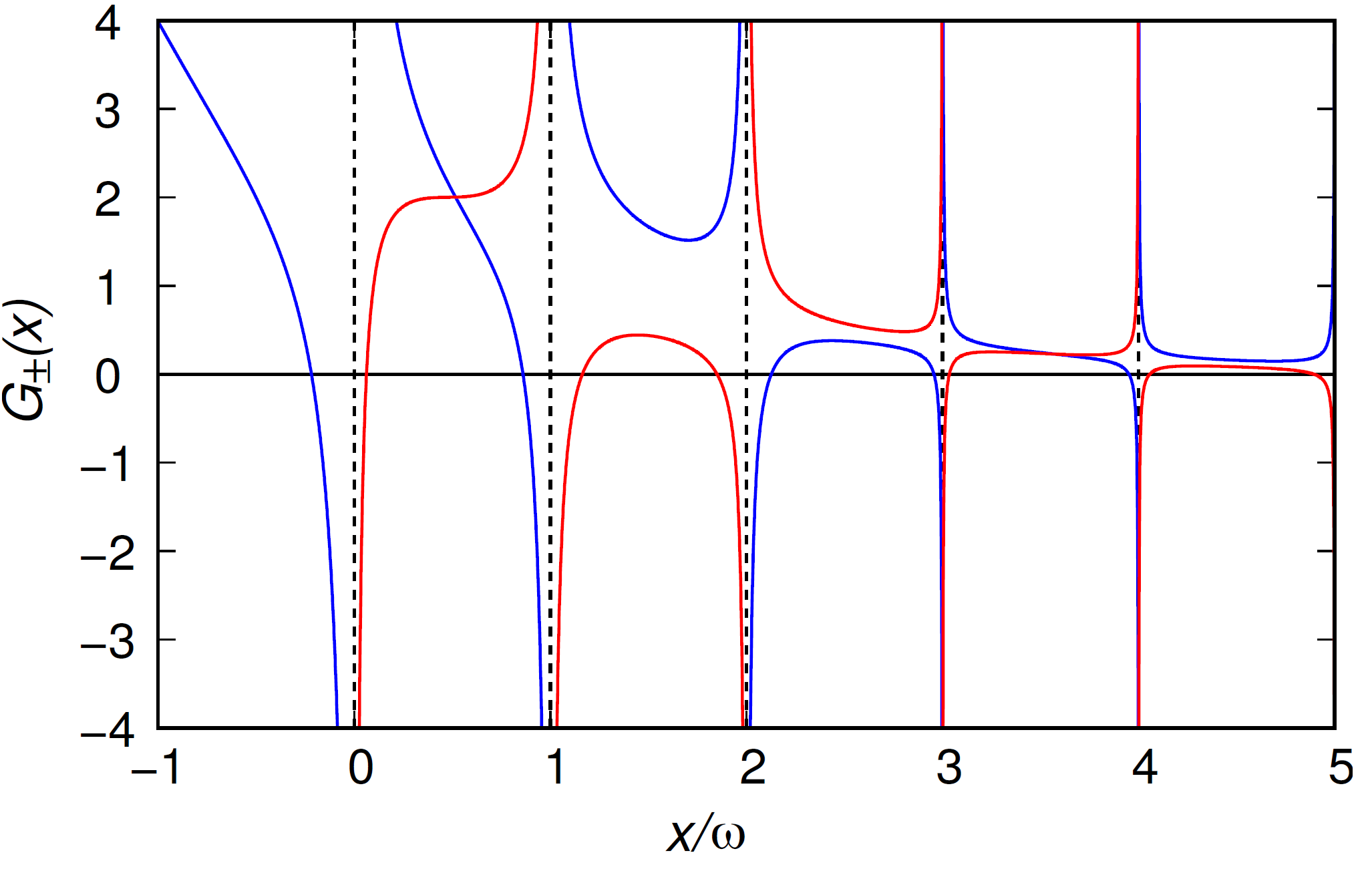}
\caption{$G_+(x)$ (\textcolor{red}{red}) and $G_-(x)$ (\textcolor{blue}{blue}) for $\omega_c=1$, $\omega_q=0.8$ and $g=0.7$. (This figure is used with permission from \cite{Braak_Integrability_2011}. See Copyright Permissions.)}
\label{Fig:Background-Rabi-Gfunction}
\end{figure}
%%%%%%%%%%%%%%%%%%%%%%%%%%%%%%%%%%%%%%%%%%%%%
The Rabi model describes the interaction of a two level system with a single bosonic mode. This rather simple model is used to describe various phenomena in physics, including light-matter coupling in cavity-QED and circuit-QED systems, magnetic resonance in solid state systems and trapped ions. Despite being introduced in 1937 \cite{Rabi_Space_1937}, about 80 years ago, spectrum and eigenfunctions of the Rabi model have been only known through numerical diagonalization. In 2011, after 74 years, an analytical solution for the spectrum of this problem was introduced by D. Braak for the first time \cite{Braak_Integrability_2011}. In this section, we briefly review this analytical solution and compare the resulting spectrum with that of the JC model.

Consider the Rabi Hamiltonian
\begin{align}
\frac{\hat{\mathcal{H}}_{\text{Rabi}}}{\hbar}\equiv\frac{\omega_q}{2}\hat{\sigma}^z+\omega_c\hat{a}^{\dag}\hat{a}+g\left(\hat{\sigma}^-+\hat{\sigma}^+\right)\left(\hat{a}+\hat{a}^{\dag}\right),
\label{eqn:Background-Rabi-H_(Rabi)}
\end{align}
where in contrast to the JC model the counter rotating terms are fully kept. Due to presence of these counter rotating terms it is not possible to find an additional conserved quantity besides the energy. This caused the common belief that the Rabi model is not integrable \cite{Graham_Quantum_1984, Bonci_Quantum_1991, Fukuo_Jaynes_1998, Emary_Quantum_2003}. Note however that the continuous $\mathbb{U}(1)$ symmetry of the JC model has not completely vanished, but rather reduced down into a discrete $\mathbb{Z}_2$ symmetry. This $\mathbb{Z}_2$ symmetry allows a decomposition of the full Hilbert space of the problem into two subspaces with opposite parities $\mathbb{H}_{\text{Rabi}}=\mathbb{H}_{\text{Rabi},+}\oplus \mathbb{H}_{\text{Rabi},-}$. It turns out that the $\mathbb{Z}_2$ symmetry is sufficient for integrability of the Rabi model \cite{Braak_Integrability_2011}. 
  
Here, we review the main results for the Rabi spectrum, whose derivation can be found in \cite{Braak_Integrability_2011}. The regular eigenvalues ($\omega_q\neq 0$) are given by roots of a transcendental function $G_{\pm}(x)$ defined as
\begin{align}
G_{\pm}(x)\equiv\sum\limits_{n=0}^{\infty}K_n(x)\left[1\mp\frac{1}{x-n}\frac{\omega_q}{2\omega_c}\right]\left(\frac{g}{\omega_c}\right)^n,
\label{eqn:Background-Rabi-Def of G(x)}
\end{align}
where the coefficients $K_n(x)$ are defined via the recursion relation
\begin{align}
nK_n(x)=f_{n-1}(x)K_{n-1}(x)-K_{n-2}(x),
\label{eqn:Background-Rabi-Def of K_n(x)}
\end{align}
with initial condition $K_0(x)=1$ and $K_1(x)=f_0(x)$. Moreover, the coefficients $f_n(x)$ are given as
\begin{align}
f_n(x)=\frac{2g}{\omega_c}+\frac{\omega_c}{2g}\left[n-x+\frac{1}{n-x}\left(\frac{\omega_q}{2\omega_c}\right)^2\right].
\label{eqn:Background-Rabi-Def of f_n(x)}
\end{align}
Based on Eq.~(\ref{eqn:Background-Rabi-Def of G(x)}), the functions $G_{\pm}(x)$ have simple poles at integer values of $x$, which are poles of the uncoupled cavity mode (Fig.~\ref{Fig:Background-Rabi-Gfunction}). The spectrum of the Rabi model is related to the roots $G_{\pm}(x)(x_{n,\pm})=0$ as
\begin{align}
\frac{E_{n,\pm}}{\hbar}=\left[x_{n,\pm}-\left(\frac{g}{\omega_c}\right)^2\right]\omega_c.
\label{eqn:Background-Rabi-E_n,pm}
\end{align}

\begin{figure}[t!]
\centering
\subfloat[\label{subfig:Background-Rabi-RabiSpectrum}]{%
\includegraphics[scale=0.35]{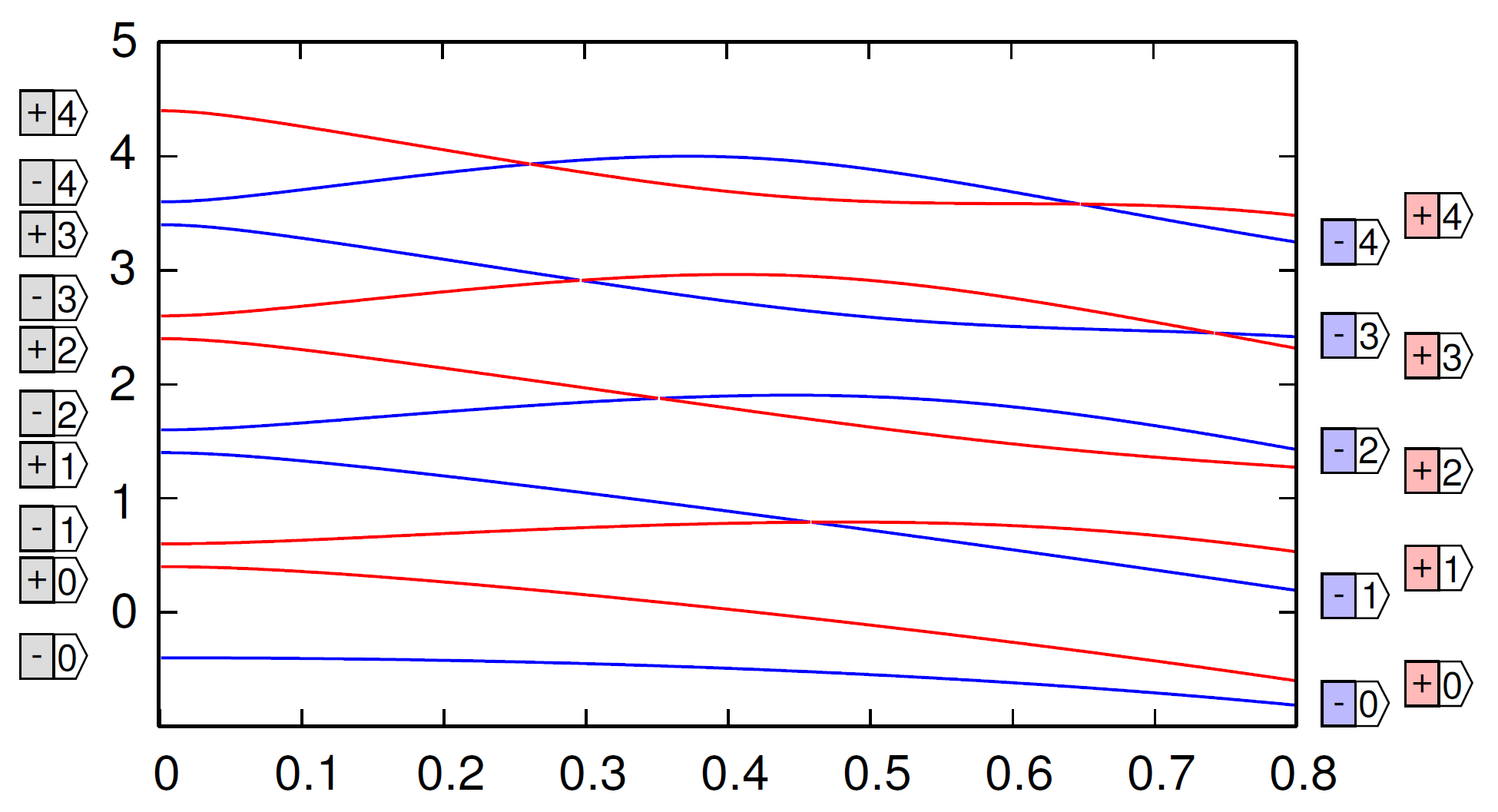}%
}\hfill
\subfloat[\label{subfig:Background-Rabi-JCVsRabiSpectrum}]{%
\includegraphics[scale=0.35]{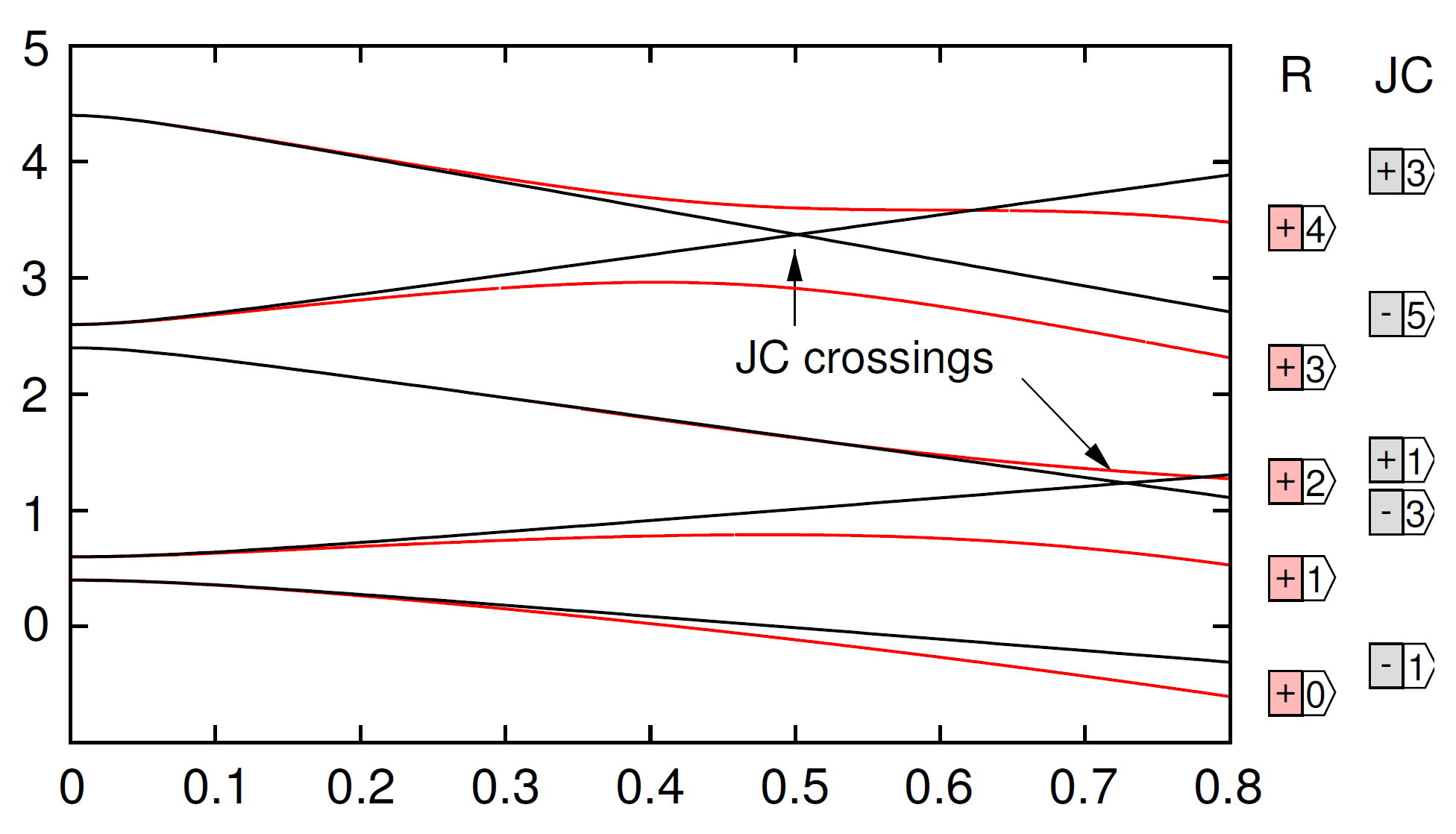}%
}
\caption{a) Spectrum of the Rabi model for $\omega_c= 1$, $\omega_q=0.8$ and $0 \leq g \leq 0.8$
with positive (\textcolor{red}{red}) and negative (\textcolor{blue}{blue}) parity. The $\mathbb{Z}_2$ symmetry partitions the Hilbert space into two subspaces. In each of these supspaces the eigenstates are labeled with increasing numbers. Importantly, we observe that level crossings of states with different parity is allowed but does not take place for states having the same parity. b) Comparison of Rabi (\textcolor{red}{red}) and JC (black) spectrums for the same parameters as in a). Interestingly, the JC model predicts fictitious level crossings between states with the same parity. (These figures are used with permission from \cite{Braak_Integrability_2011}. See Copyright Permissions.)}
\label{Fig:Background-JC Spectrum}
\end{figure}

The Rabi spectrum, including the first few levels, is shown in Fig.~\ref{subfig:Background-Rabi-RabiSpectrum} \cite{Braak_Integrability_2011} for $\omega_c=1$, $\omega_q=0.8$ as a function of coupling $0\leq g\leq 0.8$. An important observation is that level crossings of eigenstates with different parity is possible, but is not allowed for states having the same parity. On the other hand, the JC model is only valid for weak coupling and predicts fake level crossing between same parity states. A comparison of Rabi and JC spectrums is given in Fig.~\ref{subfig:Background-Rabi-JCVsRabiSpectrum} \cite{Braak_Integrability_2011}.
%%%%%%%%%%%%%%% Fig:Rabi Spectrum %%%%%%%%%%%%
%%%%%%%%%%%%%%%%%%%%%%%%%%%%%%%%%%%%
\section{Open QED systems}
\label{Sec:Background-openQED}

Up to here, we have only discussed the quantum theory of an ideal isolated QED system without taking into account the interplay between the system and its surrounding environment. In any real QED system, the electromagnetic fields and/or qubits are coupled to external environments. 
In this section, we review a few common techniques that incorporate the openness of a quantum optical system. The starting point of all these techniques is to separate the global quantum system into a subsystem of interest and an irrelevant environment, whose detailed dynamics is not important. The environment is conventionally considered to be in a thermal state, which manifests itself as a fluctuating force acting on the relevant system and leads to dissipation, decoherence and damping that influences the quantum evolution the internal degrees of the system.  
In Sec.~\ref{SubSec:Background-Lindblad}, we derive the Lindblad master equation, which is based on a density matrix formalism. In Sec.~\ref{SubSec:Background-In-Out}, we review the derivation of input-output formalism.  

\subsection{Lindblad master equation}
\label{SubSec:Background-Lindblad}

Lindblad master equation \cite{Kossakowski_Quantum_1972, Lindblad_Generators_1976, Gorini_Completely_1976} describes the non-unitary evolution of the reduced density matrix of a system coupled to an environment. This dynamical map is completely positive and trace-preserving regardless of initial condition. In the following, we first review the generic derivation of the master equation for a system linearly coupled to multiple environments. As an example, we narrow down the result to the case of a single bosonic mode linearly coupled to a thermal bath.  

Consider dividing the universe into two sectors: system and bath. The dynamics of the universe is governed by the full Hamiltonian 
\begin{align}
\hat{\mathcal{H}}=\hat{\mathcal{H}}_s+\hat{\mathcal{H}}_b+\hat{\mathcal{H}}_{sb},
\end{align}
where $s$ and $b$ refer to the system and the bath, respectively. To simplify the dynamics we choose to work in the interaction picture:
\begin{subequations}
\begin{align}
&\hat{\tilde{\mathcal{H}}}\equiv e^{\frac{i}{\hbar}(\hat{\mathcal{H}}_s+\hat{\mathcal{H}}_b)t}\hat{\mathcal{H}}e^{-\frac{i}{\hbar}(\hat{\mathcal{H}}_s+\hat{\mathcal{H}}_b)t},
\label{eqn:Background-Lindblad-H_I}\\
&\hat{\tilde{\mathcal{\rho}}}\equiv e^{\frac{i}{\hbar}(\hat{\mathcal{H}}_s+\hat{\mathcal{H}}_b)t}\hat{\mathcal{\rho}}e^{-\frac{i}{\hbar}(\hat{\mathcal{H}}_s+\hat{\mathcal{H}}_b)t},
\label{eqn:Background-Lindblad-rho_I}
\end{align}
\end{subequations}
where tilde operators belong to the frame that rotates with the free Hamiltonian. Moreover, the density operator $\hat{\rho}$ describes the state of the total system in
the product Hilbert space $\mathbb{H}= \mathbb{H}_s 
\otimes \mathbb{H}_b$. The relevant information about the system of interest can be obtained by tracing out bath degrees of freedom to obtain the reduced density matrix $\hat{\tilde{\rho}}_s \equiv \Tr_b(\hat{\tilde{\rho}})$. 

We start by the von-Neumann equation
in the interaction picture
\begin{align}
\hat{\dot{\tilde{\rho}}}=-\frac{i}{\hbar}[\hat{\tilde{\mathcal{H}}}_{sb},\hat{\tilde{\rho}}].
\label{eqn:Background-Lindblad-VonNeumann Eq1}
\end{align}
The formal solution to Eq.~(\ref{eqn:Background-Lindblad-VonNeumann Eq1}) is obtained as
\begin{align}
\hat{\tilde{\rho}}(t)=\hat{\tilde{\rho}}(0)-\frac{i}{\hbar}\int_{0}^{t}dt'[\hat{\tilde{\mathcal{H}}}_{sb}(t'),\hat{\tilde{\rho}}(t')].
\label{eqn:Background-Lindblad-VonNeumann Form Sol}
\end{align}
Substituting the integral solution~(\ref{eqn:Background-Lindblad-VonNeumann Form Sol}) into the original Eq.~(\ref{eqn:Background-Lindblad-VonNeumann Eq1}) we find 
\begin{align}
\hat{\dot{\tilde{\rho}}}(t)=-\frac{i}{\hbar}[\hat{\tilde{\mathcal{H}}}_{sb},\hat{\tilde{\rho}}(0)]-\frac{1}{\hbar^2}\int_{0}^{t} dt'[\hat{\tilde{\mathcal{H}}}_{sb}(t),[\hat{\tilde{\mathcal{H}}}_{sb}(t'),\hat{\tilde{\rho}}(t')]].
\label{eqn:Background-Lindblad-VonNeumann Eq2}
\end{align}
The next step to trace Eq.~(\ref{eqn:Background-Lindblad-VonNeumann Eq2}) over bath degrees of freedom. In order to make analytical progress, two approximations are commonly made. The first is called Born approximation and assumes that the coupling between the system and the bath is
weak, such that the density matrix of the bath $\hat{\tilde{\rho}}_b$ is barely changed by the interaction. As a result,
state of the full system at time $t$ can be approximated by a tensor product
\begin{align}
\hat{\tilde{\rho}}(t)=\hat{\tilde{\rho}}_s(t)\otimes \hat{\tilde{\rho}}_b(0).
\label{eqn:Background-Lindblad-Born App}
\end{align}
Second, note that Eq.~(\ref{eqn:Background-Lindblad-VonNeumann Eq2}) is non-Markovian, i.e., it has a memory of earlier density matrix $\hat{\tilde{\rho}}(t')$ for $t'<t$. The memory makes the analytical solution intractable and the numerical simulation very challenging. Markov approximation removes the dependence on earlier states upon replacing $\hat{\tilde{\rho}}(t')$ by $\hat{\tilde{\rho}}(t)$ and sending the upper limit for the $t'$ integral to infinity. The rationale behind this approximation is again that the system bath coupling is very weak. Applying these approximations and taking the trace over bath degrees of freedom we obtain
\begin{align}
\hat{\dot{\tilde{\rho}}}_s(t)=-\frac{1}{\hbar^2}\int_{0}^{\infty} dt'\Tr_b[\hat{\tilde{\mathcal{H}}}_{sb}(t),[\hat{\tilde{\mathcal{H}}}_{sb}(t-t'),\hat{\tilde{\rho}}_s(t)\otimes\hat{\tilde{\rho}}_b(0)]],
\label{eqn:Background-Lindblad-Langevin Eq1}
\end{align}
where due to the cyclic property of the trace we replaced 
\begin{align}
\Tr_b[\hat{\tilde{\mathcal{H}}}_{sb},\hat{\tilde{\rho}}(0)]=0.
\end{align}

Equation~(\ref{eqn:Background-Lindblad-Langevin Eq1}) is called the quantum Langevin equation and is the most generic form of the evolution for the reduced density matrix of a system weakly coupled to a bath. Considering a specific form for the coupling helps simplify this equation further. We assume a linear system bath coupling as
\begin{align}
\hat{\mathcal{H}}_{sb}(t)=\hbar\sum\limits_{\alpha} \hat{\mathcal{S}}_{\alpha}(t)\otimes\hat{\mathcal{B}}_{\alpha}(t),
\label{eqn:Background-Lindblad-SysBath Coup}
\end{align}
where $\hat{\mathcal{S}}_{\alpha}$ and $\hat{\mathcal{B}}_{\alpha}$ are system and bath operators and $\alpha$ is the label for each potential bath. Next, we insert Eq.~(\ref{eqn:Background-Lindblad-SysBath Coup}) into the Langevin equation~(\ref{eqn:Background-Lindblad-Langevin Eq1}) and define bath correlation functions
\begin{align}
C_{\alpha\beta}(t')\equiv \Tr_b\{\hat{\tilde{\rho}}_b(0)\hat{\tilde{\mathcal{B}}}_{\alpha}(t)\hat{\tilde{\mathcal{B}}}_{\beta}(t-t')\}=\delta_{\alpha\beta}\braket{\hat{\tilde{\mathcal{B}}}_{\alpha}(t')\hat{\tilde{\mathcal{B}}}_{\beta}(0)},
\end{align}
where we neglect cross correlation between distinct baths. The Langevin Eq.~(\ref{eqn:Background-Lindblad-Langevin Eq1}) becomes
\begin{align}
\begin{split}
\hat{\dot{\tilde{\rho}}}_s(t)=-\sum\limits_{\alpha}\int_{0}^{\infty} dt'\left\{C_{\alpha\alpha}(t')\left[\hat{\tilde{\mathcal{S}}}_{\alpha}(t)\hat{\tilde{\mathcal{S}}}_{\alpha}(t-t')\hat{\tilde{\rho}}_s(t)-\hat{\tilde{\mathcal{S}}}_{\alpha}(t-t')\hat{\tilde{\rho}}_s(t)\hat{\tilde{\mathcal{S}}}_{\alpha}(t)\right]\right.\\
+\left.C_{\alpha\alpha}(-t')\left[\hat{\tilde{\rho}}_s(t)\hat{\tilde{\mathcal{S}}}_{\alpha}(t-t')\hat{\tilde{\mathcal{S}}}_{\alpha}(t)-\hat{\tilde{\mathcal{S}}}_{\alpha}(t)\hat{\tilde{\rho}}_s(t)\hat{\tilde{\mathcal{S}}}_{\alpha}(t-t')\right]\right\}.
\end{split}
\end{align}
We expand the system operator in terms of eigenstates of the system Hamiltonian as 
\begin{align}
\hat{\tilde{\mathcal{S}}}_{\alpha}(t-t')=\sum\limits_{mn} \hat{\tilde{\mathcal{S}}}_{\alpha}^{mn}(t)e^{-i\Delta_{mn}t'},
\end{align}
where $\Delta_{mn}\equiv \omega_m-\omega_n$ is the transition frequency between energy eigenstates $m$ and $n$ of the system. Substituting this into Eq.~(\ref{eqn:Background-Lindblad-Langevin Eq1}) and rewriting in terms of bath power spectrum
\begin{align}
C_{\alpha\beta}(\omega)\equiv \int_{0}^{\infty} dt' C_{\alpha\beta}(t)e^{i\omega t'},
\label{eqn:Background-Lindblad-Def of C_albet(om)}
\end{align}
we obtain 
\begin{align}
\begin{split}
\hat{\dot{\tilde{\rho}}}_s(t)=-\sum\limits_{\alpha}\sum\limits_{mnm'n'}\left\{C_{\alpha\alpha}(-\Delta_{mn})\left[\hat{\tilde{\mathcal{S}}}_{\alpha}^{m'n'}(t)\hat{\tilde{\mathcal{S}}}_{\alpha}^{mn}(t)\hat{\tilde{\rho}}_s(t)-\hat{\tilde{\mathcal{S}}}_{\alpha}^{mn}(t)\hat{\tilde{\rho}}_s(t)\hat{\tilde{\mathcal{S}}}_{\alpha}^{m'n'}(t)\right]\right.\\
+\left.C_{\alpha\alpha}(\Delta_{mn})\left[\hat{\tilde{\rho}}_s(t)\hat{\tilde{\mathcal{S}}}_{\alpha}^{mn}(t)\hat{\tilde{\mathcal{S}}}_{\alpha}^{m'n'}(t)-\hat{\tilde{\mathcal{S}}}_{\alpha}^{m'n'}(t)\hat{\tilde{\rho}}_s(t)\hat{\tilde{\mathcal{S}}}_{\alpha}^{mn}(t)\right]\right\}.
\end{split}
\label{eqn:Background-Lindblad-RedfieldBloch Eq}
\end{align}
Equation~(\ref{eqn:Background-Lindblad-RedfieldBloch Eq}) is the most general form of the Redfield-Bloch equation \cite{Breuer_Theory_2002}, which is a generalization of the Lindblad master equation. Every Redfield-Bloch master equation can be transformed into a Lindblad type master equation by applying the secular approximation, where only resonant system-bath interactions are kept. Both Redfield-Bloch and Lindblad master equations are trace-preserving. However Redfield-Bloch equations, in contrast to Lindblad, do not necessarily preserve the positivity of the density matrix.

To progress the analytical discussion further, we focus on the insightful case of a single mode cavity weakly coupled to a thermal bath. Then, the system and bath operators $\hat{\tilde{\mathcal{S}}}_{\alpha}$ and $\hat{\tilde{\mathcal{B}}}_{\alpha}$ are replaced as
\begin{subequations}
\begin{align}
&\hat{\tilde{\mathcal{B}}}(t)=\sum\limits_{k}g_k\left(\hat{b}_{k}e^{-i\omega_k t}+\hat{b}_{k}^{\dag}e^{+i\omega_k t}\right),
\label{eqn:Background-Lindblad-Ex B(t)}\\
&\hat{\tilde{\mathcal{S}}}(t)=\hat{a}e^{-i\omega_c t}+\hat{a}^{\dag}e^{+i\omega_c t}.
\label{eqn:Background-Lindblad-Ex S(t)}
\end{align}
\end{subequations}
In order to derive the Lindblad master equation, we first insert Eq.~(\ref{eqn:Background-Lindblad-Ex S(t)}) into the Redfield-Bloch Eq.~(\ref{eqn:Background-Lindblad-RedfieldBloch Eq}) and keep only the resonant contributions $\hat{a}^{\dag}\hat{a}$ and $\hat{a}\hat{a}^{\dag}$. The result can be written in the following compact form
\begin{align}
\hat{\dot{\tilde{\rho}}}_s(t)=-i[\hat{\mathcal{H}}_{LS},\hat{\tilde{\rho}}_s(t)]+G_{\tilde{\mathcal{B}}\tilde{\mathcal{B}}}(\omega_c)\mathcal{D}[\hat{a}]\hat{\tilde{\rho}}(t)+G_{\tilde{\mathcal{B}}\tilde{\mathcal{B}}}(-\omega_c)\mathcal{D}[\hat{a}^{\dag}]\hat{\tilde{\rho}}(t),
\label{eqn:Background-Lindblad-Lindblad Eq}
\end{align}
where $\hat{\mathcal{H}}_{LS}$ is the Lamb-shift contribution, $\tilde{G}_{\tilde{\mathcal{B}}\tilde{\mathcal{B}}}(\omega)$ is the bilateral bath correlation function and $\mathcal{D}[\hat{A}]$ is the collapse superoperator that acts as a dissipation channel via system operator $\hat{A}$. These quantities are defined as
\begin{subequations}
\begin{align}
&\hat{\mathcal{H}}_{LS}\equiv \frac{C_{\tilde{\mathcal{B}}\tilde{\mathcal{B}}}(\omega)-C_{\tilde{\mathcal{B}}\tilde{\mathcal{B}}}^*(\omega)}{2i}\hat{\tilde{a}}^{\dag}\hat{\tilde{a}}\equiv \Delta_{LS} \hat{\tilde{a}}^{\dag}\hat{\tilde{a}}
\label{eqn:Background-Lindblad-Def of H_LS}\\
&G_{\tilde{\mathcal{B}}\tilde{\mathcal{B}}}(\omega)\equiv C_{\tilde{\mathcal{B}}\tilde{\mathcal{B}}}(\omega)+C_{\tilde{\mathcal{B}}\tilde{\mathcal{B}}}^*(\omega)=\int_{-\infty}^{+\infty}dt' \braket{\hat{\tilde{\mathcal{B}}}(t')\hat{\tilde{\mathcal{B}}}(0)}e^{i\omega t'}
\label{eqn:Background-Lindblad-Def of G_BB(Om)}\\
&\mathcal{D}[\hat{A}]\hat{\rho}\equiv \hat{A}\hat{\rho}\hat{A}^{\dag}-\frac{1}{2}\left\{\hat{A}^{\dag}\hat{A},\hat{\rho}\right\}
\label{eqn:Background-Lindblad-Def of D[a]}
\end{align}
\end{subequations}
Assuming that the bath is in a thermal distribution, we can calculate the correlation function $G_{\tilde{\mathcal{B}}\tilde{\mathcal{B}}}(\omega)$ explicitly. First, we expand the expectation value $\braket{\hat{\tilde{\mathcal{B}}}(t')\hat{\tilde{\mathcal{B}}}(0)}$ as
\begin{align}
\braket{\hat{\tilde{\mathcal{B}}}(t')\hat{\tilde{\mathcal{B}}}(0)}\equiv\sum\limits_{kk'}g_kg_k^*\left(\braket{\hat{b}_k\hat{b}_{k'}^{\dag}}e^{-i\omega_k t'}+\braket{\hat{b}_k^{\dag}\hat{b}_{k'}}e^{i\omega_k t'}\right).
\label{eqn:Background-Lindblad-<B(t')B(0)> 1}
\end{align}
For a thermal distribution, we have
\begin{align}
&\braket{\hat{b}_k\hat{b}_{k'}^{\dag}}=\left[1+\bar{n}(\omega_k,T)\right]\delta_{kk'},\\
&\braket{\hat{b}_k^{\dag}\hat{b}_{k'}}=\bar{n}(\omega_k,T)\delta_{kk'},
\end{align}
where $\bar{n}(\omega_k,T)$ is the mean photon number at frequency $\omega_k$ and temperature $T$ and is given by the Bose-Einstein distribution \cite{Bose_Planck_1924} as
\begin{align}
\bar{n}(\omega_k,T)\equiv\frac{1}{e^{\frac{\hbar\omega_k}{k_BT}}-1}.
\label{eqn:Background-Lindblad-BoseEins Dist}
\end{align}
Next, we replace the discrete sums over $k$  in Eq.~(\ref{eqn:Background-Lindblad-<B(t')B(0)> 1}) as $\sum\limits_k\to\int d\omega D_b(\omega)$, with $D_b(\omega)$ being the density of states, to obtain 
\begin{align}
\braket{\hat{\tilde{\mathcal{B}}}(t')\hat{\tilde{\mathcal{B}}}(0)}=\int_0^{\infty}d\omega D_b(\omega)|g(\omega)|^2\left\{[1+\bar{n}(\omega,T)]e^{-i\omega t'}+\bar{n}(\omega,T)e^{i\omega t'}\right\},
\label{eqn:Background-Lindblad-<B(t')B(0)> 2}
\end{align}
Substituting Eq.~(\ref{eqn:Background-Lindblad-<B(t')B(0)> 2}) into expression~(\ref{eqn:Background-Lindblad-Def of G_BB(Om)}) for the correlation function and applying Markov approximation, we obtain the correlation functions as
\begin{subequations}
\begin{align}
&G_{\tilde{\mathcal{B}}\tilde{\mathcal{B}}}(\omega_c)=2\pi D_b(\omega_c)|g(\omega_c)|^2[1+\bar{n}(\omega_c,T)],\\
&G_{\tilde{\mathcal{B}}\tilde{\mathcal{B}}}(-\omega_c)=2\pi D_b(\omega_c)|g(\omega_c)|^2n(\omega_c,T).
\end{align}
\end{subequations}
Using the Fermi's golden rule \cite{Fermi_Nuclear_1950}, we are able to relate the correlation function to the damping rate as
\begin{align}
\kappa(\omega_c)\equiv 2\pi D_b(\omega_c)|g(\omega_c)|^2.
\end{align}
Therefore, we obtain the Lindblad master equation for a single bosonic mode coupled to a thermal reservoir as
\begin{align}
\hat{\dot{\tilde{\rho}}}_s(t)=-i[\hat{\mathcal{H}}_{LS},\hat{\tilde{\rho}}_s(t)]+[1+\bar{n}(\omega_c,T)]\kappa(\omega_c)\mathcal{D}[\hat{a}]\hat{\tilde{\rho}}(t)+\bar{n}(\omega_c,T)\kappa(\omega_c)\mathcal{D}[\hat{a}^{\dag}]\hat{\tilde{\rho}}(t).
\label{eqn:Background-Lindblad-Lindblad Eq}
\end{align}
Equation~(\ref{eqn:Background-Lindblad-Lindblad Eq}) describes the effective evolution of the reduced density matrix of a cavity up to Markov and rotating wave approximation. Based on this result, the net effect of the environment on the cavity is both downward and upward temperature dependent transition rates. This equation is a standard result in quantum optics, while employing the full Redfield-Bloch equation becomes essential only in the ultrastrong coupling regime, where collective excitations are influenced by the bath as well rather than just individual constituents of the system.
\subsection{Input-output formalism}
\label{SubSec:Background-In-Out}
Lindblad master equation is derived by tracing out the bath degrees of freedom completely, which results in effective decay channels for the system. However, this formalism is not applicable to describe transmission type experiments where an incident pulse is applied to the system and we want to measure the output response of the system to the input field. In this section, we review the derivation of input-output formalism for cavity-QED, which was first proposed by Collet and Gardiner in 1984 \cite{Gardiner_Quantum_2004}. Contrary to Lindblad formalism, which is in Schrodinger picture, the input-output formalism is based on Heisenberg equations of motion. The distinguishing feature of this formalism from many other treatments of open quantum systems is keeping the bath degrees of freedom instead of tracing them out completely.

We start by the following Hamiltonian
\begin{align}
\hat{\mathcal{H}}=\hat{\mathcal{H}}_s+\hat{\mathcal{H}}_b+\hat{\mathcal{H}}_{sb},
\label{eqn:Background-InOut-H}
\end{align}
where $\hat{\mathcal{H}}_s$, $\hat{\mathcal{H}}_b$ and $\hat{\mathcal{H}}_{sb}$ are the system, bath and system-bath Hamiltonians. We keep the system Hamiltonian generic, while focus on the dynamics of one of its modes that is weakly coupled to the bath. Therefore, we write
\begin{align}
&\hat{\mathcal{H}}_{b}=\sum\limits_{k}\hbar\omega_k\hat{b}_k^{\dag}\hat{b}_k,
\label{eqn:Background-InOut-H_b}\\
&\hat{\mathcal{H}}_{sb}=\sum\limits_{k}\hbar \left(g_k\hat{a}^{\dag}\hat{b}_k+g_k^*\hat{b}_k^{\dag}\hat{a}\right),
\label{eqn:Background-InOut-H_sb}
\end{align}
where $k$ labels the bath degrees of freedom and $g_k$ is the weak system bath coupling that allows for rotating wave approximation used in $\hat{\mathcal{H}}_{sb}$. Moreover, $\hat{a}$ and $\hat{b}_k$ are bosonic operators obeying
\begin{align}
[\hat{a},\hat{a}^{\dag}]=1, \quad [\hat{b}_k,\hat{b}_{k'}^{\dag}]=\delta_{kk'}.
\label{eqn:Background-InOut-Comm Rels}
\end{align}

The Heisenberg equations of motion for the bosonic fields are obtained as
\begin{subequations}
\begin{align}
&\hat{\dot{a}}(t)=\frac{i}{\hbar}[\hat{\mathcal{H}}(t),\hat{a}(t)]=\frac{i}{\hbar}[\hat{\mathcal{H}}_s(t),\hat{a}(t)]-\sum\limits_{k}ig_k\hat{b}_k(t)
\label{eqn:Background-InOut-dot(a)}\\
&\hat{\dot{b}}_k(t)=\frac{i}{\hbar}[\hat{\mathcal{H}},\hat{b}_k]=-i\omega_k\hat{b}_k(t)-ig_k^*\hat{a}(t)
\label{eqn:Background-InOut-dot(b)_k}
\end{align}
\end{subequations}
Equation~(\ref{eqn:Background-InOut-dot(b)_k}) that describes the evolution of the bath modes $\hat{b}_k(t)$ is linear and hence can be solved explicitly as
\begin{align}
\hat{b}_k(t)=\hat{b}_k(t_i)e^{-i\omega_k (t-t_i)}-ig_k^*\int_{t_i}^{t}dt'\hat{a}(t')e^{-i\omega_k(t-t')},
\label{eqn:Background-InOut-Sol1 b_k}
\end{align}
where $t_i<t$ is some initial time in the past. The first term is the free evolution of the bath mode, while the second term arises due to coupling to the cavity mode. Substituting Eq.~(\ref{eqn:Background-InOut-Sol1 b_k}) into Eq.~(\ref{eqn:Background-InOut-dot(a)}) for the cavity mode we obtain
\begin{align}
\hat{\dot{a}}(t)=\frac{i}{\hbar}[\hat{\mathcal{H}}_s(t),\hat{a}(t)]-\int_{t_i}^{t}dt'\mathcal{K}(t-t')\left[\hat{a}(t')e^{i\omega_c(t-t')}\right]-\sum\limits_{k}ig_k e^{-i\omega_k t}\hat{b}_k(t_i).
\label{eqn:Background-InOut-dot(a) with K}
\end{align}
The Kernel $\mathcal{K}$ encodes the memory of the cavity mode $\hat{a}$ being coupled to its earlier states and is given by
\begin{align}
\mathcal{K}(t-t')=\sum\limits_k |g_k|^2 e^{-i(\omega_k-\omega_c)(t-t')}.
\label{eqn:Background-InOut-Def of K(t-t')}
\end{align}
Moreover, we have multiplied $\hat{a}(t')$ in the square bracket by an additional exponential factor $e^{i\omega_c(t-t')}$ to make it slowly varying.

To make anlytical progress, we simplify the memory term by Markov approximation. First, note that the Kernel can be reexpressed as
\begin{align}
\sum\limits_k |g_k|^2 e^{-i(\omega_k-\omega_c)(t-t')}=\int_{-\infty}^{\infty}\frac{d\omega}{2\pi}\left[\sum\limits_k 2\pi|g_k|^2\delta(\omega_c-\omega_k+\omega)\right]e^{-i\omega(t-t')},
\label{eqn:Background-InOut-FermGold trick}
\end{align}
where we have inserted an additional identity as integral over a dirac $\delta$-function. The motivation for this reformulation is that the term in the square bracket can be interpreted as the decay rate from the Fermi's golden rule \cite{Fermi_Nuclear_1950}
\begin{align}
\sum\limits_k 2\pi|g_k|^2\delta(\omega_c-\omega_k+\omega)=\kappa_c(\omega_c+\omega).
\label{eqn:Background-InOut-FermGold rule}
\end{align}
Next, we apply Markov approximation which assumes that the decay rate is not frequency dependent, i.e. $\kappa_c(\omega+\omega_c)\approx \kappa_c$. Therefore, the Kernel can be approximated as
\begin{align}
\mathcal{K}(t-t')\approx \kappa_c \delta(t-t').
\label{eqn:Background-InOut-App K}
\end{align}
Employing the indentity $\int_{t_i}^{t}dt'\delta(t-t')=1/2$ we find an effective equation for the cavity mode as
\begin{align}
\hat{\dot{a}}(t)=\frac{i}{\hbar}[\hat{\mathcal{H}}_s(t),\hat{a}(t)]-\frac{\kappa_c}{2}\hat{a}(t)-\sum\limits_{k}ig_k e^{-i\omega_k t}\hat{b}_k(0),
\label{eqn:Background-InOut-dot(a) with diss}
\end{align}
where the first term is the free evolution of the cavity mode due to $\hat{\mathcal{H}}_s$, the second term, within
the Markov approximation, has become a linear
dissipation term for the cavity mode, and finally the last term is the input field that moves towards the cavity.

Next, we simplify the input field further by assuming that the coupling $g_k$ varies very smoothly around the frequency of interest $\omega_c$, such that we can relate the dissipation $\kappa_c$ and bath density of states as
\begin{align}
&\kappa_c=2\pi g^2 D_b,\\
&D_b\equiv \sum\limits_k \delta(\omega_c-\omega_k).
\end{align}  
Defining the input field as
\begin{align}
\hat{b}_{\text{in}}(t)\equiv\frac{1}{\sqrt{2 \pi D_b}}\sum\limits_k \hat{b}_k(t_i)e^{-i\omega_k(t-t_i)},
\label{eqn:Background-InOut-Def of b_in}
\end{align}
we are able to rewrite Eq.~(\ref{eqn:Background-InOut-dot(a) with diss}) as
\begin{align}
\hat{\dot{a}}(t)=\frac{i}{\hbar}[\hat{\mathcal{H}}_s(t),\hat{a}(t)]-\frac{\kappa_c}{2}\hat{a}(t)-\sqrt{\kappa_c}\hat{b}_{in}(t).
\label{eqn:Background-InOut-dot(a) with b_in}
\end{align}

Equation~(\ref{eqn:Background-InOut-dot(a) with b_in}) provides the response of the cavity mode to an input field $\hat{b}_{in}(t)$. In most experiments, we do not have access to measure directly the intracavity field. Instead, the dynamics of the cavity is usually inferred from analyzing the output field $\hat{b}_{out}(t)$. If the cavity does not respond the to the input field at all, then the output field will be the same as the input field that is reflected from the end mirrors of the cavity. On the other hand, if the mirrors are partially transparent that allows the input field to penetrate the cavity, then the output field will contain information about the intracavity dynamics. 

In order to find a relation between the input and the output fields, we can write an alternative solution to Eq.~(\ref{eqn:Background-InOut-dot(b)_k}) in terms of a final time, $t<t_f$, as
\begin{align}
\hat{b}_k(t)=\hat{b}_k(t_f)e^{-i\omega_k (t-t_f)}+ig_k^*\int_{t}^{t_f}dt'\hat{a}(t')e^{-i\omega_k(t-t')},
\label{eqn:Background-InOut-Sol2 b_k}
\end{align}
where now the evolution is backwards from the future time $t_f$ to $t$. Defining the output field as
\begin{align}
\hat{b}_{\text{out}}(t)\equiv\frac{1}{\sqrt{2 \pi D_b}}\sum\limits_k \hat{b}_k(t_f)e^{-i\omega_k(t-t_f)},
\label{eqn:Background-InOut-Def of b_out}
\end{align}
and going through the same derivation, we obtain another equation of motion for $\hat{a}(t)$ as
\begin{align}
\hat{\dot{a}}(t)=\frac{i}{\hbar}[\hat{\mathcal{H}}_s(t),\hat{a}(t)]+\frac{\kappa_c}{2}\hat{a}(t)-\sqrt{\kappa_c}\hat{b}_{out}(t),
\label{eqn:Background-InOut-dot(a) with b_out}
\end{align}
where the crucial difference is that by reversing the direction of time damping becomes gain and vice versa. Eventually, by subtracting Eqs.~(\ref{eqn:Background-InOut-dot(a) with b_out}) and~(\ref{eqn:Background-InOut-dot(a) with b_in}) we obtain the desired input-output relation as
\begin{align}
\hat{b}_{\text{out}}(t)=\hat{b}_{\text{in}}(t)+\sqrt{\kappa}_c\hat{a}(t).
\label{eqn:Background-InOut-InOut Rel}
\end{align}
The input-output relation~(\ref{eqn:Background-InOut-InOut Rel}) states that, up to rotating-wave and Markov approximations, the output field is just the reflected input field plus an additional cavity response that is the field being radiated through the partially transparent cavity mirrors. In order to use these relations, first we have to solve for the cavity response, i.e. $\hat{a}(t)$, to a known input field $\hat{b}_{in}(t)$ from Eq.~(\ref{eqn:Background-InOut-dot(a) with b_in}) and then use Eq.~(\ref{eqn:Background-InOut-InOut Rel}) to obtain the output response $\hat{b}_{out}(t)$.

To illustrate this formalism, we apply it on a single mode cavity with system Hamiltonian
\begin{align}
\hat{\mathcal{H}}_s\equiv \hbar\omega_c\hat{a}^{\dag}\hat{a}.
\label{eqn:Background-InOut-Ex Hs}
\end{align}
For this system, Eq.~(\ref{eqn:Background-InOut-dot(a) with b_in}) simplifies to
\begin{align}
\hat{\dot{a}}(t)=-i\omega_c\hat{a}-\frac{\kappa_c}{2}\hat{a}(t)-\sqrt{\kappa_c}\hat{b}_{in}(t),
\label{eqn:Background-InOut-Ex dot(a) with b_in}
\end{align}
which is a linear ODE with constant coefficients and hence can be solved exactly via Laplace transform as
\begin{align}
\hat{\tilde{a}}(s)=\frac{\hat{a}(0)}{s+i\omega_c+\kappa_c/2}-\frac{\sqrt{\kappa_c}\hat{\tilde{b}}_{in}(s)}{s+i\omega_c+\kappa_c/2}.
\label{eqn:Background-InOut-a(s) Sol}
\end{align}
The first term in the solution~(\ref{eqn:Background-InOut-a(s) Sol}) is the cavity response to its initial condition that is a damped oscillatory solution. The second contribution gives the steady state response of the cavity to the input field. Taking Laplace transform of the input-output relation~(\ref{eqn:Background-InOut-a(s) Sol}) and substituting it into Eq.~(\ref{eqn:Background-InOut-Ex b_out(s) Sol}) we obtain the output response in the Laplace domain
\begin{align}
\hat{\tilde{b}}_{out}(s)=\frac{\sqrt{\kappa_c}}{s+i\omega_c+\kappa_c/2}\hat{a}(0)+\frac{s+i\omega_c-\kappa_c/2}{s+i\omega_c+\kappa_c/2}\hat{\tilde{b}}_{in}(s).
\label{eqn:Background-InOut-Ex b_out(s) Sol}
\end{align}
Therefore, the reflection coefficient from the cavity mirrors can be obtained as
\begin{align}
R_c(s)\equiv\frac{\hat{\tilde{b}}_{\text{out}}(s)}{\hat{\tilde{b}}_{\text{in}}(s)}=\frac{s+i\omega_c-\kappa_c/2}{s+i\omega_c+\kappa_c/2}.
\label{eqn:Background-InOut-Ex Def of R(s)}
\end{align}
\section{Summary}

In this chapter, we provided a brief summary of the necessary theoretical foundation of the field of cavity and circuit quantum electrodynamics. The presented material here will be used in and is essential to understanding the work in the subsequent chapters. In particular, our Heisenberg-Langevin formalism, discussed in chapter~\ref{Ch:NonMarkovian}, is indeed a generalization of input-output formalism without adopting rotating-wave, two-level, Born or Markov approximations.
\chapter{Non-Markovian dynamics of a superconducting qubit in an open multimode resonator\label{Ch:NonMarkovian}}

In this chapter, we study the dynamics of a transmon qubit that is capacitively coupled to an open multimode superconducting resonator. Our effective equations are derived by eliminating resonator degrees of freedom while encoding their effect in the Green's function of the electromagnetic background. We account for the dissipation of the resonator exactly by employing a spectral representation for the Green's function in terms of a set of non-Hermitian modes and show that it is possible to derive effective  Heisenberg-Langevin equations without resorting to the rotating wave, two level, Born or Markov approximations. A well-behaved time domain perturbation theory is derived to systematically account for the nonlinearity of the transmon. We apply this method to the problem of spontaneous emission, capturing accurately the non-Markovian features of the qubit dynamics, valid for any qubit-resonator coupling strength.
\section{Introduction}
Superconducting circuits are of interest for gate based quantum information processing \cite{Devoret_Implementing_2004, Blais_Quantum-Information_2007, Devoret_Superconducting_2013} and for fundamental studies of collective quantum phenomena away from equilibrium \cite{Houck_On-chip_2012, Schmidt_Circuit_2013, LeHur_Many-body_2016}. In these circuits, Josephson junctions provide the nonlinearity required to define a qubit or a pseudo-spin degree of freedom, and low loss microwave waveguides and resonators provide a convenient linear environment to mediate interactions between Josephson junctions \cite{Majer_Coupling_2007, Sillanpaa_Coherent_2007, Filipp_Preparation_2011, Filipp_Multimode_2011, Loo_Photon-mediated_2013, Shankar_Autonomously_2013, Kimchi-Schwartz_Stabilizing_2016}, act as Purcell filters \cite{Houck_Controlling_2008, Jeffrey_Fast_2014, Bronn_Broadband_2015} or as suitable access ports for efficient state preparation and readout.
% \cite{relevant}
Fabrication capabilities have reached a stage where coherent interactions between multiple qubits occur through a waveguide \cite{Loo_Photon-mediated_2013}, active coupling elements \cite{Roushan_Chiral_2016} or cavity arrays \cite{McKay_High-Contrast_2015}, while allowing manipulation and readout of individual qubits in the circuit.  In addition, experiments started deliberately probing regimes featuring very high qubit coupling strengths \cite{Niemczyk_Circuit_2010, Diaz_Observation_2010, Todorov_Ultrastrong_2010} or setups where multimode effects cannot be avoided \cite{Sundaresan_Beyond_2015}. Accurate modeling of these complex circuits has not only become important for designing such circuits, e.g. to avoid cross talk and filter out the electromagnetic environment, but also for the fundamental question of the collective quantum dynamics of qubits \cite{Mlynek_Observation_2014}. In this work, we introduce a first principles Heisenberg-Langevin framework that accounts for such complexity.

%%%%%%%%%% Fig of cQEDOpenSymbolic  %%%%%%%
\begin{figure}[t!]
\centering
\subfloat[\label{subfig:cQEDopenSybolic}]{%
\includegraphics[scale=1.2]{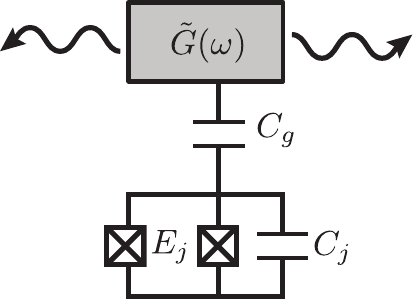}%
} 
\subfloat[\label{subfig:cQEDopenSybolicSpider}]{%
\includegraphics[scale=1.2]{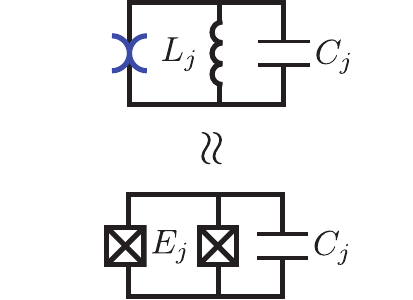}%
} 
\caption{a) Transmon qubit linearly (capacitively) coupled to an open harmonic electromagnetic background, i.e. a multimode superconducting resonator, characterized by Green's function $\tilde{G}(\omega)$. b) Separation of linear and anharmonic parts of the Josephson potential.}
\label{Fig:cQEDopenSybolic}
\end{figure}
%%%%%%%%%%%%%%%%%%%%%%%%%%%%%%%%%%%%%%%%%%%%

The inadequacy of the standard Cavity QED models based on the interaction of a pseudo-spin degree of freedom with a single cavity mode was recognized early on \cite{Houck_Controlling_2008}. In principle, the Rabi model could straightforwardly be extended to include many cavity electromagnetic modes and the remaining qubit transitions (See Sec. III of \cite{Malekakhlagh_Origin_2016}), but this does not provide a computationally viable approach for several reasons. Firstly, we do not know of a systematic approach for the truncation of this multimode multilevel system. Secondly, the truncation itself will depend strongly on the spectral range that is being probed in a given experiment (typically around a transition frequency of the qubit), and the effective model for a given frequency would have to accurately describe the resonator loss in a broad frequency range. It is then unclear whether the Markov approximation would be sufficient to describe such losses. 

Multimode effects come to the fore in the accurate computation of the effective Purcell decay of a qubit \cite{Houck_Controlling_2008} or the photon-mediated effective exchange interaction between qubits in the dispersive regime \cite{Filipp_Multimode_2011}, where the perturbation theory is divergent. A phenomenological semiclassical approach to the accurate modeling of Purcell loss has been suggested~\cite{Houck_Controlling_2008}, based on the availability of the effective impedance seen by the qubit. A full quantum model that incorporates the effective impedance of the linear part of the circuit at its core was later presented~\cite{Nigg_BlackBox_2012}. This approach correctly recognizes that a better behaved perturbation theory in the nonlinearity can be developed if the hybridization of the qubit with the linear multimode environment is taken into account at the outset \cite{Bourassa_Josephson_2012}. Incorporating the dressing of the modes into the basis that is used to expand the nonlinearity gives then rise to self- and cross-Kerr interactions between hybridized modes. This basis however does not account for the open nature of the resonator. Qubit loss is then extracted from the poles of the linear circuit impedance at the qubit port, $Z(\omega)$. This quantity can in principle be measured or obtained from a simulation of the classical Maxwell equations. Finding the poles of $Z(\omega)$ through Foster's theorem introduces potential numerical complications \cite{solgun_blackbox_2014}. Moreover, the interplay of the qubit nonlinearity and dissipation is not addressed within Rayleigh-Schr\"odinger perturbation theory. An exact treatment of dissipation is important for the calculation of multimode Purcell rates of qubits as well as the dynamics of driven dissipative qubit networks \cite{aron_photon-mediated_2016}.

The difficulty with incorporating dissipation on equal footing with energetics in open systems is symptomatic of more general issues in the quantization of radiation in finite inhomogeneous media. One of the earliest thorough treatments of this problem \cite{Glauber_Quantum_1991} proposes to use a complete set of states in the unbounded space including the finite body as a scattering object. This ``modes of the universe'' approach \cite{lang_why_1973, ching_quasinormal-mode_1998} is well-defined but has an impractical aspect: one has to deal with a continuum of modes, and as a consequence simple properties characterizing the scatterer itself (e.g. its resonance frequencies and widths) are not effectively utilized. Several methods have been proposed since then to address this shortcoming, which discussed quantization using quasi-modes (resonances) of the finite-sized open resonator \cite{dalton_quasi_1999, lamprecht_quantized_1999, dutra_quantized_2000, hackenbroich_field_2002}. Usually, these methods treat the atomic degree of freedom as a two-level system and use the rotating wave and the Markov approximations. 

In the present work, rather than using a Hamiltonian description, we derive an effective Heisenberg-Langevin equation to describe the dynamics of a transmon qubit \cite{Koch_Charge_2007} capacitively coupled to an open multimode resonator (See Fig.~\ref{subfig:cQEDopenSybolic}). Our treatment illustrates a general framework that does not rely on the Markov, rotating wave or two level approximations. We show that the electromagnetic degrees of freedom of the entire circuit can be integrated out and appear in the equation of motion through the classical electromagnetic Green's function (GF) corresponding to the Maxwell operator and the associated boundary conditions. A spectral representation of the GF in terms of a complete set of non-Hermitian modes \cite{Tureci_SelfConsistent_2006, martin_claassen_constant_2010} accounts for dissipative effects from first principles. This requires the solution of a boundary-value problem of the Maxwell operator {\it only} in the finite domain of the resonator. Our main result is the effective equation of motion~(\ref{eqn:Eff Dyn before trace}), which is a Heisenberg-Langevin \cite{Scully_Quantum_1997, Gardiner_Quantum_2004, Carmichael_Statistical_1998} integro-differential equation for the phase operator of the transmon. Outgoing fields, which may be desired to calculate the homodyne field at the input of an amplifier chain, can be conveniently related through the GF to the qubit phase operator.   

As an immediate application, we use the effective Heisenberg-Langevin equation of motion to study spontaneous emission. The spontaneous emission of a two level system in a finite polarizable medium was calculated \cite{Dung_Spontaneous_2000} in the Schr\"odinger-picture in the spirit of Wigner-Weisskopf theory \cite{Scully_Quantum_1997}. These calculations are based on a radiation field quantization procedure which incorporates continuity and boundary conditions corresponding to the finite dielectric \cite{Matloob_Electromagnetic_1995, Gruner_Green-Function_1996}, but only focus on separable geometries where the GF can be calculated semianalytically. A generalization of this methodology to an arbitrary geometry \cite{Krimer_Route_2014} uses an expansion of the GF in terms of a set of non-Hermitian modes for the appropriate boundary value problem \cite{Tureci_SelfConsistent_2006, martin_claassen_constant_2010}. This approach is able to consistently account for multimode effects where the atom-field coupling strength is of the order of the free spectral range of the cavity \cite{Meiser_Superstrong_2006, Krimer_Route_2014, Sundaresan_Beyond_2015} for which the atom is found to emit narrow pulses at the cavity roundtrip period \cite{Krimer_Route_2014}. A drawback of these previous calculations performed in the Schr\"odinger picture is that without the rotating wave approximation, no truncation scheme has been proposed so far to reduce the infinite hierarchy of equations to a tractable Hilbert space dimension. The employment of the rotating wave approximation breaks this infinite hierarchy through the existence of a conserved excitation number. The Heisenberg-Langevin method introduced here is valid for arbitrary light-matter coupling, and therefore can access the dynamics accurately where the rotating-wave approximation is not valid.

In summary, our microscopic treatment of the openness is one essential difference between our study and previous works on the collective excitations of circuit-QED systems with a localized Josephson nonlinearity \cite{Wallquist_Selective_2006, Bourassa_Josephson_2012, Nigg_BlackBox_2012, Leib_Networks_2012}. In our work, the lifetime of the collective excitations arises from a proper treatment of the resonator boundary conditions \cite{Clerk_Introduction_2010}. The harmonic theory of the coupled transmon-resonator system is exactly solvable via Laplace transform. Transmon qubits typically operate in a weakly nonlinear regime, where charge dispersion is negligible \cite{Koch_Charge_2007}. We treat the Josephson anharmonicity on top of the non-Hermitian linear theory (See Fig~\ref{subfig:cQEDopenSybolicSpider}) using multi-scale perturbation theory (MSPT) \cite{Bender_Advanced_1999, Nayfeh_Nonlinear_2008, Strogatz_Nonlinear_2014}. First, it resolves the anomaly of secular contributions in conventional time-domain perturbation theories via a resummation \cite{Bender_Advanced_1999, Nayfeh_Nonlinear_2008, Strogatz_Nonlinear_2014}. While this perturbation theory is equivalent to the Rayleigh-Schr\"odinger perturbation theory when the electromagnetic environment is closed, it allows a systematic expansion even when the environment is open and the dynamics is non-unitary. Second, we account for the self-Kerr and cross-Kerr interactions \cite{Drummond_Quantum_1980} between the collective non-Hermitian excitations extending \cite{Bourassa_Josephson_2012, Nigg_BlackBox_2012}. Third, treating the transmon qubit as a weakly nonlinear bosonic degree of freedom allows us to include the linear coupling to the environment non-perturbatively. This is unlike the dispersive limit treatment of the light-matter coupling as a perturbation \cite{Boissonneault_Dispersive_2009}. Therefore, the effective equation of motion is valid for all experimentally accessible coupling strengths \cite{Thompson_Observation_1992, Strong_Wallraff_2004, Reithmaier_Strong_2004, Anappara_Signatures_2009, Niemczyk_Circuit_2010, Diaz_Observation_2010, Todorov_Ultrastrong_2010,  Sundaresan_Beyond_2015}. 

We finally present a perturbative procedure to reduce the computational complexity of the solution of Eq.~(\ref{eqn:Eff Dyn before trace}), originating from the enormous Hilbert space size, when the qubit is weakly anharmonic. Electromagnetic degrees of freedom can then be perturbatively traced out resulting in an effective equation of motion~(\ref{eqn:NumSim-QuDuffingOscMemReduced}) in the qubit Hilbert space \textit{only}, which makes its numerical simulation tractable.

The paper is organized as follows: In Sec.~\ref{Sec:Toy Model}, we introduce a toy model to familiarize the reader with the main ideas and notation. In Sec.~\ref{Sec:Eff Dyn Of Transmon}, we present an ab initio effective Heisenberg picture dynamics for the transmon qubit. The derivation for this effective model has been discussed in detail in Apps.~\ref{App:Quantum EOM} and \ref{App:Eff Dyn of transmon}. In Sec.~\ref{Sec:Lin SE Theory}, we study linear theory of spontaneous emission. In Sec.~\ref{Sec:PertCor}, we employ quantum multi-scale perturbation theory to investigate the effective dynamics beyond linear approximation. The details of multi-scale calculations are presented in App.~\ref{App:MSPT}. In Sec.~\ref{Sec:NumSimul} we compare these results with the pure numerical simulation. We summarize the main results of this paper in Sec.~\ref{Sec:Conclusion}.
\section{Toy model}
\label{Sec:Toy Model}

In this section, we discuss a toy model that captures the basic elements of the effective equation (Eq.~(\ref{eqn:Eff Dyn before trace})), which we derive in full microscopic detail in Sec.~\ref{Sec:Eff Dyn Of Transmon}. This will also allow us to introduce the notation and concepts relevant to the rest of this paper, in the context of a tractable and well-known model. We consider the single-mode Cavity QED model, consisting of a nonlinear quantum oscillator (qubit) that couples linearly to a single bosonic degree of freedom representing the cavity mode (Fig.~\ref{Fig:cQEDopenSybolic}). This mode itself is coupled to a continuum set of bosons playing the role of the waveguide modes. When the nonlinear oscillator is truncated to the lowest two levels, this reduces to the standard open Rabi Model, which is generally studied using Master equation \cite{Ridolfo_Photon_2012} or stochastic Schr\"odinger equation \cite{Loic_Quantum_2014} approaches. Here we will discuss a Heisenberg-picture approach to arrive at an equation of motion for qubit quadratures. The Hamiltonian for the toy model is ($\hbar=1$)
\begin{align}
\begin{split}
\hat{\mathcal{H}} &\equiv \frac{\omega_j}{4}\left(\hat{\mathcal{X}}_j^2+\hat{\mathcal{Y}}_j^2\right)+\frac{\omega_j}{2}U(\hat{\mathcal{X}}_j)\\
&+\frac{\omega_c}{4}\left(\hat{\mathcal{X}}_c^2+\hat{\mathcal{Y}}_c^2\right)+g\hat{\mathcal{Y}}_j\hat{\mathcal{Y}}_c\\
&+\sum\limits_{b}\left[\frac{\omega_b}{4}\left(\hat{\mathcal{X}}_b^2+\hat{\mathcal{Y}}_b^2\right)+g_b\hat{\mathcal{Y}}_c\hat{\mathcal{Y}}_b\right],
\end{split}
\label{eqn:ToyModel-H}
\end{align}
where $\omega_j$, $\omega_c$ and $\omega_b$ are bare oscillation frequencies of qubit, the cavity and the bath modes, respectively. We have defined the canonically conjugate variables
\begin{align}
\hat{\mathcal{X}}_l\equiv(\hat{a}_l+\hat{a}_l^{\dag}), \ \hat{\mathcal{Y}}_l\equiv -i(\hat{a}_l-\hat{a}_l^{\dag}),
\label{Eq:ToyModel-Def of X&Y}
\end{align}
where $\hat{a}_{l}$ represent the boson annihilation operator of sector $l\equiv j,c,b$. Furthermore, $g$ and $g_b$ are qubit-cavity and cavity-bath couplings. $U(\hat{\mathcal{X}}_j)$ represents the nonlinear part of the potential shown in Fig.~\ref{subfig:cQEDopenSybolicSpider} with a blue spider symbol.

The remainder of this section is structured as follows. In Sec.~\ref{Sec:ToyModel-Eff Dyn}, we eliminate the cavity and bath degrees of freedom to obtain an effective Heisenberg-Langevin equation of motion for the qubit. We dedicate Sec.~\ref{Sec:ToyModel-LinearTheory} to the resulting characteristic function describing the hybridized modes of the linear theory. 
\subsection{Effective dynamics of the qubit}
\label{Sec:ToyModel-Eff Dyn}

In this subsection, we derive the equations of motion for the Hamiltonian~(\ref{eqn:ToyModel-H}). We first integrate out the bath degrees of freedom via Markov approximation to obtain an effective dissipation for the cavity. Then, we eliminate the degrees of freedom of the leaky cavity mode to arrive at an effective equation of motion for the qubit, expressed in terms of the GF of the cavity. The Heisenberg equations of motion are found as
\begin{subequations}
\begin{align}
&\hat{\dot{\mathcal{X}}}_j(t)=\omega_j\hat{\mathcal{Y}}_j(t)+2g\hat{\mathcal{Y}}_c(t),
\label{eqn:ToyModel-dotX_j}\\
&\hat{\dot{\mathcal{Y}}}_j(t)=-\omega_j\left\{\hat{\mathcal{X}}_j(t)+U'[\hat{\mathcal{X}}_j(t)]\right\},
\label{eqn:ToyModel-dotY_j}\\
&\hat{\dot{\mathcal{X}}}_c(t)=\omega_c\hat{\mathcal{Y}}_c(t)+2g\hat{\mathcal{Y}}_j(t)+\sum
\limits_{b} 2g_b\hat{\mathcal{Y}}_b(t),
\label{eqn:ToyModel-dotX_c}\\
&\hat{\dot{\mathcal{Y}}}_c(t)=-\omega_c\hat{\mathcal{X}}_c(t),
\label{eqn:ToyModel-dotY_c}\\
&\hat{\dot{\mathcal{X}}}_b(t)=\omega_b\hat{\mathcal{Y}}_b(t)+2g_b\hat{\mathcal{Y}}_c(t)
\label{eqn:ToyModel-dotX_b}\\
&\hat{\dot{\mathcal{Y}}}_b(t)=-\omega_b\hat{\mathcal{X}}_b(t),
\label{eqn:ToyModel-dotY_b}
\end{align}
\end{subequations}
where $U'[\hat{\mathcal{X}}_j]\equiv dU/d\hat{\mathcal{X}}_j$. Eliminating $\hat{\mathcal{Y}}_{j,c,b}(t)$ using Eqs.~(\ref{eqn:ToyModel-dotY_j}), (\ref{eqn:ToyModel-dotY_c}) and (\ref{eqn:ToyModel-dotY_b}) first, and integrating out the bath degree of freedom via Markov approximation \cite{Scully_Quantum_1997, Walls_Quantum_2008} we obtain effective equations for the qubit and cavity as 
\begin{subequations}
\begin{align}
\hat{\ddot{\mathcal{X}}}_j(t)+\omega_j^2\left\{\hat{\mathcal{X}}_j(t)+U'[\hat{\mathcal{X}}_j(t)]\right\}=-2g\omega_c\hat{\mathcal{X}}_c(t),
\label{eqn:ToyModel-ddotX_j}
\end{align}
\begin{align}
\begin{split}
&\hat{\ddot{\mathcal{X}}}_c(t)+2\kappa_c\hat{\dot{\mathcal{X}}}_c(t)+\omega_c^2\hat{\mathcal{X}}_c(t)\\
&=-2g\omega_j\left\{\hat{\mathcal{X}}_j(t)+U'[\hat{\mathcal{X}}_j(t)]\right\}-\hat{f}_{B}(t),
\end{split}
\label{eqn:ToyModel-ddotX_c}
\end{align}
\end{subequations}
where $2\kappa_c$ is the effective dissipation \cite{Senitzky_Dissipation_1960, Caldeira_Influence_1981, Clerk_Introduction_2010} and $\hat{f}_B(t)$ is the noise operator of the bath seen by the cavity
\begin{align}
\hat{f}_B(t)=\sum\limits_{b}2g_b\left[\omega_b\hat{\mathcal{X}}_b(0)\cos(\omega_b t)+\hat{\dot{\mathcal{X}}}_b(0)\sin(\omega_b t)\right].
\label{eqn:ToyModel-f_B(t)}
\end{align}
 
Note that Eq.~(\ref{eqn:ToyModel-ddotX_c}) is a linear non-homogoneous ODE in terms of $\hat{\mathcal{X}}_c(t)$. Therefore, it is possible to find its general solution in terms of its impulse response, i.e. the GF of the associated classical cavity oscillator:
\begin{align}
\ddot{G}_c(t,t')+2\kappa_c\dot{G}_c(t,t')+\omega_c^2G_c(t,t')=-\delta(t-t').
\label{eqn:ToyModel-Def of G(t,t')}
\end{align}
Following the Fourier transform conventions
\begin{subequations}
\begin{align}
&\tilde{G}_c(\omega)\equiv \int_{-\infty}^{\infty}dtG_c(t,t')e^{i\omega(t-t')},\\
&G_c(t,t')\equiv \int_{-\infty}^{\infty}\frac{d\omega}{2\pi}\tilde{G}_c(\omega)e^{-i\omega(t-t')},
\end{align}
\end{subequations}
we obtain an algebraic solution for $\tilde{G}_c(\omega)$ as
\begin{align}
\tilde{G}_c(\omega)=\frac{1}{(\omega-\omega_C)(\omega+\omega_C^*)},
\label{eqn:ToyModel-Sol of G(Om)}
\end{align}
with $\omega_C\equiv \nu_c-i\kappa_c$ and $\nu_c\equiv\sqrt{\omega_c^2-\kappa_c^2}$. Taking the inverse Fourier transform of Eq.~(\ref{eqn:ToyModel-Sol of G(Om)}) we find the single mode GF of the cavity oscillator
\begin{align}
G_c(t,t')=-\frac{1}{\nu_c}\sin\left[\nu_c(t-t')\right]e^{-\kappa_c(t-t')}\Theta(t-t'),
\label{eqn:ToyModel-Sol of G(t,t')}
\end{align}   
where since the poles of $\tilde{G}_c(\omega)$ reside in the lower-half of the complex $\omega$-plane, $G_c(t,t')$ is retarded (causal) and $\Theta(t)$ stands for the Heaviside step function \cite{Abramowitz_Handbook_1964}.

Then, the general solution to Eq.~(\ref{eqn:ToyModel-ddotX_c}) can be expressed in terms of $G_c(t,t')$ as \cite{Morse_Methods_1953}
\begin{align}
\begin{split}
&\hat{\mathcal{X}}_c(t)=2g\omega_j\int_{0}^{t}dt'G_c(t,t')\left\{\hat{\mathcal{X}}_j(t')+U'[\hat{\mathcal{X}}_j(t')]\right\}\\
&+\left.\left(\partial_{t'}+2\kappa_c\right)G_c(t,t')\right|_{t'=0}\hat{\mathcal{X}}_c(0)-G_c(t,0)\hat{\dot{\mathcal{X}}}_c(0)\\
&+\int_{0}^{t}dt'G_c(t,t')\hat{f}_B(t').
\end{split}
\label{eqn:ToyModel-Gen Sol of Xc}
\end{align}
Substituting Eq.~(\ref{eqn:ToyModel-Gen Sol of Xc}) into the RHS of Eq.~(\ref{eqn:ToyModel-ddotX_j}) and defining 
\begin{subequations}
\begin{align}
&\mathcal{K}(t)\equiv 4g^2\frac{\omega_c}{\omega_j}G_c(t,0),
\label{eqn:ToyModel-Def of K(t)}\\
&\mathcal{D}(t)\equiv -2g\omega_c G_c(t,0),
\label{eqn:ToyModel-Def of D(t)}\\
&\mathcal{I}(\omega)\equiv-2g\omega_c\tilde{G}_c(\omega),
\label{eqn:ToyModel-Def of I(om)}
\end{align}
\end{subequations}
we find the effective dynamics of the nonlinear oscillator in terms of $\hat{\mathcal{X}}_j(t)$ as
\begin{align}
\begin{split}
&\hat{\ddot{\mathcal{X}}}_j(t)+\omega_j^2\left\{\hat{\mathcal{X}}_j(t)+U'[\hat{\mathcal{X}}_j(t)]\right\}=\\
&-\int_{0}^{t}dt'\mathcal{K}(t-t')\omega_j^2\left\{\hat{\mathcal{X}}_j(t')+U'[\hat{\mathcal{X}}_j(t')]\right\}\\
&+\int_{0}^{t}dt'\mathcal{D}(t-t')\hat{f}_B(t')\\
&+\int_{-\infty}^{\infty}\frac{d\omega}{2\pi}
\mathcal{I}(\omega)\left[(i\omega+2\kappa_c)\hat{\mathcal{X}}_c(0)-\hat{\dot{\mathcal{X}}}_c(0)\right]e^{-i\omega t}.
\end{split}
\label{eqn:ToyModel-Eff Dyn}
\end{align}
The LHS of Eq.~(\ref{eqn:ToyModel-Eff Dyn}) is the free dynamics of the qubit. The first term on the RHS includes the memory of all past events encoded in the memory kernel $\mathcal{K}(t)$. The second term incorporates the influence of bath noise on qubit dynamics and plays the role of a drive term. Finally, the last term captures the effect of the initial operator conditions of the cavity. Note that even though Eq.~(\ref{eqn:ToyModel-Eff Dyn}) is an effective equation for the qubit, all operators act on the full Hilbert space of the qubit and the cavity.
\subsection{Linear theory}
\label{Sec:ToyModel-LinearTheory}
In the absence of the nonlinearity, i.e. $U[\hat{\mathcal{X}}_j]=0$, Eq.~(\ref{eqn:ToyModel-Eff Dyn}) is a linear integro-differential equation that can be solved exactly via unilateral Laplace transform
\begin{align}
\tilde{f}(s)\equiv \int_{0}^{\infty}dt e^{-st}f(t), 
\label{eqn:Def of Laplace}
\end{align} 
since the memory integral on the RHS appears as a convolution between the kernel $\mathcal{K}(t)$ and earlier values of $\hat{\mathcal{X}}_j(t')$ for $0<t'<t$. Employing the convolution identity
\begin{align}
\mathfrak{L}\left\{\int_{0}^{t}dt' \mathcal{K}(t-t')\hat{\mathcal{X}}_j(t')\right\}=\tilde{\mathcal{K}}(s)\hat{\tilde{\mathcal{X}}}_j(s),
\end{align}
we find that the Laplace solution to Eq.~(\ref{eqn:ToyModel-Eff Dyn}) takes the general form 
\begin{align}
\hat{\tilde{\mathcal{X}}}_j(s)=\frac{\hat{\mathcal{N}}_j(s)}{D_j(s)},
\label{eqn:ToyModel-FormSol of mathcal(X)_j}
\end{align}
where the numerator
\begin{align}
\hat{\mathcal{N}}_j(s)=s\hat{\mathcal{X}}_j(0)+\hat{\dot{\mathcal{X}}}_j(0)-\frac{2g\omega_c\left[(s+2\kappa_c)\hat{\mathcal{X}}_c(0)+\hat{\dot{\mathcal{X}}}_c(0)-\hat{\tilde{f}}_B(s)\right]}{s^2+2\kappa_c s+\omega_c^2},
\label{eqn:ToyModel-N_j(s)}
\end{align}
contains the information regarding the initial conditions and the noise operator. The characteristic function $D_j(s)$ is defined as
\begin{align}
D_j(s)\equiv s^2+\omega_j^2\left[1+\tilde{\mathcal{K}}(s)\right]=s^2+\omega_j^2-\frac{4g^2\omega_j\omega_c}{s^2+2\kappa_cs+\omega_c^2},
\label{eqn:ToyModel-D_j(s)}
\end{align}
which is the denominator of the algebraic Laplace solution~(\ref{eqn:ToyModel-FormSol of mathcal(X)_j}). Therefore, its roots determine the complex resonances of the coupled system. The poles of $D_j(s)$ are, on the other hand, the bare complex frequencies of the dissipative cavity oscillator found before, $z_c\equiv -i\omega_C$. Therefore, $D_j(s)$ can always be represented formally as
\begin{align}
D_j(s)=(s-p_j)(s-p_j^*)\frac{(s-p_c)(s-p_c^*)}{(s-z_c)(s-z_c^*)},
\label{eqn:ToyModel-Formal Rep of D(s)}
\end{align}
where $p_j$ and $p_c$ are the qubit-like and cavity-like poles such that for $g\to 0$  we get $p_j\to -i\omega_j$ and $p_c\to -i\omega_C\equiv z_c$. In writing Eq.~(\ref{eqn:ToyModel-Formal Rep of D(s)}), we have used the fact that the roots of a polynomial with real coefficients come in complex conjugate pairs.

%%%%%%%%%%%% Fig of Poles Weak Coupling%%%%%%%%%%
\begin{figure}
\centering
\subfloat[\label{subfig:ToyModelRWA}]{%
\includegraphics[scale=0.35]{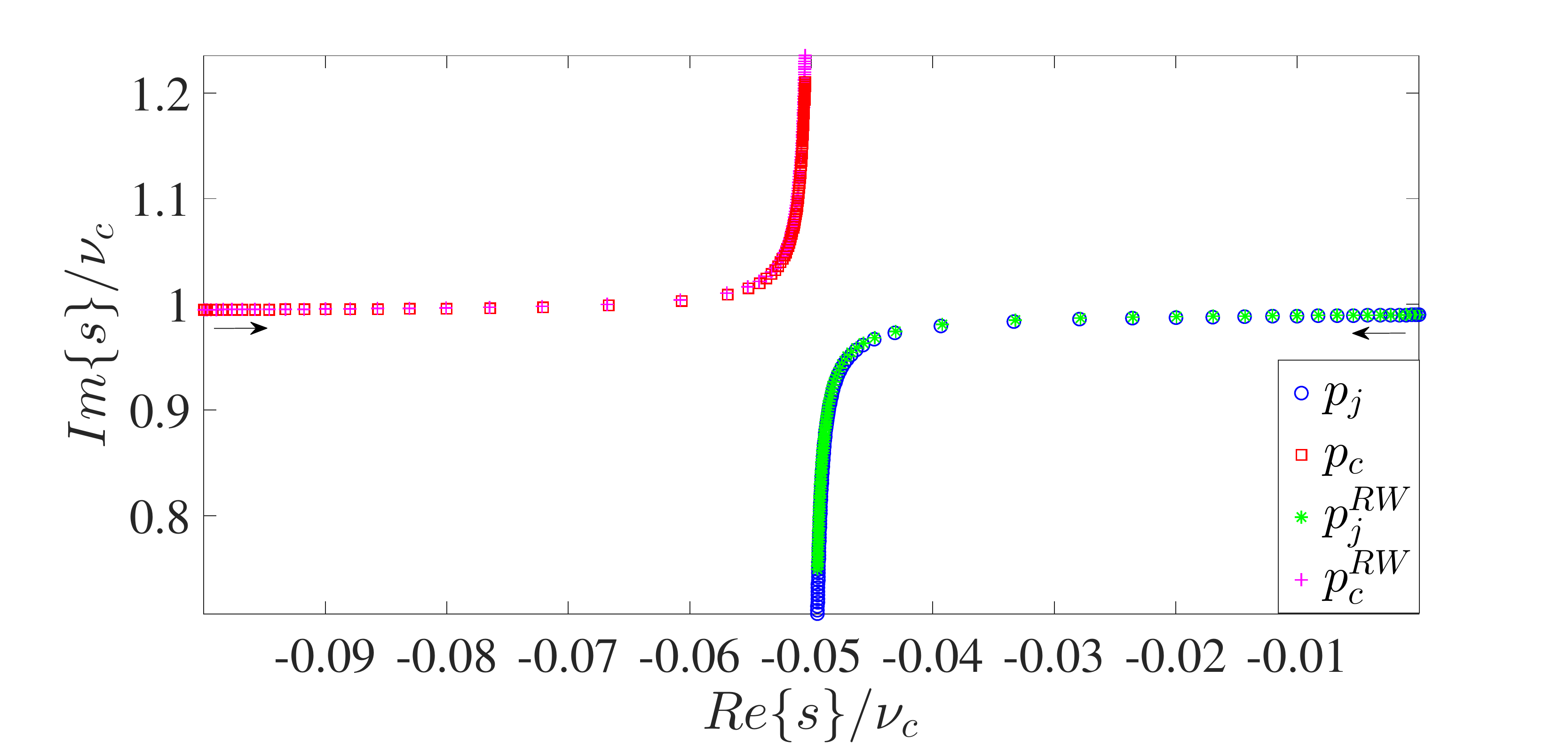}%
}\hfill 
\subfloat[\label{subfig:ToyModelRWADeltaPc}]{%
\includegraphics[scale=0.42]{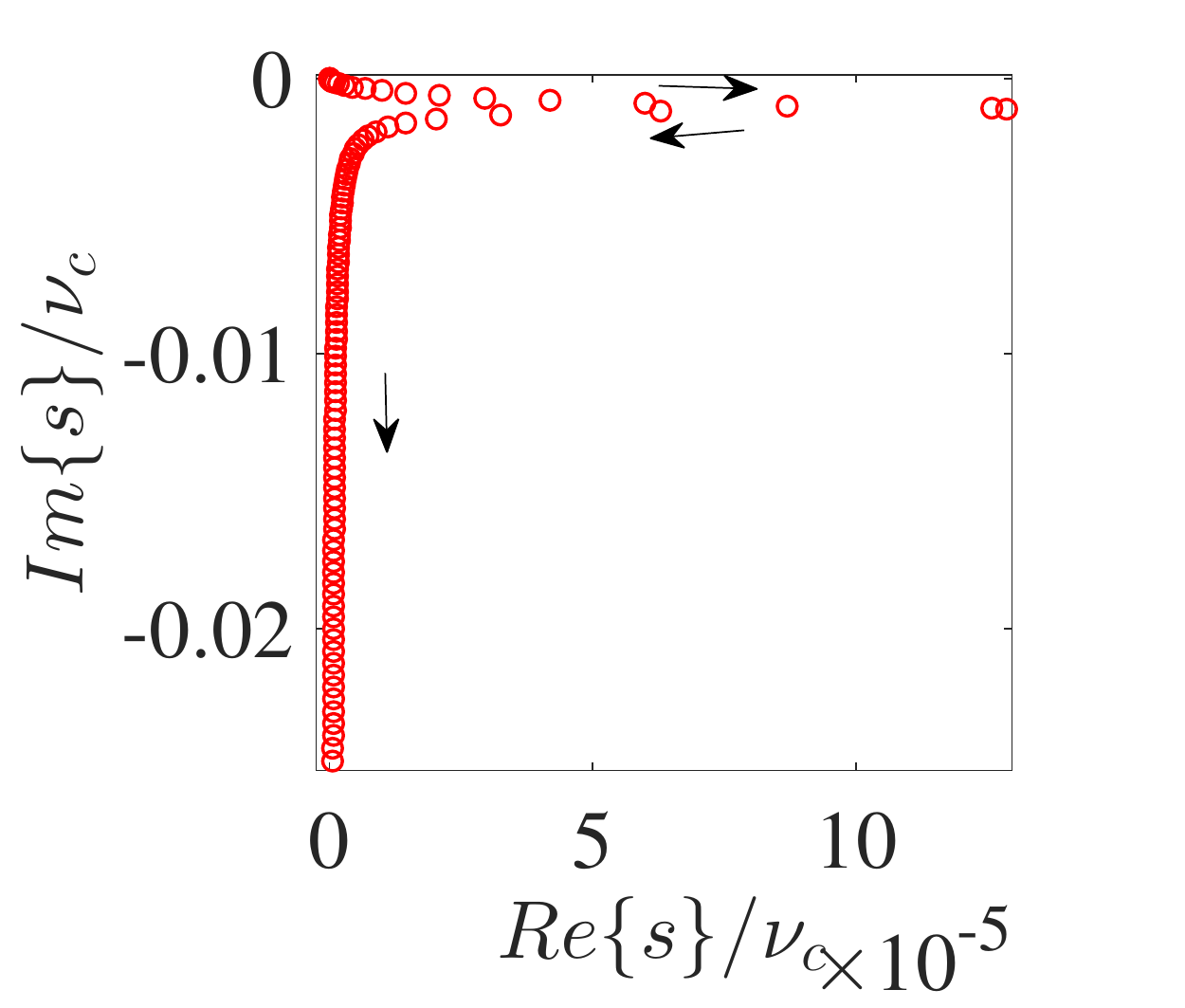}%
} 
\subfloat[\label{subfig:ToyModelRWADeltaPj}]{%
\includegraphics[scale=0.42]{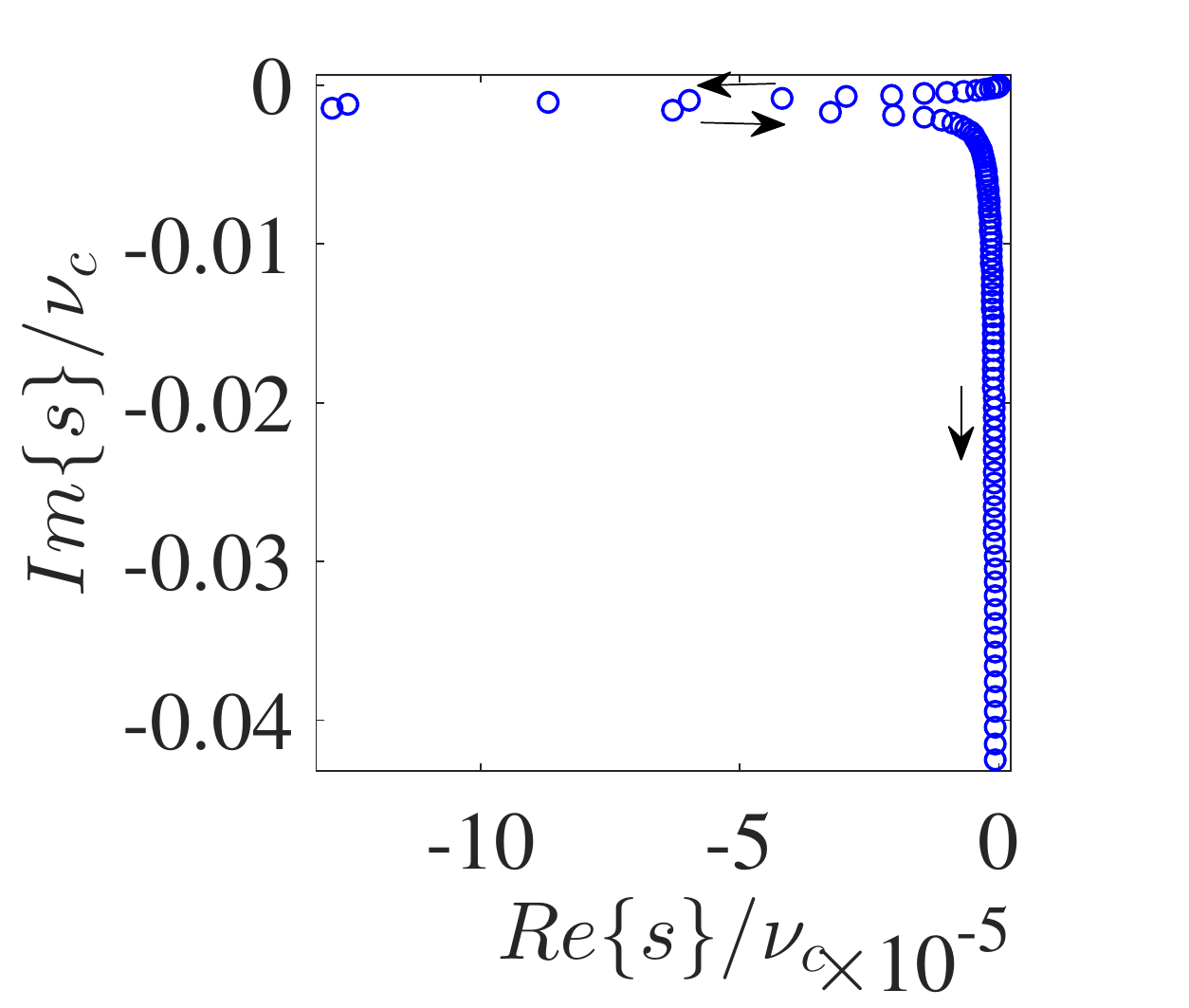}%
} 
\caption{a) Hybridized poles of the linear theory, $p_j$ and $p_c$, obtained from Eqs.~(\ref{eqn:ToyModel-D_j(s)}) and (\ref{eqn:ToyModel-RW Dq(s)}) for the resonant case  $\omega_j=\nu_c^-$, $\kappa_c=0.1\nu_c$ as a function of $g\in [0,0.5\omega_j]$ with increment $\Delta g=0.005\omega_j$. The blue circles and green stars show the qubit-like pole $p_j$ with and without RW, respectively. Similarly, the red squares and purple crosses show the cavity-like pole $p_c$. b) and c) represent the difference $\Delta p_{j,c}\equiv p_{j,c}-p_{j,c}^{RW}$ between the two solutions. The black arrows show the direction of increase in $g$.}
\label{Fig:ToyModelRWA}
\end{figure}
%%%%%%%%%%%%%%%%%%%%%%%%%%%%%%%%%%%%%%%%%%%%%%%

It is worth emphasizing that our toy model avoids the rotating wave (RW) approximation. This approximation is known to break down in the ultrastrong coupling regime \cite{Bourassa_Ultrastrong_2009, Anappara_Signatures_2009, Niemczyk_Circuit_2010, Diaz_Observation_2010, Todorov_Ultrastrong_2010}. In order to understand its consequence and make a quantitative comparison, we have to find how the RW approximation modifies $D_j(s)$. Note that by applying the RW approximation, only the coupling Hamiltonian in Eq.~(\ref{eqn:ToyModel-H}) transforms as
\begin{align}
\hat{\mathcal{Y}}_j\hat{\mathcal{Y}}_c\underset{\text{RW}}{\longrightarrow}\frac{1}{2}\left(\hat{\mathcal{X}}_j\hat{\mathcal{X}}_c+\hat{\mathcal{Y}}_j\hat{\mathcal{Y}}_c\right).
\label{eqn:ToyModel-RW Dq(s)}
\end{align}
Then, the modified equations of motion for $\hat{\mathcal{X}}_j(t)$ and $\hat{\mathcal{X}}_c(t)$ read
\begin{subequations}
\begin{align}
&\hat{\ddot{\mathcal{X}}}_j(t)+\left(\omega_j^2+g^2\right)\hat{\mathcal{X}}_j(t)=-g(\omega_j+\omega_c)\hat{\mathcal{X}}_c(t),
\label{eqn:ToyModelRWA-ddotX_j}\\
&\hat{\ddot{\mathcal{X}}}_c(t)+2\kappa_c\hat{\dot{\mathcal{X}}}_c(t)+\left(\omega_c^2+g^2\right)\hat{\mathcal{X}}_c(t)=-g(\omega_j+\omega_c)\hat{\mathcal{X}}_j(t)-\hat{f}_B(t).
\label{eqn:ToyModelRWA-ddotX_c}
\end{align}
\end{subequations}
Note that the form of Eqs.~(\ref{eqn:ToyModelRWA-ddotX_j}-\ref{eqn:ToyModelRWA-ddotX_c}) is the same as Eqs.~(\ref{eqn:ToyModel-ddotX_j}-\ref{eqn:ToyModel-ddotX_c}) except for the modified parameters. Therefore, following the same calculation as in Sec.~\ref{Sec:ToyModel-Eff Dyn} we find a new characteristic function $D_j^{RW}(s)$ which reads
\begin{align}
D_j^{RW}(s)&=s^2+\left(\omega_j^2+g^2\right)-\frac{g^2(\omega_j+\omega_c)^2}{s^2+2\kappa_cs+\left(\omega_c^2+g^2\right)}.
\label{eqn:ToyModel-D^RW(s)}
\end{align}

We compare the complex roots of $D_j(s)$ and $D_j^{RW}(s)$ in Fig.~\ref{Fig:ToyModelRWA} as a function of $g$. For $g=0$, the poles start from their bare values $i\omega_j$ and $i\nu_c-\kappa_c$ and the results with and without RW match exactly. As $g$ increases both theories predict that the dissipative cavity oscillator passes some of its decay rate to the qubit oscillator. This is seen in Fig.~\ref{subfig:ToyModelRWA} where the poles move towards each other in the $s$-plane while the oscillation frequency is almost unchanged. As $g$ is increased more, there is an avoided crossing and the poles resolve into two distinct frequencies. After this point, the predictions from $D_j(s)$ and $D_j^{RW}(s)$ for $p_j$ and $p_c$ deviate more significantly. This is more visible in Figs.~\ref{subfig:ToyModelRWADeltaPc} and \ref{subfig:ToyModelRWADeltaPj} that show the difference between the two solutions in the complex $s$-plane. In addition, there is a saturation of the decay rates to half of the bare decay rate of the dissipative cavity oscillator.

In summary, we have obtained the effective equation of motion~(\ref{eqn:ToyModel-Eff Dyn}) for the quadrature $\hat{\mathcal{X}}_j(t)$ of the nonlinear oscillator. This equation incorporates the effects of memory, initial conditions of the cavity and drive. It admits an exact solution via Laplace transform in the absence of nonlinearity. To lowest order, the Josephson nonlinearity is a time-domain perturbation $\propto \hat{\mathcal{X}}_j^3(t)$ in Eq.~(\ref{eqn:ToyModel-Eff Dyn}). This amounts to a quantum Duffing oscillator \cite{Bowen_Quantum_2015} coupled to a linear environment. Time-domain perturbation theory consists of an order by order solution of Eq.~(\ref{eqn:ToyModel-Eff Dyn}). A naive application leads to the appearance of resonant coupling between the solutions at successive orders. The resulting solution contains secular contributions, i.e. terms that grow unbounded in time. We present the resolution of this problem using multi-scale perturbation theory (MSPT) \cite{Bender_Advanced_1999, Nayfeh_Nonlinear_2008, Strogatz_Nonlinear_2014} in Sec.~\ref{Sec:PertCor}.

\section{Effective dynamics of a transmon qubit}
\label{Sec:Eff Dyn Of Transmon}
\begin{figure}
\centering
\includegraphics[scale=0.70]{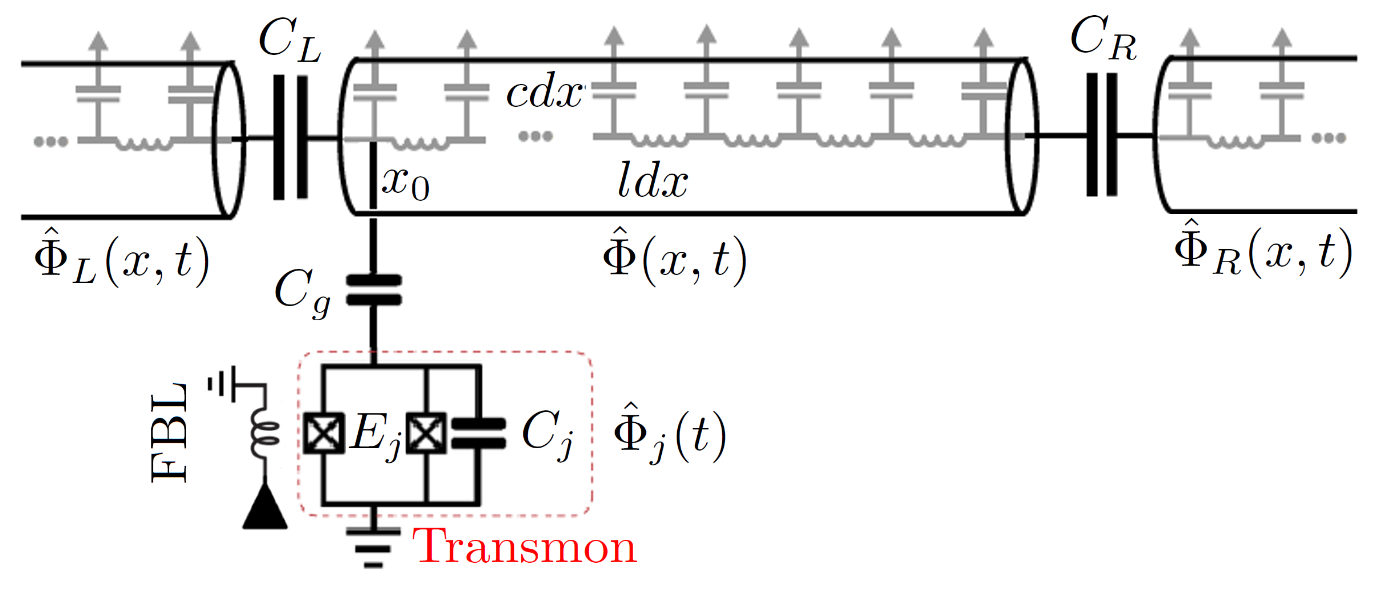}
\caption{A transmon qubit coupled to an open superconducting resonator.}
\label{Fig:cQED-open}
\end{figure}
In this section, we present a first principles calculation for the problem of a transmon qubit that couples capacitively to an open multimode resonator (see Fig.~\ref{Fig:cQED-open}). Like the toy model in Sec.~\ref{Sec:Toy Model}, this calculation relies on an effective equation of motion for the transmon qubit quadratures, in which the photonic degrees of freedom are integrated out. In contrast to the toy model where the decay rate was obtained via Markov approximation, we use a microscopic model for dissipation \cite{Senitzky_Dissipation_1960, Caldeira_Influence_1981}. We model our bath as a pair of semi-infinite waveguides capacitively coupled to each end of a resonator.

As shown in Fig.~\ref{Fig:cQED-open}, the transmon qubit is coupled to a superconducting resonator of finite length $L$ by a capacitance $C_g$. The resonator itself is coupled to the two waveguides at its ends by capacitances $C_R$ and $C_L$, respectively. For all these elements, the capacitance and inductance per length are equal and given as $c$ and $l$, correspondingly. The transmon qubit is characterized by its Josephson energy $E_j$, which is tunable by an external flux bias line (FBL) \cite{Johnson_Quantum_2010}, and its charging energy $E_c$, which is related to the capacitor $C_j$ as $E_c=e^2/(2C_j)$. The explicit circuit quantization is explained in App.~\ref{App:Quantum EOM} following a standard approach \cite{ Devoret_Quantum_1995, Clerk_Introduction_2010, Bishop_Circuit_2010, Devoret_Quantum_2014}. We describe the system in terms of flux operator $\hat{\Phi}_j(t)$ for transmon and flux fields $\hat{\Phi}(x,t)$ and $\hat{\Phi}_{R,L}(x,t)$ for the resonator and waveguides.

The dynamics for the quantum flux operators of the transmon and each resonator shown in Fig.~\ref{Fig:cQED-open} is derived in App.~\ref{App:Quantum EOM}. In what follows, we work with unitless variables 
\begin{align}
\begin{split}
\frac{x}{L}\rightarrow x,\quad
\frac{t}{\sqrt{lc}L}\rightarrow t,\quad
\sqrt{lc}L\omega\rightarrow \omega,\ 2\pi \frac{\hat{\Phi}}{\Phi_0} \rightarrow\hat{\varphi}, 
\end{split}
\label{eqn:unitless vars}
\end{align}
where $\Phi_0\equiv h/(2e)$ is the flux quantum and $1/\sqrt{lc}$ is the phase velocity. We also define unitless parameters
\begin{align}
&\chi_i\equiv\frac{C_i}{cL}, \quad i=R,L,j,g,s\\
&\mathcal{E}_{j,c}\equiv \sqrt{lc} L\frac{E_{j,c}}{\hbar}. 
\end{align}
%%%%%%% Table of variables and parameters %%%%%%%%%
\begin{table}[t!]
\centering
\begin{tabular}{c|c|c}
\textbf{Notation} & \textbf{Definition} & \textbf{Physical Meaning} \\ \hline
$\chi$ & $C/cL$ & unitless capacitance \\ \hline
$\chi_s$ & $\chi_g \chi_j/(\chi_g+\chi_j)$ & series capacitance \\ \hline
$\gamma$ & $\chi_g/(\chi_g+\chi_j)$ & capacitive ratio \\ \hline
$\chi(x,x_0)$ & $1+\chi_s\delta(x-x_0)$ & capacitance per length \\ \hline
$\mathcal{E}_{j,c}$ & $\sqrt{lc}LE_{j,c}/\hbar$ & unitless energy \\ \hline
$\omega_j$ & $\sqrt{8\mathcal{E}_c\mathcal{E}_j}$ & bare transmon frequency \\ \hline
$\epsilon$ & $\left(\mathcal{E}_c/\mathcal{E}_j\right)^{1/2}$ & nonlinearity measure \\ \hline
$\varepsilon$ & $\frac{\sqrt{2}}{6}\left(\mathcal{E}_c/\mathcal{E}_j\right)^{1/2}$ & small expansion parameter \\ \hline 
$\Phi_0$ & $h/(2e)$ & flux quantum \\ \hline
$\phi_{\text{zpf}}$ & $(2\mathcal{E}_c/\mathcal{E}_j)^{1/4}$ & zero-point fluctuation phase \\ \hline
$\hat{\Phi}(t)$ & $\int_{0}^{t}dt'\hat{V}(t)$ & flux\\ \hline
$\hat{\varphi}(t)$ & $2\pi\hat{\Phi}/\Phi_0$ & phase \\ \hline
$\hat{\phi}_j(t)$ & $\Tr_{ph}\{\hat{\rho}_{ph}(0)\hat{\varphi}_j(t)\}$ & reduced phase\\ \hline
$\hat{\mathcal{X}}(t)$ & $\hat{\varphi}(t)/\phi_{\text{zpf}}$ & unitless quadrature \\ \hline
$\hat{X}_j(t)$ & $\hat{\phi}_j(t)/\phi_{\text{zpf}}$ &  reduced unitless quadrature
\end{tabular}
\caption{Summary of definitions for some parameters and variables. Operators are denoted by a hat notation.}
\label{Tab:Def of Pars&Vars}
\end{table}
%%%%%%%%%%%%%%%%%%%%%%%%%%%%%%%%%%%%%%%%%%%%%%%%%%%
The Heisenberg equation of motion for the transmon reads
\begin{align}
\hat{\ddot{\varphi}}_j(t)+(1-\gamma)\omega_j^2\sin{[\hat{\varphi}_j(t)]}=\gamma \partial_{t}^2\hat{\varphi}(x_0,t),
\label{eqn:Transmon Dyn}
\end{align}
where $\gamma\equiv\chi_g/(\chi_g+\chi_j)$ is a capacitive  ratio, $\omega_j\equiv\sqrt{8\mathcal{E}_c\mathcal{E}_j}$ is the unitless bare transmon frequency and $x_0$ is the location of transmon. The phase field $\hat{\varphi}(x,t)$ of the resonator satisfies an inhomogeneous wave equation 
\begin{align}
\left[\partial_{x}^2-\chi(x,x_0)\partial_{t}^2\right]\hat{\varphi}(x,t)=\chi_s\omega_j^2 \sin{[\hat{\varphi}_j(t)]}\delta(x-x_0),
\label{eqn:Res Dyn}
\end{align}
where $\chi(x,x_0)=1+\chi_s\delta(x-x_0)$ is the unitless capacitance per unit length modified due to coupling to the transmon qubit, and $\chi_s\equiv\chi_g\chi_j/(\chi_g+\chi_j)$ is the unitless series capacitance of $C_j$ and $C_g$. The effect of a nonzero $\chi_s$ reflects the modification of the cavity modes due to the action of the transmon as a classical scatterer \cite{Malekakhlagh_Origin_2016}. We note that this modification is distinct from, and in addition to, the modification of the cavity modes due to the linear part of the transmon potential discussed in \cite{Nigg_BlackBox_2012}. Table~\ref{Tab:Def of Pars&Vars} lists the unitless variables and parameters used in the remainder of this paper.

The flux field in each waveguide obeys a homogeneous wave equation    
\begin{align}
\left(\partial_{x}^2-\partial_{t}^2\right)\hat{\varphi}_{R,L}(x,t)=0.
\label{eqn:Side Res Dyn}
\end{align}
The boundary conditions (BC) are derived from conservation of current at each end of the resonator as
\begin{subequations}
\begin{align}
-\left.\partial_{x}\hat{\varphi}\right|_{x=1^-}=-\left.\partial_{x}\hat{\varphi}_R\right|_{x=1^+}=\chi_R\partial_{t}^2\left[\hat{\varphi}(1^-,t)-\hat{\varphi}_R(1^+,t)\right],
\label{eqn:BC-Cons of current at 1}\\
-\left.\partial_{x}\hat{\varphi}\right|_{x=0^+}=-\left.\partial_{x}\hat{\varphi}_L\right|_{x=0^-}=\chi_L\partial_{t}^2\left[\hat{\varphi}_L(0^-,t)-\hat{\varphi}(0^+,t)\right].
\label{eqn:BC-Cons of current at 0}
\end{align}
\end{subequations}

Equations~(\ref{eqn:Transmon Dyn}-\ref{eqn:BC-Cons of current at 0}) completely describe the dynamics of a transmon qubit coupled to an open resonator. Note that according to Eq.~(\ref{eqn:Transmon Dyn}) the bare dynamics of the transmon is modified due to the force term $\gamma\partial_t^2\hat{\varphi}(x_0,t)$. Therefore, in order to find the effective dynamics for the transmon, we need to solve for $\hat{\varphi}(x,t)$ first and evaluate it at the point of connection $x=x_0$. This can be done using the {\it classical} electromagnetic GF by virtue of the homogeneous part of Eqs.~(\ref{eqn:Res Dyn},\ref{eqn:Side Res Dyn}) being linear in the quantum fields (see App.~\ref{SubApp:Def of G}). Substituting it into the LHS of Eq.~(\ref{eqn:Transmon Dyn}) and further simplifying leads to the effective dynamics for the transmon phase operator
\begin{align}
\begin{split}
&\hat{\ddot{\varphi}}_j(t)+(1-\gamma)\omega_j^2\sin{\left[\hat{\varphi}_j(t)\right]}=\\
+&\frac{d^2}{dt^2}\int_{0}^{t}dt'\mathcal{K}_0(t-t')\omega_j^2\sin{\left[\hat{\varphi}_j(t')\right]}\\
+&\int_{-\infty}^{+\infty}\frac{d\omega}{2\pi}\mathcal{D}_R(\omega)\hat{\tilde{\varphi}}_R^{inc}(1^+,\omega)e^{-i\omega t}\\
+&\int_{-\infty}^{+\infty}\frac{d\omega}{2\pi}\mathcal{D}_L(\omega)\hat{\tilde{\varphi}}_L^{inc}(0^-,\omega)e^{-i\omega t}\\
+&\int_{0^-}^{1^+}dx'\int_{-\infty}^{+\infty}\frac{d\omega}{2\pi}\mathcal{I}(x',\omega)\left[i\omega\hat{\varphi}(x',0)-\hat{\dot{\varphi}}(x',0)\right]e^{-i\omega t}.
\end{split}
\label{eqn:Eff Dyn before trace}
\end{align}
The electromagnetic GF is the basic object that appears in the various kernels constituting the above integro-differential equation:
\begin{subequations}
\begin{align}
&\mathcal{K}_n(\tau)\equiv\gamma\chi_s\int_{-\infty}^{+\infty} \frac{d\omega}{2\pi} \omega^n\tilde{G}(x_0,x_0,\omega)e^{-i\omega\tau},
\label{eqn:Def of K_n(tau)}\\
&\mathcal{D}_R(\omega)\equiv -2i\gamma\omega^3\tilde{G}(x_0,1^+,\omega),
\label{eqn:Def of D_R(om)}\\
&\mathcal{D}_L(\omega)\equiv -2i\gamma\omega^3\tilde{G}(x_0,0^-,\omega),
\label{eqn:Def of D_L(om)}\\
&\mathcal{I}(x',\omega)\equiv \gamma\omega^2\chi(x',x_0)\tilde{G}(x_0,x',\omega).
\label{eqn:Def of I(x',om)}
\end{align}
\end{subequations}
Equation~(\ref{eqn:Eff Dyn before trace}) fully describes the effective dynamics of the transmon phase operator. The various terms appearing in this equation have transparent physical interpretation. The first integral on the RHS of Eq.~(\ref{eqn:Eff Dyn before trace}) represents the retarded self-interaction of the qubit. It contains the GF in the form $\tilde{G}(x_0,x_0,\omega)$ and describes all processes in which the electromagnetic radiation is emitted from the transmon at $x=x_0$ and is scattered back again. We will see later on that this term is chiefly responsible for the spontaneous emission of the qubit. The boundary terms include only the incoming part of the waveguide phase fields. They describe the action of the electromagnetic fluctuations in the waveguides on the qubit, as described by the propagators from cavity interfaces to the qubit, $\tilde{G}(x_0,0^-,\omega)$ and $\tilde{G}(x_0,1^+,\omega)$. The phase fields $\hat{\varphi}_{L}(0^-,t)$ and $\hat{\varphi}_{R}(1^+,t)$ may contain a classical (coherent) part as well. Finally, the last integral adds up all contributions of a nonzero initial value for the electromagnetic field inside the resonator that propagates from the point $0<x'<1$ to the position of transmon $x_0$. 

The solution to the effective dynamics (\ref{eqn:Eff Dyn before trace}) requires knowledge of $\tilde{G}(x,x',\omega)$. To this end, we employ the spectral representation of the GF in terms of a set of constant flux (CF) modes \cite{Tureci_SelfConsistent_2006, Tureci_Strong_2008} 
\begin{align}
\tilde{G}(x,x',\omega)=\sum\limits_{n}\frac{\tilde{\varphi}_n(x,\omega)\bar{\tilde{\varphi}}_n^*(x',\omega)}{\omega^2-\omega_n^2(\omega)},
\label{eqn:Spec rep of G-Open}
\end{align}
where $\tilde{\varphi}_n(x,\omega)$ and $\bar{\tilde{\varphi}}_n(x,\omega)$ are the right and left eigenfunctions of the Helmholtz eigenvalue problem with outgoing BC and hence carry a constant flux when $x\to\pm \infty$. Note that in this representation, both the CF frequencies $\omega_n(\omega)$ and the CF modes $\tilde{\varphi}_n(x,\omega)$ parametrically depend on the source frequency $\omega$. The expressions for $\omega_n(\omega)$ and $\tilde{\varphi}_n(x,\omega)$ are given in App.~\ref{SubApp:Spec Rep of G-open}.  

\begin{figure}
\centering
\subfloat[\label{subfig:NHEigFreqsXrXl1Em5}]{%
\includegraphics[scale=0.45]{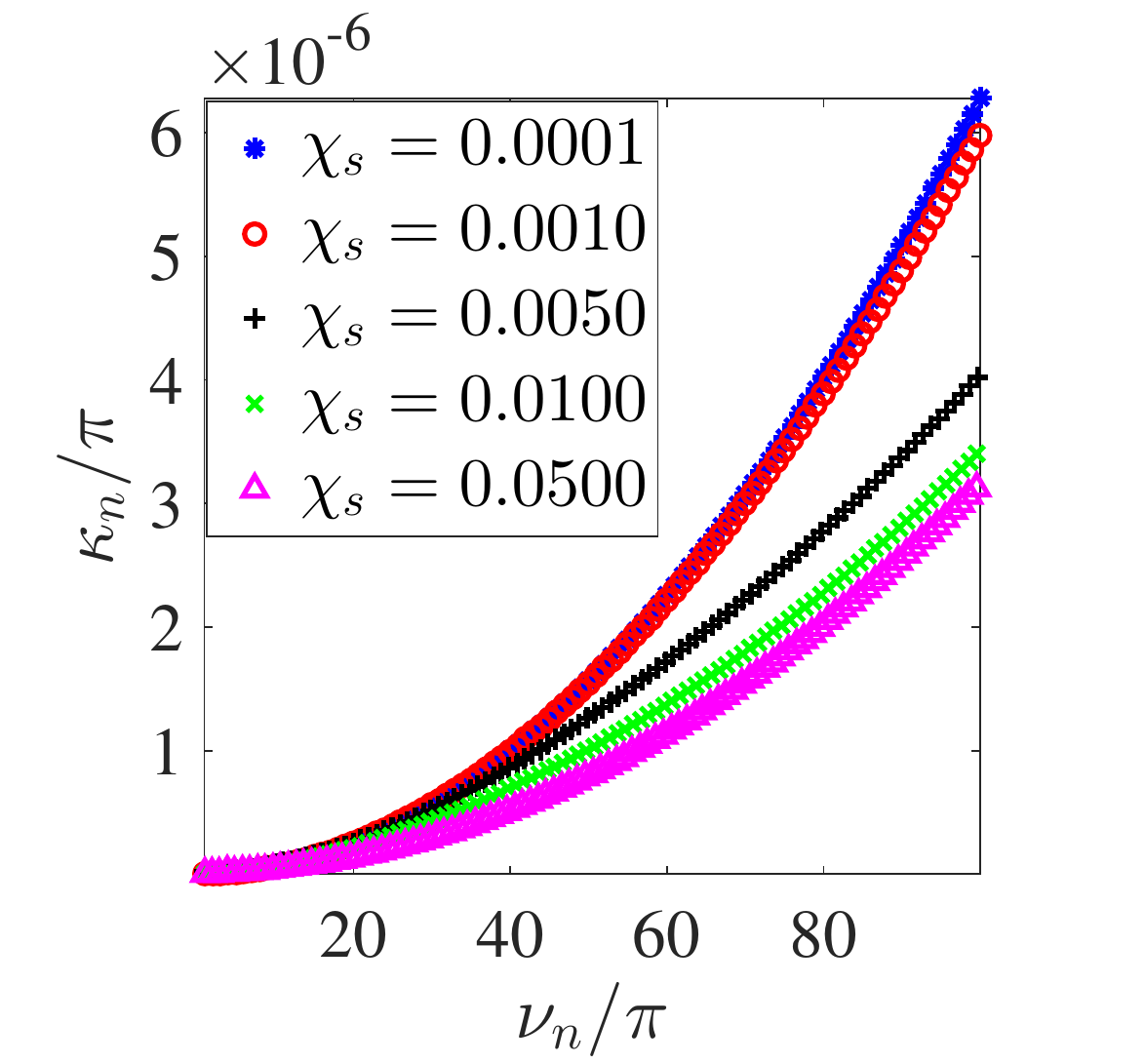}%
}
\subfloat[\label{subfig:NHEigFreqsXrXl1Em3}]{%
\includegraphics[scale=0.45]{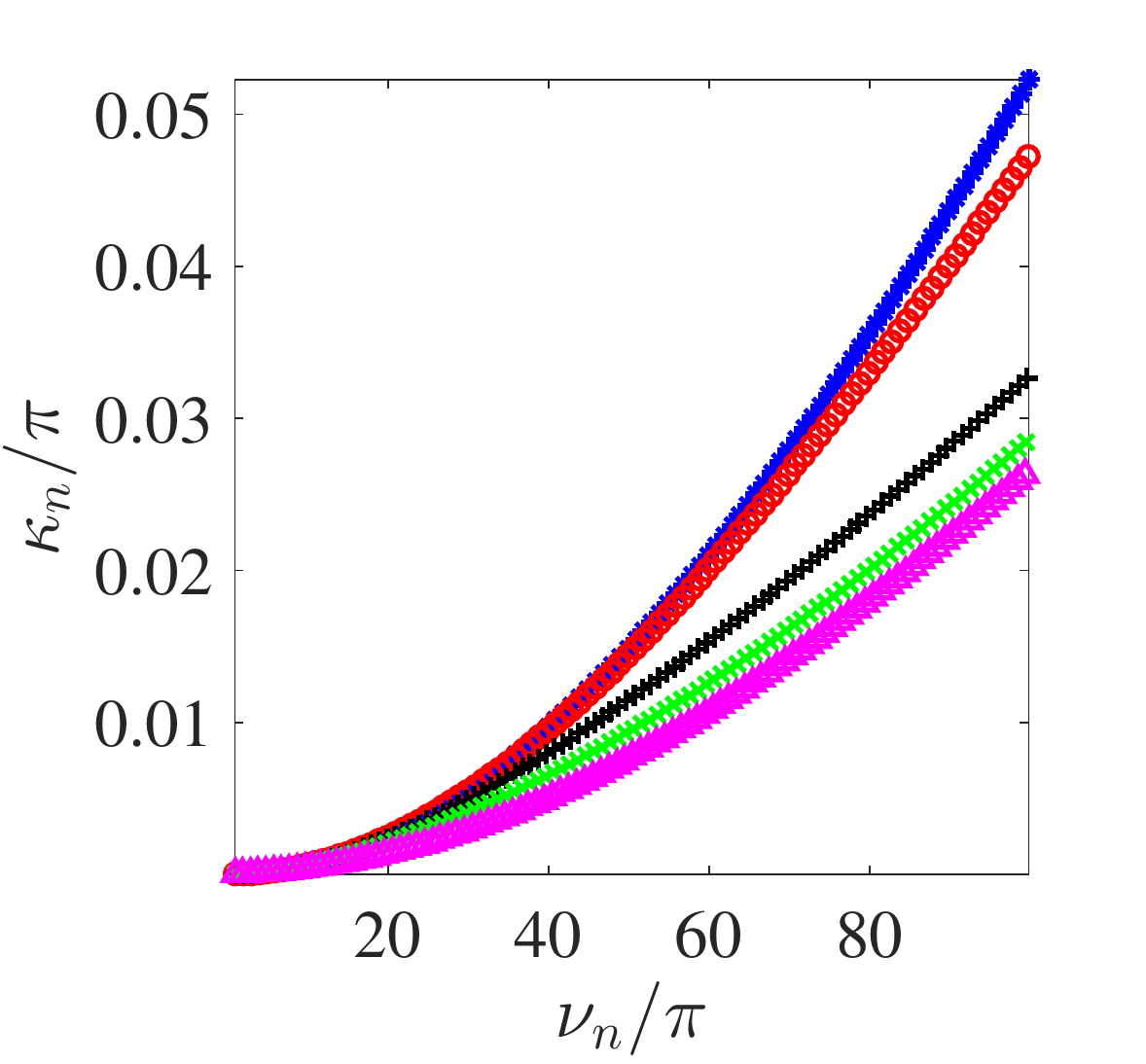}%
}\\
\subfloat[\label{subfig:NHEigFreqsXrXl1Em2}]{%
\includegraphics[scale=0.45]{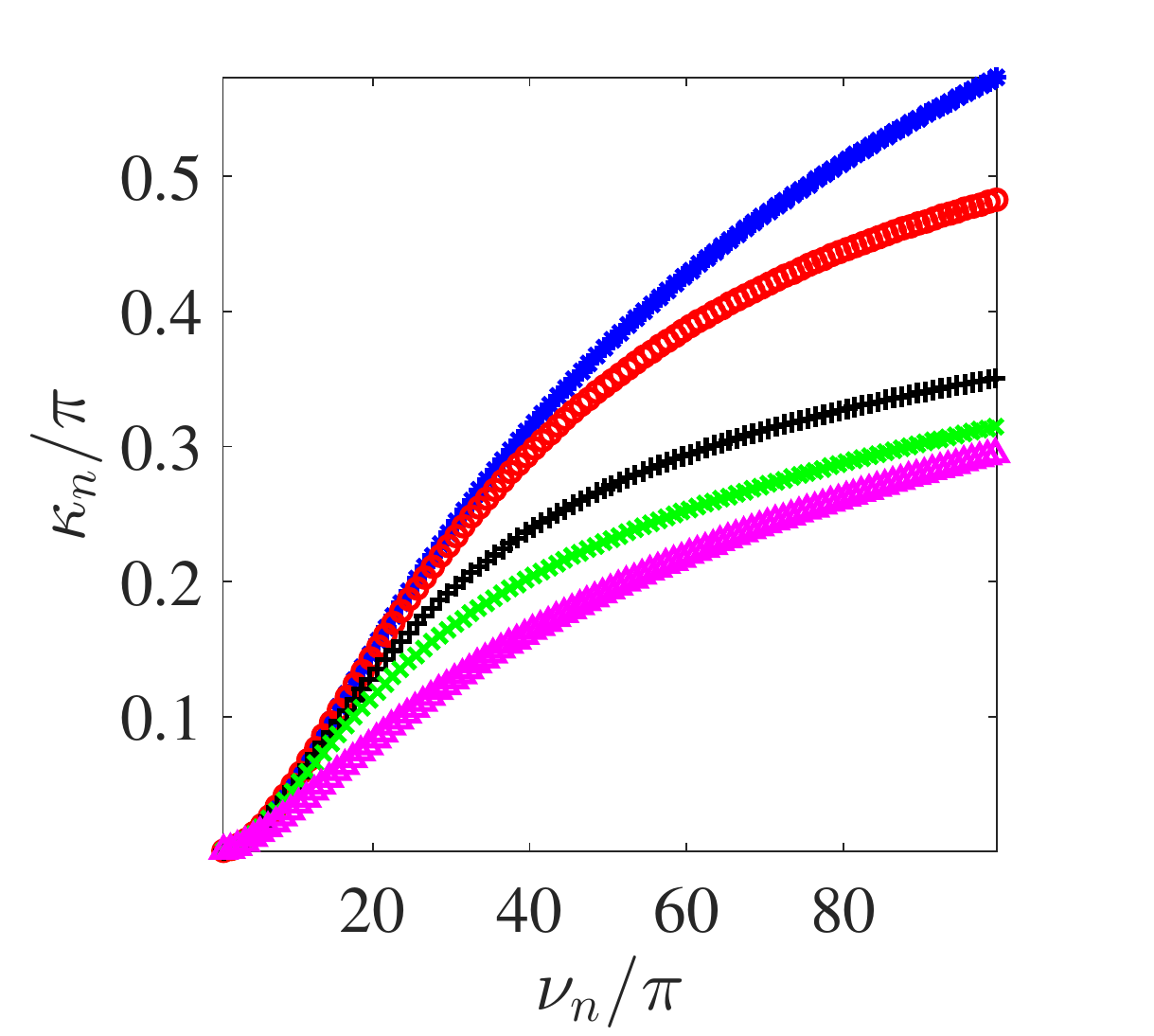}%
}
\subfloat[\label{subfig:NHEigFreqsXrXl1Em1}]{%
\includegraphics[scale=0.45]{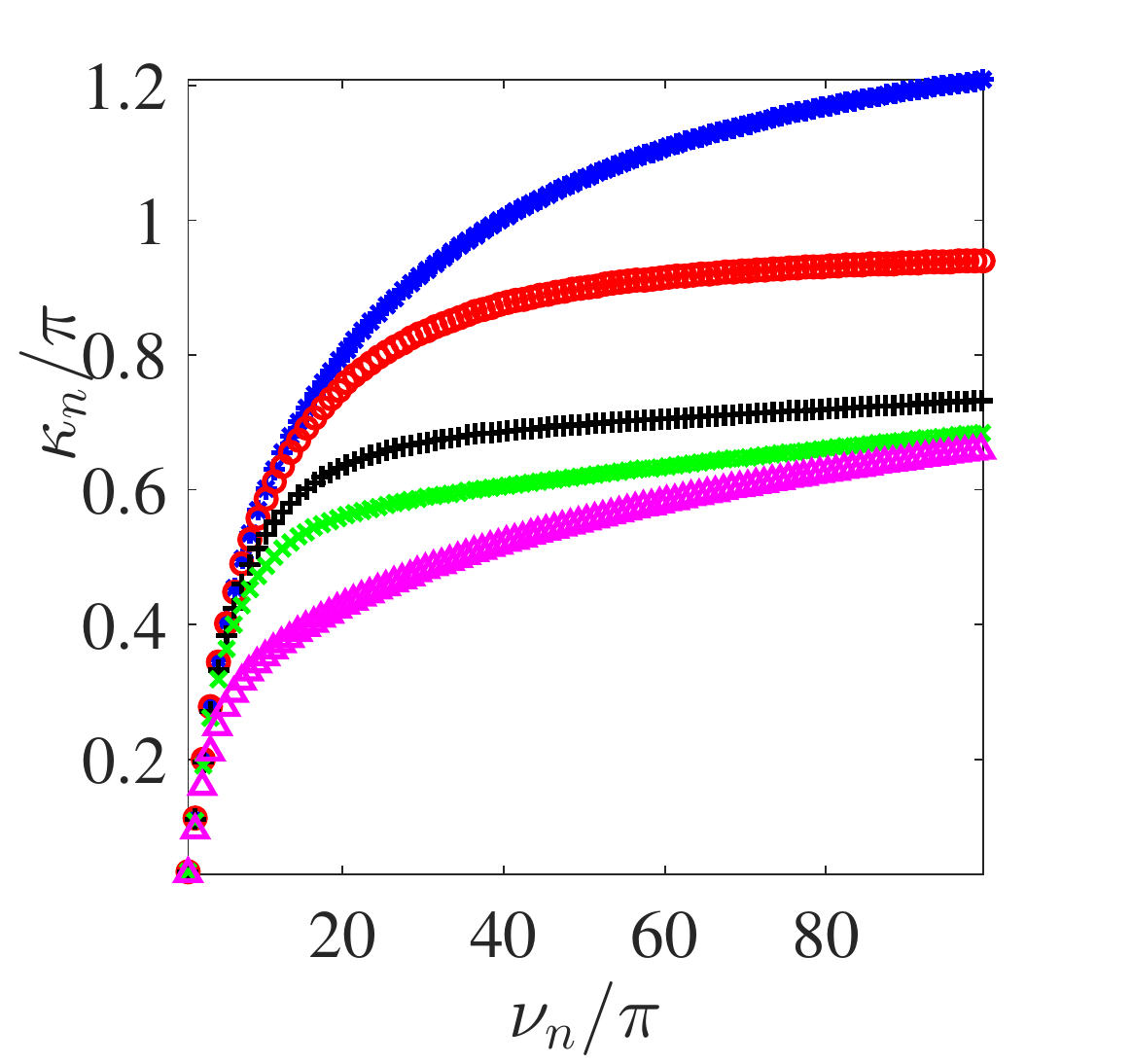}%
}
\caption{Decay rate $\kappa_n$ versus oscillation frequency $\nu_n$ for the first 100 non-Hermitian modes for $x_0=0$ and different values of $\chi_s$. a) $\chi_R=\chi_L=10^{-5}$, b) $\chi_R=\chi_L=10^{-3}$, c) $\chi_R=\chi_L=10^{-2}$ and d) $\chi_R=\chi_L=10^{-1}$.} 
\label{Fig:NHEigFreqs}
\end{figure}

The poles of the GF are the solutions to $\omega=\omega_n(\omega)$ that satisfy the transcendental equation
\begin{align}
\begin{split}
&\left[e^{2i\omega_n}-(1-2i\chi_L\omega_n)(1-2i\chi_R\omega_n)\right]\\
&+\frac{i}{2}\chi_s\omega_n[e^{2i\omega_n x_0}+(1-2i\chi_L\omega_n)]\\
&\times[e^{2i\omega_n (1-x_0)}+(1-2i\chi_R\omega_n)]=0.
\end{split}
\label{eqn:Generic NHEigfreq}
\end{align}
The solutions to Eq.~(\ref{eqn:Generic NHEigfreq}) all reside in the lower half of $\omega$-plane resulting in a finite lifetime for each mode that is characterized by the imaginary part of $\omega_n\equiv \nu_n-i\kappa_n$.  In Fig.~\ref{Fig:NHEigFreqs} we plotted the decay rate $\kappa_n$ versus the oscillation frequency $\nu_n$ of the first 100 modes for $x_0=0$ and different values of $\chi_R=\chi_L$ and $\chi_s$. There is a transition from a super-linear \cite{Houck_Controlling_2008} dependence on mode number for smaller opening to a sub-linear dependence for larger openings. Furthermore, increasing $\chi_s$ always decreases the decay rate $\kappa_n$. Intuitively, $\chi_s$ is the strength of a $\delta$-function step in the susceptibility at the position of the transmon. An increase in the average refractive index inside the resonator generally tends to redshift the cavity resonances, while decreasing their decay rate.

%%%%%%%%%%% Fig of NHEigFreqs %%%%%%%%%%%%%%%%%%%
%%%%%%%%%%%%%%%%%%%%%%%%%%%%%%%%%%%%%%%%%%%%

In summary, we have derived an effective equation of motion, Eq.~(\ref{eqn:Eff Dyn before trace}), for the transmon qubit flux operator $\hat{\varphi}_j$, in which the resonator degrees of freedom enter via the   electromagnetic GF $\tilde{G}(x,x',\omega)$ given in Eq.~(\ref{eqn:Spec rep of G-Open}). 
%%%%%%%%%%%%%%%%%%%%%%%%%%%%%%%%%%%%%%%%%%%%
\section{Spontaneous emission into a leaky resonator}
In this section, we revisit the problem of spontaneous emission \cite{Purcell_Resonance_1946, Kleppner_Inhibited_1981, Goy_Observation_1983, Hulet_Inhibited_1985, Jhe_Suppression_1987, Dung_Spontaneous_2000, Houck_Controlling_2008, Krimer_Route_2014}, where the system starts from the initial density matrix
\begin{align}
\hat{\rho}(0)=\hat{\rho}_j(0)\otimes\ket{0}_{ph}\bra{0}_{ph},
\label{eqn:SE-IC}
\end{align} 
such that the initial excitation exists in the transmon sector of Hilbert space with zero photons in the resonator and waveguides. $\hat{\rho}_j(0)$ is a general density matrix in the qubit subspace. For our numerical simulation of the spontaneous emission dynamics in terms of quadratures, we will consider $\hat{\rho}_j(0)=\ket{\Psi_j(0)}\bra{\Psi_j(0)}$ with $\ket{\Psi_j(0)}=(\ket{0}_j+\ket{1}_j)/\sqrt{2}$. The spontaneous emission was conventionally studied through the Markov approximation of the memory term which results only in a modification of the qubit-like pole. This is the Purcell modified spontaneous decay where, depending on the density of the states of the environment, the emission rate can be suppressed or enhanced \cite{Purcell_Resonance_1946, Kleppner_Inhibited_1981, Goy_Observation_1983, Hulet_Inhibited_1985, Jhe_Suppression_1987}. We extract the spontaneous decay as the real part of transmon-like pole in a full {\it multimode} calculation that is accurate for any qubit-resonator coupling strength.

A product initial density matrix like Eq.~(\ref{eqn:SE-IC}) allows us to reduce the generic dynamics significantly, since the expectation value of any operator $\hat{\mathcal{O}}(t)$ can be expressed as
\begin{align}
\Tr_{j}\Tr_{ph}\left\{\hat{\rho}_j(0)\otimes\hat{\rho}_{ph}(0)\hat{\mathcal{O}}(t)\right\}=\Tr_j\left\{\hat{\rho}_j(0)\hat{O}(t)\right\}
\end{align}
where $\hat{O}\equiv\Tr_{ph}\{\hat{\mathcal{O}}\}$ is the reduced operator in the Hilbert space of the transmon. Therefore, we define a reduced phase operator
\begin{align}
\hat{\phi}_j(t)\equiv \Tr_{ph}\{\hat{\rho}_{ph}(0)\hat{\varphi}_j(t)\}.
\label{eqn:Def of phi_j(t)}
\end{align}
In the absence of an external drive, the generic effective dynamics in Eq.~(\ref{eqn:Eff Dyn before trace}) reduces to
\begin{align}
\begin{split}
&\hat{\ddot{\phi}}_j(t)+\omega_j^2\left[1-\gamma+i\mathcal{K}_1(0)\right]\Tr_{ph}\left\{\hat{\rho}_{ph}(0)\sin{\left[\hat{\varphi}_j(t)\right]}\right\}\\
&=-\int_0^{t}dt'\mathcal{K}_2(t-t')\omega_j^2\Tr_{ph}\left\{\hat{\rho}_{ph}(0)\sin{\left[\hat{\varphi}_j(t')\right]}\right\}.
\label{eqn:NL SE Problem}
\end{split}
\end{align}
The derivation of Eq.~(\ref{eqn:NL SE Problem}) can be found in Apps.~\ref{SubApp:SE Eff Dyn} and \ref{SubApp:Spec Rep of K}.

Note that, due to the sine nonlinearity, Eq.~(\ref{eqn:NL SE Problem}) is not closed in terms of $\hat{\phi}_j(t)$. However, in the transmon regime \cite{Koch_Charge_2007}, where $\mathcal{E}_j \gg \mathcal{E}_c$, the nonlinearity in the spectrum of transmon is weak. This becomes apparent when we work with the unitless quadratures 
\begin{align}
&\hat{X}_j(t)\equiv\frac{\hat{\phi}_j(t)}{\phi_{\text{zpf}}},
\quad \hat{\mathcal{X}}_j(t)\equiv \frac{\hat{\varphi}_j(t)}{\phi_{\text{zpf}}},
\label{eqn:Def of mathcalX_j(t)}
\end{align}  
where $\phi_{\text{zpf}}\equiv (2\mathcal{E}_c/\mathcal{E}_j)^{1/4}$ is the zero-point fluctuation (zpf) phase amplitude. Then, we can expand the nonlinearity in both sides of Eq.~(\ref{eqn:NL SE Problem}) as
\begin{align}
\begin{split}
\frac{\sin{\left[\hat{\varphi}_j(t)\right]}}{\phi_{\text{zpf}}}&=\frac{\hat{\varphi}_j(t)}{\phi_{\text{zpf}}}
-\frac{\hat{\varphi}_j^3(t)}{3!\phi_{\text{zpf}}}+\mathcal{O}\left[\frac{\hat{\varphi}_j^5(t)}{\phi_{\text{zpf}}}\right]\\
&=\hat{\mathcal{X}}_j(t)-\frac{\sqrt{2}\epsilon}{6}\hat{\mathcal{X}}_j^3(t)+\mathcal{O}\left(\epsilon^2\right),
\end{split}
\label{eqn:Expansion of Sine}
\end{align}
where $\epsilon\equiv(\mathcal{E}_c/\mathcal{E}_j)^{1/2}$ appears as a measure for the strength of the nonlinearity. In experiment, the Josephson energy $\mathcal{E}_j$ can be tuned through the FBL while the charging energy $\mathcal{E}_c$ is fixed. Therefore, a higher transmon frequency $\omega_j=\sqrt{8\mathcal{E}_c\mathcal{E}_j}$ is generally associated with a smaller $\epsilon$ and hence weaker nonlinearity. 

The remainder of this section is organized as follows. In Sec.~\ref{Sec:Lin SE Theory} we study the linear theory. In Sec.~\ref{Sec:PertCor} we develop a perturbation expansion up to leading order in $\epsilon$. In Sec.~\ref{Sec:NumSimul}, we compare our analytical results with numerical simulation. Finally, in Sec.~\ref{Sec:SysOutput} we discuss the output response of the cQED system that can be probed in experiment. 
\subsection{Linear theory}
\label{Sec:Lin SE Theory}
In this subsection, we solve the linear effective dynamics and discuss hybridization of the transmon and the resonator resonances. We emphasize the importance of off-resonant modes as the coupling $\chi_g$ is increased. We next investigate the spontaneous decay rate as a function of transmon frequency $\omega_j$ and coupling $\chi_g$ and find an asymmetric dependence on $\omega_j$ in agreement with a previous experiment \cite{Houck_Controlling_2008}.

Neglecting the cubic term in Eq.~(\ref{eqn:Expansion of Sine}), the partial trace with respect to the resonator modes can be taken directly and we obtain the effective dynamics 
\begin{align}
\hat{\ddot{X}}_j(t)+\omega_j^2\left[1-\gamma+i\mathcal{K}_1(0)\right]\hat{X}_j(t)=-\int_0^{t}dt'\mathcal{K}_2(t-t')\omega_j^2\hat{X}_j(t').
\label{eqn:Lin SE Problem}
\end{align}
Then, using Laplace transform we can solve Eq.~(\ref{eqn:Lin SE Problem}) as
\begin{align}
\hat{\tilde{X}}_j(s)=\frac{s\hat{X}_j(0)+\omega_j\hat{Y}_j(0)}{D_j(s)},
\label{eqn:Sol of X_j(s)}
\end{align}
with $D_j(s)$ defined as
\begin{align}
D_j(s)\equiv s^2+\omega_j^2\left[1-\gamma+i\mathcal{K}_1(0)+\tilde{\mathcal{K}}_2(s)\right].
\label{eqn:Def of D(s)}
\end{align}
Equations~(\ref{eqn:Sol of X_j(s)}) and (\ref{eqn:Def of D(s)}) contain the solution for the reduced quadrature operator of the transmon qubit in the Laplace domain. 

In order to find the time domain solution, it is necessary to study the poles of Eq.~(\ref{eqn:Sol of X_j(s)}) and consequently the roots of $D_j(s)$. The characteristic function $D_j(s)$ can be expressed as (see App.~\ref{App:Char func D(s)}) 
\begin{align}
\begin{split}
D_j(s)=s^2+\omega_j^2+\omega_j^2\left\{-\gamma+\sum\limits_{n}M_n\frac{s\{\cos{[2\delta_{n}(x_0)]}s+\sin{[2\delta_{n}(x_0)]}\nu_n\}}{(s+\kappa_n)^2+\nu_n^2}\right\},
\label{eqn:simplified D(s)}
\end{split}
\end{align}
where $\delta_n(x)$ is the phase of the non-Hermitian eigenfunction such that $\tilde{\Phi}_n(x)=|\tilde{\Phi}_n(x)|e^{i\delta_n(x)}$. We identify the term
\begin{align}
M_n\equiv\gamma\chi_s|\tilde{\varphi}_n(x_0)|^2
\label{eqn:Def of Mn}
\end{align}
as the measure of hybridization with individual resonator modes. The form of $M_n$ in Eq.~(\ref{eqn:Def of Mn}) illustrates that the hybridization between the transmon and the resonator is bounded. This strength of hybridization is parameterized by $\gamma\chi_s$ rather than $\chi_g$. This implies that as $\chi_g$, the coupling capacitance, is increased, the qubit-resonator hybridization is limited by the internal capacitance of the qubit, $\chi_j$:
\begin{align}
\lim\limits_{\frac{\chi_g}{\chi_j}\to\infty}\gamma\chi_s=\lim\limits_{\frac{\chi_g}{\chi_j}\to\infty}\left(\frac{\chi_g}{\chi_g+\chi_j}\right)^2\chi_j=\chi_j.
\label{eqn:LinTheory-Asymptote}
\end{align}
For this reason, our numerical results below feature a saturation in hybridization as $\chi_g$ is increased.  
%%%%%%%%% Fig of Poles Weak Coupling %%%%%%%%
\begin{figure}
\centering
\subfloat[\label{subfig:Poles5ModeXrXl1Em2Weak}]{%
\includegraphics[scale=0.50]{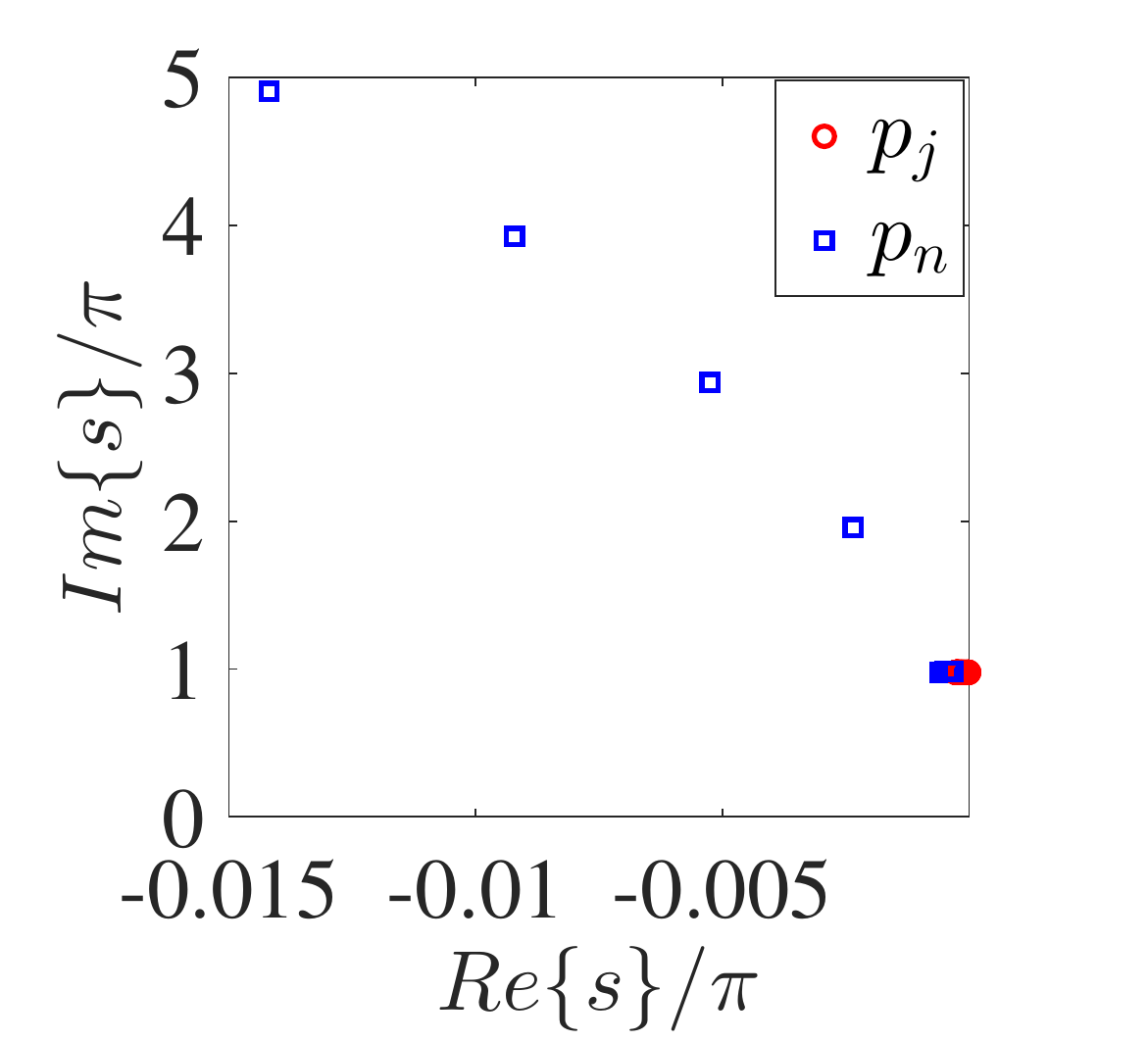}%
}
\subfloat[\label{subfig:Poles5ModeXrXl1Em2WeakZoomed}]{%
\includegraphics[scale=0.50]{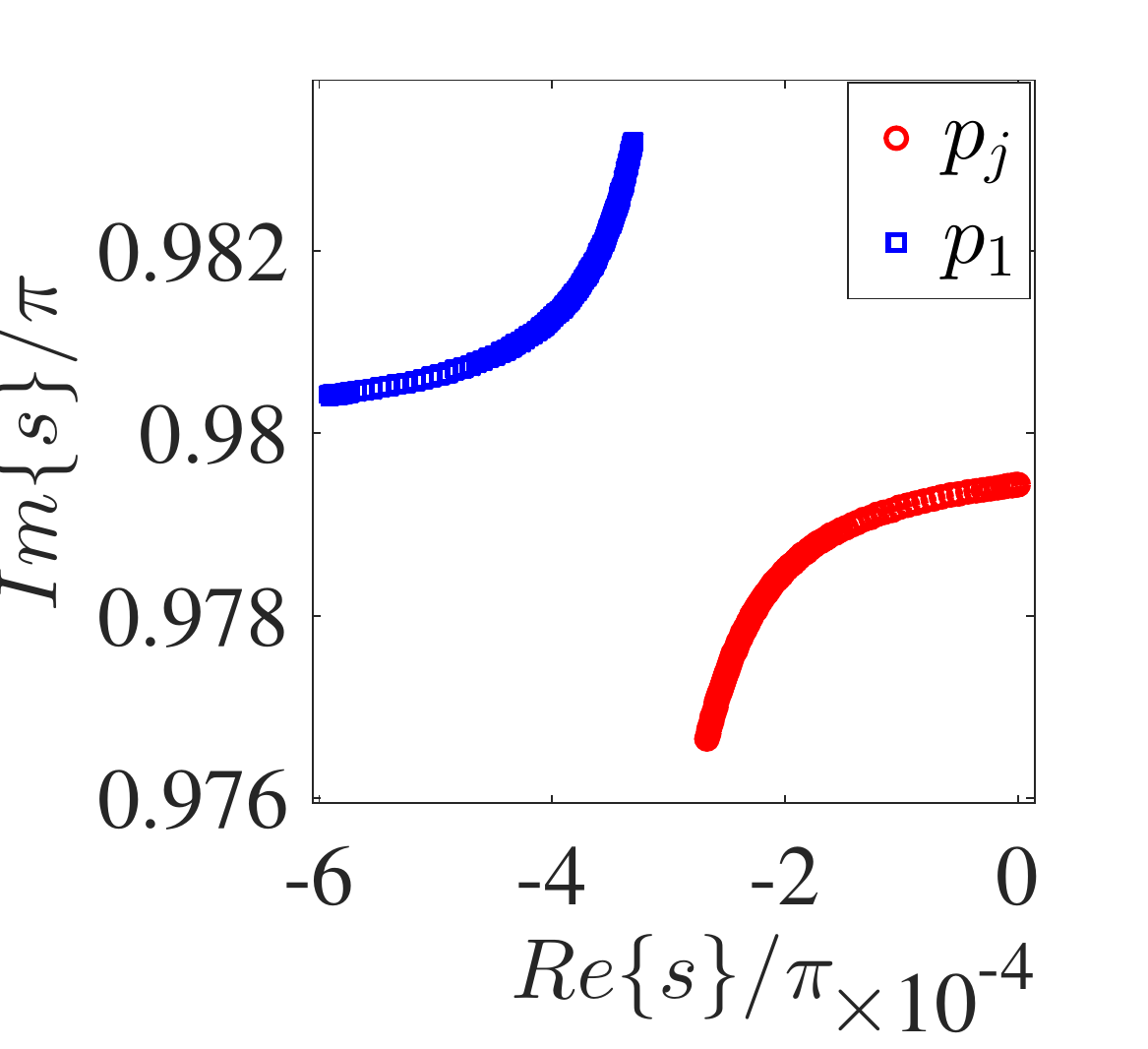}%
} 
\caption{ a) The first five hybridized poles of the resonator-qubit system, for the case where the transmon is slightly detuned below the fundamental mode, i.e. $\omega_j=\nu_1^-$. The other parameters are set as $\chi_R=\chi_L=0.01$, $\chi_j=0.05$ and $\chi_g\in [0,10^{-3}]$ with increments $\Delta\chi_g=10^{-5}$. b) Zoom-in plot of the hybridization of the most resonant modes. Hybridization of $p_1$ and $p_j$ is much stronger than that of the off-resonant poles $p_n$, $n > 1$.}
\label{Fig:Hybrid Poles Weak Coup}
\end{figure}
%%%%%%%%%%%%%%%%%%%%%%%%%%%%%%%%%%%%%%%%%%%%%%%

The roots of $D_j(s)$ are the hybridized poles of the entire system. If there is no coupling, i.e. $\chi_g=0$, then $D_j(s)=s^2+\omega_j^2=(s+i\omega_j)(s-i\omega_j)$ is the characteristic polynomial that gives the bare transmon resonance. However, for a nonzero $\chi_g$, $D_j(s)$ becomes a meromorphic function whose zeros are the hybridized resonances of the entire system, and whose poles are the bare cavity resonances. Therefore, $D_j(s)$ can be expressed as
\begin{align}
D_j(s)=(s-p_j)(s-p_j^*)\prod\limits_{m}\frac{(s-p_m)(s-p_m^*)}{(s-z_m)(s-z_m^*)}.
\label{eqn:Formal Rep of D(s)}
\end{align}
In Eq.~(\ref{eqn:Formal Rep of D(s)}), $p_j\equiv -\alpha_j-i\beta_j$ and $p_n\equiv -\alpha_n-i\beta_n$ are the zeros of $D_j(s)$ that represent the transmon-like and the $n$th resonator-like poles, accordingly. Furthermore, $z_n \equiv -i\omega_n=-\kappa_n-i\nu_n$ stands for the $n$th bare non-Hermitian resonator resonance. The notation chosen here ($p$ for poles and $z$ for zeroes) reflects the meromorphic structure of $1/D_j(s)$ which enters the solution Eq.~(\ref{eqn:Sol of X_j(s)}).

An important question concerns the convergence of $D_j(s)$ as a function of the number of the resonator modes included in the calculation. The form of $D_j(s)$ given in Eq.~(\ref{eqn:Formal Rep of D(s)}) is suitable for this discussion. Consider the factor corresponding to the $m$th resonator mode in $1/D_j(s)$. We reexpress it as
\begin{align}
\begin{split}
\frac{(s-z_m)(s-z_m^*)}{(s-p_m)(s-p_m^*)}&=\left(1-\frac{z_m-p_m}{s-p_m}\right)\left(1-\frac{z_m^*-p_m^*}{s-p_m^*}\right)\\
&=1+\mathcal{O}\left(\left|\frac{z_m-p_m}{s-p_m}\right|\right).
\end{split}
\end{align}   
The consequence of a small shift $|p_m-z_m|$ as compared to the strongly hybridized resonant mode $|p_1-z_1|$ is that it can be neglected in the expansion for $1/D_j(s)$. The relative size of these contributions is controlled by the coupling $\chi_g$. As rule  of thumb, the less hybridized a resonator pole is, the less it contributes to qubit dynamics. Ultimately, the truncation in this work is established by imposing the convergence of the numerics.

A numerical solution for the roots of Eq.~(\ref{eqn:simplified D(s)}) at weak coupling $\chi_g$ reveals that the mode resonant with the transmon is significantly shifted, with comparatively small shifts $|p_m-z_m|$ in the other resonator modes (See Fig.~\ref{Fig:Hybrid Poles Weak Coup}). At weak coupling, the hybridization of $p_j$ and $p_1$ is captured by a single resonator mode.
%%%%%%%%%%%% Fig of Poles Strong Coupling%%%%%%%%%%
\begin{figure}[t!]
\centering
\subfloat[\label{subfig:Poles1ModeXrXl1Em2Zoomed}]{%
\includegraphics[scale=0.50]{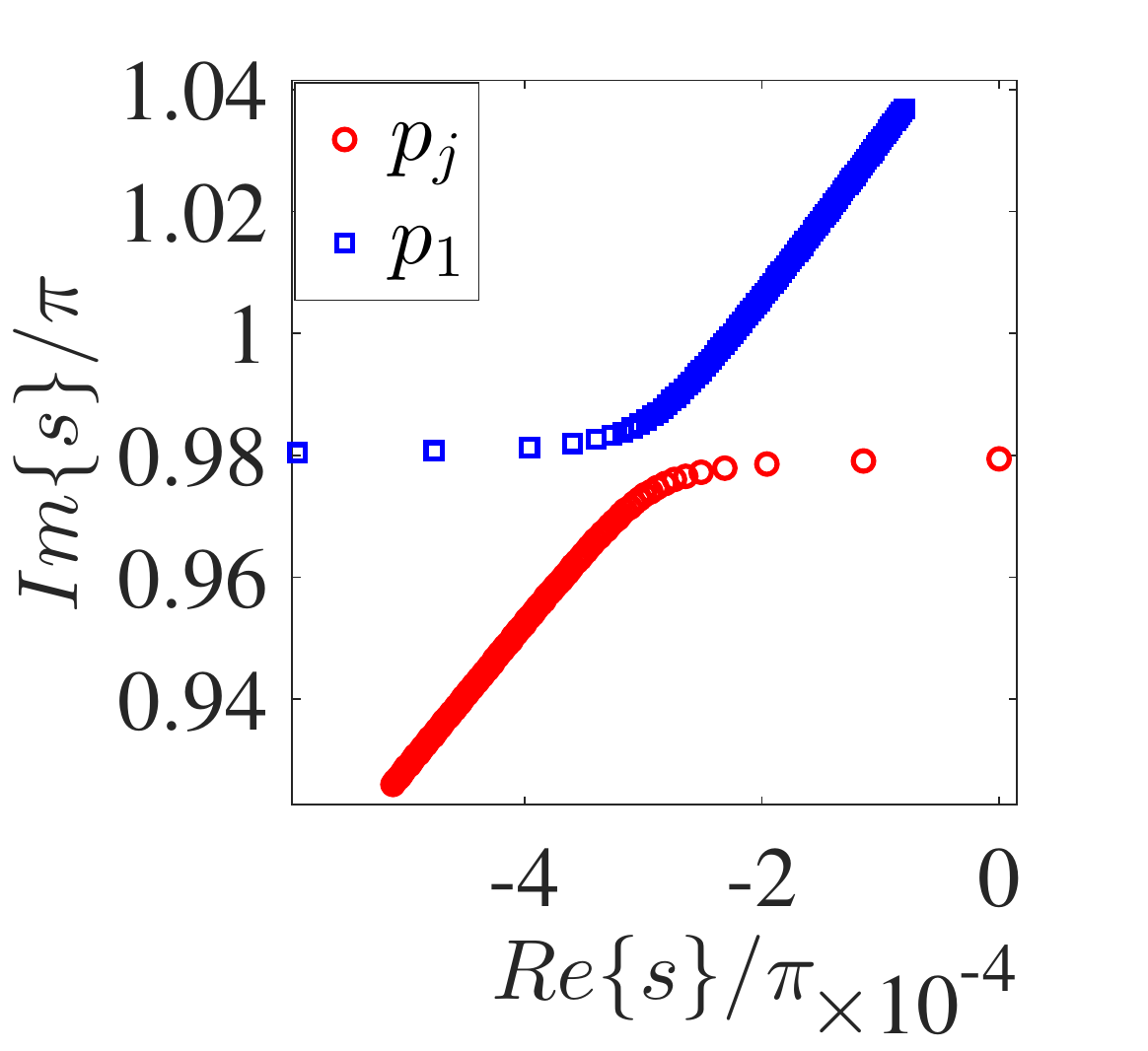}%
} 
\subfloat[\label{subfig:Poles5ModeXrXl1Em2Zoomed}]{%
\includegraphics[scale=0.50]{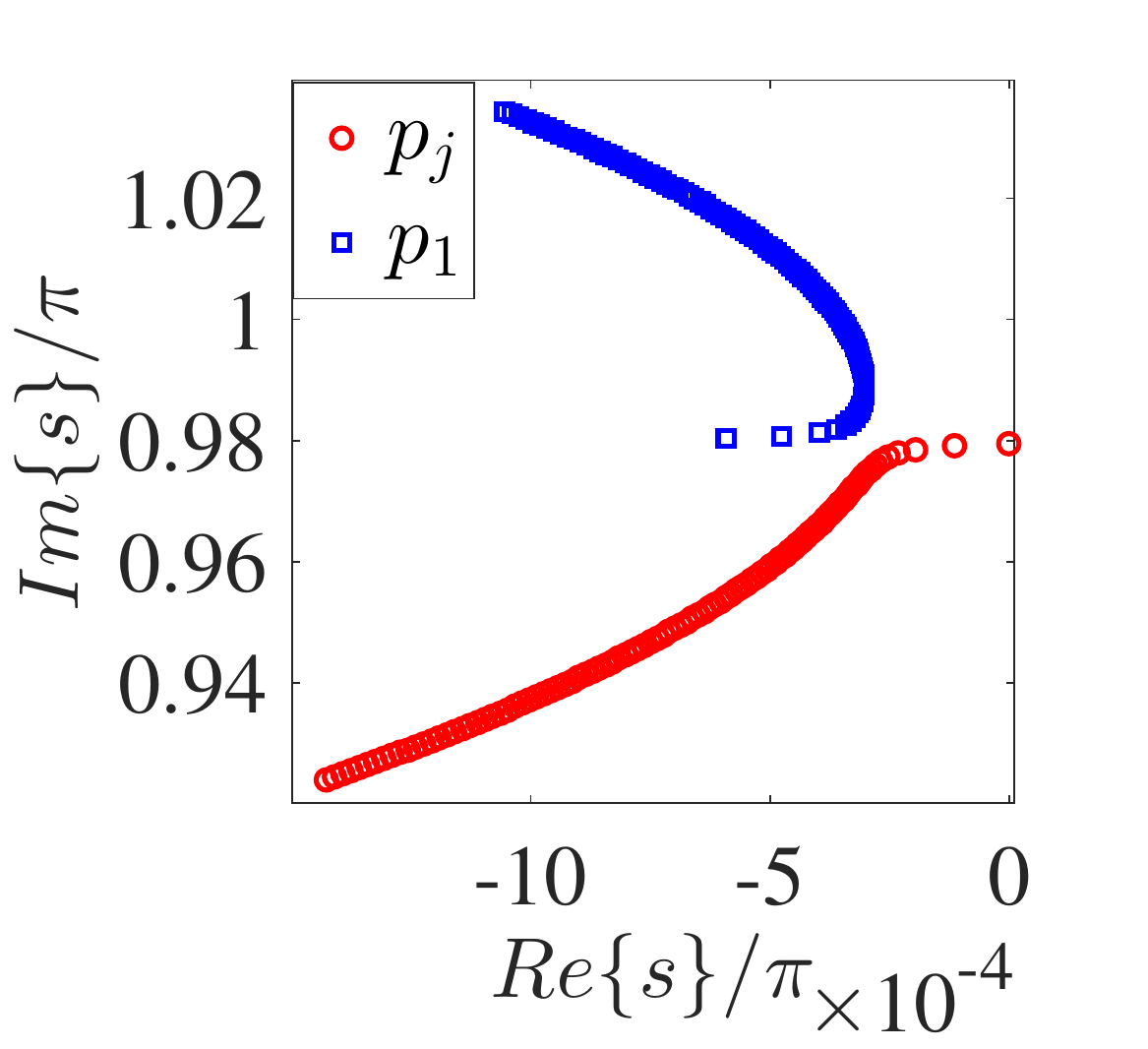}%
}\\
\subfloat[\label{subfig:Poles10ModeXrXl1Em2Zoomed}]{%
\includegraphics[scale=0.50]{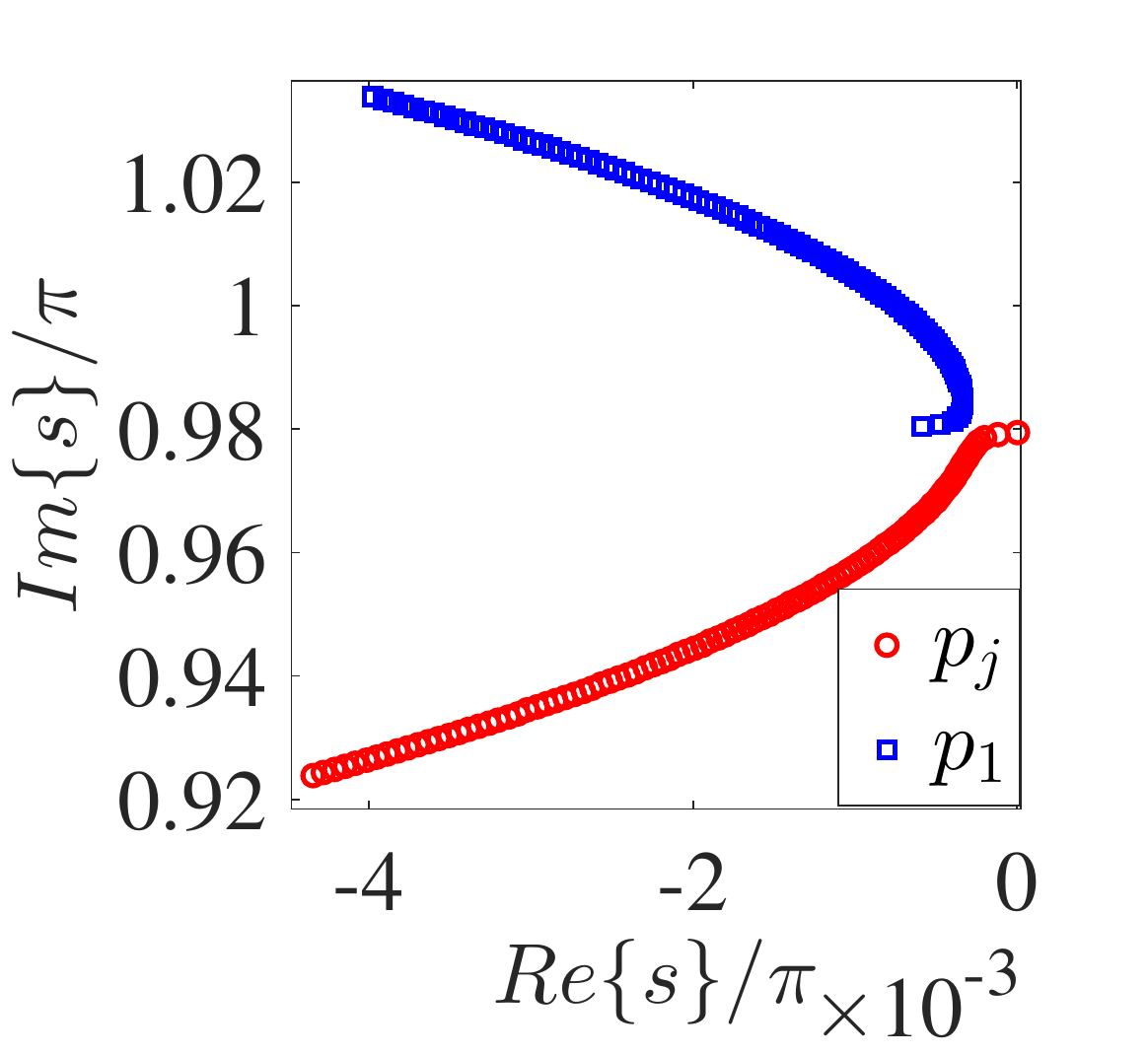}%
}
\subfloat[\label{subfig:Poles20ModeXrXl1Em2Zoomed}]{%
\includegraphics[scale=0.50]{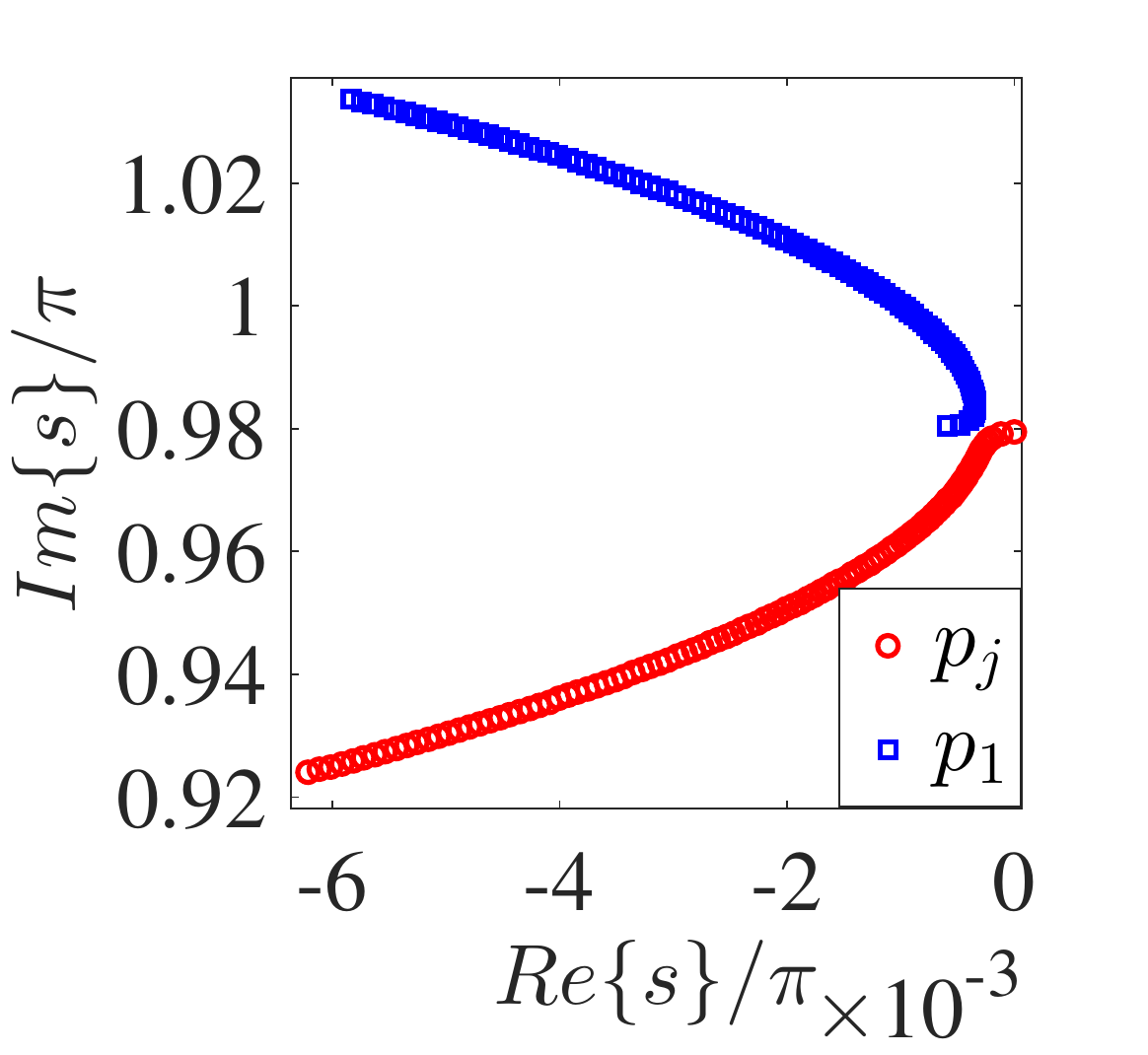}%
}
\caption{Convergence of $p_j$ and $p_1$ for the same parameters as Fig.~\ref{Fig:Hybrid Poles Weak Coup}, but for $\chi_g\in[0,0.02]$ and keeping a) 1, b) 5, c) 10 and d) 20 resonator modes in $D_j(s)$.} 
\label{Fig:Hybrid Poles Strong Coup}
\end{figure}
%%%%%%%%%%%%%%%%%%%%%%%%%%%%%%%%%%%%%%%%%%%%
Next, we plot in Fig.~\ref{Fig:Hybrid Poles Strong Coup} the effect of truncation on the response of the multimode system in a band around $s= p_j$. As the coupling $\chi_g$ is increased beyond the avoided crossing, which is also captured by the single mode truncation, the effect of off-resonant modes on $p_j$ and $p_1$ becomes significant. It is important to note that the hybridization occurs in the complex $s$-plane. On the frequency axis $\Im\{s\}$ an increase in $\chi_g$ is associated with a splitting of transmon-like and resonator-like poles. Along the decay rate axis $\Re\{s\}$ we notice that the qubit decay rate is controlled by the resonant mode at weak coupling, with noticeable enhancement of off-resonant mode contribution at strong coupling. If the truncation is not done properly in the strong coupling regime, it may result in spurious unstable roots of $D_j(s)$, i.e. $\Re\{s\}>0$, as seen in Fig.~\ref{subfig:Poles1ModeXrXl1Em2Zoomed}.

The modification of the decay rate of the transmon-like pole, henceforth identified as $\alpha_j\equiv-\Re\{p_j\}$, has an important physical significance. It describes the Purcell modification of the qubit decay (if sources for qubit decay other than the direct coupling to electromagnetic modes can be neglected). The present scheme is able to capture the  full multimode modification, that is out of the reach of conventional single-mode theories of spontaneous emission \cite{Purcell_Resonance_1946, Kleppner_Inhibited_1981, Goy_Observation_1983, Hulet_Inhibited_1985, Jhe_Suppression_1987}.

\begin{figure}
\centering
\subfloat[\label{subfig:SpEmRateXrXl1Em2Xg1Em3}]{%
\includegraphics[scale=0.50]{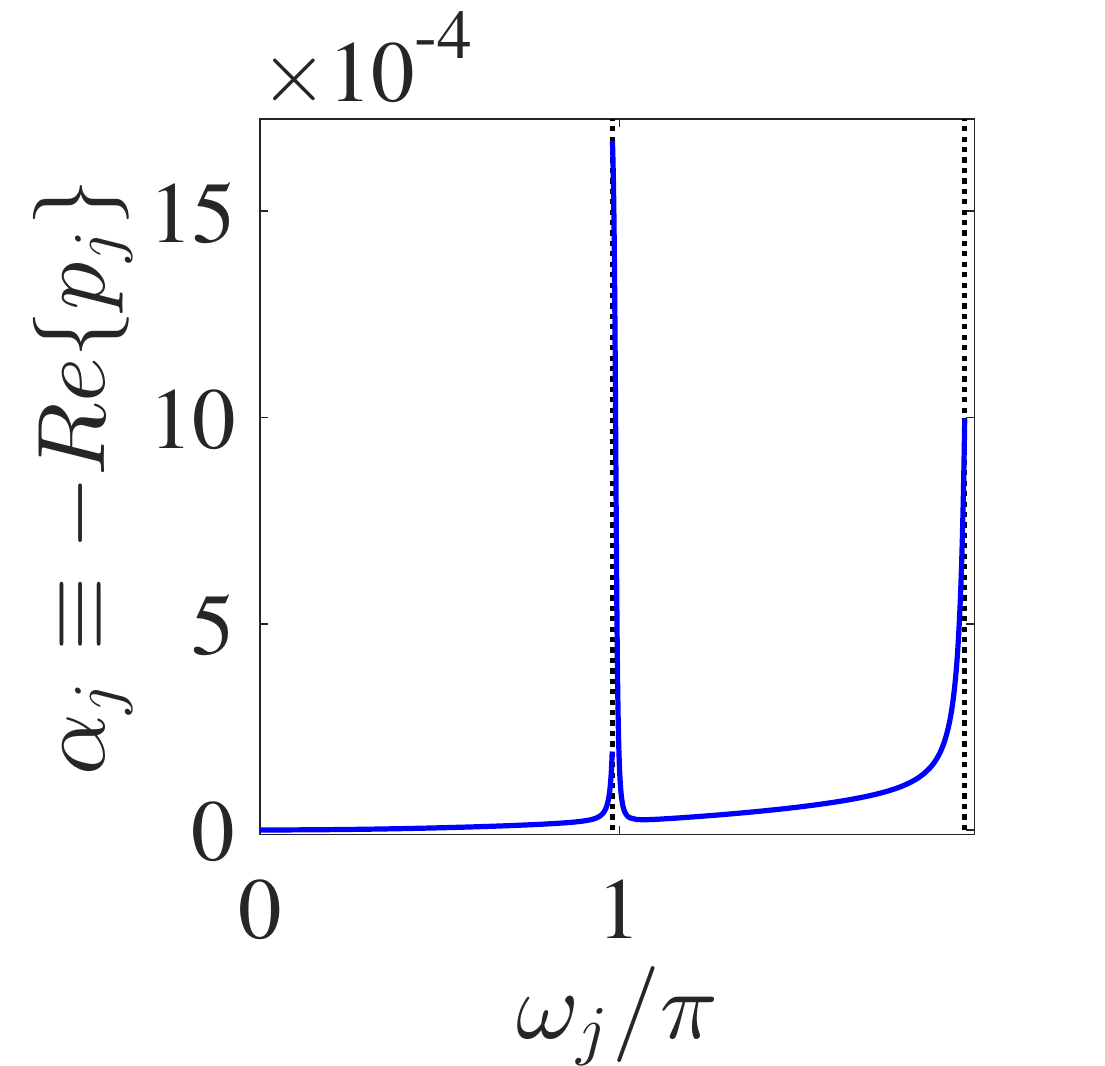}%
}
\subfloat[\label{subfig:SpEmRateXrXl1Em2Xg5Em3}]{%
\includegraphics[scale=0.50]{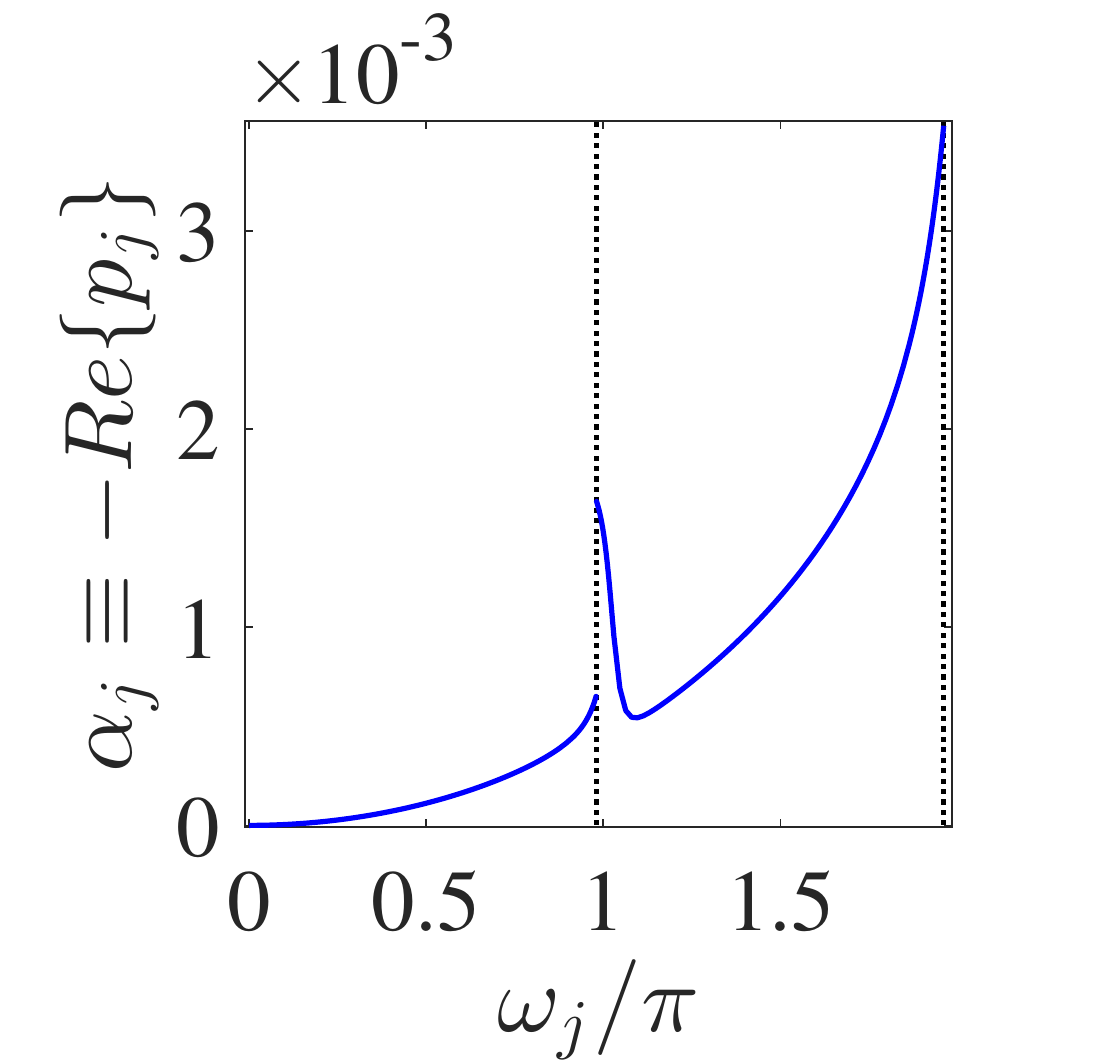}%
}
\caption{Spontaneous emission rate defined as $\alpha_j\equiv-\Re\{p_j\}$ as a function of transmon frequency $\omega_j$ for $\chi_R=\chi_L=10^{-2}$, $\chi_j=0.05$, a) $\chi_g=10^{-3}$ and b) $\chi_g=5\times 10^{-3}$. We observe that the asymmetry grows as $\chi_g$ is increased. The black vertical dotted lines show the location of resonator frequencies $\nu_n$.} 
\label{Fig:SpEmRate}
\end{figure}

At fixed $\chi_g$, we observe an asymmetry of $\alpha_j$ when the bare transmon frequency is tuned across the fundamental mode of the resonator, in agreement with a previous experiment \cite{Houck_Controlling_2008}, where a semiclassical model was employed for an accurate fit. Figure~\ref{Fig:SpEmRate} shows that near the resonator-like resonance the spontaneous decay rate is enhanced, as expected. For positive detunings spontaneous decay is significantly larger than for negative detunings, which can be traced back to an asymmetry in the resonator density of states \cite{Houck_Controlling_2008}. We find that this asymmetry grows as $\chi_g$ is increased. Note that besides a systematic inclusion of multimode effects, the presented theory of spontaneous emission goes beyond the rotating wave, Markov and two-level approximations as well.
%%%%%%%%%%%%% Fig of SpEm Rates %%%%%%%%%%
%%%%%%%%%%%%%%%%%%%%%%%%%%%%%%%%%%%%%%%%%%%%%%%

\begin{figure}
\centering
\subfloat[\label{subfig:AxXg0to05Xj005XrXl001}]{%
\includegraphics[scale=0.50]{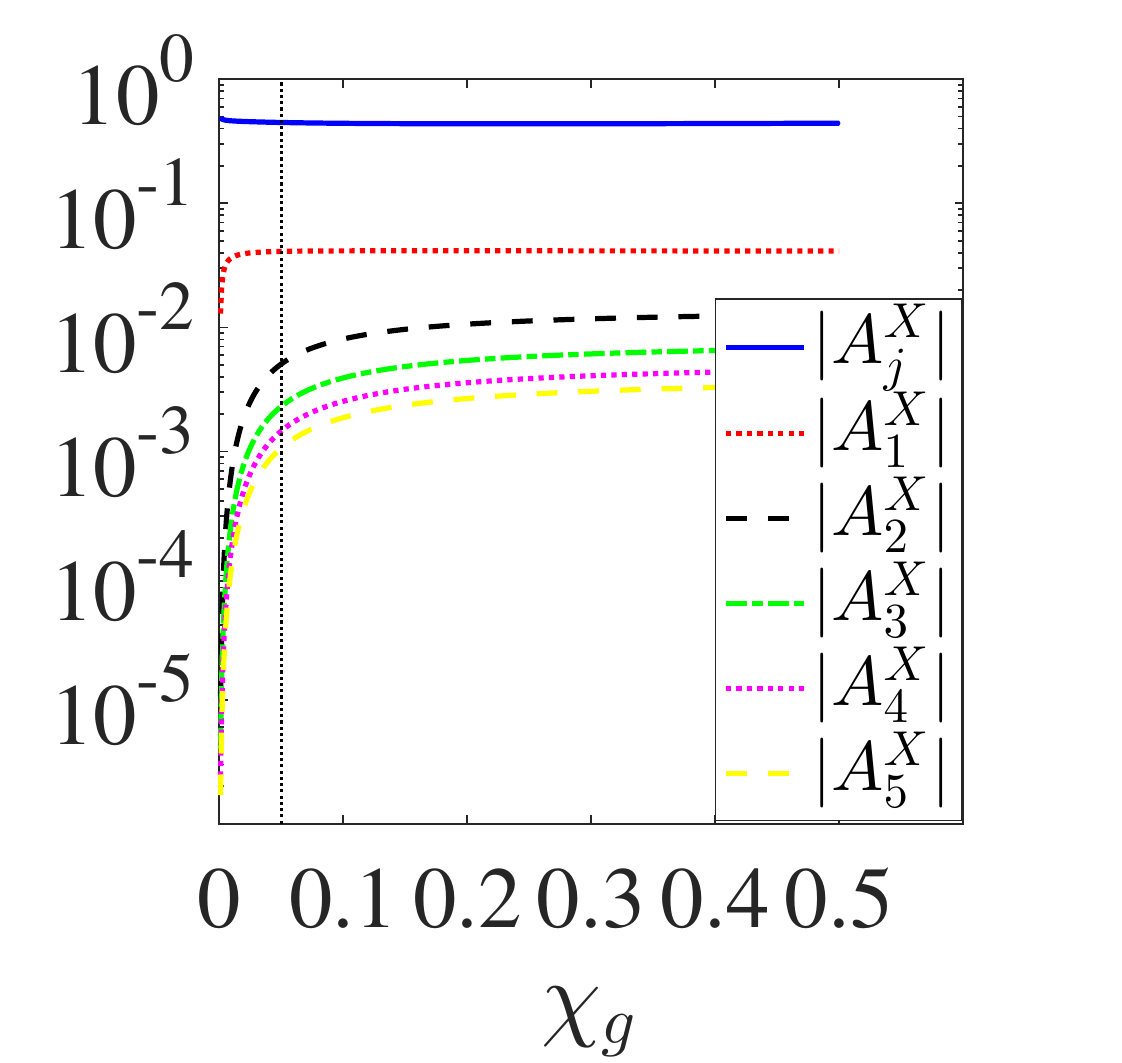}%
}
\subfloat[\label{subfig:AyXg0to05Xj005XrXl001}]{%
\includegraphics[scale=0.50]{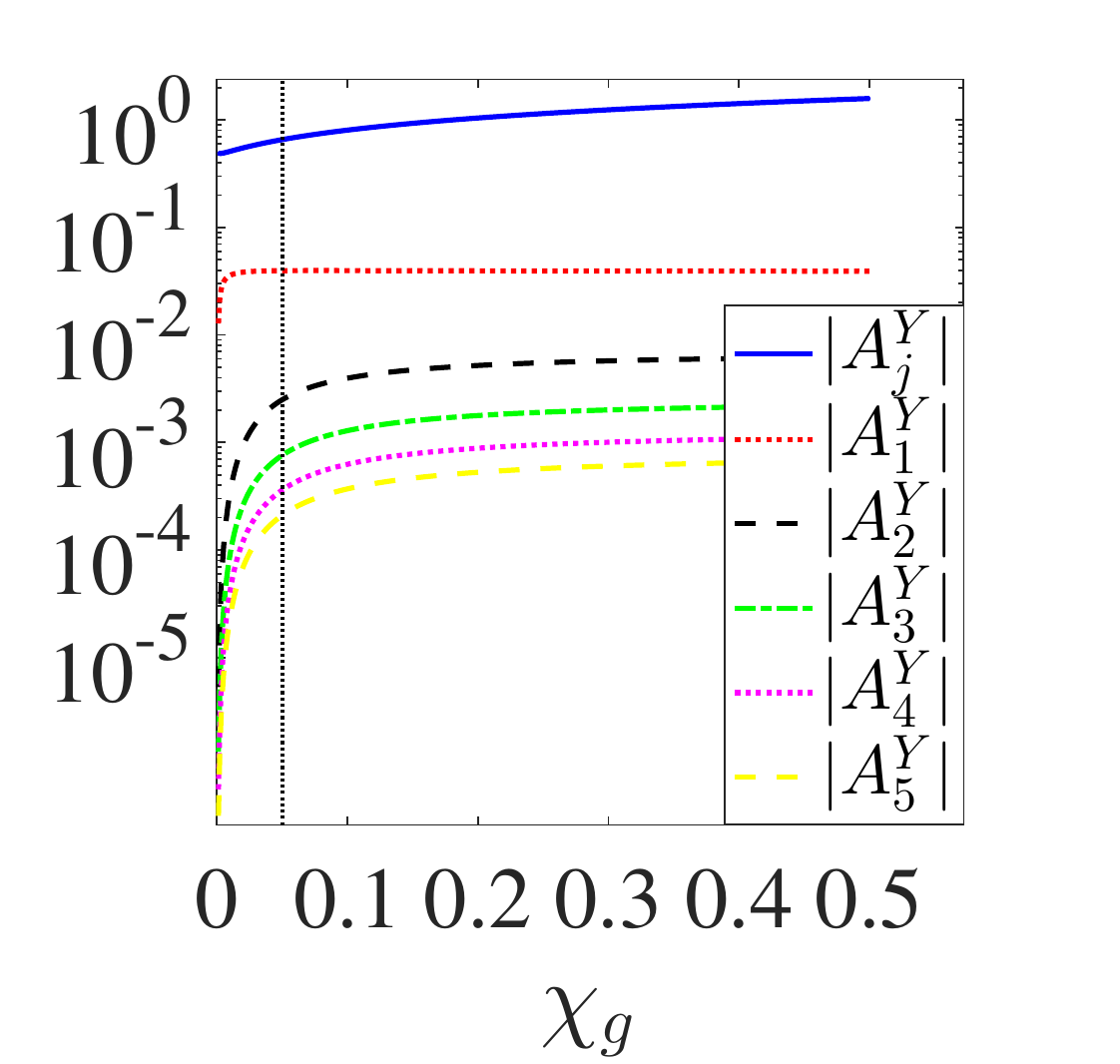}%
}
\caption{Dependence of residues defined in Eqs.~(\ref{eqn:Def of A_l^X}-\ref{eqn:Def of A_l^Y}) on $\chi_g$ for $\omega_j=\nu_1^-$, $\chi_R=\chi_L=0.01$ and $\chi_j=0.05$. The black vertical dotted line shows the value of $\chi_j$.} 
\label{Fig:AnxAnyAjxAjy}
\end{figure}

Having studied the hybridized resonances of the entire system, we are now able to provide the time-dependent solution to Eq.~(\ref{eqn:Lin SE Problem}). By substituting Eq.~(\ref{eqn:Formal Rep of D(s)}) into Eq.~(\ref{eqn:Sol of X_j(s)}) we obtain
\begin{align}
\hat{\tilde{X}}_j(s)=\left(\frac{\hat{A}_j}{s-p_j}+\sum\limits_{n}\frac{\hat{A}_n}{s-p_n}\right)+H.c.,
\label{eqn:PFE of X_j(s)}
\end{align}
from which the inverse Laplace transform is immediate
\begin{align}
\hat{X}_j(t)=\left[\left(\hat{A}_je^{p_jt}+\sum\limits_{n}\hat{A}_n e^{p_n t}\right)+H.c.\right]\Theta(t).
\label{eqn:Lin Sol X_j(t)}
\end{align}
The frequency components have operator-valued amplitudes 
\begin{subequations}
\begin{align}
&\hat{A}_j\equiv A_j^{X}\hat{X}_j(0)+A_j^{Y}\hat{Y}_j(0),
\label{eqn:Def of hat(A)_j}\\
&\hat{A}_n\equiv A_n^{X}\hat{X}_j(0)+A_n^{Y}\hat{Y}_j(0),
\label{eqn:Def of hat(A)_n}
\end{align}
\end{subequations}
with the residues given in terms of $D_j(s)$ as
\begin{subequations}
\begin{align}
&A_{j,n}^{X}\equiv \left. \left[(s-p_{j,n})\frac{s}{D_j(s)}\right]\right|_{s=p_{j,n}},
\label{eqn:Def of A_l^X}\\
&A_{j,n}^{Y}\equiv \left. \left[(s-p_{j,n})\frac{\omega_j}{D_j(s)}\right]\right|_{s=p_{j,n}}.
\label{eqn:Def of A_l^Y}
\end{align}
\end{subequations}
The dependence of $A_{j,n}^{X}$ and $A_{j,n}^{Y}$ on coupling $\chi_g$ has been studied in Fig.~\ref{Fig:AnxAnyAjxAjy}. The transmon-like amplitude (blue solid) is always dominant, and further off-resonant modes have smaller amplitudes. By increasing $\chi_g$, the resonator-like amplitude grow significantly first and reach an asymptote as predicted by Eq.~(\ref{eqn:LinTheory-Asymptote}).
%%%%%%%% Fig of Ajx Ajy Anx Any %%%%%%%%%%%%%%%%%
%%%%%%%%%%%%%%%%%%%%%%%%%%%%%%%%%%%%%%%%%%%%%%
\subsection{Perturbative corrections}
\label{Sec:PertCor}
In this section, we develop a well-behaved time-domain perturbative expansion in the transmon qubit nonlinearity as illustrated in Eq.~(\ref{eqn:Expansion of Sine}). Conventional time-domain perturbation theory is inapplicable due to the appearance of resonant coupling between the successive orders which leads to secular contributions, i.e. terms that grow unbounded in time (For a simple example see App.~\ref{SubApp:ClDuffingDiss}). A solution to this is multi-scale perturbation theory (MSPT) \cite{Bender_Advanced_1999, Nayfeh_Nonlinear_2008, Strogatz_Nonlinear_2014}, which considers multiple independent time scales and eliminates secular contributions by a resummation of the conventional perturbation series. 

The effect of the nonlinearity is to mix the hybridized modes discussed in the previous section, leading to transmon mediated self-Kerr and cross-Kerr interactions. Below, we extend MSPT to treat this problem while consistently accounting for the dissipative effects. This goes beyond the extent of Rayleigh-Schr\"odinger perturbation theory, as it will allow us to treat the energetic and dissipative scales on equal footing. 

The outcome of conventional MSPT analysis in a conservative system is frequency renormalization \cite{Bender_Advanced_1999, Bender_Multiple_1996}. We illustrate this point for a classical Duffing oscillator, which amounts to the classical theory of an isolated transmon qubit up to leading order in the nonlinearity. We outline the main steps here leaving the details to App.~\ref{SubApp:ClDuffingDiss}. Consider a classical Duffing oscillator
\begin{align}
\ddot{X}(t)+\omega^2\left[X(t)-\varepsilon X^3(t)\right]=0,
\label{eqn:PertCor-ClDuffing Osc}
\end{align}
with initial conditions $X(0)=X_0$ and $\dot{X}(0)=\omega Y_0$. Equation~(\ref{eqn:PertCor-ClDuffing Osc}) is solved order by order with the Ansatz
\begin{subequations}
\begin{align}
X(t)=x^{(0)}(t,\tau)+\varepsilon x^{(1)}(t,\tau)+\mathcal{O}(\varepsilon^2),
\label{Eq:ClDuffing Expansion of X}
\end{align}
where $\tau\equiv\varepsilon t$ is assumed to be an independent time scale such that
\begin{align}
d_t\equiv \partial_t+\varepsilon\partial_\tau+\mathcal{O}(\varepsilon^2).
\end{align} 
\end{subequations}
This additional time-scale then allows us to remove the secular term that appears in the $\mathcal{O}(\varepsilon)$ equation. This leads to a renormalization in the oscillation frequency of the $\mathcal{O}(1)$ solution as
\begin{subequations}
\begin{align}
&X^{(0)}(t)=x^{(0)}(t,\varepsilon t)=\left[a(0)e^{-i\bar{\omega}t}+c.c.\right],
\label{eqn:PertCor-ClDuffing X^(0)(t)}\\
&\bar{\omega}\equiv \left[1-\frac{3\varepsilon}{2}|a(0)|^2\right]\omega,
\label{eqn:PertCor-ClDuffing FreqRenorm}
\end{align}
\end{subequations}
where $a(0)=(X_0+iY_0)/2$.
%%%%%%%%%%%% Fig of u%%%%%%%%%%
\begin{figure}
\centering
\includegraphics[scale=0.53]{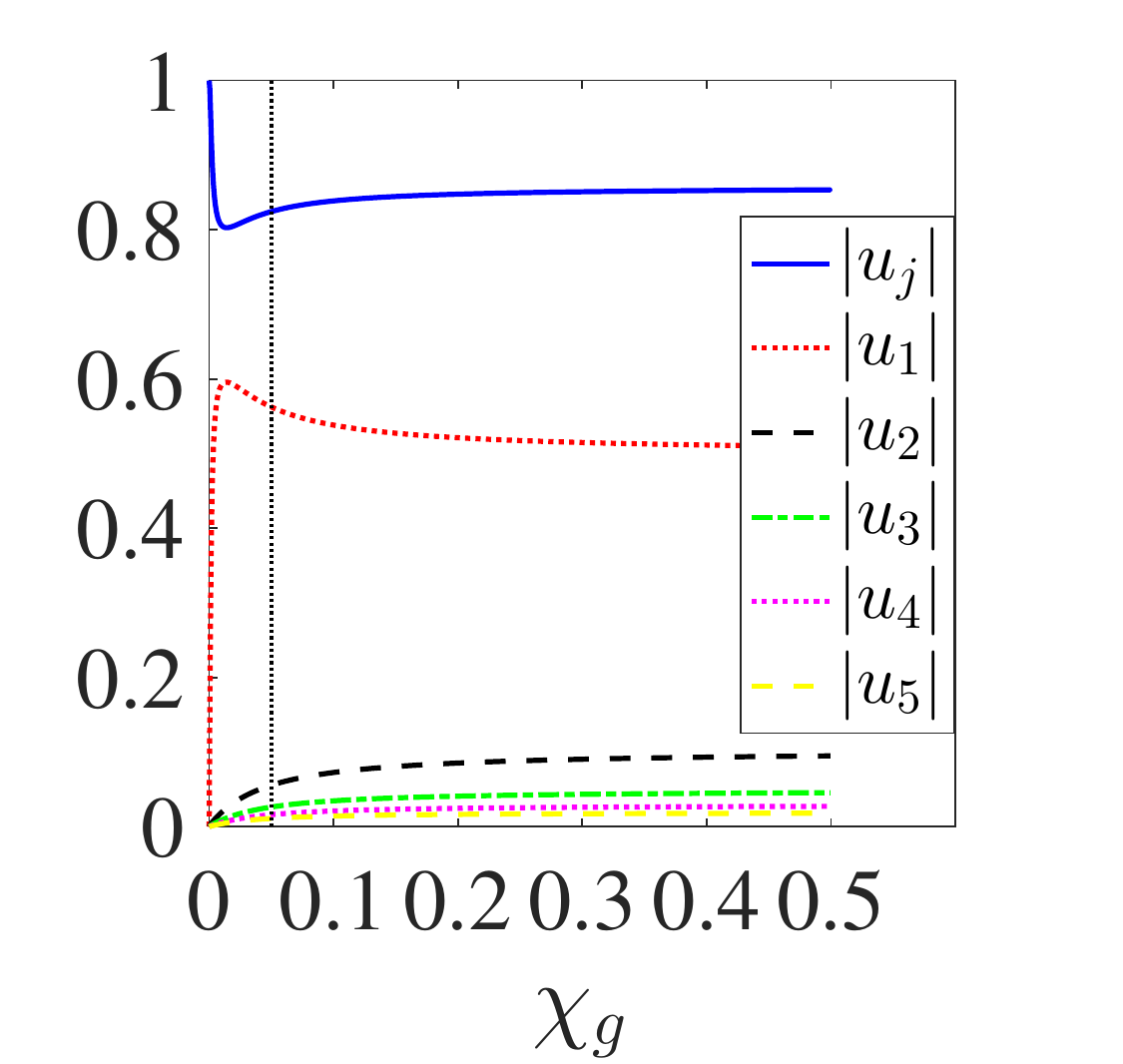} 
\caption{Hybridization coefficients $u_j$ and $u_n$ of the first five modes for the case where the transmon is infinitesimally detuned below the fundamental mode, i.e. $\omega_j=\nu_1^-$ as a function of $\chi_g\in[0,0.5]$. Other parameters are set as $\chi_R=\chi_L=0$ and $\chi_j=0.05$. The black vertical dotted line shows the value of $\chi_j$.}
\label{Fig:QuDuffQuHarm-u}
\end{figure}
%%%%%%%%%%%%%%%%%%%%%%%%%%%%%%%%%%%%%%%%%%%%%
One may wonder how this leading-order correction is modified in the presence of dissipation. Adding a small damping term $\kappa\dot{X}(t)$ to Eq.~(\ref{eqn:PertCor-ClDuffing Osc}) such that $\kappa\ll\omega$ requires a new time scale $\eta \equiv \frac{\kappa}{\omega}t$ leading to
\begin{subequations}
\begin{align}
&X^{(0)}(t)=e^{-\frac{\kappa}{2}t}\left[a(0)e^{-i\bar{\omega}t}+c.c.\right],
\label{eqn:PertCor-ClDuffingDis X^(0)(t)}\\
&\bar{\omega}\equiv\left[1-\frac{3\varepsilon}{2}|a(0)|^2e^{-\kappa t}\right]\omega.
\label{eqn:PertCor-ClDuffingDiss FreqRenorm}
\end{align}
\end{subequations}
Equations~(\ref{eqn:PertCor-ClDuffingDis X^(0)(t)}-\ref{eqn:PertCor-ClDuffingDiss FreqRenorm}) illustrate a more general fact that the dissipation modifies the frequency renormalization by a decaying envelope. This approach can be extended by introducing higher order (slower) time scales $\varepsilon^2 t, \eta^2 t$, $\eta\varepsilon t$ etc. The lowest order calculation above is valid for times short enough such that $\omega t\ll \varepsilon^{-2} , \eta^{-2} , \eta^{-1}\varepsilon ^{-1}$.

Besides the extra complexity due to non-commuting algebra of quantum mechanics, the principles of MSPT remain the same in the case of a free quantum Duffing oscillator \cite{Bender_Multiple_1996}. The Heisenberg equation of motion is identical to Eq.~(\ref{eqn:PertCor-ClDuffing Osc}) where we promote $X(t)\to \hat{X}(t)$. We obtain the $\mathcal{O}(1)$ solution (see App.~\ref{SubApp:QuDuffingNoMem}) as
\begin{subequations}
\begin{align}
\begin{split}
\hat{X}^{(0)}(t)=e^{-\frac{\kappa}{2}t}\left[\frac{\hat{a}(0)e^{-i\hat{\bar{\omega}} t}+e^{-i\hat{\bar{\omega}} t}\hat{a}(0)}{2\cos\left(\frac{3\omega}{4}\varepsilon te^{-\kappa t}\right)}+H.c.\right]
\label{eqn:QuDuffingNoMem-X^(0)(t) sol}
\end{split}
\end{align}
with an operator-valued renormalization of the frequency 
\begin{align}
&\hat{\bar{\omega}}=\left[1-\frac{3\varepsilon}{2}\hat{\mathcal{H}}(0)e^{-\kappa t}\right]\omega,
\label{eqn:QuDuffingNoMem-bar(W)}\\
&\hat{\mathcal{H}}(0)\equiv\frac{1}{2}\left[\hat{a}^{\dag}(0)\hat{a}(0)+\hat{a}(0)\hat{a}^{\dag}(0)\right].
\label{eqn:QuDuffingNoMem-H(0)}
\end{align}
\end{subequations}
The cosine that appears in the denominator of operator solution~(\ref{eqn:QuDuffingNoMem-X^(0)(t) sol}) cancels when taking the expectation values with respect to the number basis $\{\ket{n}\}$ of $\hat{\mathcal{H}}(0)$:   
\begin{align}
\bra{n-1}\hat{X}^{(0)}(t)\ket{n}=\sqrt{n}e^{-\frac{\kappa}{2}t}e^{-i\left(1-\frac{3n\varepsilon}{2}e^{-\kappa t}\right)\omega t}.
\end{align}

Having learned from these toy problems, we return to the problem of spontaneous emission which can be mapped into a quantum Duffing oscillator with $\varepsilon=\frac{\sqrt{2}}{6}\left(\mathcal{E}_c/\mathcal{E}_j\right)^{1/2}$, up to leading order in perturbation, coupled to multiple leaky quantum harmonic oscillators (see Eq.~(\ref{eqn:Expansion of Sine})). We are interested in finding an analytic expression for the shift of the hybridized poles, $p_j$ and $p_n$, that appear in the reduced dynamics of the transmon. 

The hybridized poles $p_j$ and $p_n$ are the roots of $D_j(s)$ and they are associated with the modal decomposition of the linear theory in Sec.~\ref{Sec:Lin SE Theory}. The modal decomposition can be found from the linear solution $\mathcal{X}_j(t)$ that belongs to the full Hilbert space as   
\begin{align}
\begin{split}
\hat{\mathcal{X}}_j(t)=\left(\hat{\mathcal{A}}_je^{p_jt}+\sum\limits_{n}\hat{\mathcal{A}}_n e^{p_n t}\right)+H.c.\equiv\left(u_j\hat{\bar{a}}_je^{p_jt}+\sum\limits_{n}u_n\hat{\bar{a}}_n e^{p_n t}\right)+H.c.
\end{split}
\label{eqn:Lin Sol Mathcal(X)_j(t)}
\end{align}
This is the full-Hilbert space version of Eq.~(\ref{eqn:Lin Sol X_j(t)}). It represents the unperturbed solution upon which we are building our perturbation theory. We have used bar-notation to distinguish the creation and annihilation operators in the hybridized mode basis. Furthermore, $u_j$ and $u_n$ represent the hybridization coefficients, where they determine how much the original transmon operator $\hat{\mathcal{X}}_j(t)$, is transmon-like and resonator-like. They can be obtained from a diagonalization of the linear Heisenberg-Langevin equations of motion (see App.~\ref{SubApp:QuDuffQuHarm}). The dependence of $u_j$ and $u_n$ on coupling $\chi_g$ is shown in Fig.~(\ref{Fig:QuDuffQuHarm-u}) for the case where the transmon is infinitesimally detuned below the fundamental mode of the resonator. For $\chi_g=0$, $u_j=1$ and $u_n=0$ as expected. As $\chi_g$ reaches $\chi_j$, $u_1$ is substantially increased and becomes comparable to $u_j$. By increasing $\chi_g$ further, $u_n$ for the off-resonant modes start to grow as well.

The nonlinearity acting on the transmon mixes all the unperturbed resonances through self- and cross-Kerr contributions \cite{Drummond_Quantum_1980, Nigg_BlackBox_2012, Bourassa_Josephson_2012}. Kerr shifts can be measured in a multimode cQED system \cite{Rehak_Parametric_2014, Weissl_Kerr_2015}. We therefore solve for the equations of motion of each mode. These are (see App. \ref{SubApp:QuDuffQuHarm})
\begin{align}
\begin{split}
\hat{\ddot{\bar{\mathcal{X}}}}_{l}(t)+2\alpha_{l}\hat{\dot{\bar{\mathcal{X}}}}_{l}(t)+\beta_{l}^2\left\{\hat{\bar{\mathcal{X}}}_{l}(t)-\varepsilon_{l} \left[u_j\hat{\bar{\mathcal{X}}}_j(t)+\sum\limits_n u_n\hat{\bar{\mathcal{X}}}_n(t)\right]^3\right\}=0,
\end{split}
\label{eqn:QuDuffQuHarm Osc}
\end{align}
where $\hat{\bar{X}}_l\equiv \hat{\bar{a}}_l+\hat{\bar{a}}_l^{\dag}$ is the quadrature of the $l$th mode, and $\alpha_l$ and $\beta_l$ are the decay rate and the oscillation frequency, respectively. Equation~(\ref{eqn:QuDuffQuHarm Osc}) is the leading order approximation in the inverse Q-factor of the $l$th mode, $1/Q_l\equiv\alpha_l/\beta_l$. Each hybridized mode has a distinct strength of the nonlinearity $\varepsilon_{l}\equiv\frac{\omega_j}{\beta_{l}}u_{l}\varepsilon$ for $l\equiv j,n$. In order to do MSPT, we need to introduce as many new time-scales as the number of hybridized modes, i.e. $\tau_j\equiv\varepsilon_j t$ and $\tau_n\equiv \varepsilon_n t$, and do a perturbative expansion in all of these time scales. The details of this calculation can be found in App.~\ref{SubApp:QuDuffQuHarm}. Up to lowest order in $\varepsilon$, we find operator-valued correction of $p_{j}=-\alpha_j-i\beta_j$ as
%%%%%%%%%%%% Fig of Fourier of Xj%%%%%%%%%%
\begin{figure}
\centering
\subfloat[\label{subfig:FXtXj005XrXl1Em3IC11View1}]{%
\includegraphics[scale=0.45]{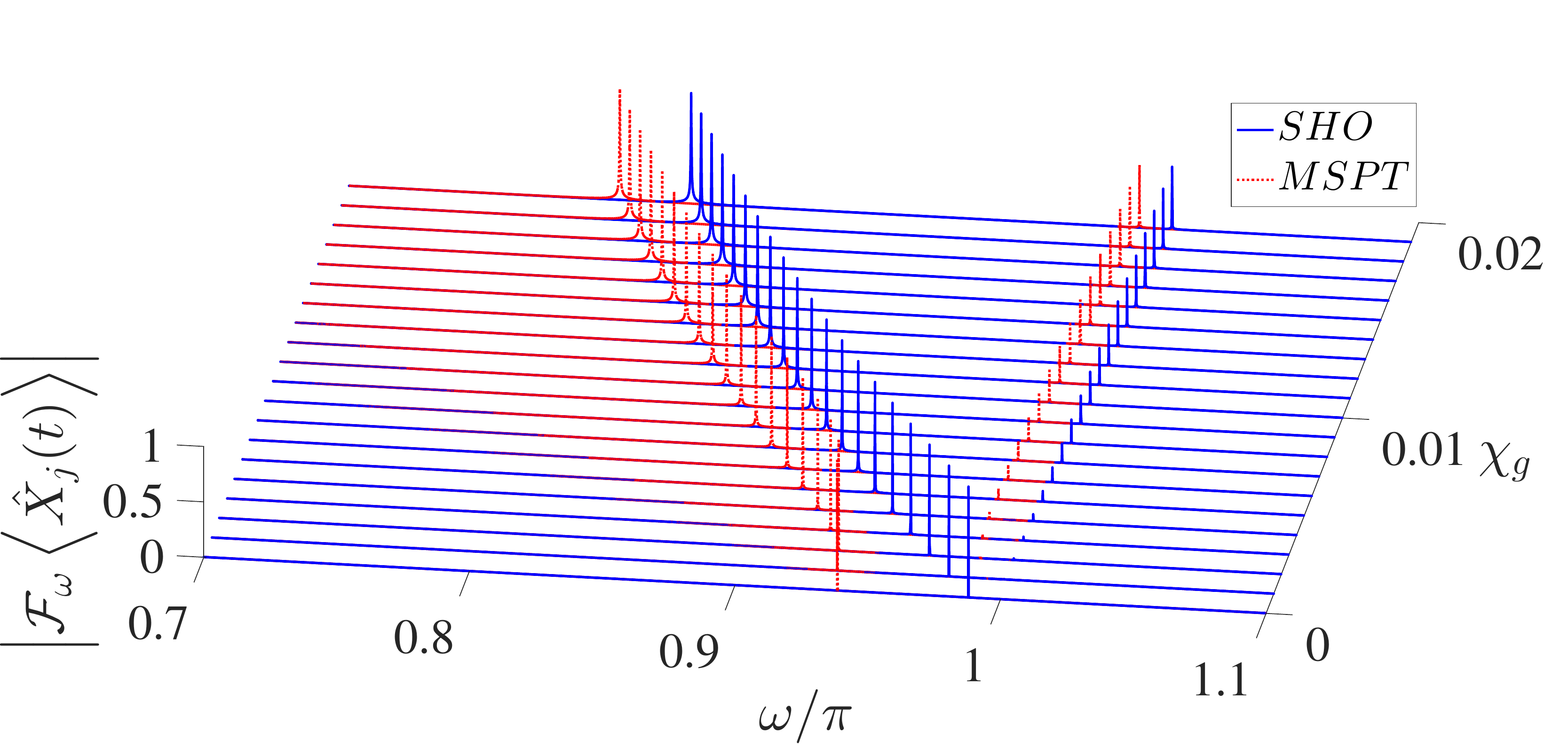}%
}\\
\subfloat[\label{subfig:FXtXj005XrXl1Em3IC11View2}]{%
\includegraphics[scale=0.45]{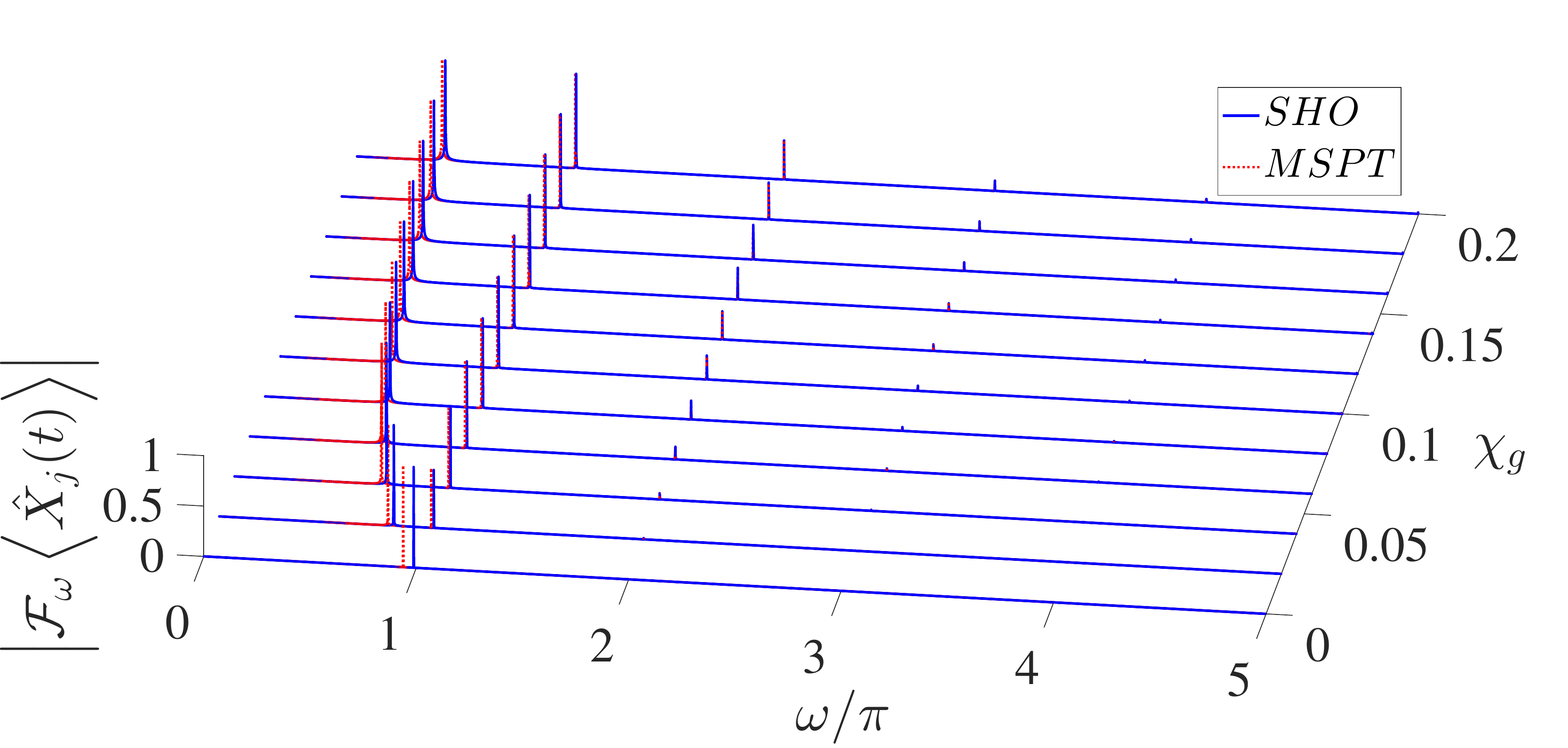}%
}  
\caption{Fourier transform of $\braket{\hat{X}_j(t)}$ from linear solution (red dashed) and MSPT (blue solid) for $\chi_j=0.05$, $\chi_R=\chi_L=0.001$, $\mathcal{E}_j=50\mathcal{E}_c$ and initial state $\ket{\Psi_j(0)}=\frac{\ket{0}_j+\ket{1}_j}{\sqrt{2}}$ as a function of $\chi_g$. The maximum value of $\left|\mathcal{F}_{\omega}\braket{\hat{X}_j(t)}\right|$ at each $\chi_g$ is set to $1$. a) $\chi_g\in[0,0.02]$, $\Delta \chi_g=0.001$. b) $\chi_g\in[0,0.2]$, $\Delta \chi_g=0.02$.}
\label{Fig:FXtXj005XrXl1Em3IC11}
\end{figure}
%%%%%%%%%%%%%%%%%%%%%%%%%%%%%%%%%%%%%%%%%%%%%
\begin{subequations}
\begin{align}
\hat{\bar{p}}_{j}=p_{j}+i\frac{3\varepsilon}{2}\omega_j\left[u_j^4\hat{\bar{\mathcal{H}}}_j(0)e^{-2\alpha_j t}+\sum\limits_{n}2u_j^2u_n^2\hat{\bar{\mathcal{H}}}_n(0)e^{-2\alpha_n t}\right],
\label{eqn:PertCorr-bar(p)_j}
\end{align}
while $p_{n}=-\alpha_n-i\beta_n$ is corrected as
\begin{align}
\begin{split}
\hat{\bar{p}}_{n}=p_{n}+i\frac{3\varepsilon}{2}\omega_j\left[u_n^4\hat{\bar{\mathcal{H}}}_n(0)e^{-2\alpha_n t}+2u_n^2u_j^2\hat{\bar{\mathcal{H}}}_j(0)e^{-2\alpha_j t}+\sum\limits_{m\neq n}2u_n^2u_m^2\hat{\bar{\mathcal{H}}}_m(0)e^{-2\alpha_m t}\right],
\end{split}
\label{eqn:PertCorr-bar(p)_n}
\end{align}
where $\hat{\bar{\mathcal{H}}}_{j}(0)$ and $\hat{\bar{\mathcal{H}}}_{n}(0)$ represent the Hamiltonians of each hybridized mode
\begin{align}
\hat{\bar{\mathcal{H}}}_{l}(0)\equiv \frac{1}{2}\left[\hat{\bar{a}}_{l}^{\dag}(0)\hat{\bar{a}}_{l}(0)+\hat{\bar{a}}_{l}(0)\hat{\bar{a}}_{l}^{\dag}(0)\right], \ l=j,n.
\label{eqn:PertCorr-Def of bar(H)_l(0)}
\end{align}
These are the generalizations of the single quantum Duffing results~(\ref{eqn:QuDuffingNoMem-bar(W)}) and (\ref{eqn:QuDuffingNoMem-H(0)}) and reduce to them as $\chi_g\to 0$ where $u_j=1$ and $u_n=0$. Each hybdridized mode is corrected due to a self-Kerr term proportional to $u_l^4$, and cross-Kerr terms proportional to $u_l^2u_{l'}^2$. Contributions of the form $u_l^2u_{l'}u_{l''}$ \cite{Bourassa_Josephson_2012} do not appear up to the lowest order in MSPT. 

In terms of Eqs.~(\ref{eqn:PertCorr-bar(p)_j}-\ref{eqn:PertCorr-bar(p)_n}), the MSPT solution reads
\end{subequations}
\begin{align}
\begin{split}
\hat{\mathcal{X}}_j^{(0)}(t)&=\frac{\hat{\mathcal{A}}_j(0)e^{\hat{\bar{p}}_j t}+e^{\hat{\bar{p}}_j t}\hat{\mathcal{A}}_j(0)}{2\cos\left(\frac{3\omega_j}{4}u_j^4\varepsilon t e^{-2\alpha_j t}\right)}+H.c.\\
&+\sum\limits_n\left[\frac{\hat{\mathcal{A}}_n(0)e^{\hat{\bar{p}}_n t}+e^{\hat{\bar{p}}_n t}\hat{\mathcal{A}}_n(0)}{2\cos\left(\frac{3\omega_j}{4}u_n^4\varepsilon t  e^{-2\alpha_n t}\right)}+H.c.\right],
\label{eqn:PertCorr-X^(0)(t) MSPT Sol}
\end{split}
\end{align}
where $\hat{\mathcal{A}}_{j,n}$ is defined in Eq.~(\ref{eqn:Lin Sol Mathcal(X)_j(t)}). In Fig.~\ref{Fig:FXtXj005XrXl1Em3IC11}, we have compared the Fourier transform of $\braket{\hat{\mathcal{X}}_j(t)}$ calculated both for the MSPT solution~(\ref{eqn:PertCorr-X^(0)(t) MSPT Sol}) and the linear solution~(\ref{eqn:Lin Sol X_j(t)}) for initial condition $\ket{\Psi(0)}=\frac{\ket{0}_j+\ket{1}_j}{\sqrt{2}}\otimes \ket{0}_{ph}$ as a function of $\chi_g$. At $\chi_g=0$, we notice the bare $\mathcal{O}(\varepsilon)$ nonlinear shift of a free Duffing oscillator as predicted by Eq.~(\ref{eqn:PertCor-ClDuffing FreqRenorm}). As $\chi_g$ is increased, the predominantly self-Kerr nonlinearity on the qubit is gradually passed as cross-Kerr contributions to the resonator modes, as observed from the frequency renormalizations~(\ref{eqn:PertCorr-bar(p)_j}) and (\ref{eqn:PertCorr-bar(p)_n}). As a result of this, interestingly, the effective nonlinear shift in the transmon resonance becomes smaller and saturates at stronger couplings. In other words, the transmon mode becomes {\it more linear} at {\it stronger} coupling $\chi_g$. This counterintuitive result can be understood from Eq.~(\ref{eqn:PertCorr-bar(p)_j}). For initial condition considered here, the last term in Eq.~(\ref{eqn:PertCorr-bar(p)_j}) vanishes, while one can see from Fig.~\ref{Fig:QuDuffQuHarm-u} that $u_j<1$ for $\chi_g>0$.

\subsection{Numerical simulation of reduced equation}
\label{Sec:NumSimul}
The purpose of this section is to compare the results from MSPT and linear theory to a pure numerical solution valid up to $\mathcal{O}(\varepsilon^2)$. A full numerical solution of the Heisenberg equation of motion~(\ref{eqn:Eff Dyn before trace}) requires matrix representation of the qubit operator $\hat{\mathcal{X}}_j(t)$ over the entire Hilbert space, which is impractical due to the exponentially growing dimension. We are therefore led to work with the reduced Eq.~(\ref{eqn:NL SE Problem}). While the nonlinear contribution in Eq.~(\ref{eqn:NL SE Problem}) cannot be traced exactly, it is possible to make progress perturbatively. We substitute the perturbative expansion Eq.~(\ref{eqn:Expansion of Sine}) into Eq.~(\ref{eqn:NL SE Problem}):
\begin{align}
\begin{split}
&\hat{\ddot{X}}_j(t)+\omega_j^2\left[1-\gamma+i\mathcal{K}_1(0)\right]\left[\hat{X}_j(t)-\varepsilon \Tr_{ph}{\{\hat{\rho}_{ph}(0)\hat{\mathcal{X}}_j^3(t)\}}\right]\\
&=-\int_{0}^{t}dt'\omega_j^2 \mathcal{K}_2(t-t')[\hat{X}_j(t')-\varepsilon \Tr_{ph}{\{\hat{\rho}_{ph}(0)\hat{\mathcal{X}}_j^3(t')\}}],
\label{eqn:NumSim-QuDuffingOscMem}
\end{split}
\end{align}
with $\varepsilon\equiv\frac{\sqrt{2}}{6}\epsilon$. If we are interested in the numerical results only up to $\mathcal{O}(\varepsilon^2)$ then the cubic term can be replaced as
\begin{align}
\begin{split}
\varepsilon\hat{\mathcal{X}}_j^3(t)=\varepsilon\left[\left.\hat{\mathcal{X}}_j(t)\right|_{\varepsilon=0}\right]^3+\mathcal{O}\left(\varepsilon^2\right).
\end{split}
\end{align}
Since we know the linear solution~(\ref{eqn:Lin Sol Mathcal(X)_j(t)}) for $ \hat{\mathcal{X}}_j(t)$ analytically, the trace can be performed directly (see App.~\ref{App:RedNumEq}). We obtain the reduced equation in the Hilbert of transmon as
\begin{align}
\begin{split}
&\hat{\ddot{X}}_j(t)+\omega_j^2\left[1-\gamma+i\mathcal{K}_1(0)\right]\left[\hat{X}_j(t)-\varepsilon\hat{X}_j^3(t)\right]\\
&=-\int_{0}^{t}dt'\omega_j^2 \mathcal{K}_2(t-t')\left[\hat{X}_j(t')-\varepsilon\hat{X}_j^3(t')\right]+\mathcal{O}(\varepsilon^2).
\label{eqn:NumSim-QuDuffingOscMemReduced}
\end{split}
\end{align}  
Solving the integro-differential Eq.~(\ref{eqn:NumSim-QuDuffingOscMemReduced}) numerically is a challenging task, since the memory integral on the RHS requires the knowledge of all results for $t'<t$. Therefore, simulation time for Eq.~(\ref{eqn:NumSim-QuDuffingOscMemReduced}) grows polynomially with $t$. The beauty of the Laplace transform in the linear case is that it turns a memory contribution into an algebraic form. However, it is inapplicable to Eq.~(\ref{eqn:NumSim-QuDuffingOscMemReduced}). 
%%%%%%%%%%% Fig of TimeDynamics %%%%%%%%%%%%%%%%%%%
\begin{figure}
\subfloat[\label{subfig:XtQuDuffingEps003Xg0XrXl0005}]{%
\includegraphics[scale=0.50]{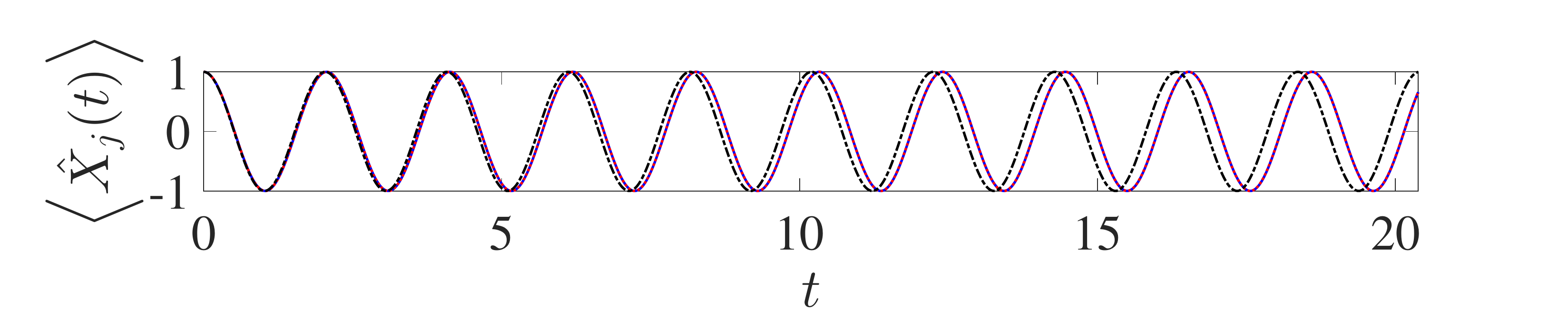}%
}\\ 
\subfloat[\label{subfig:XtQuDuffingEps003Xg001XrXl0005}]{%
\includegraphics[scale=0.50]{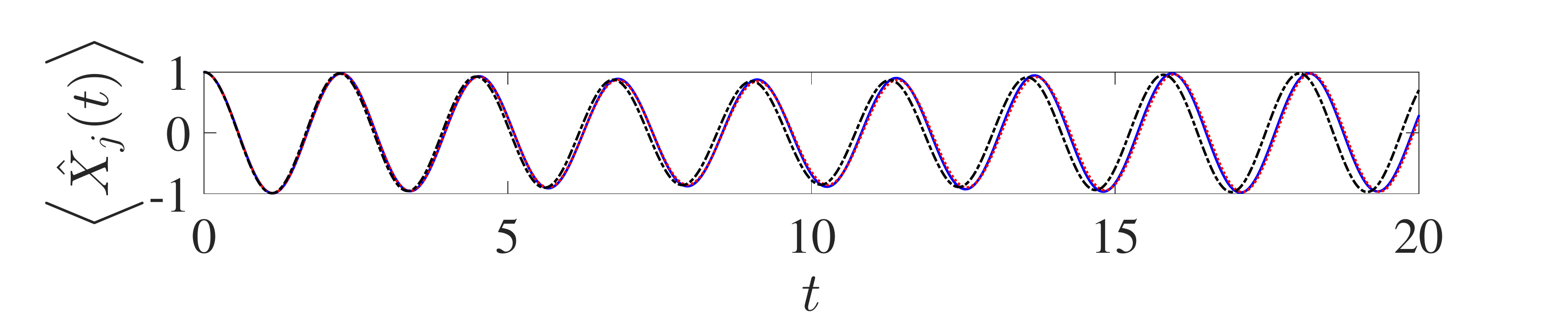}%
}
\\
\subfloat[\label{subfig:XtQuDuffingEps003Xg01XrXl0005}]{%
\includegraphics[scale=0.50]{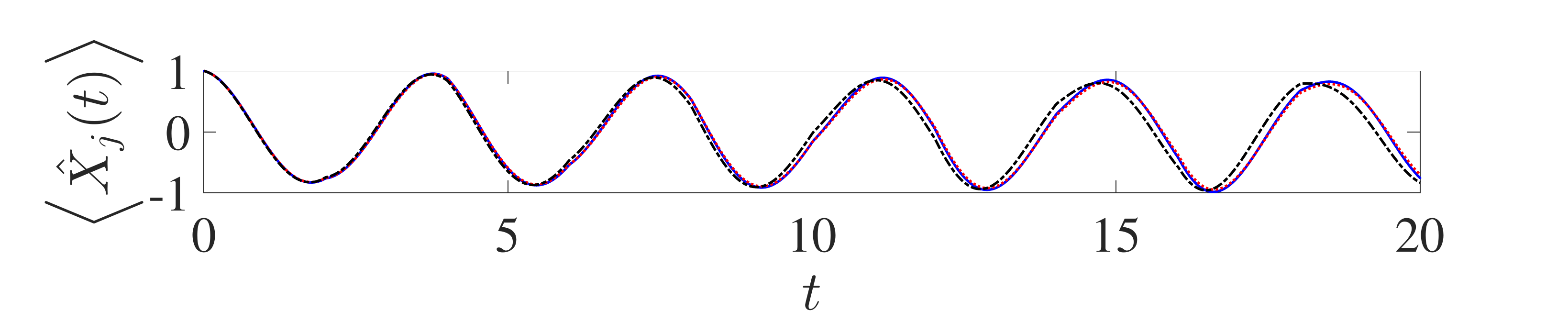}%
}
\\
\subfloat[\label{subfig:XtQuDuffingEps003Xg02XrXl0005}]{%
\includegraphics[scale=0.50]{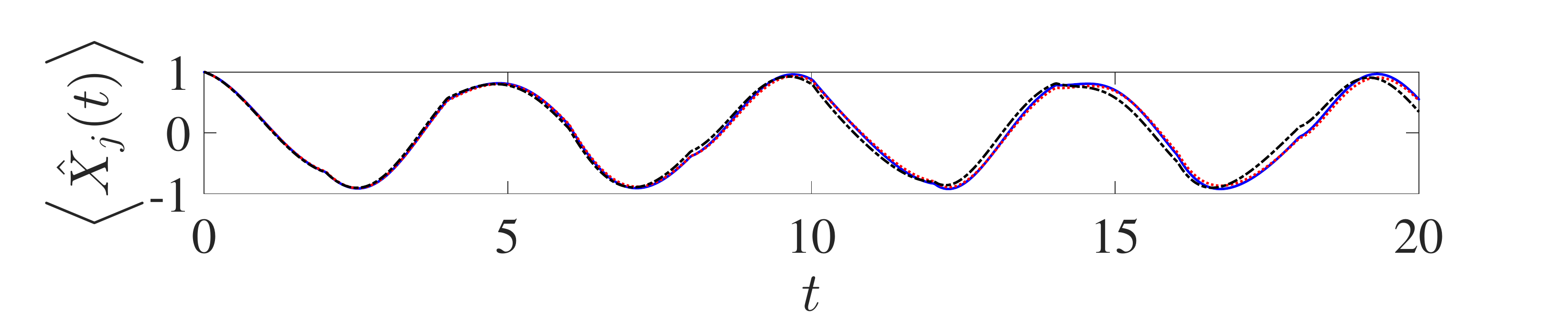}%
}
\caption{Comparison of short-time dynamics between the results from linear theory (black dash-dot), MSPT (red dotted) and numerical (blue solid) of $\braket{\hat{X}_j(t)}$ for the same parameters as in Fig.~(\ref{Fig:FXtXj005XrXl1Em3IC11}) and for a) $\chi_g=0$, b) $\chi_g=0.01$, c) $\chi_g=0.1$ and d) $\chi_g=0.2$. The oscillation frequency and decay rate of the most dominant pole (transmon-like) are controlled by the hybridization strength. For a) where $\chi_g=0$, there is no dissipation and the transmon is isolated. The decay rate increases with $\chi_g$ such that the Q-factor for the transmon-like resonance reaches $Q_j\equiv\beta_j/\alpha_j \approx 625.3$ in Fig. d).} 
\label{Fig:TimeDynAnalVsODE45}
\end{figure}
%%%%%%%%%%%%%%%%%%%%%%%%%%%%%%%%%%%%%%%%%%%%

In Fig.~\ref{Fig:TimeDynAnalVsODE45}, we compared the numerical results to both linear and MSPT solutions up to 10 resonator round-trip times and for different values of $\chi_g$. For $\chi_g=0$, the transmon is decoupled and behaves as a free Duffing oscillator. This corresponds to the first row in Fig.~(\ref{subfig:FXtXj005XrXl1Em3IC11View1}) where there is only one frequency component and MSPT provides the correction given in Eq.~(\ref{eqn:QuDuffingNoMem-bar(W)}). As we observe in Fig.~\ref{subfig:XtQuDuffingEps003Xg0XrXl0005} the MSPT results lie on top of the numerics, while the linear solution shows a visible lag by the 10th round-trip. Increasing $\chi_g$ further, brings more frequency components into play. As we observe in Fig.~\ref{Fig:FXtXj005XrXl1Em3IC11}, for $\chi_g=0.01$ the most resonant mode of the resonator has a non-negligible $u_1$. Therefore, we expect to observe weak beating in the dynamics between this mode and the dominant transmon-like resonance, which is shown in Fig.~\ref{subfig:XtQuDuffingEps003Xg001XrXl0005}. Figures~\ref{subfig:XtQuDuffingEps003Xg01XrXl0005} and \ref{subfig:XtQuDuffingEps003Xg02XrXl0005} show stronger couplings where many resonator modes are active and a more complex beating is observed. In all these cases, the MSPT results follow the pure numerical results more closely than the linear solution confirming the improvement provided by perturbation theory. 
\subsection{System output}
\label{Sec:SysOutput}

%%%%%%%%%%%% Fig of Fourier of Xcav%%%%%%%%%%
\begin{figure}[h!]
\centering
\subfloat[\label{subfig:FXcavXj005XrXl1Em3IC11View1}]{%
\includegraphics[scale=0.45]{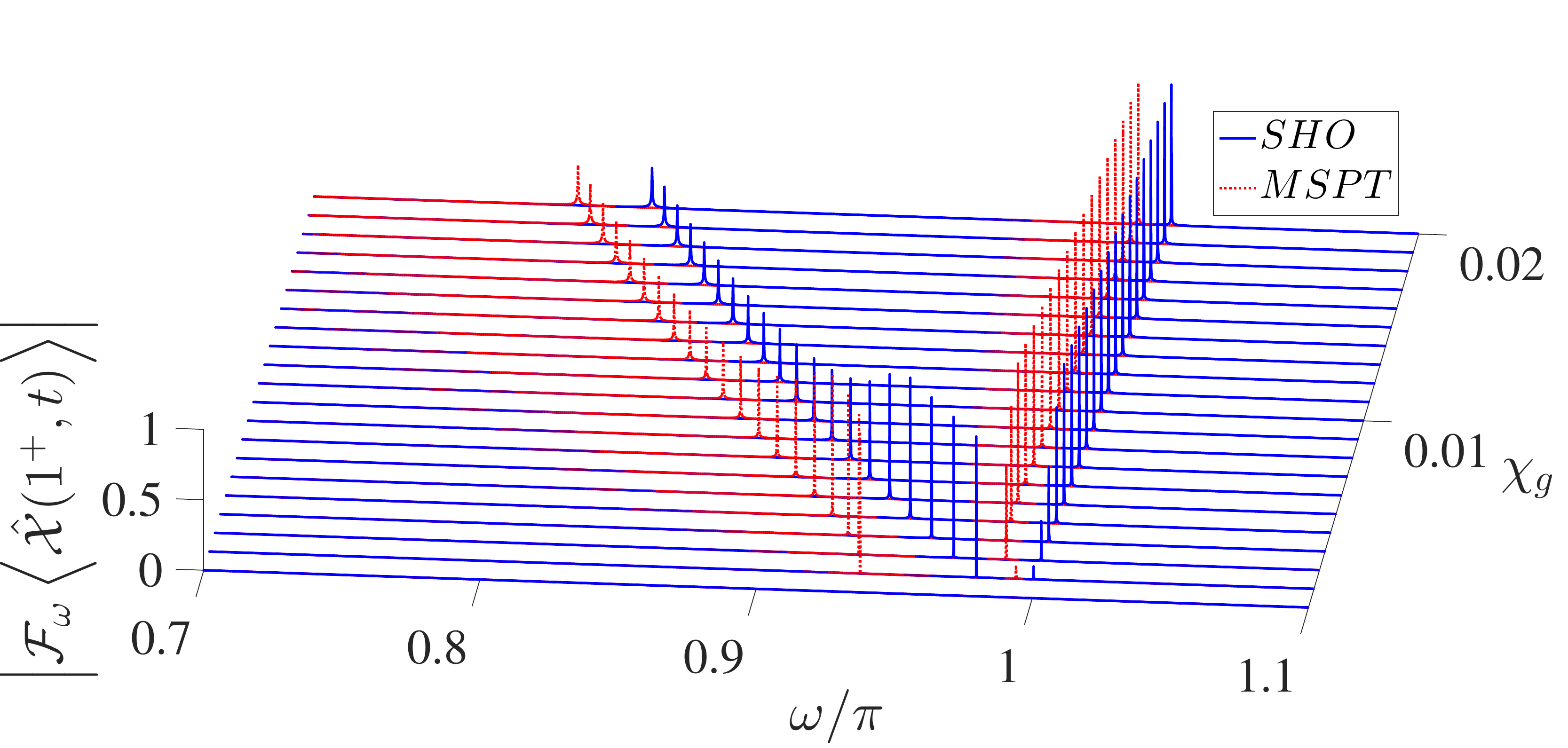}%
}\\ 
\subfloat[\label{subfig:FXcavXj005XrXl1Em3IC11View2}]{%
\includegraphics[scale=0.45]{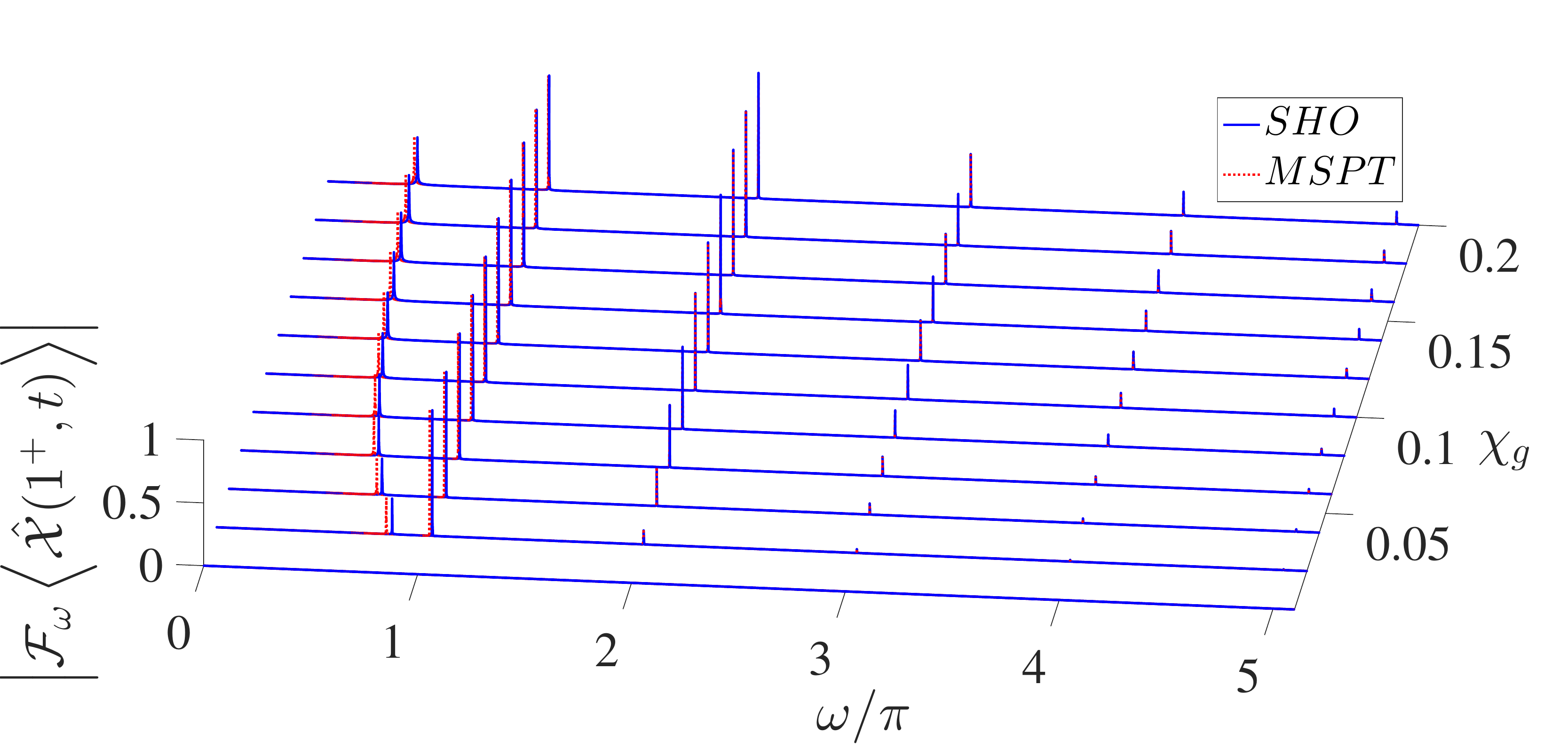}%
}  
\caption{Fourier transform of $\braket{\hat{\mathcal{X}}(1^+,t)}$ for the linear solution (red dashed) and the MSPT (blue solid) for the same parameters as in Fig.~\ref{Fig:FXtXj005XrXl1Em3IC11}. The maximum value of $\left|\mathcal{F}_{\omega}\braket{\hat{X}_j(t)}\right|$ at each $\chi_g$ is set to $1$.}
\label{Fig:FXcavXj005XrXl1Em3IC11}
\end{figure}
%%%%%%%%%%%%%%%%%%%%%%%%%%%%%%%%%%%%%%%%%%%%%
Up to this point, we studied the dynamics of the spontaneous emission problem in terms of one of the quadratures of the transmon qubit, i.e. $\braket{\hat{\mathcal{X}}_j(t)}$. In a typical experimental setup however, the measuarable quantities are the quadratures of the field outside the resonator \cite{Clerk_Introduction_2010}. We devote this section to the computation of these quantities. 

The expression of the fields $\hat{\varphi}(x,t)$ can be directly inferred from the solution of the inhomogeneous wave Eq.~(\ref{eqn:Res Dyn}) using the impulse response (GF) defined in Eq.~\ref{Eq:Def of G(x,t|x0,t0)}. We note that this holds irrespective of whether one is solving for the classical or as is the case here, for the quantum fields. Taking the expectation value of this solution (App.~\ref{SubApp:Eff Dyn of transmon}) with respect to the initial density matrix~(\ref{eqn:SE-IC}) we find
\begin{align}
\braket{\hat{\varphi}(x,t)}=\chi_s\omega_j^2\int_{0}^{t}dt' G(x,t|x_0,t')\braket{\sin[\hat{\varphi}_j(t')]}.
\label{eqn:<varphi(x,t)>-SysOutput}
\end{align} 
Dividing both sides by $\phi_{\text{zpf}}$ and keeping the lowest order we obtain the resonator response as
\begin{align}
\braket{\hat{\mathcal{X}}^{(0)}(x,t)}=\chi_s\omega_j^2\int_{0}^{t}dt' G(x,t|x_0,t')\braket{\hat{\mathcal{X}}_j^{(0)}(t')},
\label{eqn:<varphi(x,t)>-SysOutput}
\end{align} 
where $\hat{\mathcal{X}}_j^{(0)}(t)$ is the lowest order MSPT solution~(\ref{eqn:PertCorr-X^(0)(t) MSPT Sol}), which takes into account the frequency correction to $\mathcal{O}(\varepsilon)$. Taking the Laplace transform decouples the convolution 
\begin{align}
\braket{\hat{\tilde{\mathcal{X}}}^{(0)}(x,s)}=\chi_s\omega_j^2\tilde{G}(x,x_0,s)\braket{\hat{\tilde{\mathcal{X}}}_j^{(0)}(s)},
\label{eqn:<varphi(x,t)>-SysOutput}
\end{align} 
which indicates that the resonator response is filtered by the GF. 

Figure~\ref{Fig:FXcavXj005XrXl1Em3IC11} shows the field outside the right end of the resonator, $\braket{\hat{\tilde{\mathcal{X}}}(x=1^+,s=i\omega)}$, in both linear and lowest order MSPT approximations. This quadrature can be measured via heterodyne detection \cite{Bishop_Nonlinear_2009}. Note that the hybridized resonances are the same as those of $\braket{\hat{\mathcal{X}}_j(t)}$ shown in Fig.~\ref{Fig:FXtXj005XrXl1Em3IC11}. What changes is the relative strength of the residues. The GF has poles at the bare cavity resonances and therefore the more hybridized a pole is, the smaller its residue becomes. 
\section{Conclusion}
\label{Sec:Conclusion}
In this chapter, we introduced a new approach for studying the effective non-Markovian Heisenberg equation of motion of a transmon qubit coupled to an open multimode resonator beyond rotating wave and two level approximations. The main motivation to go beyond a two level representation lies in the fact that a transmon is a weakly nonlinear oscillator. Furthermore, the information regarding the electromagnetic environment is encoded in a single function, i.e. the electromagnetic GF. As a result, the opening of the resonator is taken into account analytically, in contrast to the Lindblad formalism where the decay rates enter only phenomenologically.

We applied this theory to the problem of spontaneous emission as the simplest possible example. The weak nonlinearity of the transmon allowed us to solve for the dynamics perturbatively in terms of $(\mathcal{E}_c/\mathcal{E}_j)^{1/2}$ which appears as a measure of nonlinearity. Neglecting the nonlinearity, the transmon acts as a simple harmonic oscillator and the resulting linear theory is exactly solvable via Laplace transform. By employing Laplace transform, we avoided Markov approximation and therefore accounted for the exact hybridization of transmon and resonator resonances. Up to leading nonzero order, the transmon acts as a quantum Duffing oscillator. Due to the hybridization, the nonlinearity of the transmon introduces both self-Kerr and cross-Kerr corrections to all hybdridized modes of the linear theory. Using MSPT, we were able to obtain closed form solutions in Heisenberg picture that do not suffer from secular behavior. A direct numerical solution confirmed the improvement provided by the perturbation theory over the harmonic theory. Surprisingly, we also learned that the linear theory becomes more accurate for stronger coupling since the nonlinearity is suppressed in the qubit-like resonance due to being shared between many hybdridized modes. 

The theory developed here illustrates how far one can go without the concept of photons. Many phenomena in the domain of quantum electrodynamics, such as spontaneous or stimulated emission and resonance fluorescence, have accurate semiclassical explanations in which the electric field is treated classically while the atoms obey the laws of quantum mechanics. For instance, the rate of spontaneous emission can be related to the local density of electromagnetic modes in the weak coupling limit. While it is now well understood that the electromagnetic fluctuations are necessary to start the spontaneous emission process \cite{Haake_Delay-time_1981}, it is important to ask to what extent the quantum nature of electromagnetic field effects the qubit dynamics \cite{scully_concept_1972}. We find here that although the electromagnetic degrees of freedom are integrated out and the dynamics can systematically be reduced to the Hilbert space of the transmon, the quantum state of the electromagnetic environment reappears in the initial and boundary conditions when computing observables.  
\chapter{Cut-off free circuit quantum electrodynamics}
\label{Ch-OriginOfA2}

Any discrete quantum system coupled to the continuum of modes experiences radiative decay and shift of its energy levels. When coupled to a cavity, these quantities can be significantly changed with respect to their values in vacuum. Conventionally, this modification was accurately incorporated by including only the closest resonant mode of the cavity. In the circuit quantum electrodynamics architecture, where the coupling strengths can be substantial, these rates are strongly influenced by far off-resonant modes. A multimode calculation accounting for the infinite set of cavity modes leads to a divergent result unless an artificial cutoff is imposed manually. It has so far not been identified what the source of this divergence is. We show that by placing an atom into a cavity, the electromagnetic modal structure of the cavity is modified. This modification can be understood as the atom acts as a scattering center such that bends the electromagnetic environment around itself. In cavity QED, this scattering phenomenon is captured by the appearance of a diamagnetic term, known as the $A^2$ contribution, due to the atom. Although in atomic cavity QED, the resulting modification in the eigenmodes is negligible, in recent superconducting circuit realizations, such corrections can be observable and may have qualitative implications. By treating the contribution of $A^2$ exactly, we account for the gauge invariance of interaction and conservation of current in these circuits as a result. We show here that unless gauge invariance is respected, any attempt at the calculation of circuit QED quantities is bound to diverge. Moreover, we revisit the canonical quantization procedure of a circuit QED system consisting of a single superconducting transmon qubit coupled to a multimode superconducting microwave resonator, where we introduce a complete set of modes that properly conserves the current in the entire structure. An effective multimode Rabi  model is derived with coefficients that are given in terms of circuit parameters. Finally, we apply our Heisenberg-Langevin approach to the calculation of a finite spontaneous emission rate and the Lamb shift that is free of cutoff.   

\section{Introduction}

An atom-like degree of freedom coupled to the continuum of electromagnetic (EM) modes spontaneously decays. When the atom is confined in a resonator, the emission rate can be modified with respect to its value in free space, depending on the EM local density of states at the atomic position \cite{Kleppner_Inhibited_1981, Goy_Observation_1983, Hulet_Inhibited_1985, Jhe_Suppression_1987}, which is called the Purcell effect \cite{Purcell_Resonance_1946}. An accompanying effect is the Lamb shift, a radiative level shift first observed in the microwave spectroscopy of the hydrogen $^2P_{1/2} -~^2S_{1/2}$ transition \cite{Lamb_Fine_1947}. These quantities have been experimentally accurately characterized for superconducting Josephson junction (JJ) based qubits coupled to coplanar transmission lines \cite{Fragner_Resolving_2008, Houck_Controlling_2008} and three-dimensional resonators \cite{Nigg_BlackBox_2012}. In the dispersive regime where a qubit with transition frequency $\omega_j$ is far-detuned from the nearest resonant cavity mode (frequency $\nu_r$, loss $\kappa_r$), the Purcell decay rate is $\gamma_P = (g/\delta)^2 \kappa_r$ and the Lamb shift is $\Delta_L = g^2 / \delta$. These well-known approximate estimates are often used in analyzing qubit state read-out, hence we employ them to benchmark our results. Here $g$ denotes the coupling between the qubit and the cavity mode and $\delta=\omega_j - \nu_r$ denotes their detuning \cite{Boissonneault_Dispersive_2009}. However, for large couplings accessible in circuit QED, the single-mode approximation is often inaccurate \cite{Fragner_Resolving_2008, Houck_Controlling_2008}. In particular, due to particular boundary conditions imposed by the capacitive coupling of a resonator to external waveguides, the qubit relaxation time is limited by the EM modes that are far-detuned from the qubit frequency \cite{Houck_Controlling_2008}. Similarly the measured Lamb shift in the dispersive regime can only be accurately fit with an extended Jaynes-Cummings (JC) model including several modes and qubit levels \cite{Fragner_Resolving_2008}. A generalization of the Purcell rate can be found 
\begin{equation}
\gamma_P=\sum_n(g_n/\delta_n)^2 \, \kappa_n,
\label{eqn:Multimode Purcell Rate} 
\end{equation}
where $g_n$ is the coupling strength to mode $n$ and $\delta_n = \omega_j - \nu_n$ is a detuning from resonator mode $n$ with frequency $\omega_n$ and decay rate $\kappa_n$. This expression is divergent without imposing a high-frequency cutoff \cite{Houck_Controlling_2008}. Divergences appear as well in the Lamb shift and other vacuum-induced phenomena, e.g. photon-mediated qubit-qubit interactions \cite{Filipp_Multimode_2011}. These divergences are neither specific to the dispersive limit nor to the calculational scheme used to compute QED quantities. This issue is well-known for the Lamb shift \cite{Lamb_Fine_1947}, but less noted for the spontaneous emission rate. Indeed, free space spontaneous emission rate diverges as well, as we show in App.~\ref{App:WignerWeisskopf}. The finite result by Wigner and Weisskopf \cite{Weisskopf_Berechnung_1930, Scully_Quantum_1997} is due to Markov approximation which filters out the ultraviolet divergence. Recent generalizations of the Wigner-Weisskopf approach imposes an artificial cut-off to obtain a finite result \cite{Krimer_Route_2014}. No satisfactory theoretical explanation has been given for these divergences. 
%%%%%%%%%%%%%%%%%%%%%%%%%%%%%%%%%%%%%%%%

In single mode realization of cavity QED (CQED), a single atom coupled to a small high-Q electromagnetic resonator can be well-described by a model wherein the matter is described by a single atomic transition, and its coupling to one of the modes of the resonator can saturate this transition before other modes are populated \cite{Haroche_Exploring_2013}. A plethora of fundamental physical phenomena and their recent applications in quantum information science has been explored and vigorously pursued with a superconducting circuit-based realization of this setup \cite{Blais_Cavity_2004, Wallraff_Strong_2004, Chiorescu_Coherent_2004, Devoret_Implementing_2004, Girvin_Circuit_2009, Devoret_Superconducting_2013}. In such systems, the existence of the atom leads to a modification in the cavity modal structure due to Rayleigh-like scattering. Such corrections are unobservably small in atomic CQED unless a special cavity structure is chosen. In recent realizations of circuit quantum electrodynamics (cQED), however, such corrections may have observable consequences which we discuss in this chapter. 

A well known manifestation of the aforementioned scattering corrections is the so-called $A^2$ term in CQED literature. There has been a lively debate in recent years \cite{Nataf_No-Go_2010, Viehmann_Superradiant_2011, Hayn_Superradiant_2012, Vukics_Adequacy_2012, Baksic_Superradiant_2013, Zhang_Quantum_2014} about the impact of this term on synthetic realizations of the single mode superradiant phase transition \cite{Hepp_Superradiant_1973, Wang_Phase_1973, Carmichael_Higher_1973} when instead of one, $N$ identical non-interacting quantum dipoles are coupled with an identical strength to a single cavity mode. This particular instability of the electromagnetic vacuum has originally been discussed \cite{Hepp_Superradiant_1973, Wang_Phase_1973, Carmichael_Higher_1973} within the context of the single mode version of the Dicke model \cite{Dicke_Coherence_1954} where the $A^2$ term was not included. Subsequent work shortly thereafter \cite{Rzazewski_Phase_1975, Knight_Are_1978, Bialynicki-Birula_No-Go_1979} pointed out that the $A^2$ term rules out such a transition. Recent theoretical work on superconducting realizations of the Dicke Model \cite{Nataf_No-Go_2010} has challenged the validity of such "no-go" theorems \cite{Bialynicki-Birula_No-Go_1979}. Leaving this contentious matter aside~\cite{Viehmann_Superradiant_2011, Ciuti_Comment_2012}, we note here that the $A^2$ term is a gauge-dependent object, and specifically appears in the Coulomb gauge description of the single mode atomic CQED. However, the scattering corrections due to the existence of an atom in a cavity are physical and measurable, and hence not dependent on the choice of gauge. In fact, recent realizations \cite{Sundaresan_Beyond_2015} of the multimode strong coupling regime in a very long coplanar waveguide cavity, as well as cQED systems in the ultra-strong coupling regime \cite{Niemczyk_Circuit_2010, Peropadre_Switchable_2010} provide settings where such corrections may be observable. 

We show that respecting the gauge-inavriance of interactions in a cQED system translates as a modified capacitance per unit length for the resonator which is locally altered at the position of the qubit. This impurity scattering term is typically neglected \cite{Koch_Time-reversal_2010, Nunnenkamp_Synthetic_2011, Schmidt_Circuit_2013, Devoret_Quantum_2014} in the derivation of the quantized Hamiltonian for the multimode regime of cQED \cite{Krimer_Route_2014}. However, we show that, within the framework of circuit QED~\cite{Devoret_Quantum_1995}, finite expressions arise when gauge invariance of the circuit is respected. We focus here on a transmon qubit \cite{Koch_Charge_2007} coupled to an open transmission-line resonator, but our results should be valid for other types of one-dimensional open EM environments. 

The structure of this chapter is as follows: In Sec.~\ref{Sec:A^2-GaugeInvRabiModel}, we revisit the quantization of a closed cQED system consisting of a transmon qubit coupled to a closed superconducting coplanar resonator. We discuss how the qubit changes the propagation properties of the resonator and how as a result this modifies its eigenmodes and eigenfrequencies. We show in particular that this new basis is the one which properly fulfills current conservation law at the point of connection to the qubit. In terms of this current conserving basis we derive a renormalized Rabi model that respects gauge invariance of the circuit. In Sec.~\ref{Sec:A^2-OpenGen} we show how the quantization procedure can be generalized to an open cavity, one that is connected capacitively to external waveguides. In Sec.~\ref{Sec:A2-ConvergentPurcellLamb}, we show that finite values for QED quantities are obtained if gauge invariance of the circuit is fully incorporated. In particular, we revisit the dispersive limit calculation of Purcell decay rate and the Lamb shift in terms of our gauge invariant Rabi model and provide finite values for such quantities. More importantly, we apply our Heisenberg-Langevin framework on calculating these QED quantities and show its improvements with respect to the dispersive JC calculation. Finally, in Sec.~\ref{Sec:A^2-Discussion}, we briefly discuss the comparison to the case of atomic CQED, where we point out that including the $A^2$ term in the Hamiltonian will lead to the same type of modification in the modes of a cavity. A more detailed calculation is provided in App.~\ref{App:DerivationOfHamCQED}.

\section{Gauge Invariant Rabi Model}
\label{Sec:A^2-GaugeInvRabiModel}

\begin{figure}[t!]
\centering
\subfloat[\label{Fig:cQEDClosed-Schematic}]{%
\includegraphics[scale=0.23]{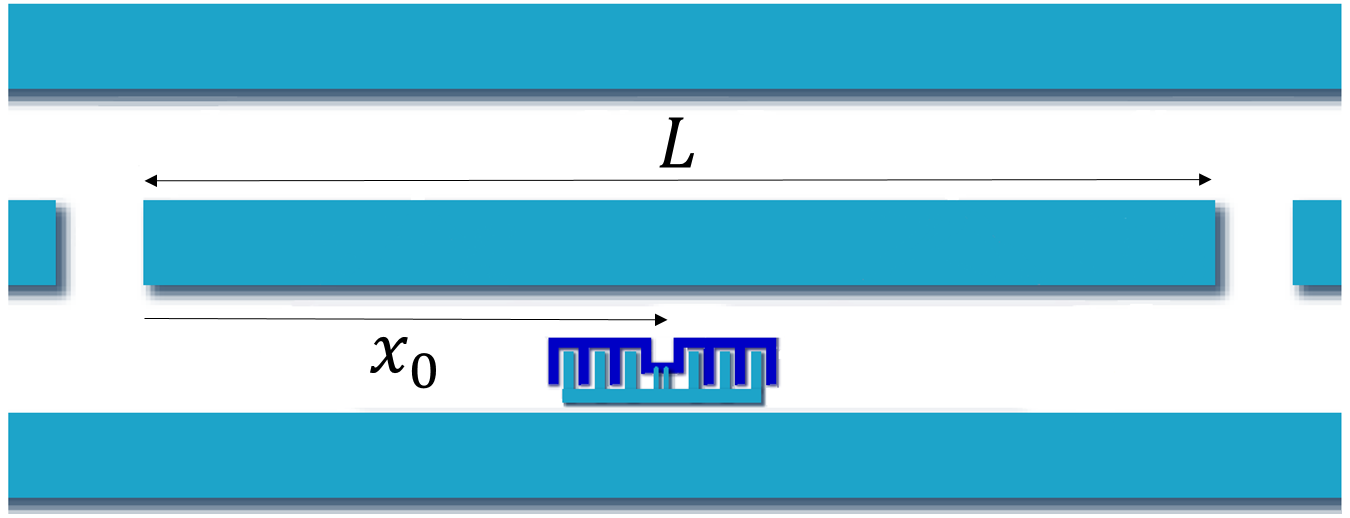}%
}\hfill
\subfloat[\label{Fig:cQEDOpen-EqCircuit}]{%
\includegraphics[scale=0.42]{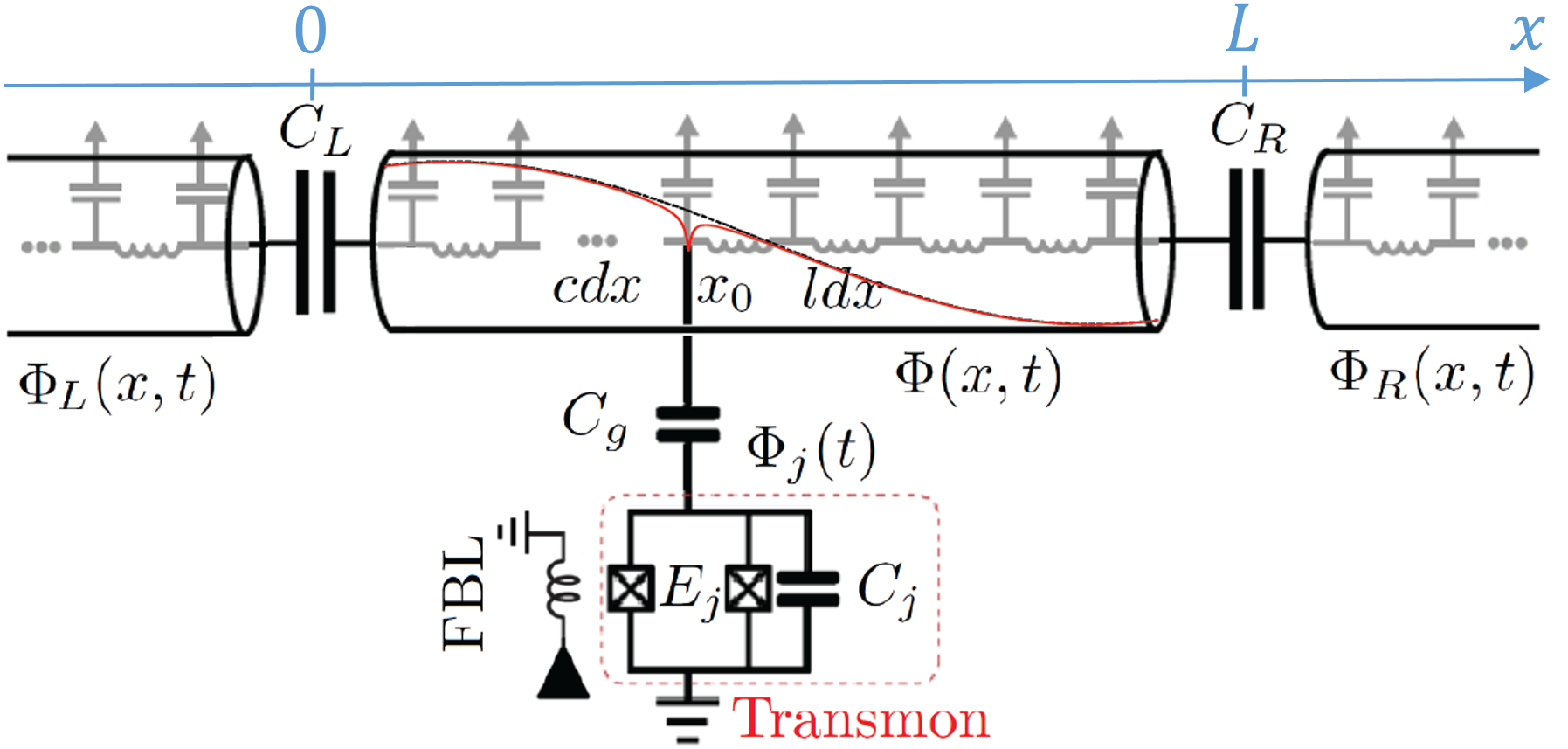}%
}
\caption{A transmon qubit coupled to an open superconducting resonator. a) Device view. b) Equivalent circuit. Transmon is described by the flux variable $\Phi_j(t)$, while resonator is assigned a continuous flux field $\Phi(x,t)$. The black dashed line is a cartoon of the fundamental bare mode of the resonator, while the red solid curve represents the modified resonator mode.}
\label{Fig:A2-cQEDclosed} 
\end{figure}

In this section, we discuss the derivation of a Rabi model that respects the gauge invariance and hence conservation of current at the point of connection to the qubit. We consider a common cQED design \cite{Koch_Charge_2007} consisting of a transmon qubit capacitively coupled to a coplanar resonator that is coupled at both ends to semi-infinite waveguides (\Fig{Fig:A2-cQEDclosed}). Here, we assume that $C_{L,R} = 0$, which corresponds to closed (perfectly reflecting) boundary conditions at $x = 0, L$. The discussion of the open case is given in Sec.~\ref{Sec:A^2-OpenGen}.  
%%%%%%%%%%%%%%%Figure of circuit%%%%%%%%%%%%%%%%%%
%%%%%%%%%%%%%%%%%%%%%%%%%%%%%%%%%%%%%%%%%%%%%
\subsection{Classical Hamiltonian and CC basis}
\label{Sec:A2-ClassHam&CCBasis}

We assign flux variables to each node in the circuit, $\Phi_n (t) = \int^t d\tau \, V_n(\tau)$, with $V_n(t)$ being the instantaneous voltage at node $n$ with respect to the ground node \cite{Devoret_Quantum_1995, Devoret_Quantum_2014}. Fixing the ground amounts to a particular gauge choice \cite{Devoret_Quantum_1995}. For the connection geometry in Fig~\ref{Fig:cQEDOpen-EqCircuit}, the light-matter interaction derives from the energy on the coupling capacitor in the dipole approximation, $T_\text{int} = \frac{1}{2} C_g [\dot{\Phi}(x_0,t) - \dot{\Phi}_j(t)]^2$, with $x_0$ being the qubit position. If from the three terms in its expansion, $T_\text{EM} =  \frac{1}{2} C_g  \dot{\Phi}^2(x_0,t)$, $T_ \text{EM-JJ} = - C_g \dot{\Phi}(x_0,t) \cdot \dot{\Phi}_j(t)$ and $T_\text{JJ} = \frac{1}{2} C_g  \dot{\Phi}_j^2(t)$, only the direct interaction $T_ \text{EM-JJ}$ is kept, a multimode JC model in terms of circuit parameters can be derived, but gives rise to a diverging Purcell rate~(\ref{eqn:Multimode Purcell Rate}). Keeping only the direct interaction $T_ \text{EM-JJ}$ violates gauge invariance. We find that inclusion of all terms, in particular $T_\text{EM}$, equivalent to the diamagnetic $A^2$ term in the minimal coupling Hamiltonian $(\vec{p}_e - e \vec{A})^2/2m_e$ \cite{Malekakhlagh_Origin_2016}, is essential to make all studied QED observables finite.

The complete derivation of the classical Hamiltonian from the classical Lagrangian is given in App.~\ref{App:DerivationOfcQEDHam}. The Hamiltonian for the closed case ($C_{R,L}=0$) reads 
\begin{align}
\begin{split}
\mathcal{H}&=\underbrace{\frac{Q_j^2}{2C_j}-E_J\cos\left(2\pi\frac{\Phi_j}{\Phi_0}\right)}_{\mathcal{H}_A}\\
&+\underbrace{\int_{0}^{L}dx \left[\frac{\rho^2(x,t)}{2c(x,x_0)}+\frac{1}{2l}\left(\frac{\partial \Phi(x,t)}{\partial x}\right)^2\right]}_{\mathcal{H}_C^{mod}} \\
&+\underbrace{\gamma Q_j \int_{0}^{L}dx \frac{\rho(x,t)}{c(x,x_0)} \delta(x-x_0)}_{\mathcal{H}_{int}},
\end{split}
\label{eqn:A2-ModcQEDHam}
\end{align}
where $\mathcal{H}_A$, $\mathcal{H}_C^{\text{mod}}$ and $\mathcal{H}_{\text{int}}$ are the transmon, cavity and interaction Hamiltonians, accordingly. The notation used here follows the canonical approach to quantization of superconducting electrical circuits \cite{Bishop_Circuit_2010, Devoret_Quantum_2014}, briefly reviewed for completeness at the beginning of sec.~\ref{SubSec:Background-circuit quantization}. We translate the results of this section in terms of unitless phase and number operators, defined in table~(\ref{Tab:Def of Pars&Vars}), when we compare our Heisenberg-Langevin formalism to dispersive JC in Sec.~\ref{Sec:A2-ConvergentPurcellLamb}. In Hamiltonian~(\ref{eqn:A2-ModcQEDHam}), the canonical variables $\Phi_j$ and $Q_j$ represent the flux and charge of the transmon qubit, respectively. In a similar manner, the canonical fields $\Phi(x,t)$ and $\rho(x,t)$ are the flux field and charge density field of the transmission line. Furthermore, $\Phi_0\equiv\frac{h}{2e}$ is the flux quantum and $\gamma\equiv \frac{C_g}{C_g+C_j}$ is a capacitive ratio. 

There is a crucial difference between the Hamiltonian we have found here, with respect to earlier treatments \cite{Koch_Charge_2007, Devoret_Quantum_2014}. We do include the modification in the resonator's capacitance per length at the qubit connection point $x_0$, $c(x,x0) = c + C_s \delta(x-x_0)$ where $C_s$ is the series capacitance of $C_j$ and $C_g$ given as $\frac{C_jC_g}{C_j+C_g}$. The Dirac $\delta$-function is the result of treating the qubit as a point object with respect to the resonator, whereas a more realistic model would replace that with a smooth function discussed in App.~\ref{App:FiniteSizeTrans}. As we see shortly, the $\delta$-function appearing in the denominator will not cause any issues in the quantization procedure, since the charge density $\rho(x,t)$ also contains the appropriate information regarding this point object so that $\frac{\rho(x,t)}{c(x,x_0)}$ turns out to be a continuous function in $x$. Once we understand how this correction influences the photonic mode structure of the resonator, we will move on to use that information in the quantization procedure. 

In order to arrive at a second quantized Hamiltonian, we quantize each sector, separately. The Hamiltonian equations of motion derived from $\mathcal{H}_{C}^{mod}$ that includes the impurity scattering term is given by (See App.~\ref{App:DerivationOfcQEDHam}):
\begin{align}
\partial_t \Phi(x,t)&=\frac{\rho(x,t)}{c(x,x_0)}, \\
\partial_t \rho(x,t)&=\frac{1}{l}\partial_x^2 \Phi(x,t).
\end{align}
The solution to these linear equations can be written in terms of the Fourier transform $\Phi(x,t) = \frac{1}{2\pi} \int_{-\infty}^{+\infty} dt \, e^{-i\omega t} \tilde{\Phi}(x,\omega)$, where $\tilde{\Phi}(x,\omega)$ is the solution of the 1D Helmholtz equation
\begin{align}
\left[\partial_x^2 +lc(x,x_0)\omega^2\right]\tilde{\Phi}(x,\omega)=0.
\label{Eq:Modified Wave Equation-Closed Case_body}
\end{align}

We look for solutions that carry zero current across the boundaries, implemented by Neumann-type boundary conditions $\partial_x \tilde{\Phi} (x)|_{x=0,L} = 0$. A solution then exists only at discrete and real values $\omega=\omega_n$. The Dirac $\delta$-function hidden in $c(x,x_0)$ can be translated into discontinuity in $\partial_x \tilde{\Phi}(x)$ which is proportional to the current $\tilde{I}(x,t)=-\frac{1}{l}\partial_x \tilde{\Phi}(x,t)$ that enters and exits the point of connection to the transmon
\begin{align}
-\frac{1}{l}\left.\partial_x \tilde{\Phi}(x,\omega)\right]_{x_0^-}^{x_0^+}=C_s\omega^2 \tilde{\Phi}(x_0,\omega),
\end{align}
where the right hand side is the current that enters $C_j$ through $C_g$ , therefore the series capacitance $C_s$. This condition amounts to the conservation of current at the point of connection to the qubit and thus it is appropriate to call the set of eigenmodes satisfying this condition the {\it current-conserving (CC) basis}. 
\begin{figure}
\centering
\subfloat[\label{subfig:Eigfreqsx001}]{%
\centerline{\includegraphics[scale=0.24]{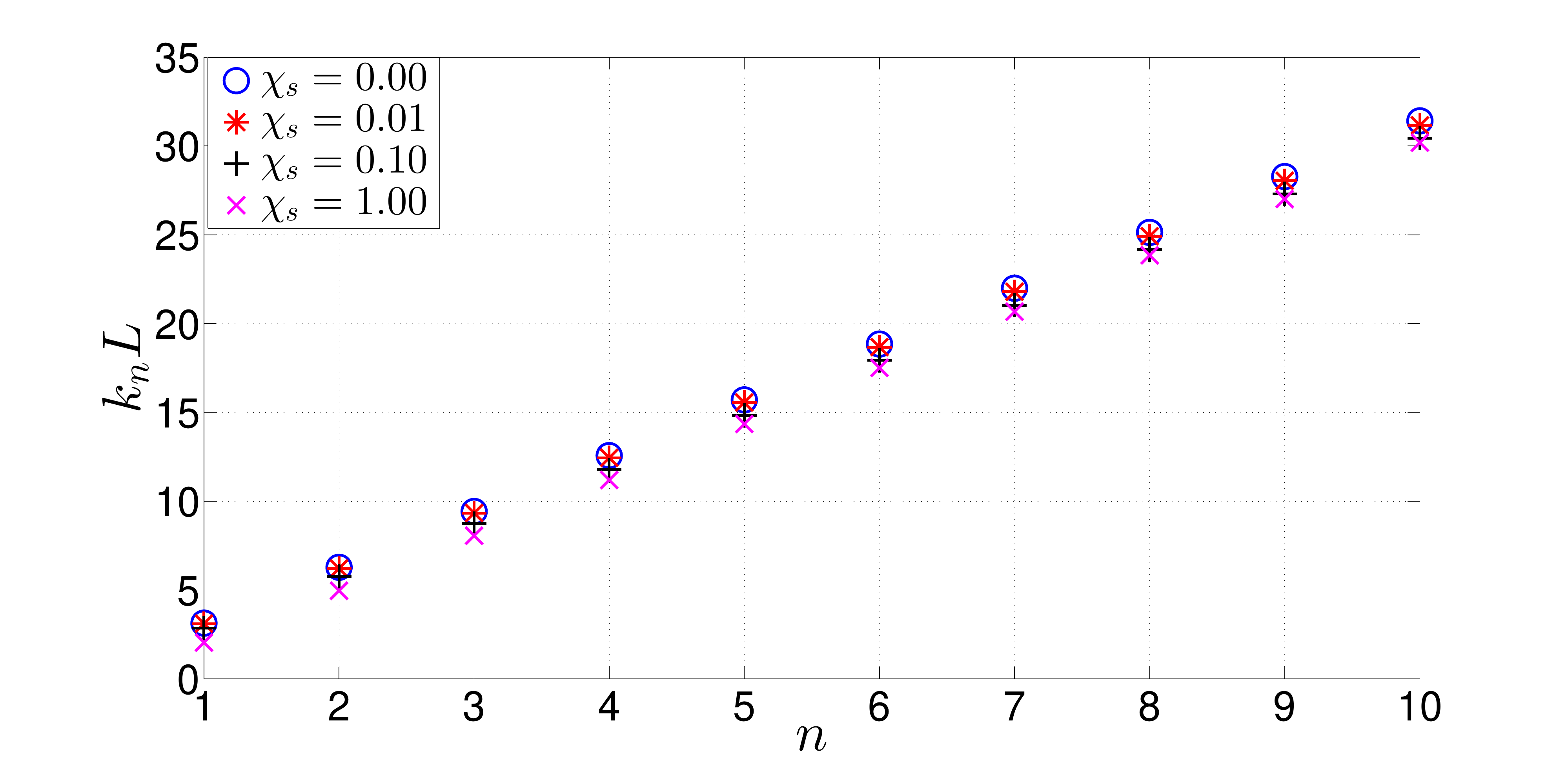}%
}}\hfill
\centering
\subfloat[\label{subfig:LevelSpacingchi0001}]{%
\includegraphics[scale=0.15]{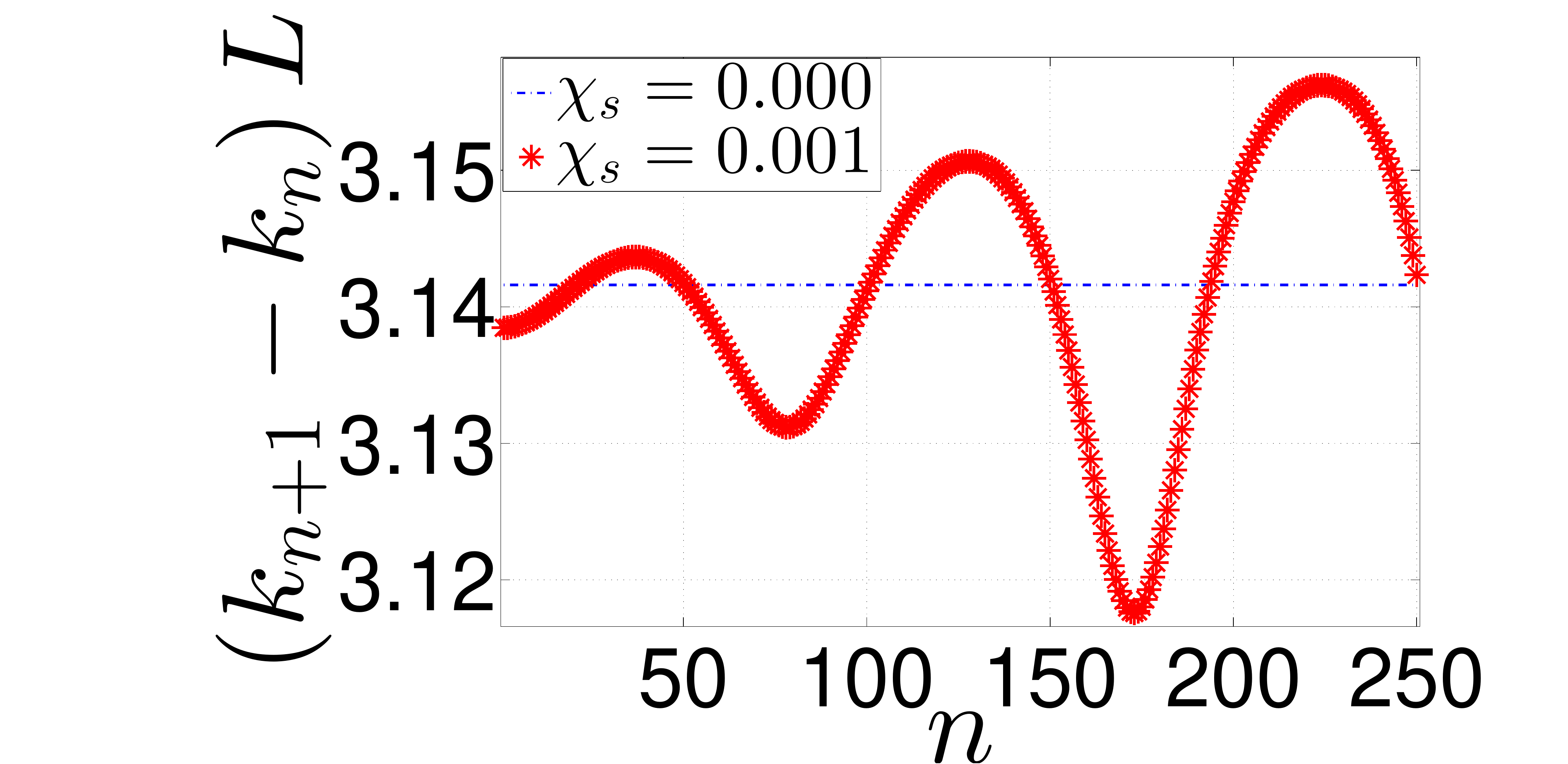}%
}
\centering
\subfloat[\label{subfig:LevelSpacingchi001}]{%
\includegraphics[scale=0.15]{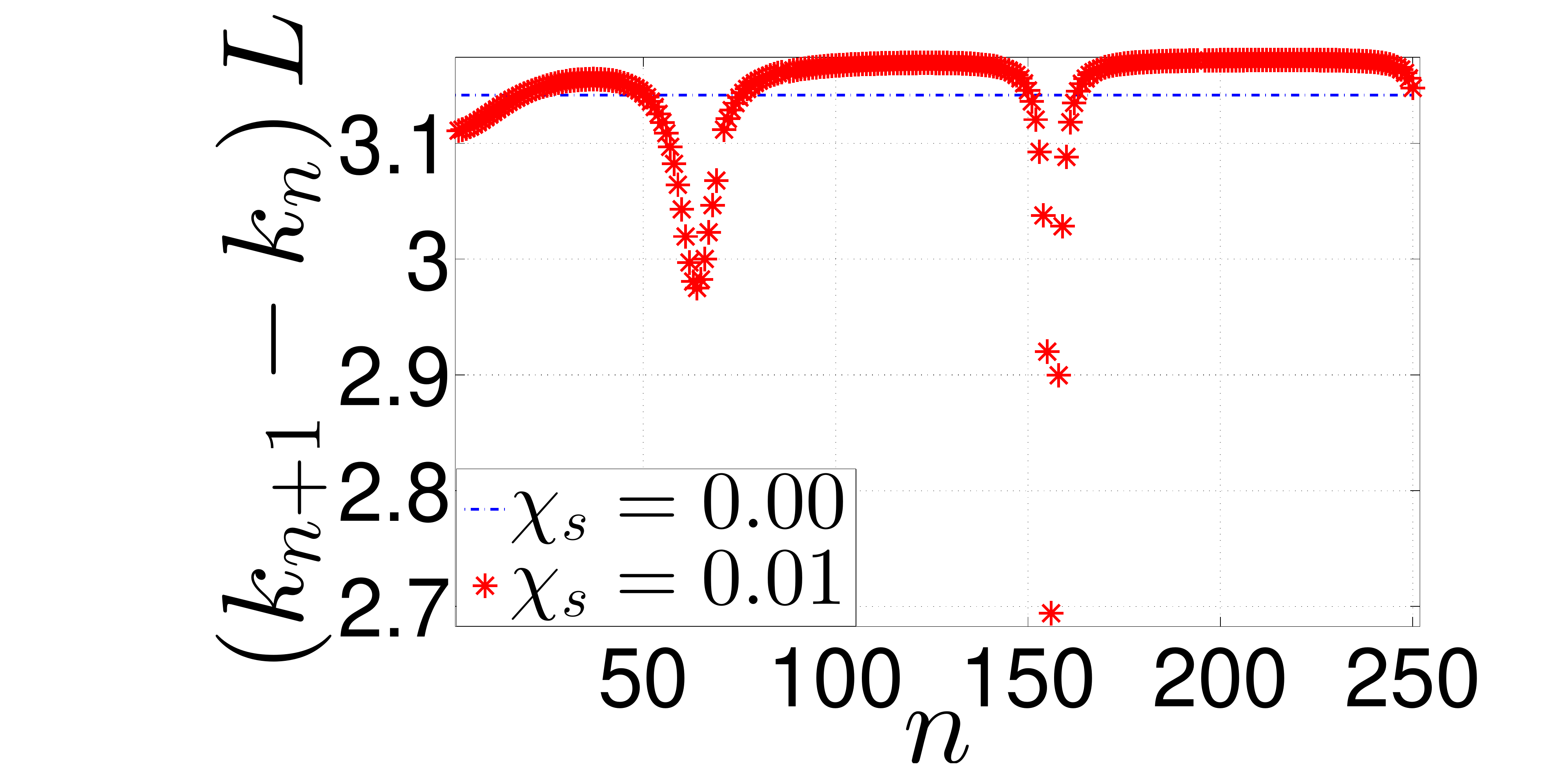}%
}
\caption{a) The first 10 modified resonances for different values of $\chi_s$ and $x_0=L/100$. Higher modes with larger $\chi_s$ experience a larger shift in frequency. b,c) Normalized level-spacing for $\chi_s=0.001$, and $0.01$ respectively. The blue dashed line shows the constant level spacing for $\chi_s=0$.}
\label{fig: eigenfrequencies for different values of chi_s}
\end{figure}
The solution of the above-stated Neumann problem gives the CC eigenfrequencies $\omega_n$ through the transcendental equation
\begin{align}
\sin(k_nL)+\chi_s k_nL\cos(k_n x_0)\cos[k_n(L-x_0)]=0.
\label{Eq: Closed CC Eigenfrequencies_body}
\end{align}
In Eq.~(\ref{Eq: Closed CC Eigenfrequencies_body}), $k_n L= \frac{\omega_n}{v_p}L=\sqrt{lc}\omega_n L$ is the normalized eigenfrequency and $\chi_s=C_s/cL$ is a unitless measure of the transmon-induced modification in eigenfrequencies/eigenstates compared to the conventional cosine basis. The CC eigenfunctions are given by
\begin{align}
\tilde{\Phi}_n(x)=\mathcal{N}_n
\begin{cases}
\cos{\left[k_n(L-x_0)\right]}\cos{(k_n x)}&0<x<x_0\\
\cos{(k_n x_0)}\cos{[k_n(L-x)]} &x_0<x<L
\end{cases}.
\label{Eq: Closed CC Eigenmodes_ body}
\end{align}
The normalization constant $\mathcal{N}_n$ has to be set by the orthogonality relations that can be found directly from the modified wave equation~(\ref{Eq:Modified Wave Equation-Closed Case_body}) as
\begin{align}
&\int_0^L dx \frac{c(x,x_0)}{c} \tilde{\Phi}_n(x) \tilde{\Phi}_m(x) = L\delta_{mn},
\label{Eq:Orthogonality of Phi _body}\\
&\int _{0}^{L}dx \partial_x\tilde{\Phi}_m(x)\partial_x\tilde{\Phi}_n(x)=k_m k_n L \delta_{mn}.
\label{Eq:Orthogonality of d/dx Phi _body}
\end{align}

Based on these results, eigenfrequencies are not only sensitive to $\chi_s$, but also to the point of connection $x_0$. In order to understand this modification better, first we have plotted the normalized eigenfrequencies in Fig. \ref{subfig:Eigfreqsx001} for different values of $\chi_s$ and the case where qubit is connected very closely to one of the ends, i.e. $x_0=0.01 L$. This is a standard location for fabricating a qubit \cite{Sundaresan_Beyond_2015} to attain a strong coupling strength between the resonator modes and the qubit, since the electromagnetic energy concentration is generally highest near the ends. In this figure, the blue circles representing the eigenfrequencies for $\chi_s=0$ are located at $n\pi$. For $\chi_s \neq 0$, all lower CC eigenfrequencies are red-shifted with respect to the $\chi_s=0$ solutions and by going to higher mode number and higher $\chi_s$, the deviation becomes more visible. In a larger scale however, the behavior of $\chi_s \neq 0$ eigenvalues are non-monotonic and most notably, display a dispersion in frequency. For better visibility of this periodic behavior, in Figs \ref{subfig:LevelSpacingchi0001}-\ref{subfig:LevelSpacingchi001} we have compared the level spacing of CC modes for different values of $\chi_s$ to the constant level spacing of unmodified cosine modes. This behavior is determined by the position of the qubit connection point $x_0$ and is easy to understand. Since $x_0=L/100$ sits at the local minima of modes 50, 150, 250 and so on, we expect a periodic behavior in the values of CC eigenfrequencies where within some portion of that period set by $\chi_s$, CC solutions are less than the $\chi_s=0$ solutions and vice versa in the remaining portion. 
\begin{figure}
\centering
\subfloat[\label{subfig:mode1x001}]{%
\includegraphics[scale=0.35]{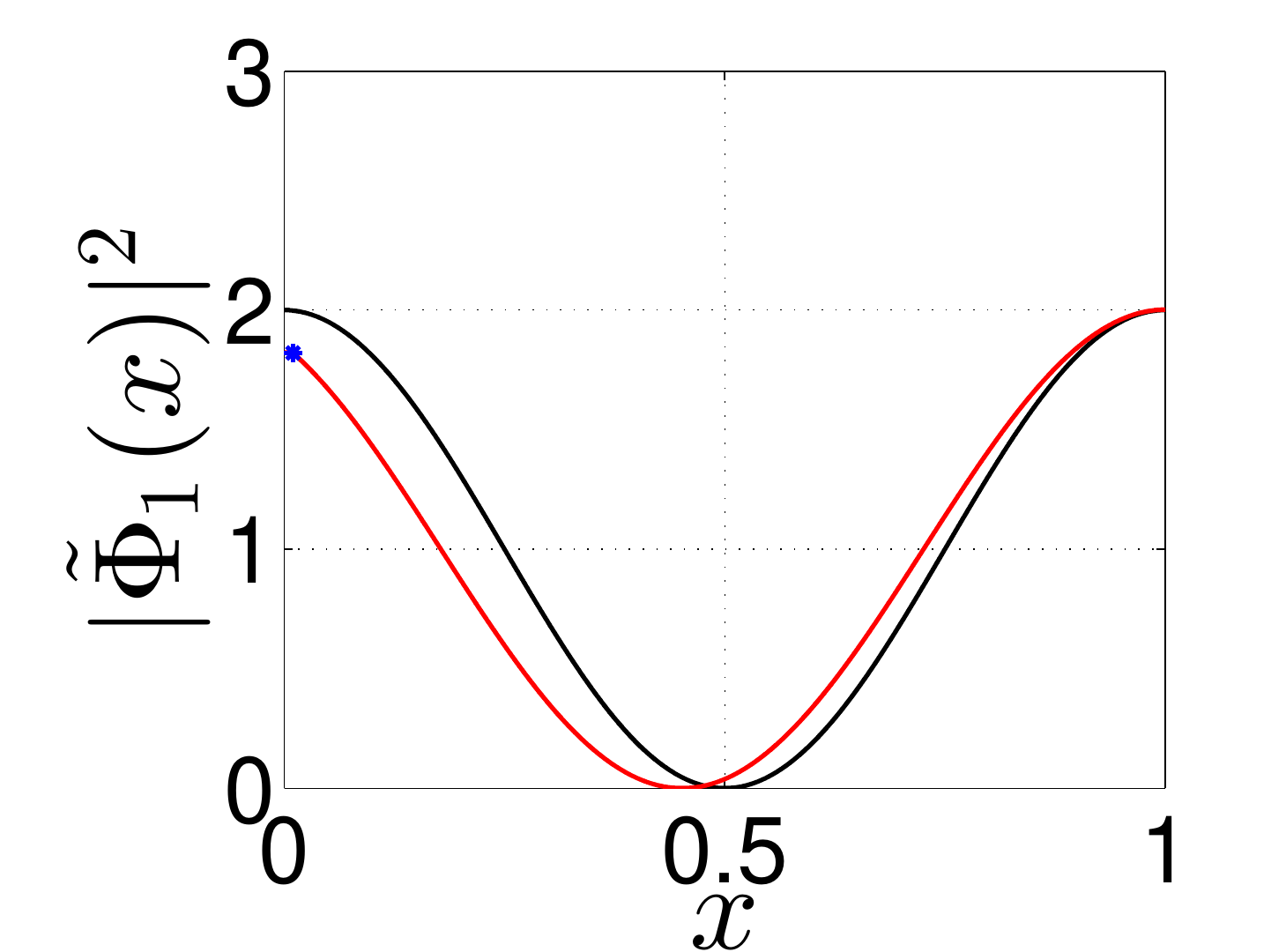}%
}
\subfloat[\label{subfig:mode2x001}]{%
\includegraphics[scale=0.35]{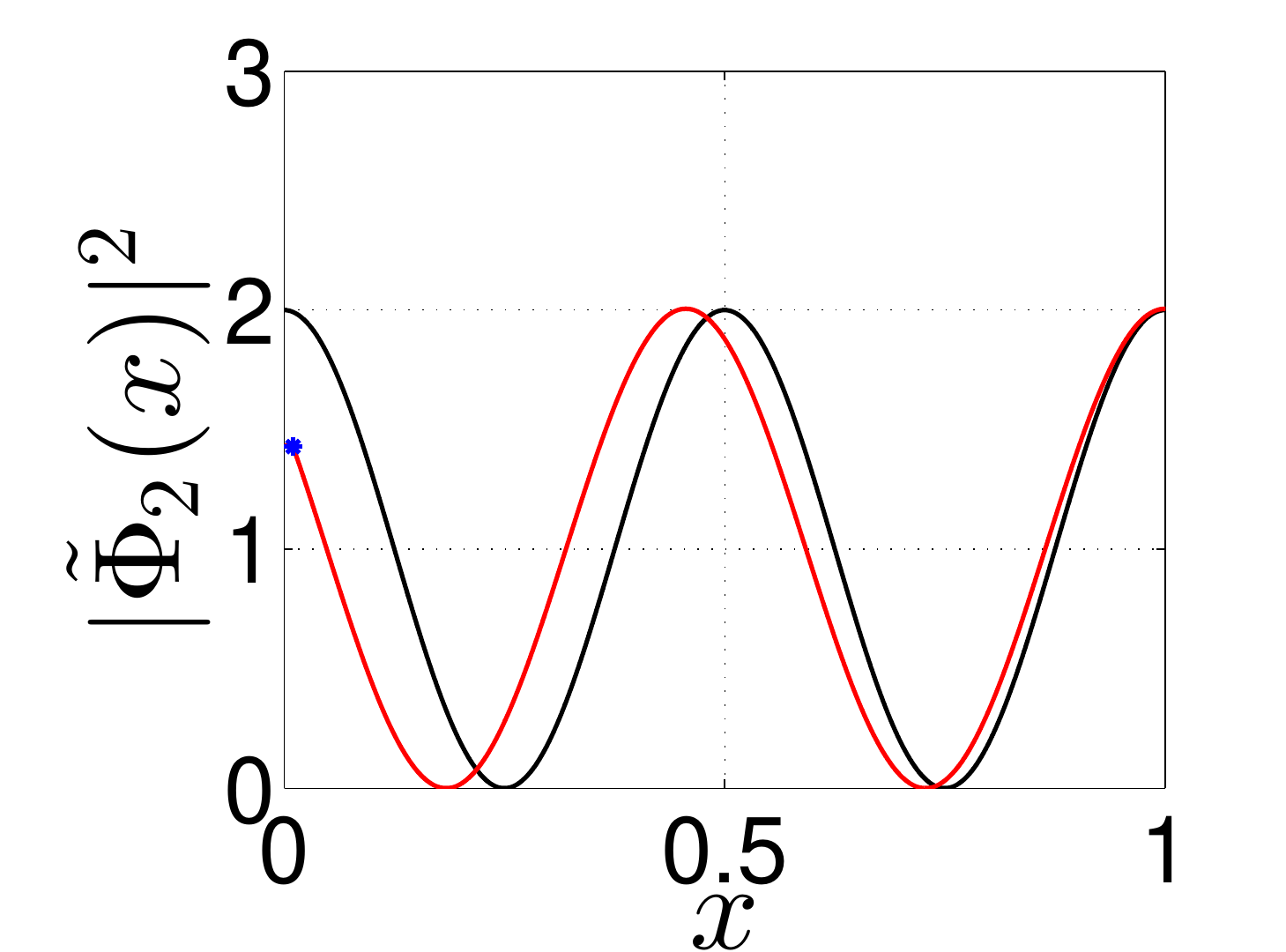}%
}\\
\subfloat[\label{subfig:mode3x001}]{%
\includegraphics[scale=0.35]{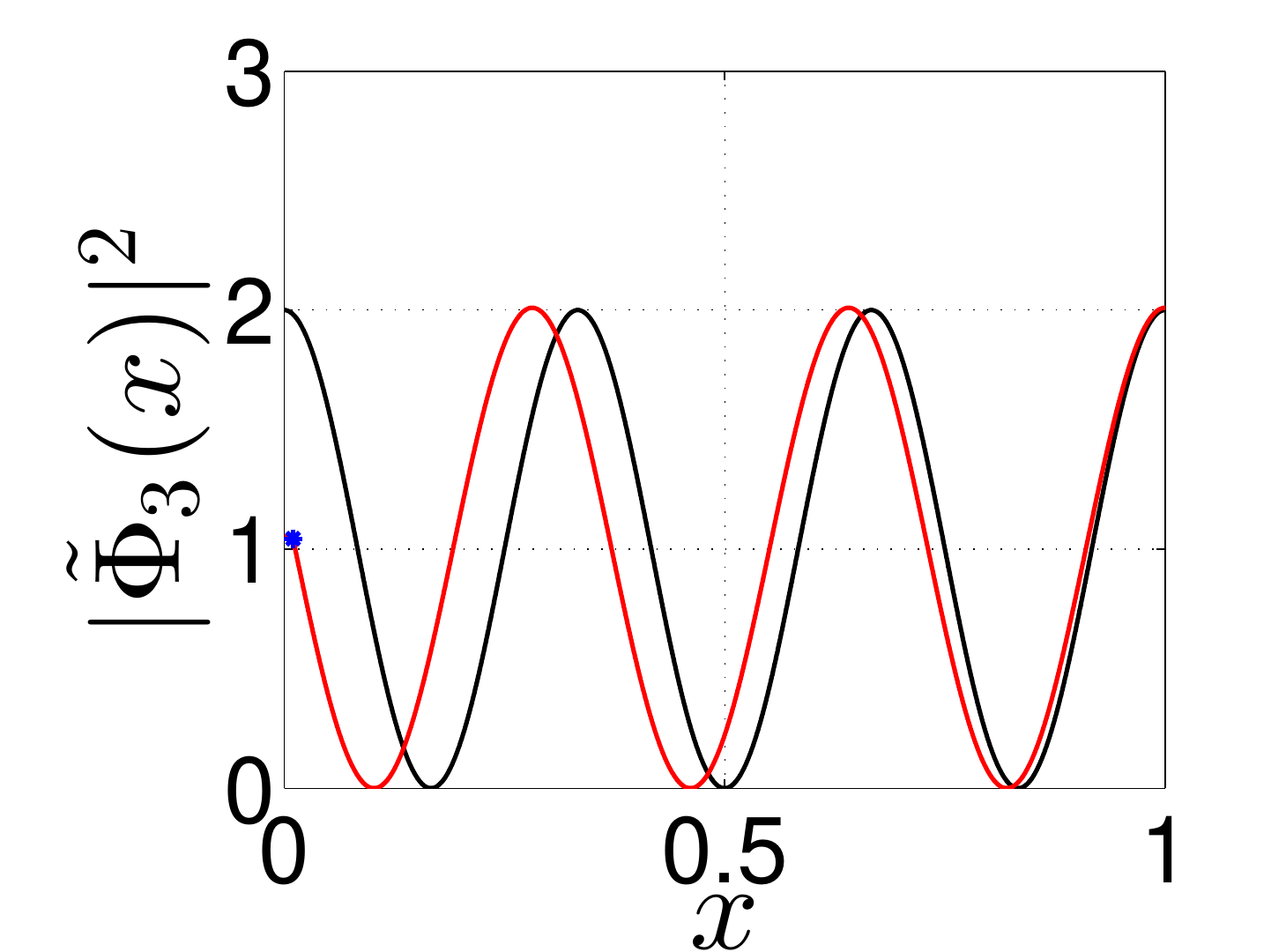}%
}
\subfloat[\label{subfig:mode4x001}]{%
\includegraphics[scale=0.35]{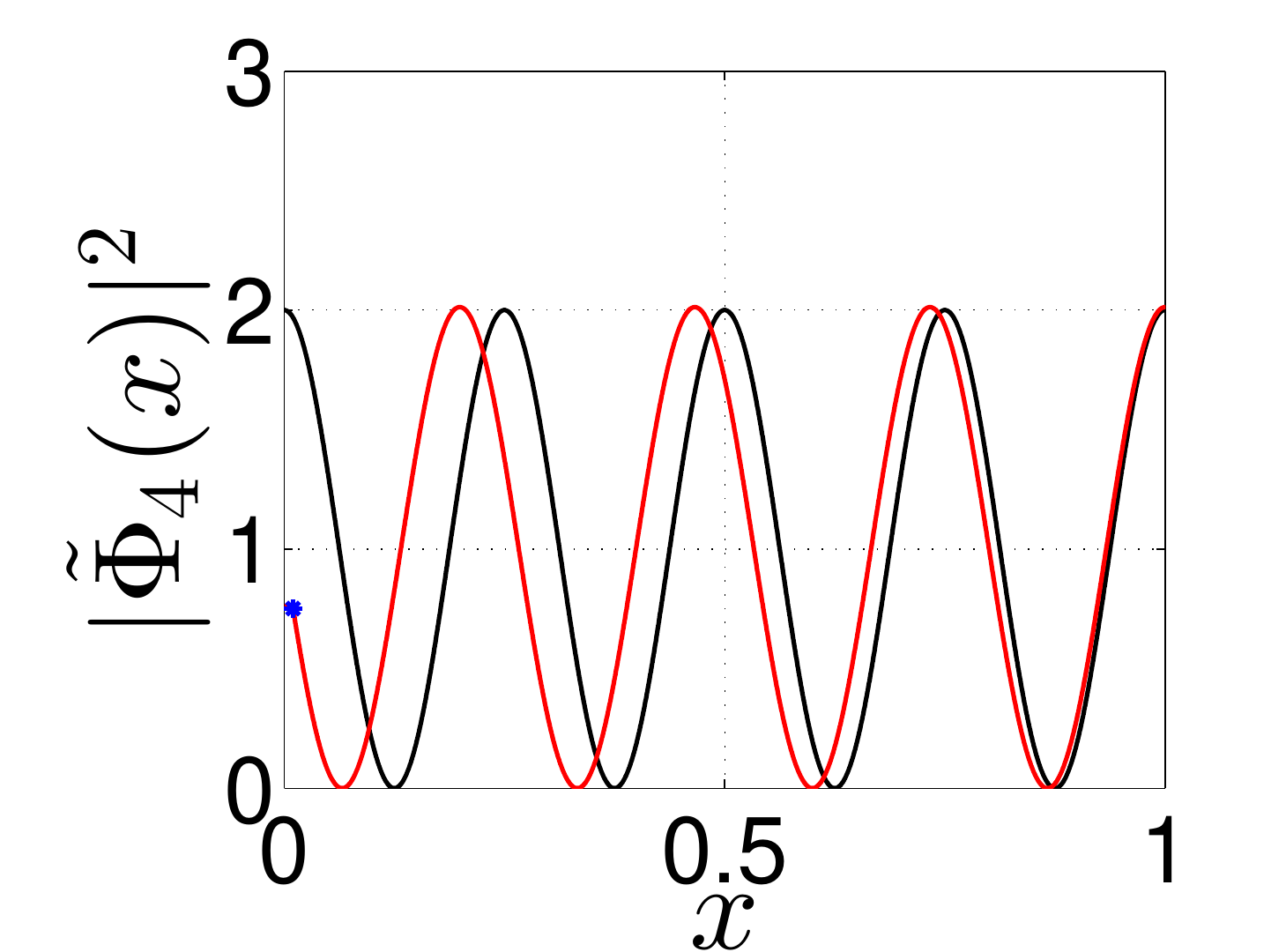}%
}
\caption{ Normalized energy density of the first 4 modes for $x_0=L/100$ and $\chi_s=0.1$. The black curve shows cosine modes while the red curve represents CC modes. The blue star shows where the qubit is connected.}
\label{fig:Eigx001}
\end{figure}
Moreover, we show the spatial dependence of the first four modes in Fig. \ref{fig:Eigx001}. The amplitude of the CC eigenmodes at the qubit location are consistently less than that for the unmodified cosine eigenmodes. This has an important implication such that the actual coupling strengths of the qubit to these modes are below the ones predicted by the $\chi_s=0$ modes as we see shortly.

\begin{figure}[t!]
\centering
\subfloat[\label{subfig:Eigfreqsx050}]{%
\centerline{\includegraphics[scale=0.24]{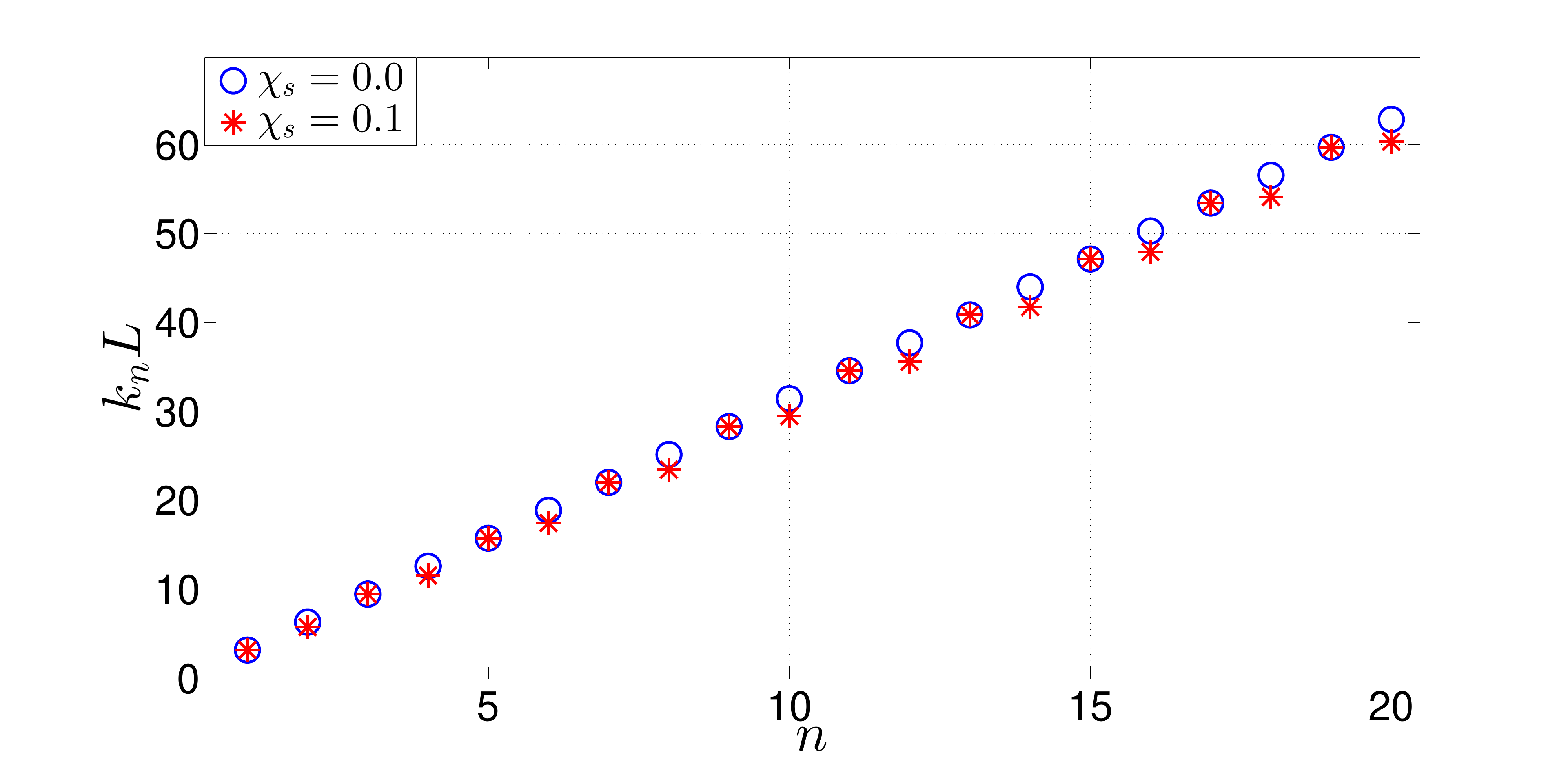}%
}}\hfill
\subfloat[\label{subfig:mode1x050}]{%
\includegraphics[scale=0.29]{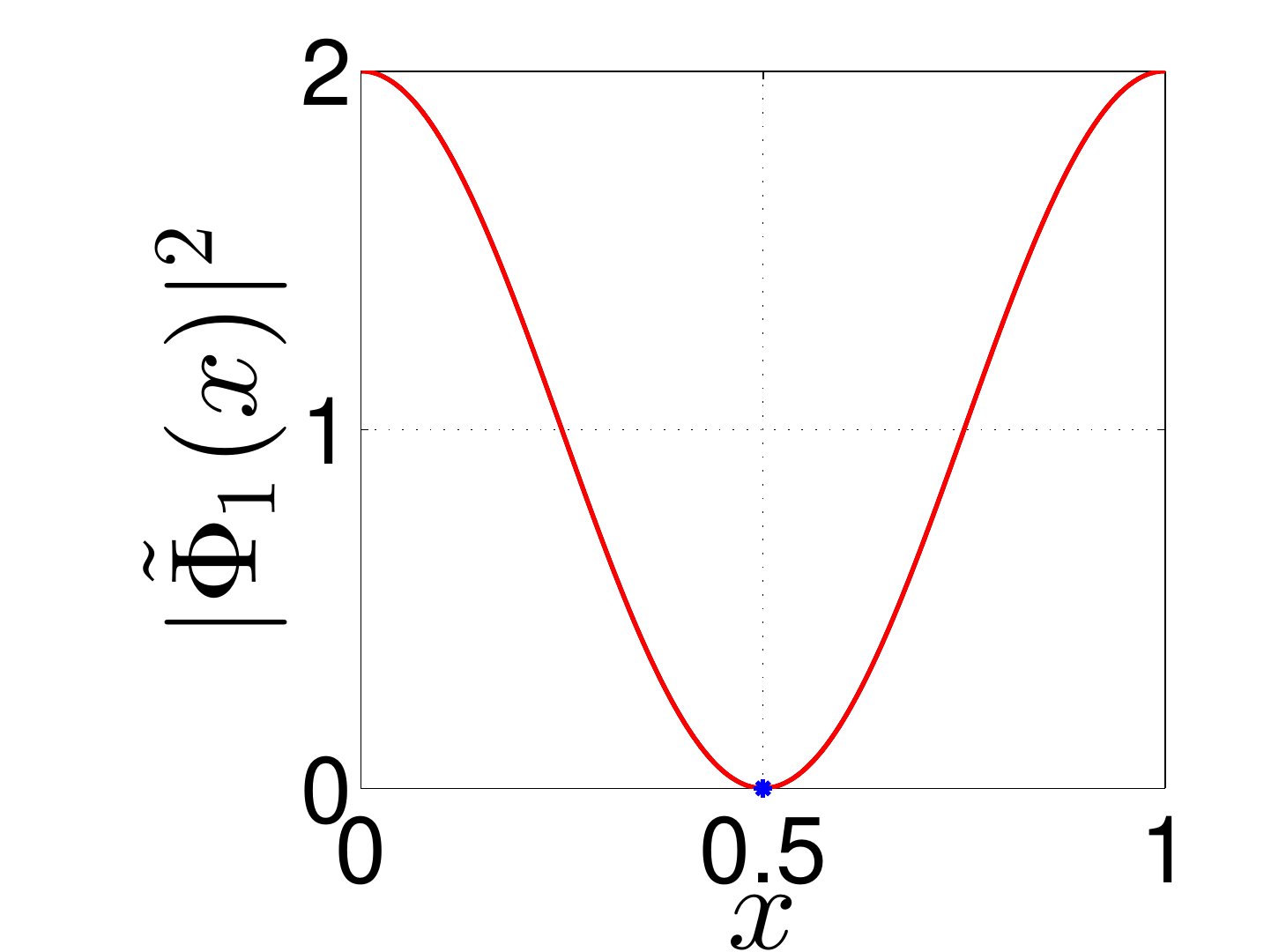}%
}
\subfloat[\label{subfig:mode2x050}]{%
\includegraphics[scale=0.29]{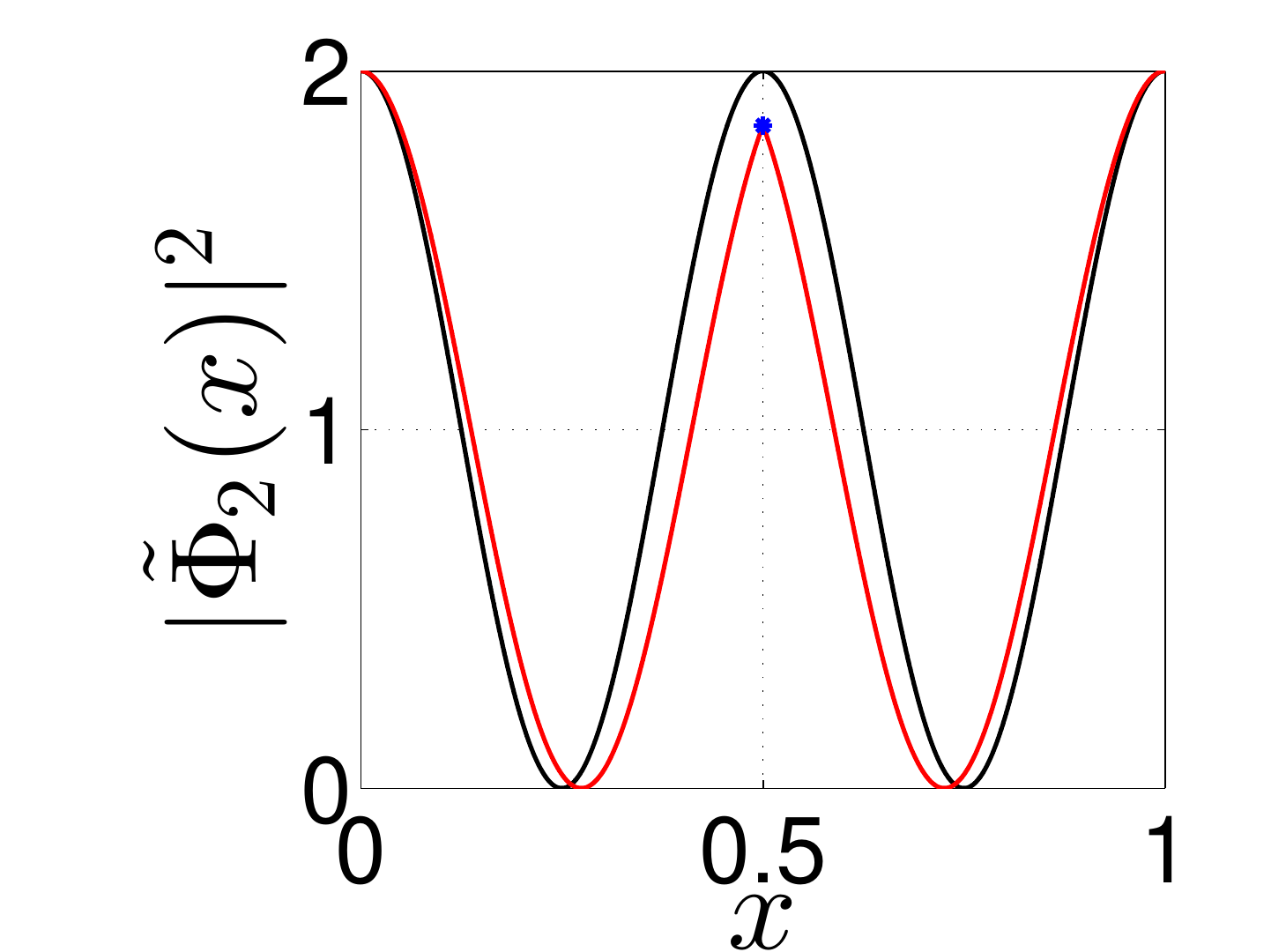}%
}\hfill
\subfloat[\label{subfig:mode3x050}]{%
\includegraphics[scale=0.29]{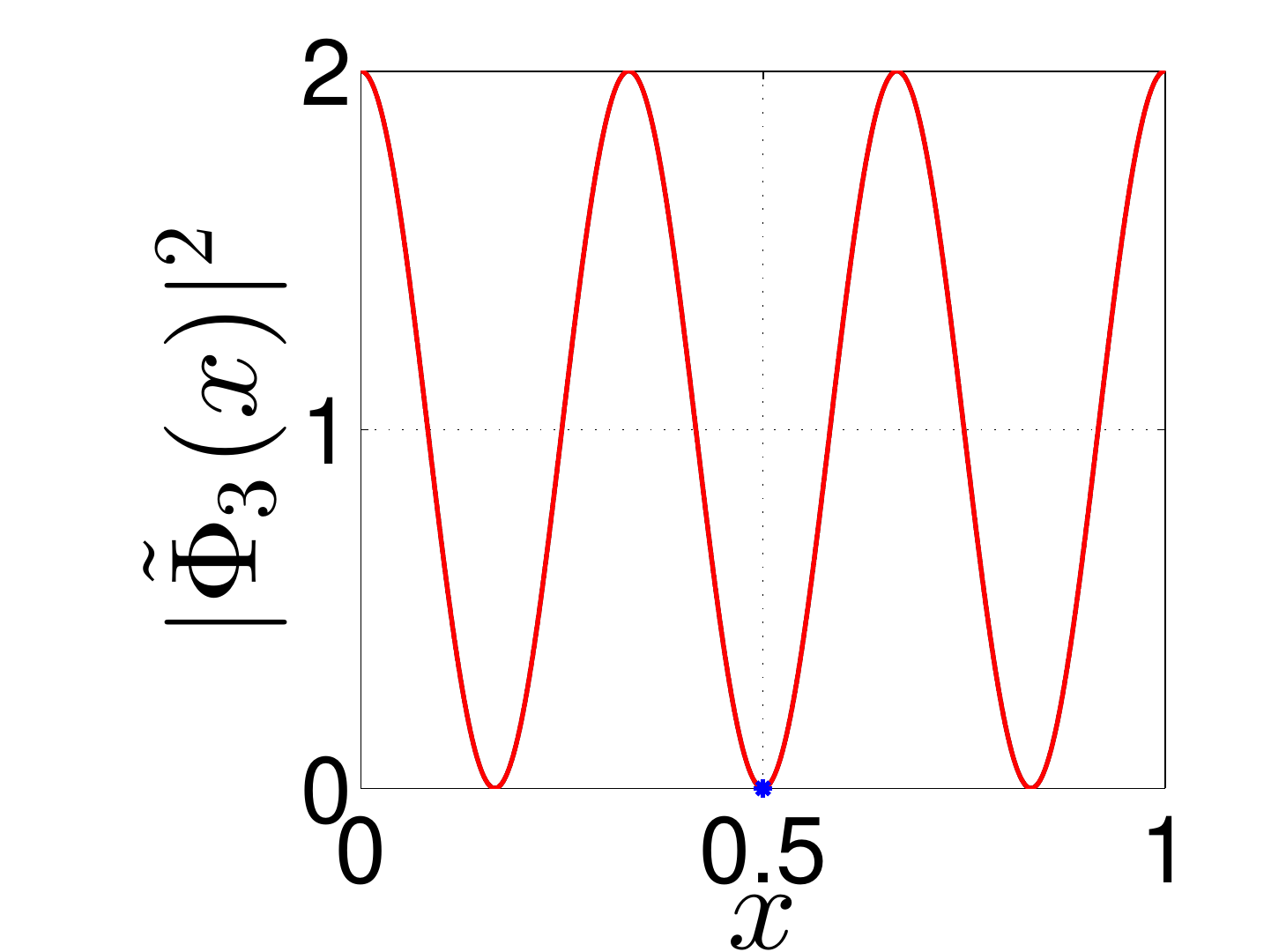}%
}
\subfloat[\label{subfig:mode4x050}]{%
\includegraphics[scale=0.29]{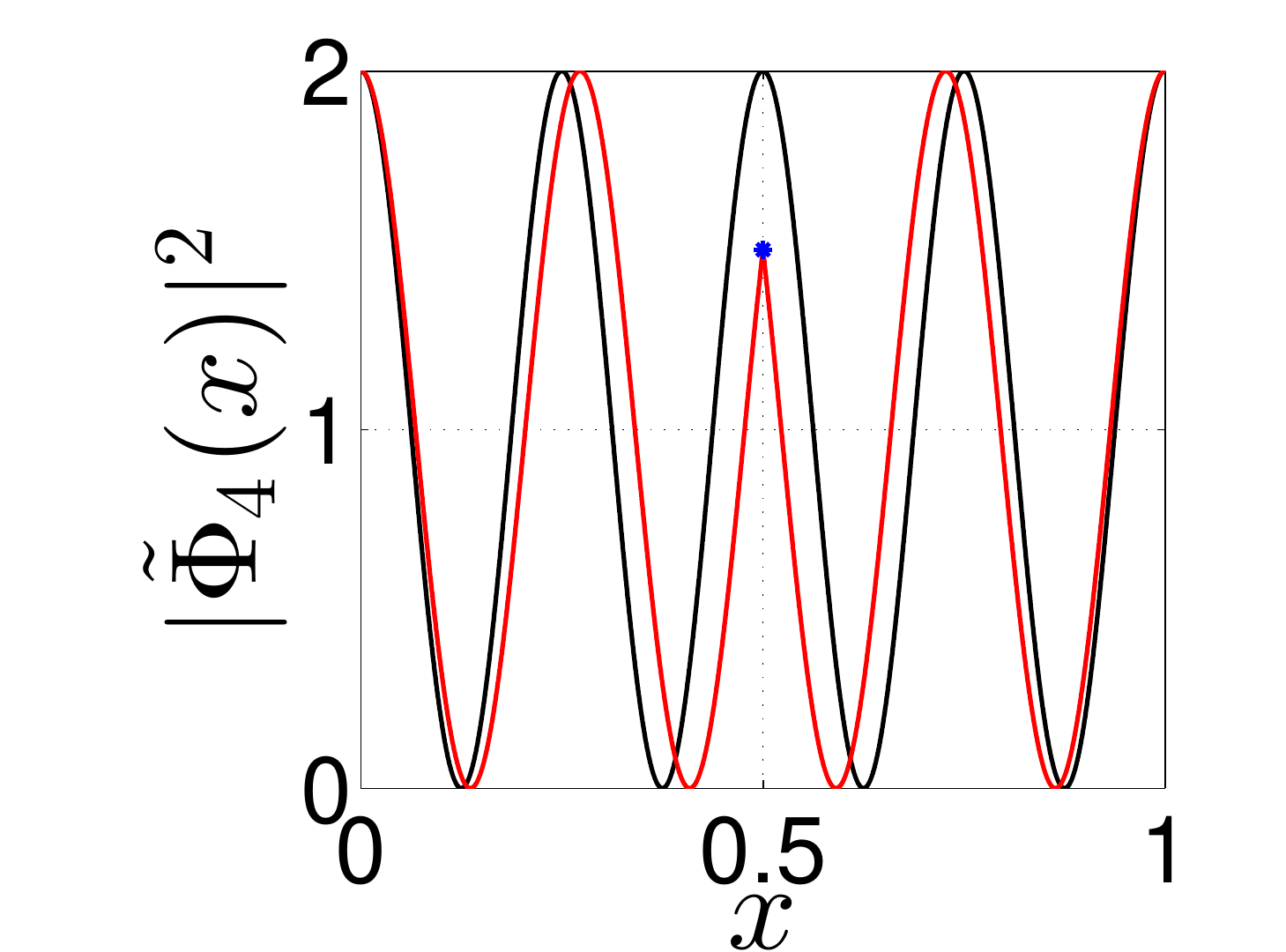}%
}
\caption{Closed-boundary CC modes for $x_0=0.5L$ (a)- First 20 Eigenfrequencies for $\chi_s=0.1$ b-e) Normalized energy density of the first 4 modes for $\chi_s=0.1$. The black curve shows cosine modes while the red ones represent CC modes. The blue star shows where the qubit is connected.}
\label{fig:Eigx050}
\end{figure}

\begin{figure}[t!]
\centering
\subfloat[\label{subfig:Eigfreqsx025}]{%
\centerline{\includegraphics[scale=0.24]{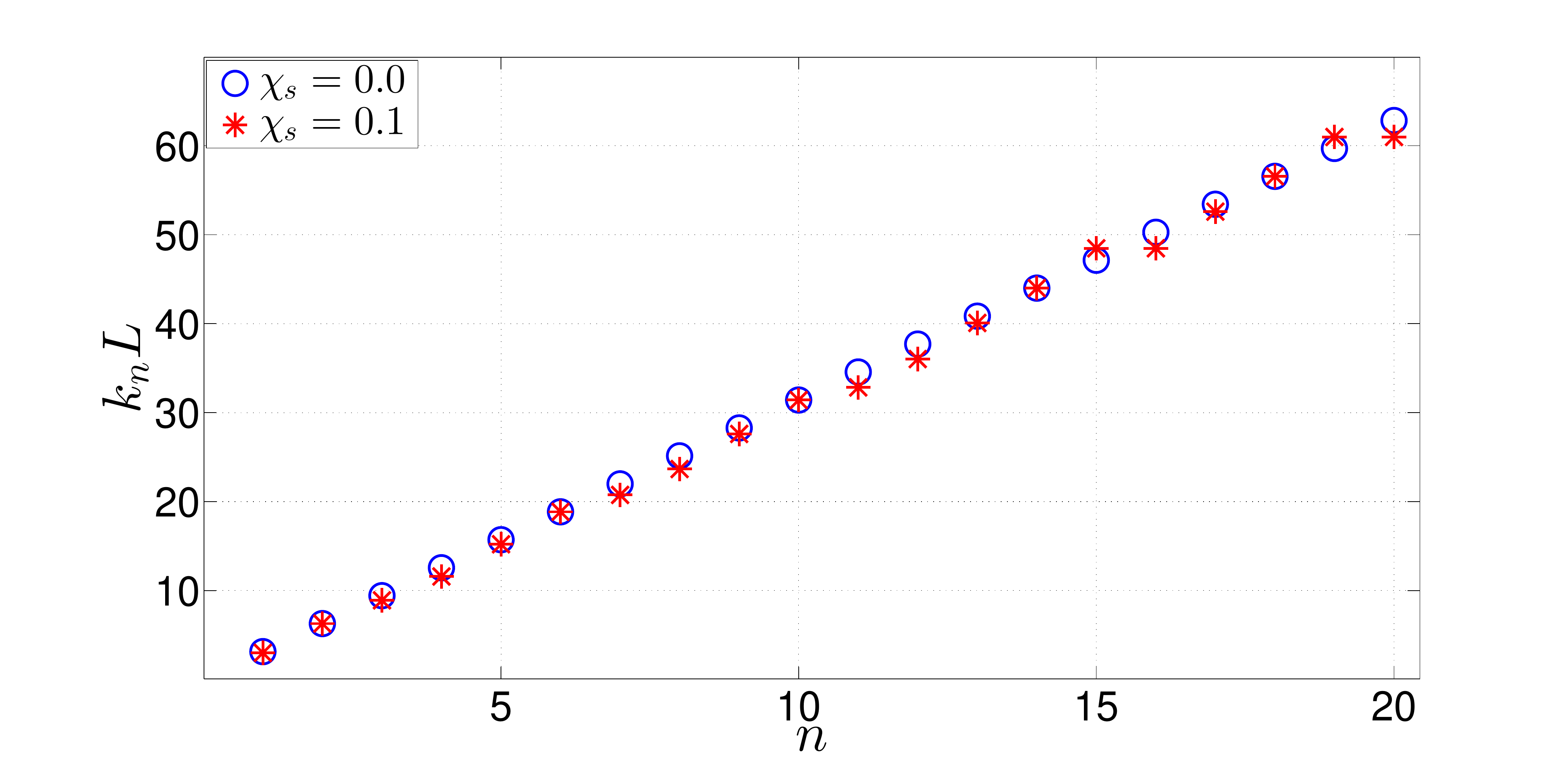}%
}}\hfill
\subfloat[\label{subfig:mode1x025}]{%
\includegraphics[scale=0.29]{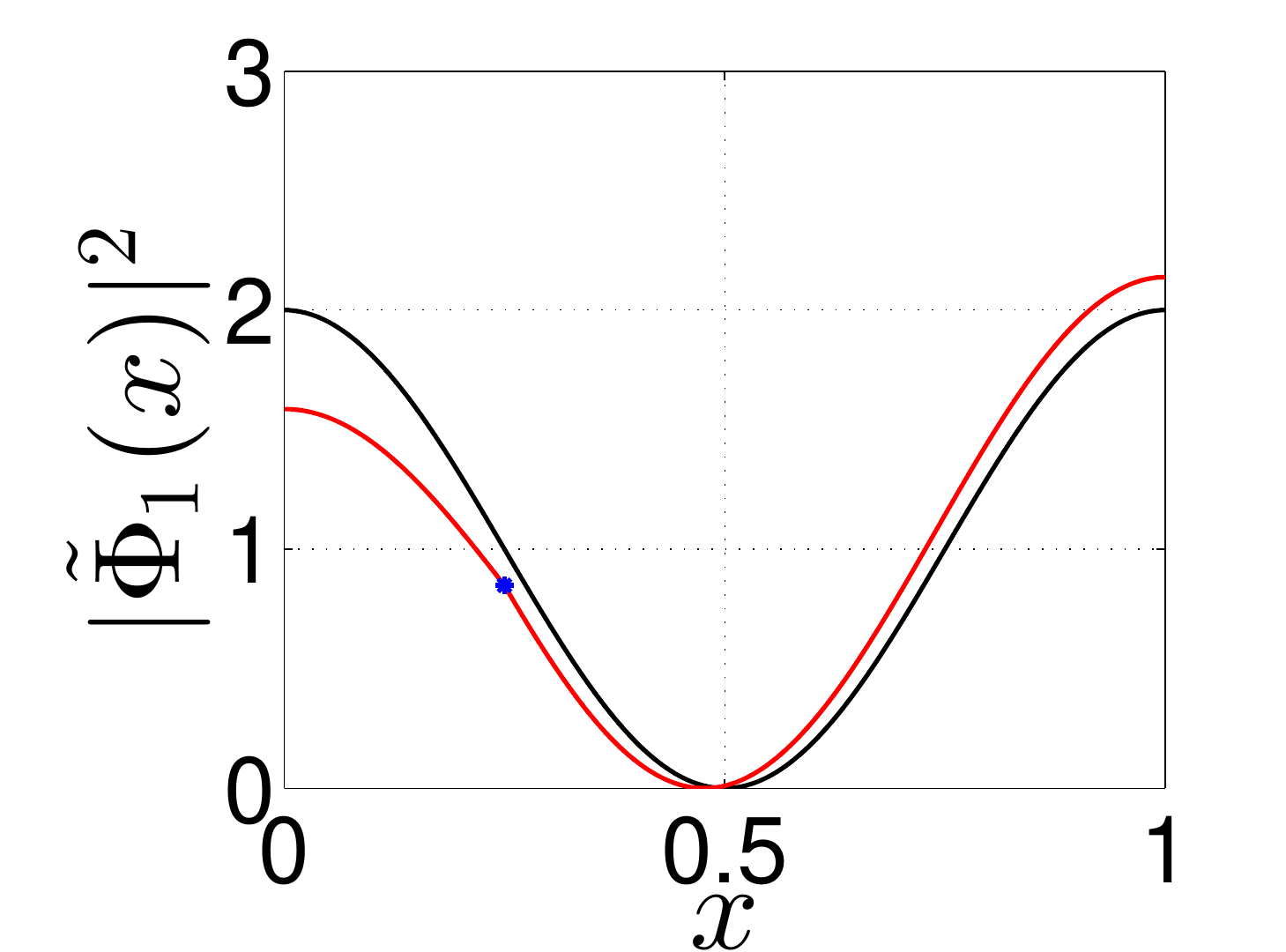}%
}
\subfloat[\label{subfig:mode2x025}]{%
\includegraphics[scale=0.29]{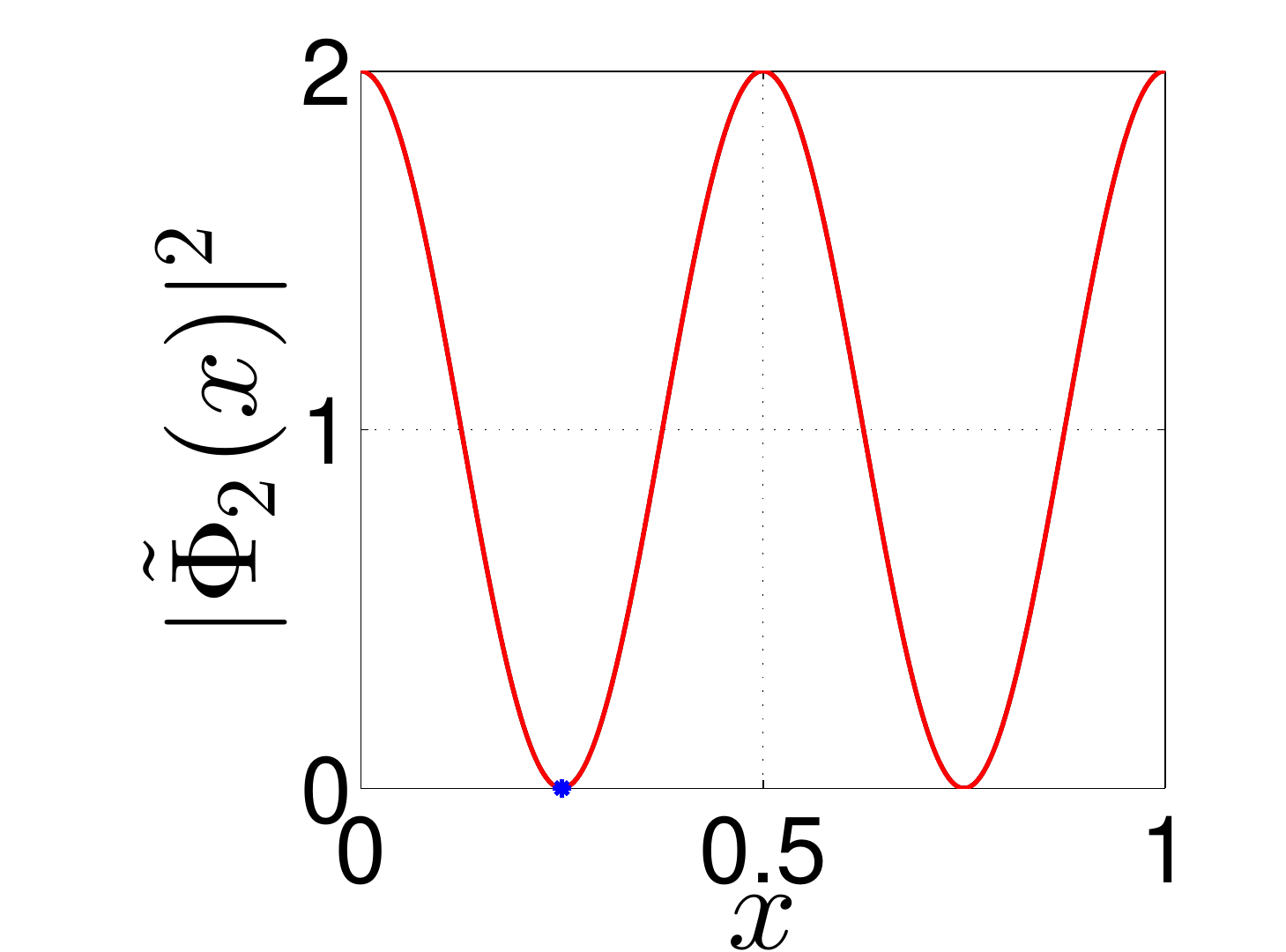}%
}\hfill
\subfloat[\label{subfig:mode3x025}]{%
\includegraphics[scale=0.29]{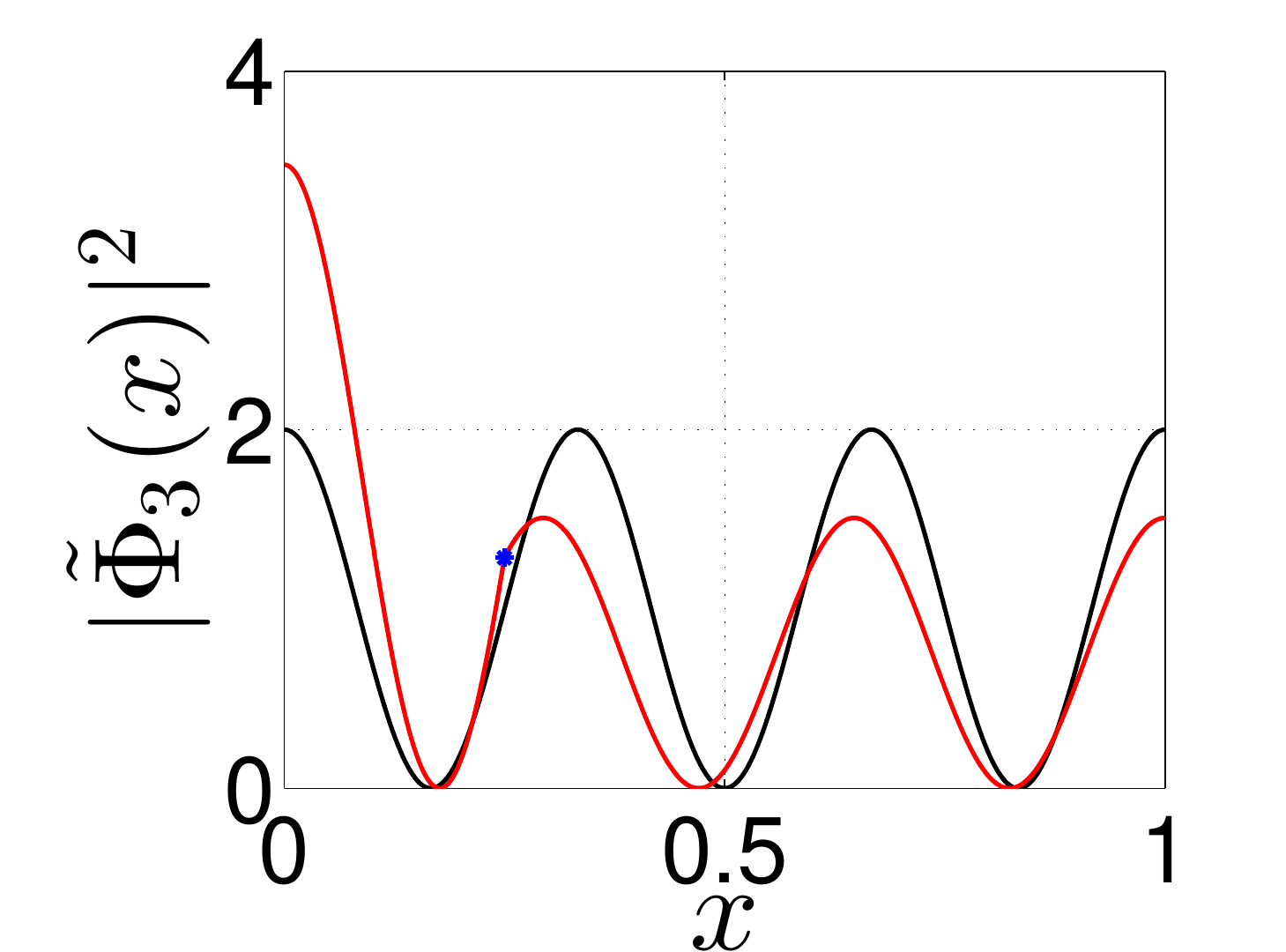}%
}
\subfloat[\label{subfig:mode4x025}]{%
\includegraphics[scale=0.29]{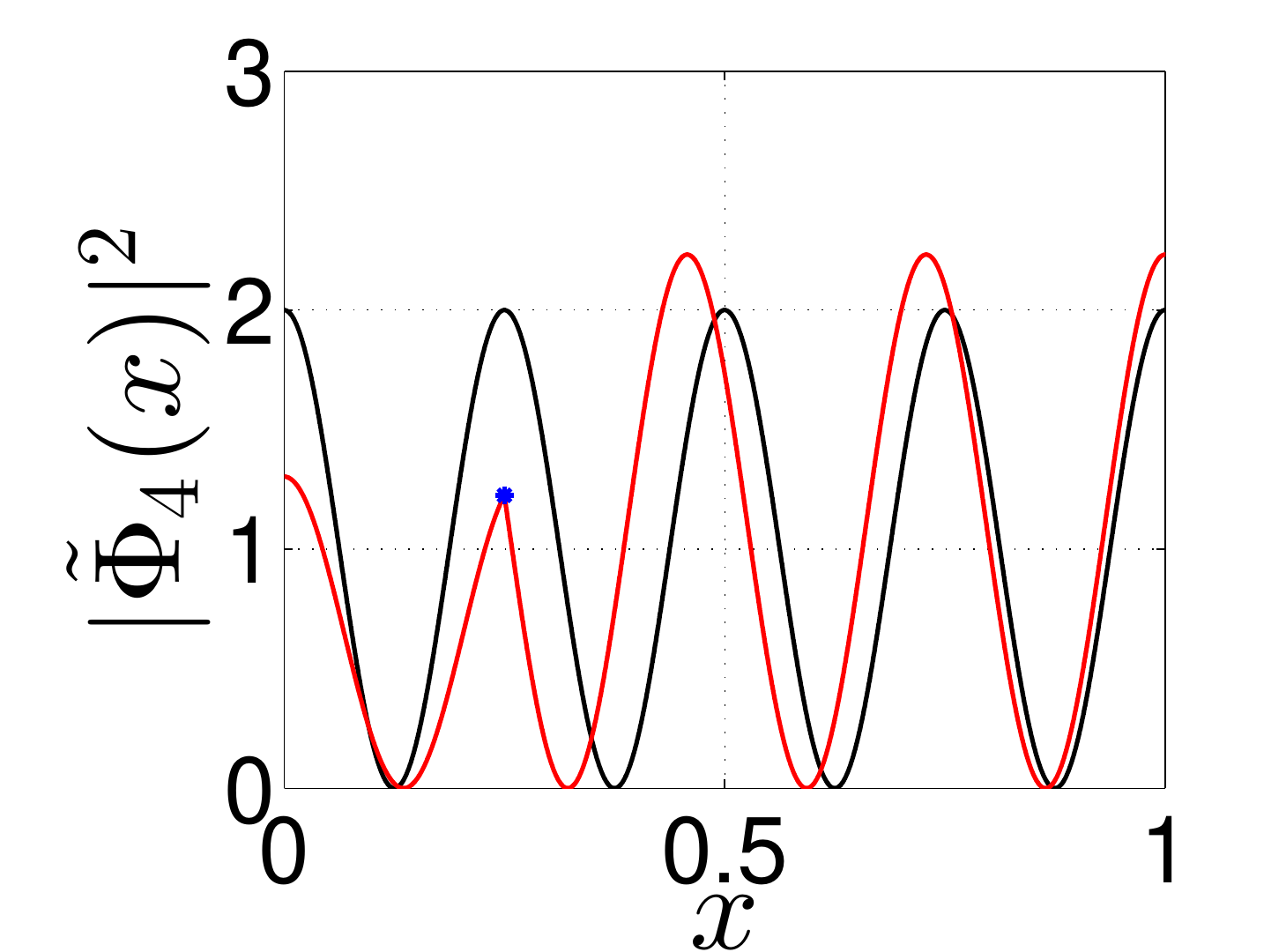}%
}
\caption{Closed-boundary CC modes for $x_0=L/4$ (a)- First 20 Eigenfrequencies b-e) Normalized energy density of the first 4 modes for $\chi_s=0.1$. The black curve shows cosine modes while the red ones represent CC modes. The blue star shows where the qubit is connected.}
\label{fig:Eigx025}
\end{figure}

Next we study the dependence of these CC modes on the qubit location $x_0$, where we consider two different cases $x_0=L/2$ and $x_0=L/4$ in Figs. \ref{fig:Eigx050} and \ref{fig:Eigx025}, consequently.
%%%%%%%%%%% Fig:Eigenmodesx050%%%%%%%%%%%%%
%%%%%%%%%%%%%%%%%%%%%%%%%%%%%%%%%%%%%%%%%%%%%
%%%%%%%%%%% Fig:Eigenmodesx025%%%%%%%%%%%%%
%%%%%%%%%%%%%%%%%%%%%%%%%%%%%%%%%%%%%%%%%%%%%
For $x_0=L/2$ we observe that all even numbered CC modes are unperturbed, while for odd numbered CC modes, both eigenmodes and eigenfrequencies are found to be less than the cosine modes. The reason for invariance of even numbered modes is that originally qubit sits on a local minimum of the photonic energy density and therefore does not interact with these modes. This behavior is not specific to $x_0=L/2$. Generally, if the qubit is placed at $x_0=\frac{L}{n},n\in \mathbb{N}$, then there is a periodicity in the mode structure such that every $n$ modes remain unperturbed. This is for example observed in Fig. \ref{fig:Eigx025} where $ x_0=L/4 $ and thus modes indexed as $4n-2,n\in\mathbb{N}$ are unchanged.

\subsection{Canonical quantization}
In this section, we discuss the quantization of $\mathcal{H}_{C}^{\text{mod}}$ in terms of the CC basis (See App.~\ref{App:CanonicalQuant} for more details). The conjugate quantum fields of the resonator can be expanded in terms of CC basis as
\begin{align}
&\hat{\Phi}(x,t)=\sum\limits_n \left(\frac{\hbar}{2\omega_n cL}\right)^{\frac{1}{2}}\left[\hat{a}_n(t) + \hat{a}_n^{\dagger}(t)\right]\tilde{\Phi}_n(x),\\
&\hat{\rho}(x,t)=-i\sum\limits_n \left(\frac{\hbar \omega_n}{2cL}\right)^{\frac{1}{2}}\left[\hat{a}_n(t) - \hat{a}_n^{\dagger}(t)\right]c(x,x_0)\tilde{\Phi}_n(x). 
\end{align}
Substituting these expressions into $\hat{\mathcal{H}}_C^{mod}$ and employing orthogonality conditions~(\ref{Eq:Orthogonality of Phi _body}) and (\ref{Eq:Orthogonality of d/dx Phi _body}), one finds a diagonal representation of $\hat{\mathcal{H}}_C^{\text{mod}}$ 
\begin{align}
\hat{\mathcal{H}}_C^{mod}=\sum\limits_n \hbar \omega_n \left( \hat{a}_n^{\dagger} \hat{a}_n + \frac{1}{2} \right),
\end{align}
as a sum over the energy of each normal CC mode.

In a similar manner, the qubit flux $\hat{\Phi}_j$ and charge $\hat{Q}_j$ operators can be represented in terms of eigenmodes of transmon Hamiltonian as
\begin{align}
&\hat{\Phi}_j(t)=\sum\limits_{m,n}\bra{m}\hat{\Phi}_j(0)\ket{n}\hat{P}_{mn}(t),
\label{eqn:A2-Spec Rep Of Phi_j 1}\\
&\hat{Q}_j(t)=\sum\limits_{m,n}\bra{m}\hat{Q}_j(0)\ket{n}\hat{P}_{mn}(t),
\label{eqn:A2-Spec Rep Of Q_j 1}
\end{align}
where $\hat{P}_{mn}(t)$ represent a set of projection operators acting between states $\ket{m}$ and $\ket{n}$. Working in the flux basis, the eigenmodes are found through solving a Schr\"odinger equation as
\begin{align}
\left[-\frac{\hbar^2}{2C_j}\frac{d^2}{d\Phi_j^2}-E_J\cos{\left(2\pi\frac{\Phi_j}{\Phi_0}\right)}\right]\Psi_n(\Phi_j)=\hbar\Omega_n\Psi_n(\Phi_j),
\end{align}
whose solution can be characterized in terms of Mathieu functions~\cite{Koch_Charge_2007}. Due to the invariance of the Hamiltonian under flux parity transformation, the eigenmodes are either even or odd functions of $\Phi_j$ and only off-diagonal elements between states having different parities are non-zero (See App.~\ref{App:CanonicalQuant}). Consequently, flux and charge matrix elements are purely real and imaginary. Therefore, we can rewrite Eqs.~(\ref{eqn:A2-Spec Rep Of Phi_j 1}) and ~(\ref{eqn:A2-Spec Rep Of Q_j 1}) as
\begin{align}
\hat{\Phi}_j(t)&=\sum\limits_{m<n}\bra{m}\hat{\Phi}_j(0)\ket{n}\left(\hat{P}_{mn}(t)+\hat{P}_{nm}(t)\right),
\label{eqn:A2-Spec Rep Of Phi_j 2}\\
\hat{Q}_j(t)&=\sum\limits_{m<n}\bra{m}\hat{Q}_j(0)\ket{n}\left(\hat{P}_{mn}(t)-\hat{P}_{nm}(t)\right).
\label{eqn:A2-Spec Rep Of Q_j 2}
\end{align}

To arrive at a standard form, we apply the unitary transformation $\hat{a}_l\rightarrow i\hat{a}_l$ and $\hat{P}_{mn}\rightarrow i\hat{P}_{mn}$ for $m<n$, so that the second quantized representation of the Hamiltonian in its most general form can be expressed as
\begin{align}
\hat{\mathcal{H}}=\sum\limits_n \hbar\Omega_n \hat{P}_{nn}+\sum\limits_n \hbar \omega_n \hat{a}_n^{\dagger} \hat{a}_n+\sum\limits_{m<n,l} \hbar g_{mnl} \left(\hat{P}_{mn}+\hat{P}_{nm}\right)\left(\hat{a}_l+\hat{a}^{\dagger}_l\right),
\end{align}
where the $g_{mnl}$ stands for the coupling strength between mode $l$ of the resonator and the transition dipole $\hat{P}_{mn}$ and is obtained as
\begin{align}
\hbar g_{mnl} \equiv \gamma \left(\frac{\hbar \omega_l}{2cL}\right)^{\frac{1}{2}}(iQ_{j,mn})\tilde{\Phi}_l(x_0).
\end{align}

Various TRK sum rules \cite{Wang_Generalization_1999} can be developed for a transmon qubit, as discussed in detail in App.~\ref{App:TRKSumRules}. For instance, the sum of transition matrix elements of $\hat{Q}_j$ between the ground state and all the excited states obey 
\begin{align}
\sum\limits_{n>0} 2\left(E_n-E_0 \right)\left|\bra{0}\hat{Q}_j\ket{n}\right|^2=(2e)^2E_J \bra{0}\cos{\left(\frac{2\pi}{\Phi_0}\hat{\Phi}_j\right)}\ket{0}<(2e)^2E_J.
\label{Eq:Sum Rule for Q_J _body}
\end{align}  
Since all terms on the right hand side are positive, this imposes an upper bound to the strength of $Q_{j,0n}$. 

A multimode Rabi Hamiltonian can be recovered by truncating the transition matrix elements to only one relevant quasi-resonant transition term (assumed here to be the $0 \rightarrow 1$ transition):
\begin{align}
\hat{\mathcal{H}}=\frac{1}{2} \hbar \omega_{01} \hat{\sigma}^z + \sum\limits_n \hbar \omega_n \hat{a}_n^{\dagger} \hat{a}_n +\sum\limits_n \hbar g_n (\hat{\sigma}^- + \hat{\sigma}^+)(\hat{a}_n+\hat{a}^{\dagger}_n),
\end{align}
where coupling strength $g_n$ is now reduced to
\begin{align}
\hbar g_n \equiv \gamma \left(\frac{\hbar \omega_n}{2cL}\right)^{\frac{1}{2}}(iQ_{j,01})\tilde{\Phi}_n(x_0). 
\label{Eq: coupling strength gn-Body}
\end{align}
Based on Eq. \ref{Eq:Sum Rule for Q_J _body}, $Q_{j,01}$ has to satisfy
\begin{align}
|Q_{j,01}|^2<\frac{2e^2 E_J}{E_1-E_0}\approx \frac{2E_J}{\sqrt{8E_JE_C}-E_C}e^2, 
\end{align}
where we have defined the charging energy $E_C\equiv e^2/(2C_j)$.   
\begin{figure}
\centering
\subfloat[\label{subfig:couplingsx001}]{%
\centerline{\includegraphics[scale=0.30]{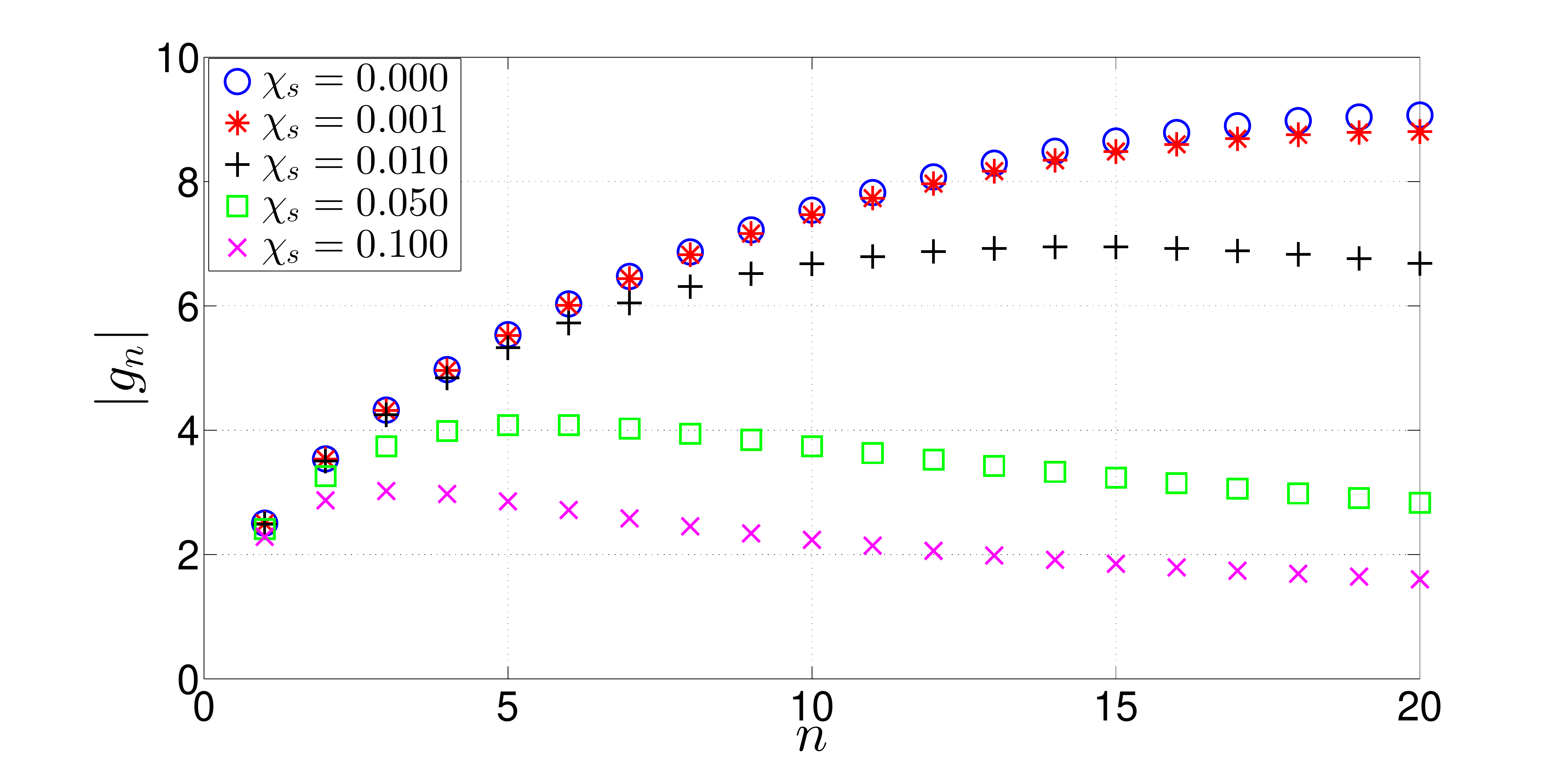}%
}}\hfill
\subfloat[\label{subfig:couplingsx001Largescale}]{%
\centerline{\includegraphics[scale=0.30]{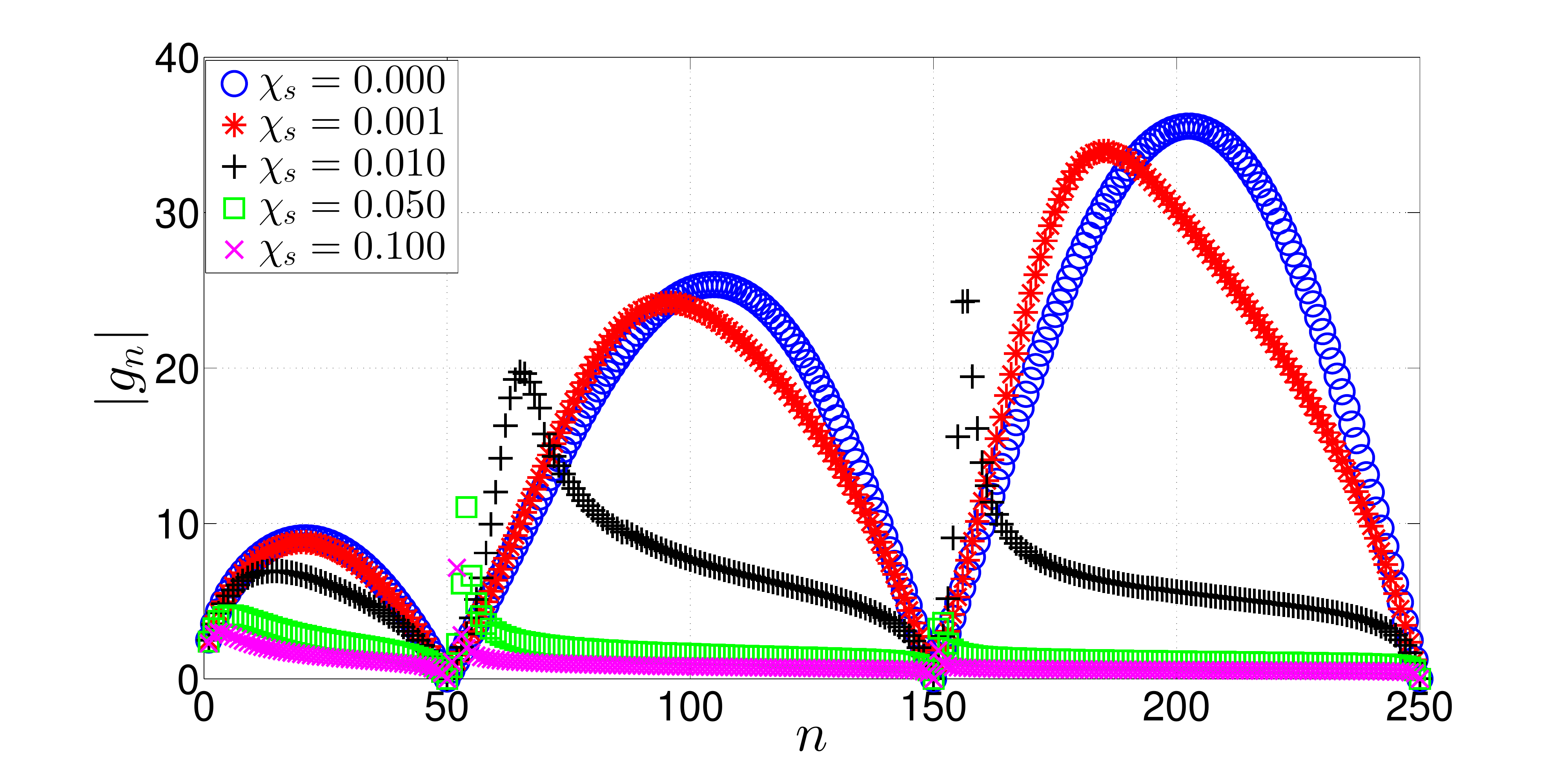}%
}}\hfill
\caption{Normalized coupling strength $g_n$ for $x_0=L/100$ a) First 20 modes b) Large scale behavior for 250 modes. In both graphs, coupling strength is normalized such that only the normalized photonic dependence is kept i.e. $g_n=(k_n L)^{\frac{1}{2}}\tilde{\Phi}_n(x_0)$}
\label{fig:couplingsx001}
\end{figure}

In order to understand how much $g_n$ can deviate in practice from its former widely used expression in terms of the unmodified ($\chi_s = 0$) modes, in Fig. \ref{fig:couplingsx001}, we have compared the results for  various values of $\chi_s$. We note that in recent experiments on an ultra-long ($\sim70$ cm) transmission line cavity \cite{Sundaresan_Beyond_2015}, $\chi_s$ was found to be  around $10^{-3}$. For shorter, more standard transmission line resonators we should expect $\chi_s \sim 0.1$ because $\chi_s \propto 1/L$. 

Importantly, CC couplings $g_n$ are very sensitive to a change in $\chi_s$ shown in Fig.~\ref{fig:couplingsx001}. For instance, in Fig. \ref{subfig:couplingsx001} which is for the common case of connecting the qubit to one end, $x_0=0.01 L$, it is observed that even for $\chi_s=0.001$ (red stars) the relative shift in the highest mode shown (mode 20) is about $3\%$. This relative change increases to $26\%$, $69\%$ and $80\%$ for $\chi_s=0.01$, $\chi_s=0.05$ and $\chi_s=0.1 $, respectively. 

These modifications, even for small $\chi_s$, are clearly observable in the multimode regime, i.e. when the qubit is resonant with a very high order mode. To study the large scale behavior of couplings, we have plotted the first 250 CC couplings $g_n$ in Fig. \ref{subfig:couplingsx001Largescale} for the same parameters. As we mentioned earlier, due to the fact that the qubit is placed at a symmetry point, we expect that with a period of 100 modes, the couplings fall down to zero. The first mode that has a local minimum at $x_0=L/100$ is mode 50 and it occurs again at modes 150, 250 and so on.  As a general rule, higher CC modes experience a larger shift in their coupling strength. Another important observation is the suppression of coupling strength as $\chi_s$ increases. The asymptotic behavior of $g_n$ is studied in App.~\ref{App:AsymptoticOfgn} where we show that $g_n\sim1/\sqrt{\omega}_n$ is suppressed at high frequencies. This is a very important observation that renders QED quantities convergent in the number of modes without the need for a cut-off. 

We note that the mechanism of this frequency cut-off is distinct from other potential suppression mechanisms which might appear at frequencies close to the superconducting gap. In other words, we have shown that there is a natural frequency cut-off even for an ideal superconductor with $\Delta_c\to\infty$. Whether the non-zero dispersion in frequencies and the modifications of coupling strengths to higher order modes is observable before reaching the gap depends on specifics of the circuit, in particular on the resonator length and the superconductor material that determines the gap energy. 
\section{Generalization to an open cavity: open-boundary CC basis}
\label{Sec:A^2-OpenGen}
In this section, We discuss the quantization procedure in an open geometry, where the resonator is coupled to two long microwave transmission lines, of length $L_L$ and $L_R$, at each side through nonzero capacitors $C_L$ and $C_R$ (Fig.~\ref{Fig:cQEDOpen-EqCircuit}). The aim of this section is to reach an open Rabi model in which the resonator is coupled to two quantum bath. In App.~\ref{App:2ndQuantOpen}, we have discussed how these nonzero capacitances alter the boundary conditions at each end, and hence the mode structure as a result as well. The resulting real eigenfrequencies of the resonator can be found from the transcendental equation
\begin{align}
\begin{split}
&+\left(1-\chi_R\chi_L (k_nL)^2\right)\sin{(k_n L)}\\
&+\left(\chi_R+\chi_L\right)k_nL\cos{(k_n L)}\\
&+\chi_s k_n L\cos{(k_n x_0)}\cos{(k_n (L-x_0))}\\
&-\chi_R\chi_s (k_n L)^2 \cos{(k_n x_0)}\sin{(k_n (L-x_0))}\\
&-\chi_L\chi_s (k_n L)^2 \sin{(k_n x_0)}\cos{(k_n (L-x_0))}\\
&+\chi_R\chi_L\chi_s (k_n L)^3 \sin{(k_n x_0)}\sin{(k_n (L-x_0))}=0,
\end{split}
\label{Eq:Open CC Eigenfrequencies_body}
\end{align}
where $\chi_{R,L}\equiv\frac{C_{R,L}}{cL}$ are normalized coupling capacitors to the left and right waveguides. Considering only the first two terms in Eq.~(\ref{Eq:Open CC Eigenfrequencies_body}), equivalent to setting $\chi_s=0 $, would lead to the well-known equation in the literature \cite{Koch_Time-reversal_2010, Schmidt_Circuit_2013, Nunnenkamp_Synthetic_2011}
\begin{align}
\tan{(k_n L)}=\frac{(\chi_R+\chi_L)k_n L}{\chi_R\chi_L (k_n L)^2-1},
\end{align}
which only describes eigenfrequencies of an isolated resonator and does not contain appropriate current conservation at the qubit location. The third term is the same modification we have found in the closed case and has a significant influence as $\chi_s$ increases, while the others represent higher order corrections and are almost negligible except for very high order modes. The real-space representation of these eigenmodes are found as
\begin{align}
\tilde{\Phi}_n(x)\propto
\begin{cases}
\tilde{\Phi}_n^{<}(x) \quad 0<x<x_0\\
\tilde{\Phi}_n^{>}(x) \quad x_0<x<L
\end{cases},
\end{align}
where $\tilde{\Phi}_n^{<}(x)$ and $\tilde{\Phi}_n^{>}(x)$ are given by
\begin{align}
\begin{split}
\tilde{\Phi}_n^{<}(x)&=\left[\cos{(k_n(L-x_0))}-\chi_Rk_nL\sin{(k_n(L-x_0))}\right]\\
&\times\left[\cos{(k_n x)}-\chi_L k_nL\sin{(k_n x)}\right],
\end{split}\\
\begin{split}
\tilde{\Phi}_n^{>}(x)&=\left[\cos{(k_n x_0)}-\chi_L k_nL\sin{(k_n x_0)}\right]\\
&\times\left[\cos{(k_n(L-x))}-\chi_Rk_nL\sin{(k_n(L-x))}\right].
\end{split}
\end{align}
The {\it open-boundary CC basis} can be shown to satisfy the modified orthogonality relations 
\begin{align}
&\int_{0}^{L} dx \frac{c_{op}(x,x_0)}{c}\tilde{\Phi}_m(x)\tilde{\Phi}_n(x)=L\delta_{mn},
\end{align}
where the capacitance per unit length $c_{op}(x,x_0)$, due to the leaky boundary is given by 
\begin{align}
c_{op}(x,x_0)=c+C_s \delta(x-x_0)+ C_R\delta(x-L^-)+C_L \delta(x-0^+).
\end{align}
The remaining orthogonality relations for the current is also modified as
\begin{align}
\begin{split}
\int _{0}^{L}dx \frac{\partial\tilde{\Phi}_m(x)}{\partial x}\frac{\partial\tilde{\Phi}_n(x)}{\partial x}&-\frac{1}{2}\left(k_m^2+k_n^2\right)L\left[\chi_R\tilde{\Phi}_m(L^-)\tilde{\Phi}_n(L^-)+\chi_L\tilde{\Phi}_m(0^+)\tilde{\Phi}_n(0^+)\right]\\
&=k_m k_n L \delta_{mn}.
\end{split}
\end{align}

Note that the same argument holds for the CC modes of the left and right transmission lines, while the exact knowledge of these modes requires assigning appropriate boundary conditions at their outer boundaries. For instance, if the side resonators are assumed to be very long, an outgoing boundary condition is a very good approximation, since the time scale by which the escaped signal bounces back and reaches the original resonator is much larger than the round-trip time of the central resonator. On the other hand, if we have a lattice \cite{Koch_Time-reversal_2010, Nunnenkamp_Synthetic_2011, Schmidt_Circuit_2013} of identical resonators each connected to a qubit and capacitively coupled to each other, then the same basis can be used for each of them. Assuming we also have the solution for the CC basis of right and left resonators as $\{\omega_{n,S},\tilde{\Phi}_{n,S}| n \in \mathbb{N}, S=\{R,L\}\}$, the quantum flux fields in each side resonator can be expanded in terms of these CC modes as
\begin{align}
\begin{split}
&\hat{\Phi}_S(x,t)=\sum \limits_{n,S} \left(\frac{\hbar}{2\omega_{n,S} cL_S}\right)^{\frac{1}{2}} \left[\hat{b}_{n,S}(t) + \hat{b}_{n,S}^{\dagger}(t) \right]\tilde{\Phi}_{n,S}(x),
\end{split}
\end{align}
where $\hat{b}_{n,S=\{R,L\}}$ are the annihilation and creation operators for the $n$th open CC mode in each side resonator. Following the quantization procedure discussed in App.~\ref{App:2ndQuantOpen}, we find the Hamiltonian in its 2nd quantized representation as
\begin{align}
\begin{split}
\hat{\mathcal{H}}&=\sum\limits_n \hbar\Omega_n \hat{P}_{nn}+\sum\limits_n \hbar\omega_n \hat{a}_n^{\dagger} \hat{a}_n+\sum\limits_{n,S=\{L,R\}} \hbar\omega_{n,S} \hat{b}_{n,S}^{\dagger}\hat{b}_{n,S}\\
&+\sum\limits_{m<n,l} \hbar g_{mnl} \left(\hat{P}_{mn}+\hat{P}_{nm}\right)\left(\hat{a}_l+\hat{a}^{\dagger}_l\right)\\
&+\sum\limits_{m,n,S=\{L,R\}} \hbar\beta_{mn,S}\left(\hat{a}_m+\hat{a}_m^{\dagger}\right)\left(\hat{b}_{n,S}+ \hat{b}_{n,S}^{\dagger}\right).
\end{split}
\label{Eq:2nd Quantized Hamiltonian_Open Case_Body}
\end{align}

In Eq.~(\ref{Eq:2nd Quantized Hamiltonian_Open Case_Body}), $\beta_{mn,S}$ stands for coupling strength of $m$-th open CC mode of the resonator to $n$-th open CC mode of the side baths and is found as
\begin{align}
&\beta_{mn,S}=\frac{C_{S}}{2c\sqrt{L}\sqrt{L_{S}}}\omega_m^{\frac{1}{2}}\omega_{n,S}^{\frac{1}{2}}\tilde{\Phi}_m(L^-)\tilde{\Phi}_{n,S}(L^+),
\label{Eq:beta_{L,R}}
\end{align}
where $C_S$ here stands for side capacitors $C_{R,L}$ and should not be confused with the series capacitance $C_s$ introduced earlier. Notice that light-matter coupling strength $g_{mnl}$ has the same form as before, but in terms of open CC eigenmodes and eigenfrequencies.

\section{Convergent multimode cQED} 
\label{Sec:A2-ConvergentPurcellLamb}

The $A^2$-term was thought to have no impact on transition frequencies in vacuum-induced effects such as the Lamb shift. Because it does not involve atomic operators, it is expected to make the same perturbative contribution to every atomic energy level, precluding observable shifts in {\it transition} frequencies \cite{Milonni_Quantum_2013}. This argument relies on perturbation theory in the $A^2$-term. We show that the diamagnetic term {\it does} have an impact when accounted for exactly to all orders through the solutions of the aforementioned transcendetal equations.   

First, we illustrate the role of modal modification on the convergence of QED quantities with a simple phenomenological model. Previously, the Purcell rate and the Lamb shift have been calculated using the Lindblad formalism in the dispersive limit \cite{Boissonneault_Dispersive_2009}. An effective multimode JC model
\begin{equation}
\frac{\hat{\mathcal{H}}_\text{JC}}{\hbar} = \frac{\omega_j}{2} \hat{\sigma}_z  + \sum_n \nu_n \hat{a}_n^\dagger \hat{a}_n + \sum_n g_n \left(\hat{\sigma}^+\hat{a}_n+\hat{\sigma}^- \hat{a}_n^{\dag}\right) 
\end{equation}
can be obtained (Sec.~\ref{Sec:A^2-GaugeInvRabiModel}) from our first principles calculation, which incorporates the modifications to the resonator modes and the qubit dynamics. Resonator losses are included through a Bloch--Redfield equivalent zero-temperature master equation for the reduced density matrix of the resonator and qubit $\hat{\dot{\rho}} = -\frac{i}{\hbar} [ \hat{\mathcal{H}}_\text{JC}, \hat{\rho} ]+\kappa_n \left( 2 \hat{a}_n \hat{\rho} \hat{a}_n^\dagger - \{ \hat{\rho}, \hat{a}_n^\dagger \hat{a}_n \} \right)$. The expressions of cavity frequencies $\nu_n$, associated losses $\kappa_n$ and modal interaction strengths $g_n$ are given in App.~\ref{SubApp:Spec Rep of G-open}. All these quantities are functions of $\chi_s$, the strength of the modification of the capacitance per unit length. 

%%%%%%%%%%% Fig of NHEigFreqs %%%%%%%%%%%%%%%%%%%
\begin{figure}[t!]
\centering
\subfloat[\label{subfig:gXrXl1Em3}]{%
\includegraphics[scale=0.45]{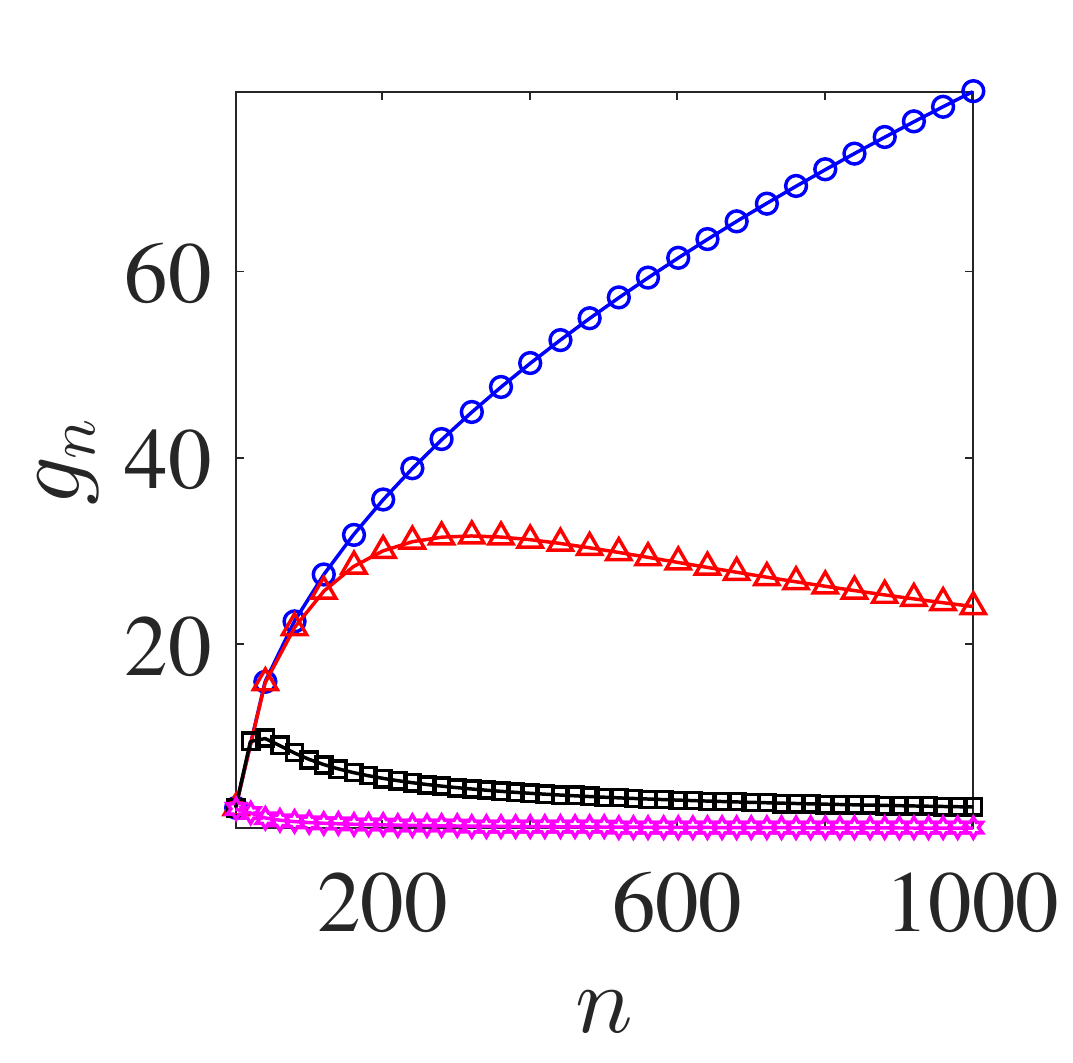}%
}
\subfloat[\label{subfig:KappaXrXl1Em3}]{%
\includegraphics[scale=0.45]{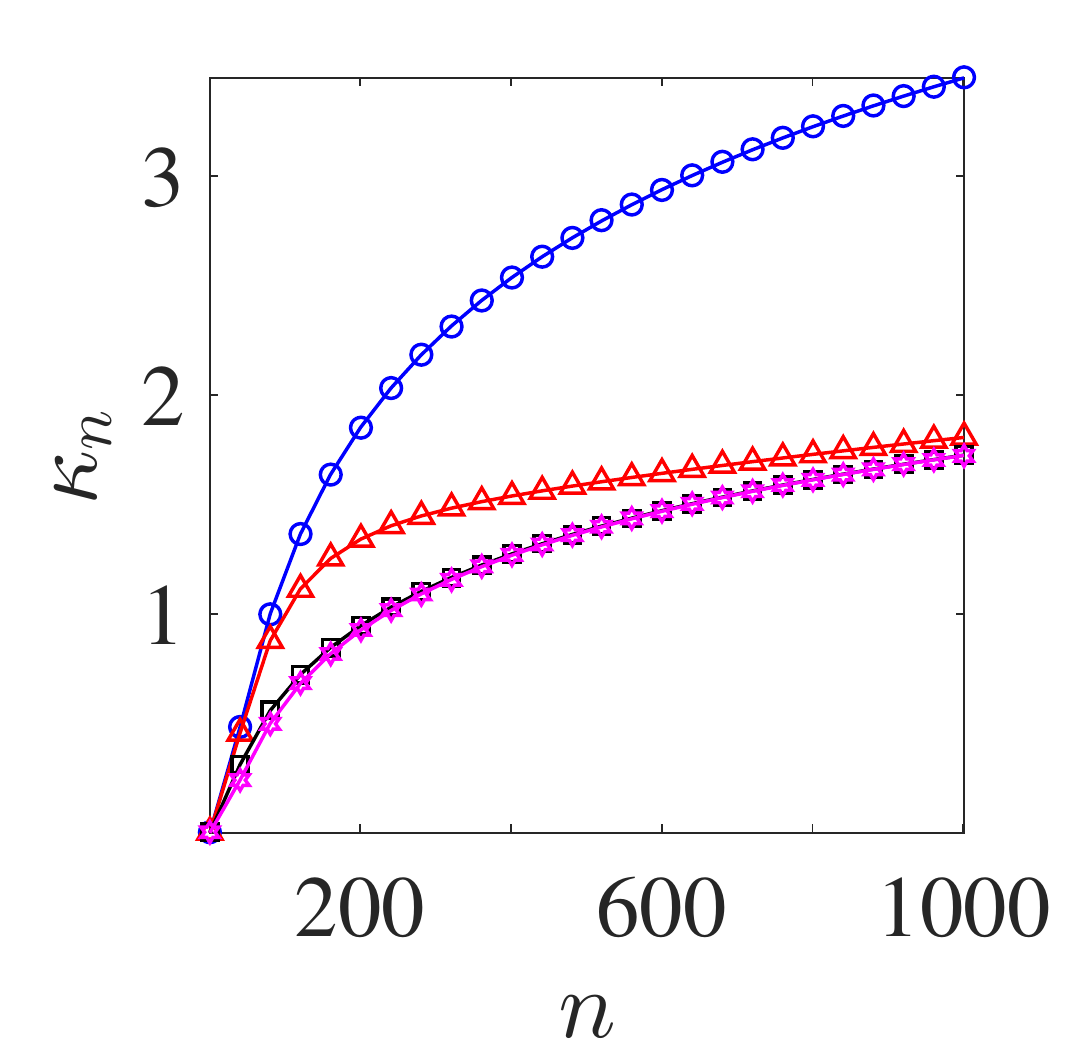}%
}\\
\subfloat[\label{subfig:LogLogSpEmRateXrXl1Em3}]{%
\includegraphics[scale=0.33]{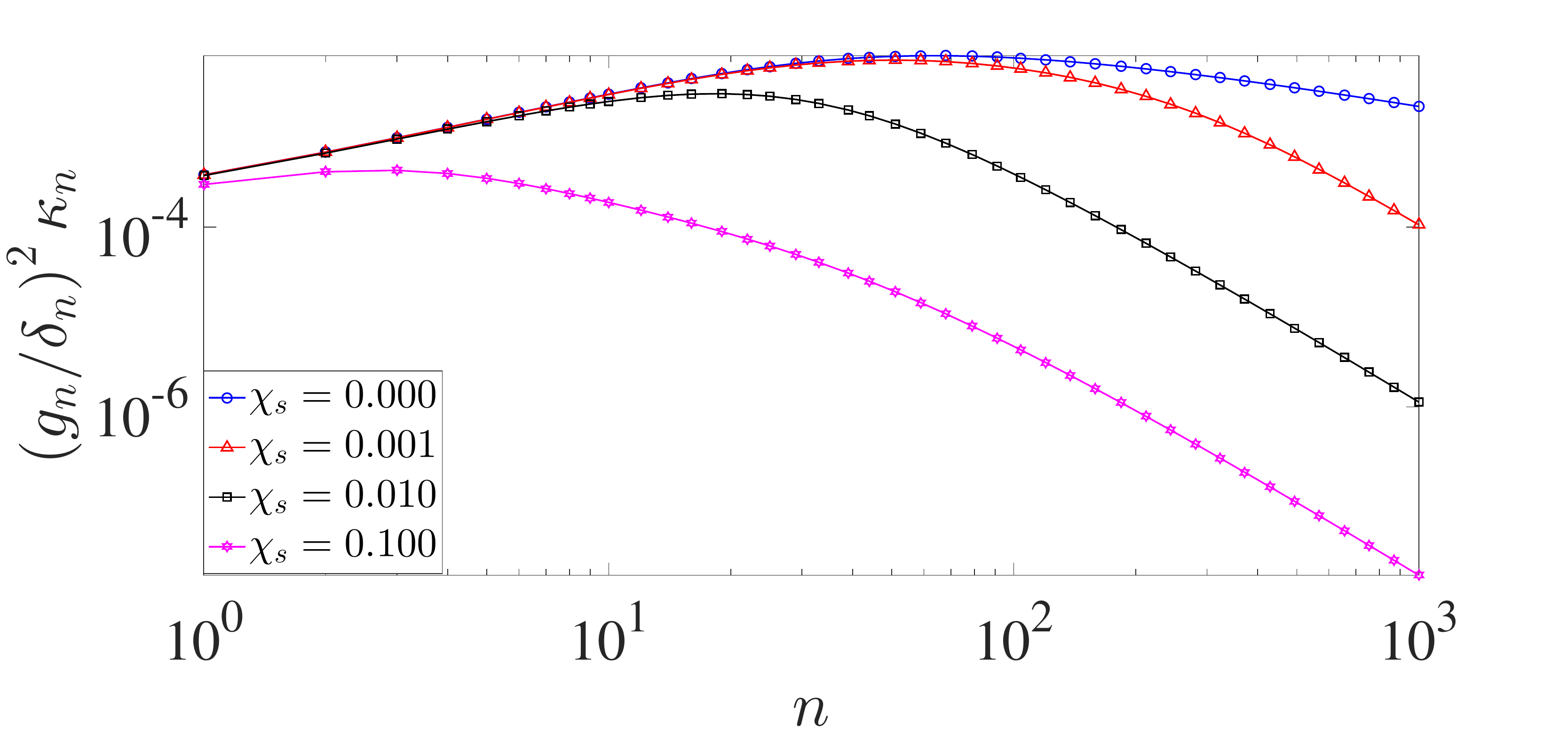}%
}
\caption{Dependence of a) coupling strength $g_n$, b) resonator decay rate $\kappa_n$ (See App.~\ref{SubApp:Spec Rep of G-open} for derivation) and c) Purcell decay rate in the dispersive regime $(g_n/\delta_n)^2\kappa_n$ on mode number $n$ for different values of $\chi_s=\{0,10^{-3},10^{-2},10^{-1}\}$. Other parameters are set as $\chi_R=\chi_L=10^{-3}$. The qubit position is chosen as $x_0=0^+$ so that only the suppression in the envelope is observed and but not the periodic oscillations observed in Fig.~\ref{subfig:couplingsx001Largescale}.} 
\label{Fig:Effect of nonzero Xs}
\end{figure}
%%%%%%%%%%%%%%%%%%%%%%%%%%%%%%%%%%%%%%%%%%%%
With the modified parameters, we revisit the convergence of the dispersive Purcell rate~(\ref{eqn:Multimode Purcell Rate}). We show again in Fig.~\ref{subfig:gXrXl1Em3} that $g_n$ is non-monotonic \cite{Malekakhlagh_Origin_2016} for any $\chi_s\neq 0$, first increasing, then turning over at a critical $\chi_s$-dependent mode $n$, decreasing as $g_n \sim 1/\sqrt{n}$ in the large-$n$ limit (See App.~\ref{App:AsymptoticOfgn}). This high frequency behavior of $g_n$ is the key to render the multimode Purcell rate finite, without an imposed cutoff. From circuit theory point of view, this can understood as the series capacitance $\chi_s$ becomes a short-circuit to ground at high frequencies, acting as a low-pass filter and suppressing mode amplitude at $x_0$. This is the cause of the power law drop of $g_n$ as $n \rightarrow \infty$. This phenomenon is not specific to the resonator geometry in \Fig{Fig:cQEDOpen-EqCircuit}. The underlying physical reason is the conservation of current at the position $x_0$ of the qubit. Moreover, eliminating the continuum degrees of freedom of the waveguides gives an effective decay rate for each mode, $\kappa_n$, which increases monotonically as $\kappa_n \sim n^{0.3}$ (\Fig{subfig:KappaXrXl1Em3}). 

As pointed out in previous studies \cite{Houck_Controlling_2008, Filipp_Multimode_2011}, for $\chi_s=0$ the resulting series \Eq{eqn:Multimode Purcell Rate} diverges. This is the case since with $g_n\sim\sqrt{n}$, $\nu_n\sim n$ and monotonically increasing $\kappa_n$ we find the asymptotic of individual terms as
\begin{align}
\left(\frac{g_n}{\omega_j-\nu_n}\right)^2\kappa_n \sim \frac{1}{n^d},
\end{align} 
with $0<d<1$ which results in a diverging series. On the other hand, for {\it any} nonzero $\chi_s$, individual terms in the sum~(\ref{eqn:Multimode Purcell Rate}) attain a universal power law $\sim n^{-2.7}$ (\Fig{subfig:LogLogSpEmRateXrXl1Em3}), which guarantees convergence. Note that the power law dependence of $\kappa_n$ and $g_n$, albeit universal with respect to $\chi_s$, are specific to the chosen circuit topology.

\begin{figure}[t!]
\centering
\subfloat[\label{subfig:SpEmRateXrXl1Em3Xj5Em2Xg1Em3}]{%
\includegraphics[scale=0.45]{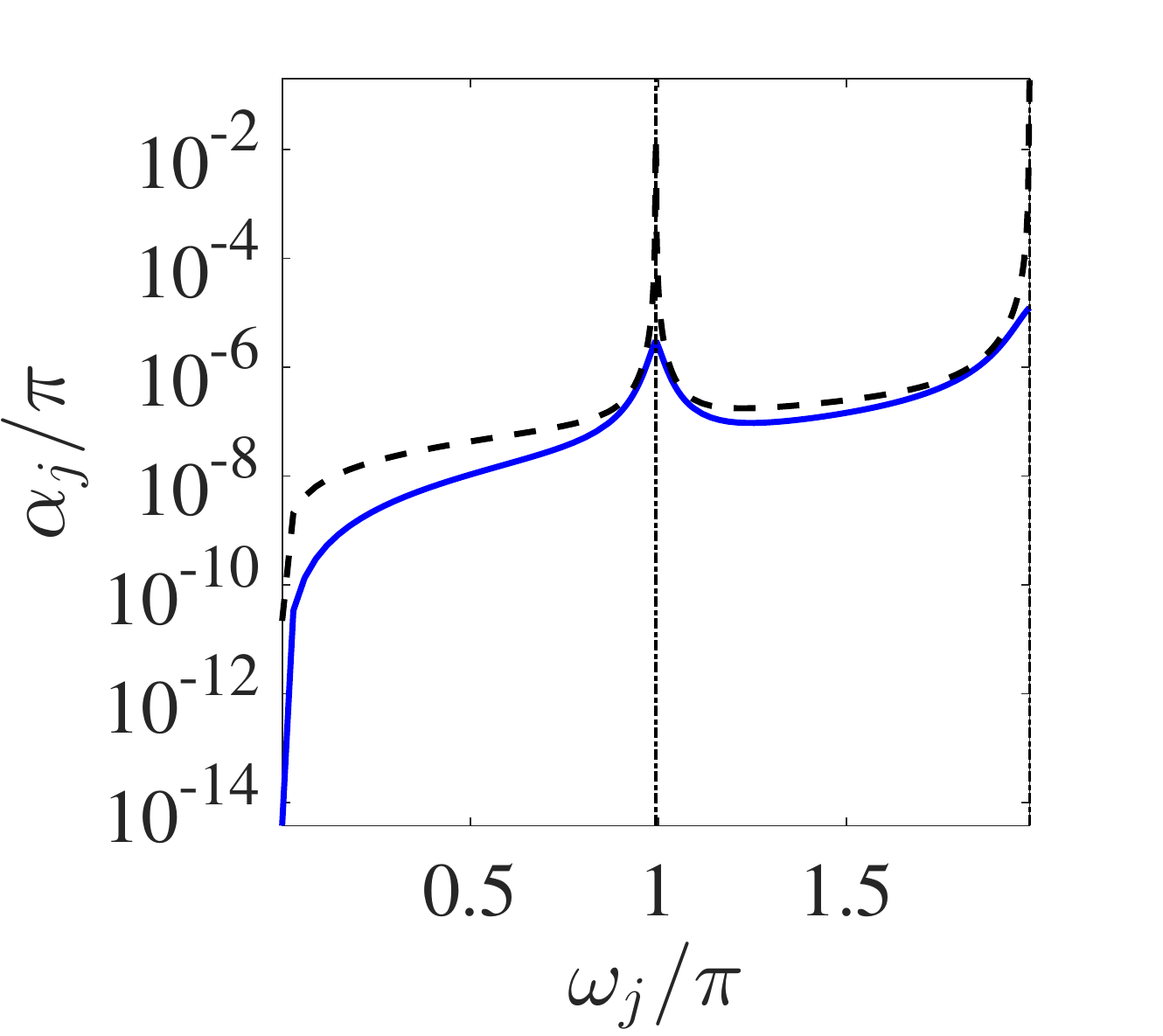}%
}
\subfloat[\label{subfig:SpEmRateXrXl1Em3Xj5Em2Xg1Em1}]{%
\includegraphics[scale=0.45]{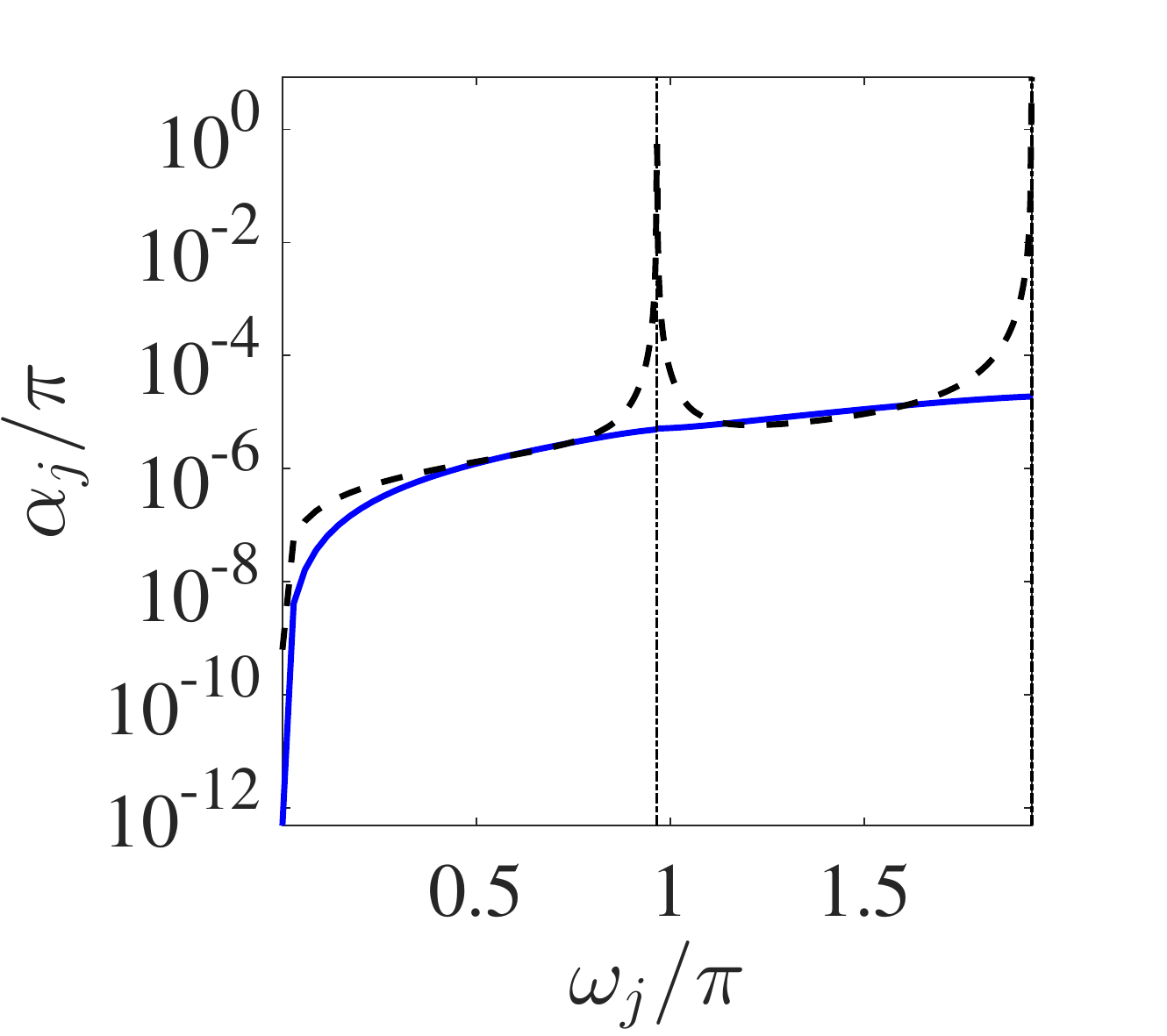}%
}\\
\subfloat[\label{subfig:LmbShftXrXl1Em3Xj5Em2Xg1Em3}]{%
\includegraphics[scale=0.45]{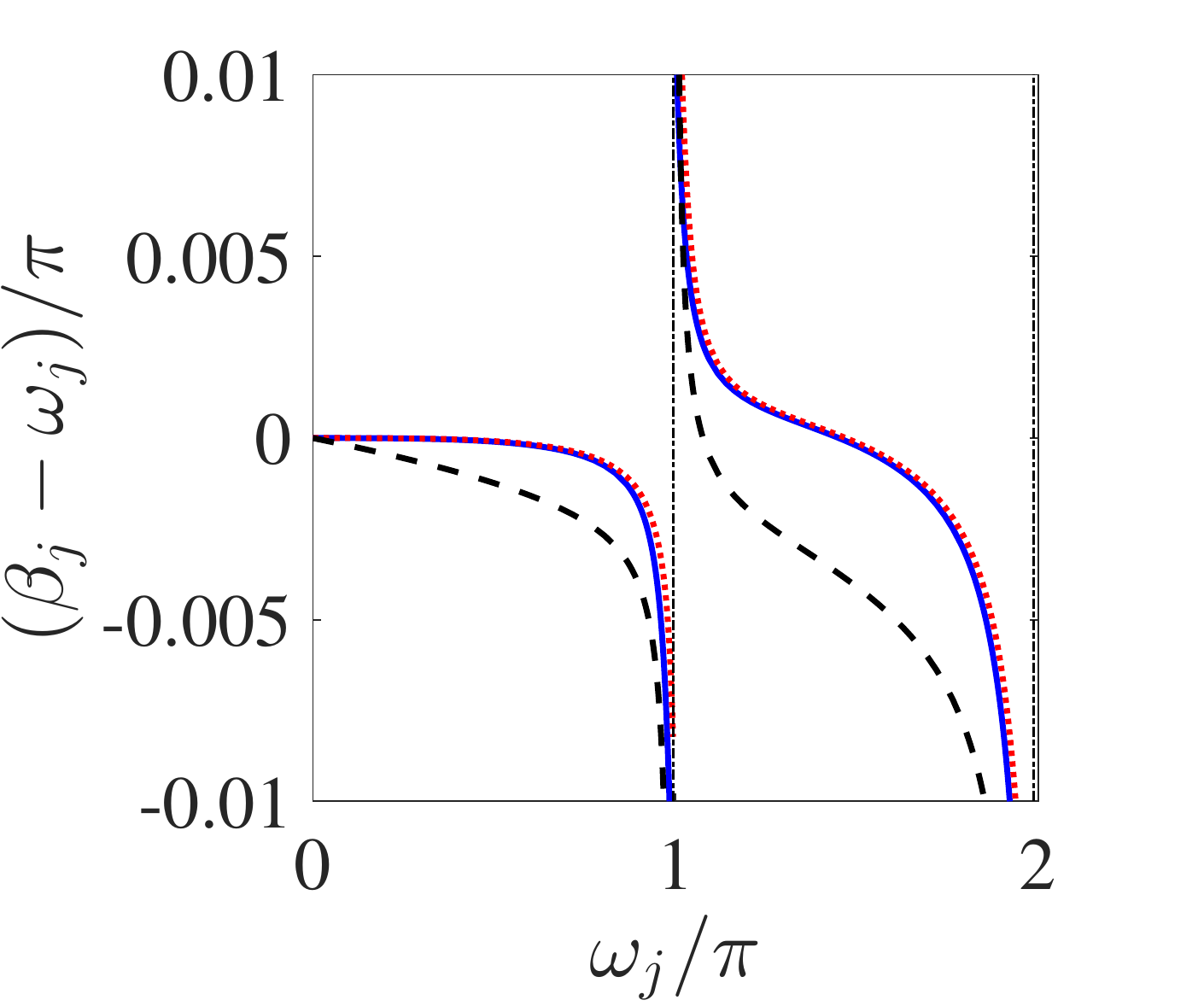}%
}
\subfloat[\label{subfig:LmbShftXrXl1Em3Xj5Em2Xg1Em1}]{%
\includegraphics[scale=0.45]{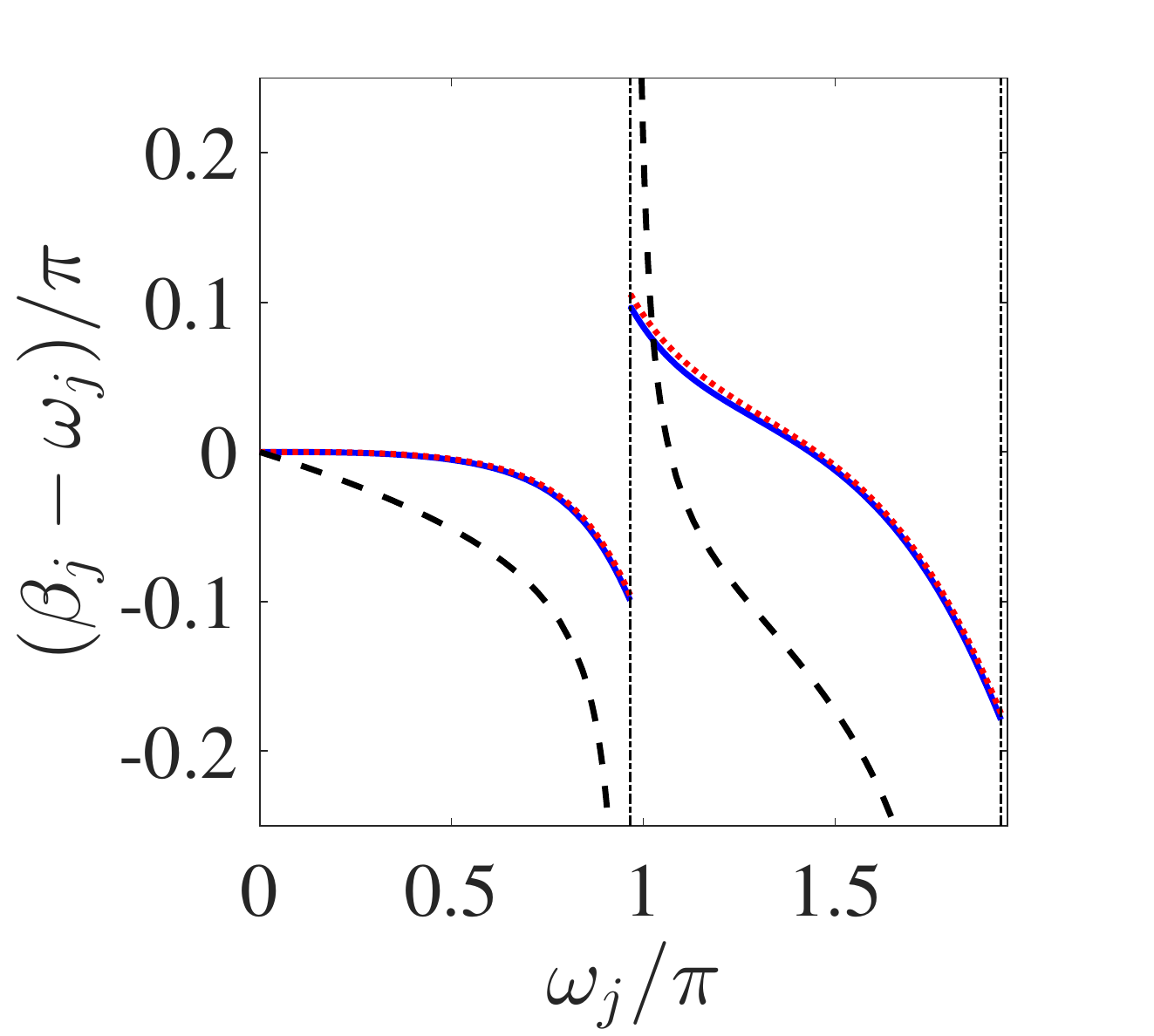}%
}
\caption{Comparison of a,b) spontaneous decay rate between the linear theory (blue solid) and the dispersive limit result $\gamma_P$ (black dashed) as a function of $\omega_j$. c,d) Lamb shift between the linear theory (blue solid), leading order perturbation (red dotted) and the dispersive limit result $\Delta_L$ (black dashed). a,c) $\chi_g=0.001$ and b,d) $\chi_g=0.1$. Both values of $\chi_g$ are in strong coupling regime, i.e. $g_1/\alpha_j\gg 1$. However, $\chi_g=0.1$ ($g_1/\nu_1=0.1033$) reaches ultrastrong coupling \cite{Niemczyk_Circuit_2010}, where multimode effects are non-negligible. The nonlinearity is set as $\epsilon=0.1$, while other parameters are $\chi_R=\chi_L=10^{-3}$ and $\chi_j=0.05$. The vertical dash-dotted black line shows the position of the fundamental frequency of the resonator.} 
\label{Fig:SpEmRate&LambShift}
\end{figure}

Although we showed that the expression~(\ref{eqn:Multimode Purcell Rate}) for the Purcell decay rate converges, this estimate is only valid in the dispersive regime $g_n \ll |\omega_j-\nu_n|$. The estimates for the Purcell decay rate and the Lamb shift will deviate substantially from the exact result for a range of order $g_n$ around each cavity resonance, diverging as the qubit frequency approaches the resonance (See Fig.~\ref{Fig:SpEmRate&LambShift}). This fictitious divergence can in principle be cured by solving the full multimode Master equation. Even if computational challenges relating to the long-time dynamics in such a large Hilbert space can be addressed, the resulting rate would still be subject to the TLA, RWA, Born and Markov approximations, casting a priori an uncertainty on its reliability.

To overcome the limitations of the dispersive limit JC model, we next apply our Heisenberg-Langevin formalism that has been discussed in chapter~.\ref{Ch:NonMarkovian} \cite{Malekakhlagh_NonMarkovian_2016}. According to this, the transmon qubit and the resonator are described by a set of coupled differential equations in Heisenberg picture as
\begin{align}
&\hat{\ddot{\varphi}}_j(t)+ ( 1- \gamma) \omega_j^2\sin{[\hat{\varphi}_j(t)]}  = \gamma\partial_{t}^2\hat{\varphi}(x_0,t),  \label{eqn:TransDyn} \\
&\left[\partial_{x}^2-\chi(x,x_0)\partial_{t}^2\right]\hat{\varphi}(x,t)  = \chi_s\omega_j^2 \sin{[\hat{\varphi}_j(t)]}\delta(x-x_0).
\label{eqn:ResDyn}
\end{align}
Here $\hat{\varphi}_j(t)$ and $\hat{\varphi}(x,t)$ are dimensionless phase operators for the JJ and the resonator-waveguide system, respectively. These two inhomogeneous equations show that the flux field at $x_0$ drives the dynamics of the JJ [\Eq{eqn:TransDyn}], while the JJ acts as a source driving the EM fields [\Eq{eqn:ResDyn}]. In addition, the fields are subject to continuity conditions at the ends of the resonator $x=0,1$ (in units of $L$). This is consistent \cite{Malekakhlagh_Origin_2016} with Kirchhoff's laws for current conservation. 

In chapter~\ref{Ch:NonMarkovian}, we explained in detail how we obtain the effective dynamics for the qubit and provided a systematic solution via perturbation theory. Here, we just mention the important steps. Equation~\ref{eqn:ResDyn} can be solved in the Fourier domain, where 
$\hat{\tilde{\varphi}}(x,\omega) = \int_{-\infty}^{\infty} dt \, \hat{\varphi}(x,t) e^{-i\omega t}$ can be expanded in terms of the basis $\tilde{\varphi}_n (x,\omega)$ that solves the generalized eigenvalue problem 
\begin{align}
\left[\partial_x^2 + \chi(x,x_0) \, \omega^2 \right] \tilde{\varphi}_n (x,\omega)=0,
\end{align}
subject to continuity conditions at the ends of the resonator, i.e. 
\begin{align}
&-\partial_x \tilde{\varphi}_n (1^-,\omega)= \chi_{R} \omega^2[ \tilde{\varphi}_n (1^-,\omega)-\tilde{\varphi}_n (1^+,\omega)],\\ 
&-\partial_x \tilde{\varphi}_n (0^+,\omega)= \chi_{L} \omega^2 [\tilde{\varphi}_n (0^-,\omega)-\tilde{\varphi}_n (0^+,\omega)],
\end{align}
which models the coupling to the waveguides and associated loss. The Dirac $\delta$-function in $\chi(x,x_0)$ leads to the discontinuity
\begin{equation}
-\left.\partial_x \tilde{\varphi}_n(x)\right]_{x_0^-}^{x_0^+} = \chi_s \omega_n^2 \tilde{\varphi}_n(x_0),
\label{eqn:Current-Conservation} 
\end{equation}
amounting to current conservation at the qubit position, resulting in a modified, current-conserving (CC) basis \cite{Malekakhlagh_Origin_2016}. These modifications in the spectrum of the transmission line resonator impact the qubit dynamics that is driven by resonator fluctuations.

\begin{figure}[t!]
\centering
\includegraphics[scale=0.30]{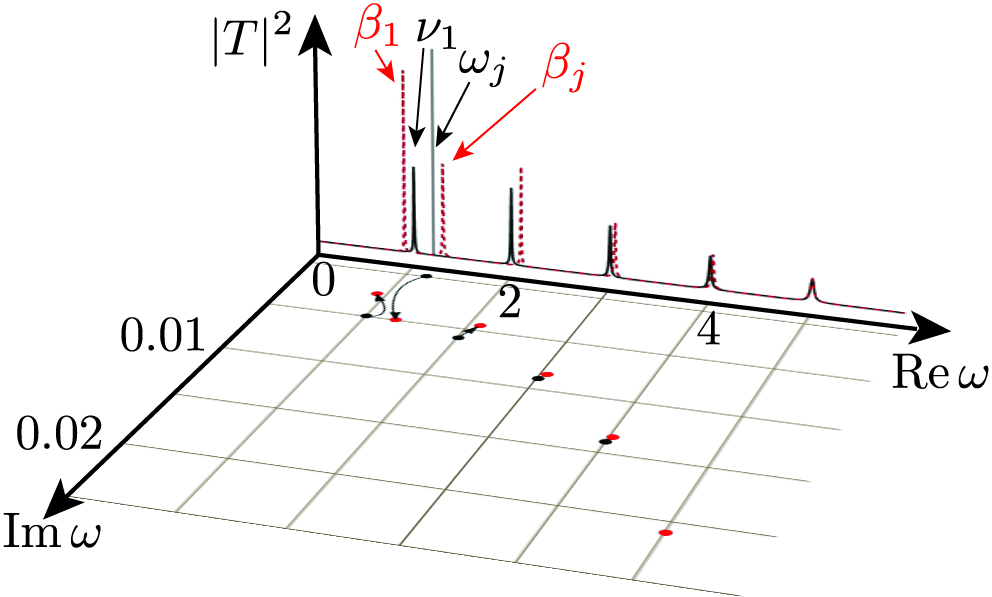}
\caption{A schematic demonstration of the calculation of the Purcell decay rate and the  Lamb shift using our Heisenberg-Langevin framework. The transmission  $|T|^2$ is shown versus the real frequency for the bare resonator modes (solid black curves). Capacitive coupling to the qubit, whose transition frequency $\omega_j$ is slightly above the fundamental resonator frequency $\nu_1$, gives rise to hybridized modes (dashed red curves). Alternatively, one may study the positions of these resonances in the complex frequency plane, where the bare resonator and qubit poles (black points) are displaced into hybridized resonator-like and qubit-like resonances (red points). The Purcell decay and the Lamb shift are obtained as the displacement of the qubit-like pole. The bare (hybridized) complex frequencies are the poles (zeros) of the characteristic function $D_j(s)$.}
\label{Fig:A2-PolesSFLabeled}
\end{figure}

Using our formalism, we can provide an improved analytic result that is uniformly valid in the transmon frequency, and is not limited by the aforementioned approximations. Electromagnetic degrees of freedom can be integrated out by solving \Eq{eqn:ResDyn} exactly, plugging into \Eq{eqn:TransDyn} and tracing over the photonic Hilbert space. To lowest order in the transmon nonlinearity $\epsilon = (E_c/E_j)^{1/2}$, where $E_c$ and $E_j$ are the charging and Josephson energy, respectively, the effective equation for the qubit is \cite{Malekakhlagh_NonMarkovian_2016}
\begin{align}
\hat{\ddot{X}}_j(t)+\omega_j^2\left[1-\gamma+i\mathcal{K}_1(0)\right]\hat{X}_j(t)=- \omega_j^2 \int_0^{t}dt' \mathcal{K}_2(t-t') \hat{X}_j(t'),
\label{eqn:LinSEProblem}
\end{align}
where $\hat{X}_j (t) =  \Tr_{ph}\{\hat{\rho}_{ph}(0)\hat{\varphi}_j(t)\}/\phi_{\text{zpf}}$ is the reduced flux operator traced over the photonic degrees of freedom and $\phi_{\text{zpf}}\equiv(\sqrt{2}\epsilon)^{1/2}$ is the magnitude of the zero-point phase fluctuations. This delay equation features the memory kernels  
\begin{align}
\mathcal{K}_n(\tau)\equiv\gamma\chi_s\int_{-\infty}^{+\infty} \frac{d\omega}{2\pi} \, \omega^n \, \tilde{G} (x_0,x_0,\omega)e^{-i\omega\tau},
\end{align}
where $\tilde{G}(x,x',\omega)$ is the classical EM Green's function defined by 
\begin{align}
\left[\partial_{x}^2+\chi(x,x_0)\omega^2\right] \tilde{G}(x,x',\omega)=\delta(x-x'),
\end{align} 
implying that $\tilde{G} (x,x',\omega)$ is the amplitude of the flux field created at $x$ by a transmon oscillating with a frequency $\omega$ at $x'$ \cite{Malekakhlagh_NonMarkovian_2016}. The term on the right hand side of \Eq{eqn:LinSEProblem} is therefore proportional to the fluctuating current driving the qubit at time $t$, that was excited by itself at an earlier time $t'$. This Green's function correctly encodes the modification of the capacitance per length. Equation~(\ref{eqn:LinSEProblem}) can be solved exactly in the Laplace domain
\begin{equation}
\hat{\tilde{X}}_j(s)=\frac{s\hat{X}_j(0)+ \hat{\dot{X}}_j(0)}{D_j(s)},
\label{eqn:Sol of X_j(s)}
\end{equation}
where $\tilde{h}(s)\equiv\int_{0}^{\infty} dt \, h(t) \, e^{-st}$, with $D_j(s)$ defined as \cite{Malekakhlagh_NonMarkovian_2016}
\begin{equation}
D_j(s)\equiv s^2+\omega_j^2\left[1-\gamma+i\mathcal{K}_1(0)+\tilde{\mathcal{K}}_2(s)\right].
\label{eqn:Def of D(s)}
\end{equation}
We express the characteristic function $D_j(s)$ in meromorphic form 
\begin{equation}
D_j(s)=(s-p_j)(s-p_j^*)\prod\limits_{m}\frac{(s-p_m)(s-p_m^*)}{(s-z_m)(s-z_m^*)}.
\label{eqn:Formal Rep of D(s)}
\end{equation}
The poles of $1/D_j(s)$ are the hybridized qubit-like and resonator-like complex-valued excitation frequencies, $p_j\equiv -\alpha_j-i\beta_j$ and $p_n\equiv -\alpha_n-i\beta_n$, respectively, of the qubit-resonator system, while its zeroes $z_n \equiv -i\omega_n=-\kappa_n-i\nu_n$ correspond to bare non-Hermitian cavity resonances. The real part of the qubit-like pole, $\alpha_j$, is the Purcell loss rate, while $\beta_j - \omega_j$ is the Lamb shift, as shown in Fig.~\ref{Fig:A2-PolesSFLabeled}. More importantly, it can be shown that the characteristic function $D_j(s)$ is itself convergent, and hence so are all hybridized frequencies, for any nonzero $\chi_s$ (See App.~\ref{App:ConvOfDj}).
%%%%%%%%%%%%%% Fig:LC Osc %%%%%%%%%%%%%%
%%%%%%%%%%%%%%%%%%%%%%%%%%%%%%%%%%%%%%%%%%
%%%%%%%%%%% Fig of SpEm&LambShift %%%%%%%%%%%%%%%%%%%
%%%%%%%%%%%%%%%%%%%%%%%%%%%%%%%%%%%%%%%%%%%%

The transmon nonlinearity neglected in Eq.~(\ref{eqn:LinSEProblem}) can be reintroduced as a weak perturbation to the exactly solvable linear theory. The leading order correction to the hybridized resonances amounts to self- and cross-Kerr interactions \cite{Nigg_BlackBox_2012, Bourassa_Josephson_2012}. Using multi scale perturbation theory \cite{Bender_Advanced_1999, Malekakhlagh_NonMarkovian_2016}, the correction to the transmon qubit-like resonance $\beta_j$ is given by
\begin{equation}
\hat{\beta}_j=\beta_j-\frac{\sqrt{2}\epsilon}{4}\omega_j\left[u_j^4\hat{\mathcal{H}}_j(0)+\sum\limits_n 2u_j^2u_n^2\hat{\mathcal{H}}_n(0)\right],
\label{eqn:PertCor}
\end{equation}
where the coefficients $u_{j,n}$ define the transformation from the hybridized to the unhybridized modes and $\hat{\mathcal{H}}_{j,n}(0)$ are the free Hamiltonian of the transmon and mode $n$, respectively. For $\chi_g\to0$, we find $u_j\to1$, $u_n=0$ and $\beta_j\to\omega_j$ such that we recover the frequency correction of free quantum Duffing oscillator $\hat{\bar{\omega}}_j=\omega_j[1-\frac{\sqrt{2}\epsilon}{4}\hat{\mathcal{H}}_j(0)]$ \cite{Bender_Multiple_1996}. We note three features of this result. Firstly, the correction is an operator and that expresses the fact that transmon levels are anharmonic. The anharmonicity can be calculated from the expectation value of a corrected quadrature operator. Secondly, by virtue of the lowest order result being convergent without a cutoff, the perturbative corrections are also convergent in the number of modes included. Finally, this result is not limited by the qubit-resonator coupling strength or the openness of the cavity. The final result is finite for all qubit frequencies, as opposed to the dispersive-limit result. The correction to the Purcell decay is higher order and forms the subject of future work.

We compared the spontaneous decay from the linear theory (\textcolor{blue}{blue solid}) to the dispersive limit estimate $\gamma_P$ in Eq.~(\ref{eqn:Multimode Purcell Rate}) (black dashed) as the transmon frequency is tuned across the fundamental mode in Figs.~\ref{subfig:SpEmRateXrXl1Em3Xj5Em2Xg1Em3}-\ref{subfig:SpEmRateXrXl1Em3Xj5Em2Xg1Em1}. 
First, the spontaneous decay is asymmetric, since there are (in)finitely many modes with frequency (larger) smaller than $\omega_j$. This feature is captured by both theories. Second, the spontaneous decay is enhanced as the qubit frequency approaches the fundamental resonator frequency. However, the dispersive limit estimate is perturbative in $g_n/\delta_n$ and hence yields a divergent result (fake kink) on resonance regardless of coupling constant, contrary to our result~\ref{eqn:Formal Rep of D(s)} which predicts a finite value even in ultrastrong coupling (Fig.~\ref{subfig:SpEmRateXrXl1Em3Xj5Em2Xg1Em1}).

In Figs.~\ref{subfig:LmbShftXrXl1Em3Xj5Em2Xg1Em3}-\ref{subfig:LmbShftXrXl1Em3Xj5Em2Xg1Em1} we compare the Lamb shift from the linear theory (blue solid) and the leading order perturbation theory (red dotted) to the dispersive multimode estimate (black dashed) $\sum_n g_n^2/\delta_n$ \cite{Boissonneault_Dispersive_2009}. Below the fundamental mode, the Lamb shift is negative due to the collective influence of all higher modes that redshifts the qubit frequency. Above the fundamental mode, there appears a competition between the hybridization with the fundamental mode and all higher modes. Close enough to the fundamental mode, the Lamb shift is positive until it changes sign, as predicted by all three curves.
\section{Discussion}
\label{Sec:A^2-Discussion}

The corrections to the spectral structure of the resonator found in Sec.~\ref{Sec:A^2-GaugeInvRabiModel} is mathematically equivalent to the scattering corrections that result from the presence of an atom in atomic CQED systems. Electromagnetic field quantization has been studied in great detail for CQED systems including single electron atoms \cite{Scully_Quantum_1997, Walls_Quantum_2008, Schleich_Quantum_2011}, multi-electron atoms \cite{Knoll_Action_1987}, and for atoms embedded in dispersive and absorptive dielectric media~\cite{Knoll_Action_1987, Glauber_Quantum_1991, Huttner_Quantization_1992, Matloob_Electromagnetic_1995, Dung_Three-Dimensional_1998, Knoll_QED_2000}. For completeness, in App.~\ref{App:DerivationOfHamCQED}, we present a full derivation of the minimal coupling Hamiltonian (neglecting electron's spin) for this system starting from a Lagrangian formalism that yields the Maxwell's equations and the Lorentz force law \cite{Knoll_Action_1987}. The term $\mathcal{L}_{int}=\mathcal{H}_{int}=\frac{1}{2}C_g\left[\dot{\Phi}_J-\dot{\Phi}(x_0,t)\right]^2$ that appears in the canonical quantization of cQED systems is mathematically equivalent to the approximate (zero-order dipole approximated) minimal coupling term $\mathcal{T}_e \approx \frac{1}{2m_e}\left[{\vec{p}_e-e\vec{A(\vec{R}_{cm},t)}}\right]^2$ appearing in CQED Hamiltonian, thus their impact on the cavity modal structure is similar. 

It could be argued that the freedom in the choice of the point of reference for the generalized fluxes i.e. the choice of ground, is analogous to the gauge freedom. However, the fact that the cavity modes are modified due to  the existence of the qubit is a property that is gauge-independent. In App.~\ref{App:DerivationOfHamCQED}, we show that in a similar manner to the discussion here, the existence of the $A^2$ term in the Coulomb gauge gives rise to modified spectral properties of a cavity. However, in atomic CQED these corrections are tiny because of the smallness of typical atomic transition dipoles and the fine structure constant. In App.~\ref{SubApp:A^2-ModEigModes}, we have also studied the reverse question and proved it is feasible to retrieve an $A^2$-like term if one naively performs the quantization by the cosine basis. This completes the similarity between cQED and CQED Hamiltonians in the lowest order (zeroth order dipole approximation) where the dimension of transmon (atom) is completely neglected compared to the cavity's wavelength.

Furthermore, the $A^2$ term kept in our calculation to enforce gauge invariance plays the role of the ``counterterm" discussed by Caldeira and Leggett to cancel an infinite frequency renormalization in their phenomenological description of interaction of a quantum system to its environment \cite{Caldeira_Quantum_1983, Leggett_Quantum_1984}. This problem had also been encountered in the quantum theory of laser radiation \cite{Schwabl_Quantum_1964}.

In summary, we have presented a framework to calculate the spontaneous decay and the Lamb shift of a transmon qubit, which are convergent in the number of resonator modes without the need for rotating-wave, two-level, Born or Markov approximations, or a high frequency cutoff. This is achieved by an \textit{ab initio} derivation of the quantum circuit equations of motion containing the $A^2$ term to enforce gauge invariance. As a result, the modes of the resonator are modified such that the light-matter coupling is suppressed at high frequencies. Formulating the cavity resonances in terms of non-Hermitian modes provides access to the spontaneous decay, the Lamb shift, and any other QED observables in a unified way.
\chapter{Summary and outlook}
\label{Ch:SummaryOutlook}

The interest in circuit-QED systems has increased significantly during the past decade as a promising platform for both quantum information processing and fundamental studies of non-equilibrium quantum phenomena\cite{Devoret_Implementing_2004, Blais_Quantum-Information_2007, Devoret_Superconducting_2013}. Quantum information applications have motivated the majority of previous theoretical studies to employ models that first appeared in earlier analoguous cavity-QED systems. These phenomenological models, like Jaynes-Cummings, use simplifying approximations that work well for cavity-QED systems, but fail to capture some experimental results on circuit-QED systems due to the different regime of parameters that such systems operate in. To resolve some of the earlier theoretical anomalies originating from these phenomenological models, we found it essential to introduce a first principles formalism based on Heisenberg-Langevin equations of motion.
 
First, even though most charge-based superconducting qubits operate in a weakly nonlinear regime \cite{Koch_Charge_2007}, majority of previous studies use a two-level approximation to describe such qubits. Treating qubits as two-level systems facilitates numerical simulation by decreasing the Hilbert size of interest, but the spin-$\frac{1}{2}$ nonlinearity is harder to deal with analytically due to the non-commuting spin algebra. Moreover, a two-level approximation completely neglects the origin of nonlinearity. Instead of two-level approximation, we have built a perturbation theory in the weak nonlinearity of the qubit. The conventional perturbation theories, like Rayleigh-Schr\"odinger, are not applicable at the level of Heisenberg equations of motion due to appearance of secular terms that grow unbounded in time. This anomaly can be resolved by employing a different perturbation scheme, called multi scale perturbation theory.

Second, stronger light-matter coupling in circuit-QED systems leads to violation of rotating-wave approximation \cite{Niemczyk_Circuit_2010}. The main motivation behind this approximation, which is applied to the Jaynes-Cummings model, is to impose a fictitious $\mathbb{U}(1)$ symmetry that renders the total number of excitation a constant of motion. Therefore, one can obtain analytical expressions for the eigenmodes of the multimode Jaynes-Cummings model. Although it has been proven that the $\mathbb{Z}_2$ symmetry of the single mode Rabi model is also sufficient for integrability \cite{Braak_Integrability_2011, Moroz_Solvability_2013}, no generalization to the multimode scenario has been provided so far. On the other hand, using our Heisenberg-Langevin formalism, we solve for the spectrum without the need for rotating wave approximation. The reason lies in the weak nonlinearity of transmon, which allows us to follow a perturbation scheme. Up to the lowest order, the transmon behaves like a quantum harmonic oscillator and the theory is exactly solvable without resorting to rotating wave approximation. Higher order corrections are then achieved using multi scale perturbation theory, which appears as self-and cross-Kerr contributions.

Third, the general line of thought to account for the openness of QED system is to extend the Hamiltonian into a Linbladian superoperator that takes into account the dissipation using Born-Makrov approximation. In our Heisenberg-Langevin formalism, we take advantage of the linear equations of motion that govern the electromagnetic degrees of freedom. These equations can always be solved exactly and the information about the electromagnetic background can be encoded in the classical electromagnetic Green's function. Then, poles of the Green's function give the \textit{exact} bare oscillation frequencies and dissipation rates of the cavity.

Fourth, in circuit-QED architecture, the light-matter coupling can be substantial \cite{Niemczyk_Circuit_2010, Bourassa_Ultrastrong_2009, Forn_Observation_2010} and it is found that QED quantities are strongly influenced by far off-resonant modes \cite{Houck_Controlling_2008, Sundaresan_Beyond_2015}. Previous multimode models lead to divergent results, while they cured this anomaly by applying an artificial cutoff \cite{Krimer_Route_2014}. In chapter~\ref{Ch-OriginOfA2} we showed that by placing the qubit into a resonator, the electromagnetic mode structure of the cavity is altered such that renders multimode quantities convergent. This was achieved by incorporating the effect of the diamagnetic $A^2$ contribution in our formalism exactly, in contrast to previous perturbative treatments of $A^2$ term.

To show the advantage of our Heisenberg-Langevin formalism, we applied it to the problem of spontaneous emission, and compared the results for the Purcell decay rate and the Lamb shift to the ones obtained from multimode Jaynes-Cummings model in the dispersive limit. Besides resolving the anomaly of multimode divergence, our theory is not limited by light-matter coupling strength or detuning, contrary to the dispersive Jaynes-Cummings results. Even though we only studied spontaneous emission, our formalism can be applied to the driven-dissipative case as well. Moreover, since we developed the theory in terms of Heisenberg operators, higher order correlation functions can be achieved immediately from the solutions for single body operators.

%%%%%%%%%%%%%%%%%%%%%%%%%%%%%%%%%%%%%%%%%%%
%%%%%%%%%%%%%%%%% Appendices %%%%%%%%%%%%%%
\appendix
\chapter{Classical Hamiltonian and modified eigenmodes of a closed cQED system}
\label{App:DerivationOfcQEDHam}
Here, we follow the quantization procedure discussed in sec.~\ref{SubSec:Background-circuit quantization} for the system shown in Fig.~\ref{Fig:cQEDOpen-EqCircuit}. We first use a discretized lumped element LC-model \cite{yurke_quantum_1984} for the microwave resonator and then take the limit where these infinitessimal elements go to zero while leaving the capacitance and inductance per length of the resonator invariant. 
\section{Discrete limit}
\subsection{Classical Lagrangian for the discretized circuit}
In terms of the generalized coordinates introduced in Appendix A, the Lagrangian for the discretized circuit can be written as the difference between kinetic capacitive energy and potential inductive energy and it reads
\begin{align}
\begin{split}
\mathcal{L}&=\frac{1}{2}C_j\dot{\Phi}_j^2+E_j\cos\left(2\pi\frac{\Phi_j}{\Phi_0}\right)\\
&+\sum\limits_{n}\left[\frac{1}{2}c\Delta x\dot{\Phi}_n^2-\frac{1}{2l\Delta x}\left(\Phi_{n+1}-\Phi_n\right)^2\right]\\
&+\frac{1}{2}C_g(\dot{\Phi}_0-\dot{\Phi}_j)^2	.
\end{split}
\end{align}
In the expression above, we have labeled the discrete nodes such that the qubit is connected to the zeroth node.
\subsection{Classical Hamiltonian for the discretized circuit}

The first step to find the Hamiltonian is to derive the conjugate variables associated with the generalized coordinate $\{\{\Phi_{n}\};\Phi_j\}$. These conjugate variables will have the dimension of charge and we represent them as $\{\{Q_{n}\};Q_j\}$. By definition, these conjugate variables read
\begin{subequations}
\begin{align}
Q_j\equiv\frac{\delta \mathcal{L}}{\delta \dot{\Phi}_j}&=(C_j+C_g)\dot{\Phi}_j-\sum\limits_{n} C_g\delta_{n0}\dot{\Phi}_n, \\
Q_n\equiv\frac{\delta \mathcal{L}}{\delta \dot{\Phi}_n}&=(c\Delta x+C_g\delta_{n0})\dot{\Phi}_n-C_g\delta_{n0}\dot{\Phi}_j.
\end{align}
\end{subequations}

The next step is to calculate the discrete Hamiltonian by a Legendre transformation 
\begin{align}
\begin{split}
\mathcal{H}&=\sum\limits_n Q_n\dot{\Phi}_n+Q_j\dot{\Phi}_j-\mathcal{L}\\
&=\frac{1}{2}(C_j+C_g)\dot{\Phi}_j^2-E_j\cos\left(2\pi\frac{\Phi_j}{\Phi_0}\right)\\
&+\sum\limits_{n}\left[\frac{1}{2}(c\Delta x+C_g\delta_{n0})\dot{\Phi}_n^2+\frac{1}{2l\Delta x}\left(\Phi_{n+1}-\Phi_n\right)^2\right]\\
&-C_g\dot{\Phi}_0\dot{\Phi}_j.
\end{split}
\end{align}

Now, we need to solve for $\dot{\Phi}_j$ and $\dot{\Phi}_n$ in terms of $Q_j$ and $Q_n$ to represent the Hamiltonian only in terms of generalized coordinates and their conjugate variables. Before proceeding further, we define the following quantities in order to simplify the calculation
\begin{subequations}
\begin{align}
&\gamma\equiv\frac{C_g}{C_g+C_j},
\label{gamma}\\
&C_s\equiv\frac{C_gC_j}{C_g+C_j},
\label{Cs}\\
&C_{s,n}\equiv c\Delta x+C_s\delta_{n0},\\
&C_{g,n}\equiv c\Delta x+C_g\delta_{n0},
\end{align}
\end{subequations}
where $C_{s}$ represents the series combination of the coupling capacitor $C_g$ and Transmon capacitor $C_j$. In terms of these new quantities we can write
\begin{subequations}
\begin{align}
&\dot{\Phi}_n=\frac{Q_n}{C_{s,n}}+\frac{\gamma\delta_{n0}Q_j}{C_{s,n}},\\
&\dot{\Phi}_j=\left(\frac{\gamma}{C_j}+\frac{\gamma^2}{C_{s,0}}\right)Q_j+\sum\limits_n \frac{\gamma\delta_{n0}}{C_{s,n}}Q_n,\\
\begin{split}
&\mathcal{H}=\frac{1}{2}\frac{C_g}{\gamma}\dot{\Phi}_j^2 -E_j\cos\left(2\pi\frac{\Phi_j}{\Phi_0}\right)\\
&+\sum\limits_{n}\left[\frac{1}{2}C_{g,n}\dot{\Phi}_n^2+\frac{1}{2l\Delta x}\left(\Phi_{n+1}-\Phi_n\right)^2\right]\\
&-C_g\dot{\Phi}_0\dot{\Phi}_j
\end{split}
\end{align}
\end{subequations}

By inserting the expressions for $\dot{\Phi}_n$ and $\dot{\Phi}_j$ into the one for the Hamiltonian one finds that
\begin{align}
\begin{split}
\mathcal{H}&=\frac{1}{2}\left[\frac{\gamma}{C_g}+\frac{\gamma^2 (C_{g,0}-\gamma C_g)}{C_{s,0}^2}\right]Q_j^2-E_j\cos\left(2\pi\frac{\Phi_j}{\Phi_0}\right)\\
&+\sum\limits_{n}\left[\frac{1}{2}\frac{C_{g,n}-\gamma C_g\delta_{n0}}{C_{s,n}^2}Q_n^2+\frac{1}{2l\Delta x}\left(\Phi_{n+1}-\Phi_n\right)^2\right]\\
&+\frac{\gamma}{C_{s,0}}Q_jQ_0.
\end{split}
\end{align}
Notice that this expression can be further simplified since $C_{g,n}$ and $C_{s,n}$ are related as $C_{g,n} -\gamma C_g \delta_{n0}=C_{s,n} $. Therefore, the final result for the discretized Hamiltonian will be
\begin{align}
\begin{split}
\mathcal{H}&=\frac{1}{2}\left(\frac{\gamma}{C_g}+\frac{\gamma^2}{C_{s,0}}\right)Q_j^2-E_j\cos\left(2\pi\frac{\Phi_j}{\Phi_0}\right)\\
&+\sum\limits_{n}\left[\frac{Q_n^2}{2C_{s,n}}+\frac{1}{2l\Delta x}\left(\Phi_{n+1}-\Phi_n\right)^2\right]\\
&+\frac{\gamma}{C_{s,0}}Q_jQ_0 ,
\end{split}
\end{align}
where the conjugate variables obey the classical Poisson bracket relations
\begin{subequations}
\begin{align}
\{\Phi_n,Q_m\}&=\delta_{mn} \\
\{\Phi_j,Q_j\}&=1, \\
\{Q_n,Q_m\}&=\{\Phi_n,\Phi_m\}=0, \\
\{Q_j,Q_j\}&=\{\Phi_j,\Phi_j\}=0.
\end{align}
\end{subequations}

\section{Continuum limit\label{SubApp:A^2-ContLimit}}
Now that we have the expressions for Lagrangian and Hamiltonian in the discrete limit, we can obtain the analogous continuous ones by simply taking the limit $\Delta x \to 0$, while keeping the capacitance and inductance per length constant. In order to do so, we find first how some of the terms change in this limit. Let us first investigate the Kronecker delta. It is quite natural to argue that
\begin{align}
\lim\limits_{\Delta x \to 0} \frac{\delta_{n0}}{\Delta x}&=\delta(x),
\label{Eq:DerOfcQEDHam-Def of delta(x)}
\end{align}
where $\delta(x)$ here represents the Dirac delta function. It can be verified by checking all the properties of a Dirac delta function:
\begin{enumerate}
\item  $\delta(x)=0 , \qquad\  x\neq0$
\item  $\delta(x) \to +\infty , \quad x \to 0$
\item  $\lim\limits_{\Delta x \to 0}\sum\limits_n \frac{\delta_{n0}}{\Delta x}\Delta x=\int_{-\epsilon}^{+\epsilon} \delta(x)\,dx\ =1$ \\
\end{enumerate}
Based on Eq.~(\ref{Eq:DerOfcQEDHam-Def of delta(x)}), it is possible to find how $C_{s,n}$ transform in the continuous case as 
\begin{align}
c(x)\equiv\lim_{\Delta x \to 0}\frac{C_{s,n}}{\Delta x}=c+C_s\delta(x). 
\label{Eq:modified capacitance per length}
\end{align}
We call this quantity \textit{modified capacitance per length} of the resonator, since it has the information regarding the position of the qubit and the way it changes the capacitance at the point of connection. Moreover, by going to continuum limit, the charge variable $Q_n$ goes to zero, since it represents the charge of infinitesimal capacitors. However, the charge density remains a finite quantity
\begin{align}
\rho(x,t)\equiv \lim_{\Delta x \to 0} \frac{Q_n(t)}{\Delta x}.
\end{align}
Finally, by definition we replace
\begin{align}
\lim_{x \to 0}\frac{\Phi_{n+1}(t)-\Phi_n(t)}{\Delta x}&=\partial_x\Phi(x,t).
\end{align}
\subsection{Classical Lagrangian and Euler-Lagrange equations of motion in the continuum limit}
Applying the limits introduced in the previous section, Lagrangian in the continuum limit reads
\begin{align}
\begin{split}
\mathcal{L}&=\frac{1}{2}(C_j+C_g)\dot{\Phi}_j^2-U_j(\Phi_j)\\
&+\int_{-L/2}^{L/2}\,dx\ \left[\frac{1}{2}[c+C_g\delta(x)][\partial_t\Phi(x,t)]^2-\frac{1}{2l}[\partial_x\Phi(x,t)]^2\right]\\
&-\int_{-L/2}^{L/2}\,dx\ C_g\delta(x)\dot{\Phi}_j\partial_t\Phi, 
\end{split}
\end{align}
where $U_j(\Phi_j)=-E_j\cos\left(2\pi\frac{\Phi_j}{\Phi_0}\right)$. Euler-Lagrange equations of motion are derived from the variational principle $\delta \mathcal{L}=0$ as
\begin{subequations}
\begin{align}
&(C_j+C_g)\ddot{\Phi}_j-C_g\int_{-L/2}^{L/2} dx \delta(x)\partial_t^2\Phi(x,t)+\frac{\partial U_j(\Phi_j)}{\partial\Phi_j}=0,
\label{Eq: unuseful E.O.M for Phi_j ddot} \\
&-\partial_x^2\Phi(x,t)+lc\partial_t^2\Phi(x,t)+lC_g\delta(x)\left[\partial_t^2\Phi(x,t)-\ddot{\Phi}_j\right]=0.
\label{Eq: unuseful E.O.M for Phi_x ddot}
\end{align}
\end{subequations}
It is helpful to rewrite these equations by first finding $\ddot\Phi_j $ from \ref{Eq: unuseful E.O.M for Phi_j ddot} and plugging into \ref{Eq: unuseful E.O.M for Phi_x ddot} as
\begin{align}
\left[\partial_x^2-lc(x)\partial_t^2\right]\Phi(x,t)=l\gamma\delta(x)\frac{\partial U_j(\Phi_j)}{\partial\Phi_j},
\end{align}
which is a wave equation with modified capacitance per length and the transmon qubit as a source on the right hand side. Therefore, the simplified equations of motion read
\begin{subequations}
\begin{align}
&\ddot{\Phi}_j+\frac{\gamma}{C_g}\frac{\partial U_j(\Phi_j)}{\partial\Phi_j}=\gamma\partial_t^2\Phi(0,t),
\label{Eq:Transmon's E.O.M} \\
&\left[\partial_x^2-lc(x)\partial_t^2\right]\Phi(x,t)=l\gamma\delta(x)\frac{\partial U_j(\Phi_j)}{\partial\Phi_j}. 
\label{Eq:Resonator's E.O.M}
\end{align}
\end{subequations}

The Dirac delta function in the wave equation \ref{Eq:Resonator's E.O.M} can be translated into discontinuity in the spatial derivative of $\Phi(x,t)$. Therefore, equation \ref{Eq:Resonator's E.O.M} can be understood as
\begin{subequations}
\begin{align}
&\left(\partial_x^2-lc\partial_t^2\right)\Phi(x,t)=0,\quad x \neq 0, \\
&\Phi(0^+,t)=\Phi(0^-,t),
\label{Eq:Continuity of Voltage} \\
&\left.\partial_x\Phi(x,t)\right]_{x=0^-}^{x=0^+}=lC_s\left.\partial_t^2\Phi(x,t)\right|_{x=0}+l\gamma \frac{\partial U_j(\Phi_j)}{\partial\Phi_j}.
\label{Eq:Discontinuity of current}
\end{align}
\end{subequations}
In terms of voltage and current,  equation \ref{Eq:Continuity of Voltage} means that voltage is spatially continuous while \ref{Eq:Discontinuity of current} means that current is not continuous at the position of the transmon, since some of the current has to go into the qubit. The two terms on the right hand side of \ref{Eq:Discontinuity of current} are proportional to the current that enters $C_j$ and the Josephson junction respectively.

\subsection{Classical Hamiltonian and Heisenberg equations of motion in the continuum limit}
Starting from our discrete Hamiltonian, we try to find the continuous Hamiltonian again by taking the limit $\Delta x\longrightarrow 0 $. We consider each term separately. First, note that
\begin{align}
\lim_{\Delta x \to 0}\left(\frac{\gamma}{C_g}+\frac{\gamma^2}{C_{s,0}}\right)=\frac{1}{C_j},
\end{align} 
Therefore, the transmon Hamiltonian becomes
\begin{align}
\frac{Q_j^2}{2C_j}-E_j\cos\left(2\pi\frac{\Phi_j}{\Phi_0}\right).
\end{align}
The Hamiltonian of the resonator transforms as
\begin{align}
\begin{split}
&\lim_{\Delta x\to0} \sum\limits_{n}\left[\frac{Q_n^2}{2c_n}+\frac{1}{2l\Delta x}\left(\Phi_{n+1}-\Phi_n\right)^2\right]\\
&=\lim\limits_{\Delta x\to0} \sum\limits_{n}\Delta x\left[\frac{1}{2}\frac{(\frac{Q_n}{\Delta x})^2}{\frac{c_n}{\Delta x}}+\frac{1}{2l}\left(\frac{\Phi_{n+1}-\Phi_n}{\Delta x}\right)^2\right]\\
&=\int_{-L/2}^{L/2}dx \left\{\frac{\rho^2(x,t)}{2c(x)}+\frac{1}{2l}\left[\partial_x \Phi(x,t)\right]^2\right\}.
\end{split}
\end{align}
Finally, the interaction term can be written as
\begin{align}
\begin{split}
\lim\limits_{\Delta x\to 0}\frac{\gamma}{C_{s,0}}Q_jQ_0=&\lim\limits_{\Delta x\to 0}\gamma Q_j \sum\limits_{n} \frac{Q_n}{C_{s,n}}\delta_{n0}\\
=&\lim\limits_{\Delta x\to 0}\gamma Q_j \sum\limits_{n} \frac{\frac{Q_n}{\Delta x}}{\frac{C_{s,n}}{\Delta x}} \frac{\delta_{n0}}{\Delta x}\Delta x\\
=&\gamma Q_j \int_{-L/2}^{L/2}dx \frac{\rho(x,t)}{c(x)} \delta(x).
\end{split}
\end{align}
Putting all the terms together, the final expression for the Hamiltonian will be
\begin{align}
\begin{split}
\mathcal{H}&=\underbrace{\frac{Q_j^2}{2C_j}-E_j\cos\left(2\pi\frac{\Phi_j}{\Phi_0}\right)}_{\mathcal{H}_{A}}\\
&+\underbrace{\int_{-L/2}^{L/2}dx\left\{\frac{\rho^2(x,t)}{2c(x)}+\frac{1}{2l}\left[\partial_x \Phi(x,t)\right]^2\right\}}_{\mathcal{H}^{mod}_C} \\
&+\underbrace{\gamma Q_j \int_{-L/2}^{L/2}dx \frac{\rho(x,t)}{c(x)} \delta(x)}_{\mathcal{H}_{int}},
\end{split}
\label{Closed-Circuit-QED Hamiltonian}
\end{align}
where the Poisson bracket relations now change to
\begin{subequations}
\begin{align}
&\{\Phi_j,Q_j\}=1, 
\label{Eq:DerOfcQEDHam-PoissonBrack j}\\
&\{\Phi(x,t),\rho(x',t)\}=\delta(x-x') .
\label{Eq:DerOfcQEDHam-PoissonBrack r}
\end{align}
\end{subequations}

Notice that in expression~(\ref{Closed-Circuit-QED Hamiltonian}) for Hamiltonian we have a Dirac delta function hidden in $ c(x) $ in the denominator of both resonator's capacitive energy and the interaction term. At the first sight, it might seem unconventional to have a Dirac delta function in the denominator. However, we will show that the charge density $\rho(x,t)$  is also proportional to $c(x)$ which makes these integrals have finite values. we know that the time dependence of an operator $O\left(\{\Phi_n\},\{Q_n\};\Phi_j,Q_j;t\right)$ is determined by
\begin{align}
\frac{dO}{dt}=\{O,H\}+\partial_t O. 
\end{align}
Using the Poisson bracket relations introduced above we obtain the following Hamilton equations of motion
\begin{subequations}
\begin{align}
&\partial_t \Phi(x,t)=\frac{\rho(x,t)}{c(x)}+\frac{\gamma \delta(x)}{c(x)} Q_j, \\
&\partial_t \rho(x,t)=\frac{1}{l}\partial_x^2 \Phi(x,t), \\
&\partial_t \Phi_j=\frac{Q_j}{C_j}+\int_{-L/2}^{L/2} dx \frac{\gamma \delta(x)}{c(x)}\rho(x,t),
\label{Eq:Transmon's Flux variable time derivative}
\\
&\partial_t Q_j=-\frac{\partial U_j(\Phi_j)}{\partial \Phi_j}=-\frac{2\pi}{\Phi_0}E_j \sin\left(2\pi \frac{\Phi_j}{\Phi_0}\right). 
\label{Eq:Transmon's Charge Variable time derivative}
\end{align}
\end{subequations}
The results here, can be generalized to a case where the transmon is connected to some arbitrary point $x_0$, where the modified capacitance per length now changes to $c(x,x_0)=c+C_s \delta(x-x_0)$. 

\section{Modified resonator eigenmodes and eigenfrequencies\label{SubApp:A^2-ModEigModes}}
Consider $\mathcal{H}_{C}^{mod}$ in Eq.~\ref{Closed-Circuit-QED Hamiltonian} which is the modified resonator Hamiltonian. The focus of this section is to find out how this modification in capacitance per length  influences the closed Hermitian eigenmodes and eigenfrequencies of the resonator. Assuming that the transmon is connected to some arbitrary point $x_0$ the Hamiltonian is given as
\begin{align}
\mathcal{H}_C^{mod}=\int_{0}^{L} dx \left\{\frac{\rho^2(x,t)}{2c(x,x_0)}+\frac{1}{2l}\left[\partial_x \Phi(x,t)\right]^2\right\}.
\end{align}

Applying the Poisson bracket relations~(\ref{Eq:DerOfcQEDHam-PoissonBrack j}-\ref{Eq:DerOfcQEDHam-PoissonBrack r}), the Hamilton equations of motion for the resonator fields read
\begin{subequations}
\begin{align}
\partial_t \Phi(x,t)&=\frac{\rho(x,t)}{c(x,x_0)}, 
\label{Eq: E.O.M for Phi from H_c mod}\\
\partial_t \rho(x,t)&=\frac{1}{l}\partial_x^2 \Phi(x,t).
\label{Eq: E.O.M for rho from H_c mod}
\end{align} 
\end{subequations}
By combining the above equations and rewriting them in Fourier representation in terms of $\tilde{\Phi}(x,\omega)=\int_{-\infty}^{+\infty} dt \Phi(x,t)e^{i\omega t}$ we obtain
\begin{align}
\left[\partial_x^2+lc(x,x_0)\omega^2\right]\tilde{\Phi}(x,\omega)=0.
\label{Eq:Modified Wave Equation-Closed Case}
\end{align}
Note that there is a Dirac delta function hidden in $c(x,x_0)$. As we mentioned earlier, this can be translated into discontinuity in $\partial_x \tilde{\Phi}(x,\omega)$ which is proportional to the current $\tilde{I}(x,\omega)=-\frac{1}{l}\partial_x \tilde{\Phi}(x,\omega)$ that enters and exits the point of connection to the transmon
\begin{align}
-\frac{1}{l}\left.\partial_x \tilde{\Phi}(x,\omega)\right]_{x_0^-}^{x_0^+} =C_s\omega^2 \tilde{\Phi}(x_0,\omega).
\end{align}

We are after a complete set of modes $\tilde{\Phi}_n(x)\equiv\tilde{\Phi}(x,\omega_n)$ where any solutions to the previous wave equation can be linearly decomposed on them. In order to find these modes, we have to solve
\begin{align}
\left(\partial_x^2+lc\omega_n^2\right)\tilde{\Phi}_n(x)=0 ,\qquad x\neq x_0 
\end{align}
with boundary conditions
\begin{subequations}
\begin{align}
&\left.\partial_x \tilde{\Phi}_n(x)\right|_{x=0}=\left.\partial_x \tilde{\Phi}_n(x)\right|_{x=L}=0, \\
&\left.\partial_x \tilde{\Phi}_n(x)\right]_{x_0^-}^{x_0^+} + lC_s\omega_n^2 \tilde{\Phi}_n(x_0)=0, \\
&\tilde{\Phi}_n(x_0^+)=\tilde{\Phi}_n(x_0^-). 
\end{align}
\end{subequations}
Applying the boundary conditions we find a transcendental equation whose roots will give the Hermitian eigenfrequencies of this closed system as 
\begin{align}
\sin(k_nL)+\chi_s k_nL\cos(k_n x_0)\cos[k_n(L-x_0)]=0. 
\label{Eq: Closed CC Eigenfrequencies}
\end{align}
In the transcendental Eq.~(\ref{Eq: Closed CC Eigenfrequencies}), $ k_n $ represents the wavevector defined as $k_n^2 \equiv lc\omega_n^2$ and the quantity $\chi_{s}\equiv \frac{C_s}{cL}$ is a unitless measure for the discontinuity of current introduced by the transmon.
Eventually, the eigenfunctions are found as
\begin{align}
\tilde{\Phi}_n(x)\propto
\begin{cases}
\cos{\left[k_n(L-x_0)\right]}\cos{(k_n x)}&0<x<x_0\\
\cos{(k_n x_0)}\cos{[k_n(L-x)]} &x_0<x<L
\end{cases},
\end{align}
where the proportionality constant is set by the orthogonality relation
\begin{align}
\int_0^L dx \frac{c(x,x_0)}{c} \tilde{\Phi}_n(x) \tilde{\Phi}_m(x) = L\delta_{mn}.
\label{Eq: Orthogonality of Phi} 
\end{align}
Another important orthogonality condition can be derived between $\{\partial_x \tilde{\Phi}_n \}$ as
\begin{align}
\int _{0}^{L}dx \partial_x\tilde{\Phi}_m(x)\partial_x\tilde{\Phi}_n(x)=k_m k_n L \delta_{mn}.
\label{Eq: Orthogonality of d/dx Phi}
\end{align}

Finally, it is instructive to show explicitly the origin of an $A^2$-like term when instead of the CC-basis the conventional cosine modes are chosen. Replacing $\rho(x,t)$ from \ref{Eq: E.O.M for Phi from H_c mod} in $\mathcal{H}_C^{mod}$ gives
\begin{align}
\begin{split}
\mathcal{H}_C^{mod}=\int_{0}^{L} dx \left\{\frac{c(x,x_0)}{2}\left[\partial_t \Phi(x,t)\right]^2+\frac{1}{2l}\left[\partial_x \Phi(x,t)\right]^2\right\}.
\end{split}
\end{align}
Substituting $c(x,x_0)=c+C_s\delta(x-x_0)$ leads to
\begin{align}
\mathcal{H}_C^{mod}=\underbrace{\int_{0}^{L} dx \left\{\frac{c}{2}\left[\partial_t \Phi(x,t)\right]^2+\frac{1}{2l}\left[\partial_x \Phi(x,t)\right]^2\right\}}_{\mathcal{H}_C}+\underbrace{\frac{1}{2}C_s \left[\partial_t \Phi(x_0,t)\right]^2}_{\mathcal{H}^{mod}}.
\end{align}  
Note that $\mathcal{H}_C$ has a diagonal representation in terms of the cosine basis \cite{Devoret_Quantum_2014}. However, by choosing this basis $\mathcal{H}^{mod}$ remains as an $A^2$-like term giving rise to intermode  interaction.

\chapter{Unitless quantum equations of motion for an open cQED system}
\label{App:Quantum EOM}

The derivation for the classical Lagrangian for the system shown in Fig.~\ref{Fig:cQED-open} has been discussed in App.~\ref{SubApp:A^2-ContLimit} for a closed cQED system. In this appendix, we review these equations for the open case and reexpress them in terms of unitless quantities. These unitless equations are then employed in chapter~\ref{Ch:NonMarkovian} to construct our Heisenberg-Langevin formalism. 

The classical Euler-Lagrange equations of motion can then be found by setting the variation of Lagrangian 
\begin{align}
\begin{split}
\mathcal{L}&=\frac{1}{2}C_j\dot{\Phi}_{j}(t)^2-U(\Phi_j(t)) \\
&+\int_{0^+}^{L^-}\,dx\left\{\frac{1}{2} c[\partial_t \Phi(x,t)]^2-\frac{1}{2l}[\partial_x \Phi(x,t)]^2\right\}\\
&+\int_{L^+}^{+\infty}\,dx\left\{\frac{1}{2} c[\partial_t \Phi_R(x,t)]^2-\frac{1}{2l}[\partial_x \Phi_R(x,t)]^2\right\}\\
&+\int_{-\infty}^{0^-}\,dx\left[\frac{1}{2} c[\partial_t \Phi_L(x,t)]^2-\frac{1}{2l}[\partial_x \Phi_L(x,t)]^2\right]\\
&+\frac{1}{2}C_L\left[\dot{\Phi}_L(0^-,t)-\dot{\Phi}(0^+,t)\right]^2\\&
+\frac{1}{2}C_R\left[\dot{\Phi}_R(L^+,t)-\dot{\Phi}(L^-,t)\right]^2\\
&+\frac{1}{2}C_g\left[\dot{\Phi}_j(t)-\dot{\Phi}(x_0,t)\right]^2,
\end{split}
\end{align}
with respect to each flux variable to zero. For the transmon and the resonator we find
\begin{align}
&\ddot{\Phi}_j+\frac{1}{C_g+C_j}\frac{\partial U_j(\Phi_j)}{\partial \Phi_j}=\gamma \partial_{t}^2\Phi(x_0,t),
\label{Eq:Transmon Dyn in terms of Phi_j}\\ 
&\partial_{x}^2\Phi(x,t)-lc(x,x_0)\partial_{t}^2\Phi(x,t) =l\gamma \delta(x-x_0)\frac{\partial U_j(\Phi_j)}{\partial \Phi_j},
\label{Eq:Resonatr-dynamics in terms of Phi}
\end{align}
where $U_j(\Phi_j)$ stands for the Josephson potential as
\begin{align}
U_j(\Phi_j)=-E_j\cos{\left(\frac{2\pi}{\Phi_0}\Phi_j\right)},
\label{Eq:Josephson Potential}
\end{align} 
and $\Phi_0\equiv \frac{h}{2e}$ is the superconducting flux quantum. Furthermore, $C_s \equiv C_gC_j/(C_g+C_j)$ is the series capacitance of $C_j$ and $C_g$ and  $\gamma\equiv C_g/(C_g+C_j)$. Moreover, $l$ and $c$ are the inductance and capacitance per length of the resonator and waveguides while $c(x,x_0)\equiv c+C_s\delta(x-x_0)$ represents the modified capacitance per length due to coupling to transmon. In addition, we find two wave equations for the flux field of the left and right waveguides as
\begin{align}
\partial_{x}^2\Phi_{R,L}(x,t)-lc\partial_{t}^2\Phi_{R,L}(x,t) =0,
\label{Eq:Side Bath dynamics}
\end{align}

The boundary conditions are derived from continuity of current at each end as
\begin{subequations}
\begin{align}
&-\frac{1}{l} \left.\partial_{x}\Phi\right|_{x=L^-}=-\frac{1}{l} \left.\partial_{x}\Phi_R\right|_{x=L^+}=C_R\partial_t^2\left[\Phi(L^-,t)-\hat{\Phi}_R(L^+,t)\right],
\label{Eq:BC-Conservation of current at 0 in terms of Phi}\\
&-\frac{1}{l} \left.\partial_{x}\Phi\right|_{x=0^+}=-\frac{1}{l} \left.\partial_{x}\Phi_L\right|_{x=0^-}=C_L\partial_t^2\left[\Phi_L(0^-,t)-\Phi(0^+,t)\right],
\label{Eq:BC-Conservation of current at 1 in terms of Phi}
\end{align}
\end{subequations}
continuity of flux at $x=x_0$
\begin{align}
\Phi(x=x_0^-,t)=\Phi(x=x_0^+,t), 
\end{align}
and conservation of current at $x=x_0$ as
\begin{align}
\left. \partial_x \Phi\right|_{x=x_0^+} - \left.\partial_x \Phi \right|_{x=x_0^-} -lC_s\partial_t^2 \Phi(x_0,t)=l\gamma \frac{\partial U_j(\Phi_j)}{\partial \Phi_j}.
\end{align}

In order to find the quantum equations of motion, we follow the common procedure of canonical quantization \cite{Devoret_Quantum_2014}: 
\begin{itemize}
\item[1)] Find the conjugate momenta $Q_n\equiv \frac{\delta \mathcal{L}}{\delta \dot{\Phi}_n}$ 
\item[2)] Find the classical Hamiltonian via a Legendre transformation as $\mathcal{H}=\sum\limits_{n}Q_n\dot{\Phi_n}-\mathcal{L}$
\item[3)] Find the Hamiltonian operator by promoting the classical conjugate variables to quantum operators such that $\{\hat{\Phi}_m,\hat{Q}_n\}=\delta_{mn}\to [\hat{\Phi}_m,\hat{Q}_n]=i\hbar\delta_{mn}$. We use a hat-notation to distinguish operators from classical variables.
\end{itemize} 

The derivation for the quantum Hamiltonian of the the closed version of this system where $C_{R,L}\to 0$ can be found in App.~\ref{SubApp:A^2-ContLimit}. Note that nonzero end capacitors $C_{R,L}$ leave the equations of motion for the resonator and waveguides unchanged, but modify the BC of the problem at $x=0,L$. The resulting equations of motion for the quantum flux operators $\hat{\Phi}_j$, $\hat{\Phi}(x,t)$ and $\hat{\Phi}_{R,L}(x,t)$ have the exact same form as the classical Euler-Lagrange equations of motion.

Next, we define unitless parameters and variables as
\begin{align}
\begin{split}
\bar{x}\equiv\frac{x}{L},\quad
\bar{t}\equiv\frac{t}{\frac{L}{v_{p}}},\quad
\bar{\omega}\equiv\frac{\omega}{v_p}L,\quad \hat{\varphi}\equiv 2\pi \frac{\hat{\Phi}}{\Phi_0}, \quad \hat{n}\equiv\frac{\hat{Q}}{2e}
\end{split}
\label{Eq:unitless vars}
\end{align}
where $v_p\equiv 1/\sqrt{lc}$ is the phase velocity of the resonator and waveguides. Furthermore, we define unitless capacitances as
\begin{align}
\chi_i\equiv\frac{C_i}{cL}, \quad i=R,L,j,g,s 
\end{align}
as well as a unitless modified capacitance per length as
\begin{align}
\chi(\bar{x},\bar{x}_0)\equiv 1+\chi_s \delta(\bar{x}-\bar{x}_0).
\label{Eq:Def of chi(x,x0)}
\end{align}
Then, the unitless equations of motion for our system are found as 
\begin{subequations}
\begin{align}
&\hat{\ddot{\varphi}}_j(\bar{t})+(1-\gamma)\bar{\omega}_j^2\sin{[\hat{\varphi}_j(\bar{t})]}=\gamma \partial_{\bar{t}}^2\hat{\varphi}(\bar{x}_0,\bar{t}),
\label{Eq:Transmon Dyn}\\
&\left[\partial_{\bar{x}}^2-\chi(\bar{x},\bar{x}_0)\partial_{\bar{t}}^2\right]\hat{\varphi}(\bar{x},\bar{t})=\chi_s\bar{\omega}_j^2 \sin{[\varphi_j(\bar{t})]}\delta(\bar{x}-\bar{x}_0),
\label{Eq:Res Dyn}\\
&\partial_{\bar{x}}^2\hat{\varphi}_{R,L}(\bar{x},\bar{t})-\partial_{\bar{t}}^2\hat{\varphi}_{R,L}(\bar{x},\bar{t}) =0,
\label{Eq:Side Res Dyn}
\end{align}
\end{subequations}
with the unitless BCs given as
\begin{subequations}
\begin{align}
&-\left.\partial_{\bar{x}}\hat{\varphi}\right|_{\bar{x}=1^-}=-\left.\partial_{\bar{x}}\hat{\varphi}_R\right|_{\bar{x}=1^+}=\chi_R\partial_{\bar{t}}^2\left[\hat{\varphi}(1^-,\bar{t})-\hat{\varphi}_R(1^+,\bar{t})\right],
\label{Eq:BC-Conservation of current at 1}\\
&-\left.\partial_{\bar{x}}\hat{\varphi}\right|_{\bar{x}=0^+}=-\left.\partial_{\bar{x}}\hat{\varphi}_L\right|_{\bar{x}=0^-}=\chi_L\partial_{\bar{t}}^2\left[\hat{\varphi}_L(0^-,\bar{t})-\hat{\varphi}(0^+,\bar{t})\right],
\label{Eq:BC-Conservation of current at 0}\\
&\hat{\varphi}(\bar{x}=\bar{x}_0^-,\bar{t})=\hat{\varphi}(\bar{x}=\bar{x}_0^+,\bar{t}),\\
&\left. \partial_{\bar{x}} \hat{\varphi}\right|_{\bar{x}=\bar{x}_0^+} - \left.\partial_{\bar{x}} \hat{\varphi} \right|_{\bar{x}=\bar{x}_0^-}
-\chi_s\partial_{\bar{t}}^2 \hat{\varphi}(\bar{x}_0,\bar{t})=\chi_s \bar{\omega}_j^2 \sin{[\varphi_j(\bar{t})]}.
\label{Eq:BC-conservation of current at x0}
\end{align}
\end{subequations}

In Eqs.~(\ref{Eq:Transmon Dyn}) and (\ref{Eq:Res Dyn}), we have defined the unitless oscillation frequency $\bar{\omega}_j$ as
\begin{align}
\bar{\omega}_j^2  \equiv lc L^2\frac{E_j}{C_j} \left(\frac{2\pi}{\Phi_0}\right)^2=8\mathcal{E}_c\mathcal{E}_j,
\label{Eq:Def of bar(W)_j}
\end{align}
where $\mathcal{E}_c$ and $\mathcal{E}_j$ stand for the unitless charging and Josephson energy given as
\begin{align}
\mathcal{E}_{j,c}\equiv \sqrt{lc} L\frac{E_{j,c}}{\hbar},\
\end{align}
with $E_c\equiv \frac{e^2}{2C_j}$. 

In what follows, we work with the unitless Eqs.~(\ref{Eq:Transmon Dyn}-\ref{Eq:Side Res Dyn}) and BCs~(\ref{Eq:BC-Conservation of current at 1}-\ref{Eq:BC-conservation of current at x0}) and drop the bars.
\chapter{Effective dynamics of the transmon via a Heisenberg picture Green's function method}
\label{App:Eff Dyn of transmon}
In order to find the effective dynamics of the transmon qubit, one has to solve for the flux field $\hat{\varphi}(x,t)$ and substitute the result back into the RHS of time evolution of the qubit given by Eq.~(\ref{Eq:Transmon Dyn}). It is possible to perform this procedure in terms of the resonator GF. In Sec.~\ref{SubApp:Def of G} we define the resonator GF. In Sec.~\ref{SubApp:Spec Rep of G-open} we study the spectral representation of the GF in terms of a suitable set of non-Hermitian modes. In Sec.~\ref{SubApp:Eff Dyn of transmon}, we discuss the derivation of the effective dynamics of transmon in terms of the resonator GF. Finally, in Secs.~\ref{SubApp:SE Eff Dyn} and \ref{SubApp:Spec Rep of K} we discuss how the generic dynamics is reduced for the problem of spontaneous emission.
\section{Definition of $G(x,t|x',t')$}
\label{SubApp:Def of G}
The resonator GF is defined as the response of the linear system of Eqs.~(\ref{Eq:Res Dyn}-\ref{Eq:Side Res Dyn}) to a $\delta$-function source in space-time as
\begin{align}
\begin{split}
\left[\partial_x^2 -\chi(x,x_0)\partial_t^2\right]&G(x,t|x_0,t_0)=\delta(x-x_0)\delta(t-t_0),
\end{split}
\label{Eq:Def of G(x,t|x0,t0)}
\end{align}
with the same BCs as Eqs.~(\ref{Eq:BC-Conservation of current at 1}-\ref{Eq:BC-conservation of current at x0}). 
Using the Fourier transform conventions
\begin{subequations}
\begin{align}
&\tilde{G}(x,x_0,\omega)=\int_{-\infty}^{\infty}dt G(x,t|x_0,t_0) e^{+i\omega(t-t_0)}, \\
&G(x,t|x_0,t_0)=\int_{-\infty}^{\infty}\frac{d\omega}{2\pi} \tilde{G}(x,x_0,\omega)e^{-i\omega(t-t_0)},
\end{align}
\end{subequations}
Eq.~(\ref{Eq:Def of G(x,t|x0,t0)}) transforms into a Helmholtz equation 
\begin{align}
\begin{split}
\left[\partial_x^2 +\omega^2\chi(x,x_0) \right]\tilde{G}(x,x_0,\omega)=\delta(x-x_0).
\end{split}
\label{Eq:Helmholtz Eq for G(x,x0,W)}
\end{align}
Moreover, the BCs are transformed by replacing $\partial_x\to \partial_x$ and $\partial_t \to -i\omega $ as
\begin{subequations}
\begin{align}
&\left.\tilde{G}\right|_{x=x_0^+}=\left.\tilde{G}\right|_{x=x_0^-}, 
\label{Eq:Cont of G(x,x0,W) at x0}\\
&\left.\partial_x \tilde{G} \right|_{x=x_0^+}-\left.\partial_x \tilde{G}\right|_{x=x_0^-}+\chi_s\omega^2\left.\tilde{G}\right|_{x=x_0}=1 ,
\label{Eq:Cont of dxG(x,x0,W) at x0}\\
&\left.\partial_x\tilde{G}\right|_{x=1^-}=\left.\partial_x \tilde{G}\right|_{x=1^+}=\chi_R \omega^2 \left(\left.\tilde{G}\right|_{x=1^-}-\left.\tilde{G}\right|_{x=1^+} \right), 
\label{Eq:Cont of dxG(x,x0,W) at 1}\\
&\left.\partial_x \tilde{G}\right|_{x=0^-}= \left.\partial_x \tilde{G}\right|_{x=0^+}=\chi_L \omega^2 \left(\left.\tilde{G}\right|_{x=0^-}-\left.\tilde{G}\right|_{x=0^+} \right). 
\label{Eq:Cont of dxG(x,x0,W) at 0}
\end{align}
\end{subequations}

Note that BCs (\ref{Eq:Cont of G(x,x0,W) at x0}-\ref{Eq:Cont of dxG(x,x0,W) at 0}) do not specify what happens to $\tilde{G}(x,x_0,\omega)$ at $x\to \pm \infty$. We model the baths by imposing outgoing BCs at infinity as
\begin{align}
\left. \partial_x \tilde{G}(x,x_0,\omega)\right|_{x\to \pm\infty}=\pm i\omega \tilde{G}(x\to\pm\infty,x_0,\omega),
\label{Eq:Outgoing BC for G(x,x0,W))}
\end{align}
which precludes any reflections from the waveguides to the resonator. 

\section{Spectral representation of GF for a closed resonator}
\label{SubApp:Spec Rep of G-closed}

It is instructive to revisit spectral representation of GF for the closed version of our system by setting $\chi_R=\chi_L=0$. This imposes Neumann BC $\partial_x \tilde{G}|_{x=0,1}=0$ and the resulting differential operator becomes Hermitian. The idea of spectral representation is to expand $\tilde{G}$ in terms of a discrete set of normal modes that obey the homogeneous wave equation
\begin{subequations}
\begin{align}
&\partial_x^2\tilde{\varphi}_n(x)+\chi(x,x_0)\omega_n^2\tilde{\varphi}_n(x)=0,\\
\label{Eq:Helmholtz Eq for Phi_n(x)}
&\left.\partial_x \tilde{\varphi}_n(x)\right|_{x=0,1}=0.
\end{align}
\end{subequations}
Then, the real valued eigenfrequencies obey the transcendental equation
\begin{align}
\sin{(\omega_n)}+\chi_s\omega_n\cos{(\omega_n x_0)}\cos{\left[\omega_n (1-x_0)\right]}=0.
\label{Eq:Hermitian Eigenfrequencies}
\end{align}
The eigenfunctions read
\begin{align}
\tilde{\varphi}_n(x)\propto
\begin{cases}
\cos{\left[\omega_n (1-x_0)\right]}\cos{(\omega_n x)},&0<x<x_0\\
\cos{(\omega_n x_0)}\cos{\left[\omega_n (1-x)\right]},&x_0<x<1
\end{cases}
\label{Eq:Sol of Phi_n(x)-closed}
\end{align}
where the normalization is fixed by the orthogonality condition
\begin{align}
\int_{0}^{1}dx\chi(x,x_0)\tilde{\varphi}_m(x)\tilde{\varphi}_n(x)=\delta_{mn}.
\label{Eq:Closed Orthogonality Condition}
\end{align}

Note that eigenfunctions of a Hermitian differential operator form a complete orthonormal basis. This allows us to deduce the spectral representation of $\tilde{G}(x,x',\omega)$ \cite{Morse_Methods_1953, Economou_Green_1984, Hassani_Mathematical_2013} as
\begin{align}
\tilde{G}(x,x',\omega)=\sum\limits_{n\in \mathbb{N}}\frac{\tilde{\varphi}_n(x)\tilde{\varphi}_n(x')}{\omega^2-\omega_n^2}=\sum\limits_{n\in \mathbb{Z}\atop n\neq 0}\frac{1}{2\omega}\frac{\tilde{\varphi}_n(x)\tilde{\varphi}_n(x')}{\omega-\omega_n},
\label{Eq:Spectral rep of G-closed}
\end{align}
where the second representation is written due to relations $\omega_{-n}=-\omega_{n}$ and $\tilde{\varphi}_{-n}(x)=\tilde{\varphi}_{n}(x)$.
\section{Spectral representation of GF for an open resonator}
\label{SubApp:Spec Rep of G-open} 
A spectral representation can also be found for the GF of an open resonator in terms of a discrete set of non-Hermitian modes that carry a constant flux away from the resonator. The Constant Flux (CF) modes \cite{Tureci_SelfConsistent_2006} have allowed a consistent formulation of the semiclassical laser theory for complex media such as random lasers \cite{Tureci_Strong_2008}. The non-Hermiticity originates from the fact that the waveguides are assumed to be infinitely long, hence no radiation that is emitted from the resonator to the waveguides can be reflected back. This results in discrete and complex-valued poles of the GF.  The CF modes satisfy the same homogeneous wave equation
\begin{align}
\partial_x^2\tilde{\varphi}_n(x,\omega)+\chi(x,x_0)\omega_n^2(\omega)\tilde{\varphi}_n(x,\omega)=0,
\label{Eq:Helmholtz Eq for Phi_n(x)}
\end{align}
but with open BCs the same as Eqs.~(\ref{Eq:Cont of G(x,x0,W) at x0}-\ref{Eq:Outgoing BC for G(x,x0,W))}). Note that the resulting CF modes $\tilde{\Phi}_n(x,\omega)$ and eigenfrequencies $\omega_n(\omega)$ parametrically depend on the source frequency $\omega$. 

Considering only an outgoing plane wave solution for the left and right waveguides based on (\ref{Eq:Outgoing BC for G(x,x0,W))}), the general solution for $\tilde{\varphi}_n(x,\omega)$ reads
\begin{align}
\tilde{\varphi}_n(x,\omega)=
\begin{cases}
A_{n}^<e^{i\omega_n(\omega) x}+B_{n}^< e^{-i\omega_n(\omega) x},&0<x<x_0\\
A_{n}^>e^{i\omega_n(\omega) x}+B_{n}^> e^{-i\omega_n(\omega) x},&x_0<x<1\\
C_ne^{i\omega x},&x>1 \\
D_ne^{-i\omega x},&x<0 \\
\end{cases}
\label{Eq:General Ansantz for Phi_n(x)}
\end{align}
Applying BCs (\ref{Eq:Cont of G(x,x0,W) at x0}-\ref{Eq:Cont of dxG(x,x0,W) at 0}) leads to a characteristic equation 
\begin{align}
\begin{split}
&\sin\left[\omega_n(\omega)\right]+(\chi_R+\chi_L)\omega_n(\omega)\left\{\cos[\omega_n(\omega)]-\frac{\omega_n(\omega)}{\omega}\sin[\omega_n(\omega)]\right\}\\
&-\chi_R\chi_L\omega_n^2(\omega)\left\{2i\frac{\omega_n(\omega)}{\omega}\cos[\omega_n(\omega)]+\left[1+\frac{\omega_n^2(\omega)}{\omega^2}\right]\sin[\omega_n(\omega)]\right\}\\
&+\chi_s\omega_n(\omega)\left\{\cos[\omega_n(\omega)x_0]-\chi_L\frac{\omega_n(\omega)}{\omega}\left\{i\omega_n(\omega)\cos[\omega_n(\omega)x_0]+\omega\sin[\omega_n(\omega)x_0]\right\}\right\}\\
&\times\left\{\cos[\omega_n(\omega)(1-x_0)]-\chi_R\frac{\omega_n(\omega)}{\omega}\left\{i\omega_n(\omega)\cos[\omega_n(\omega)(1-x_0)]+\omega\sin[\omega_n(\omega)(1-x_0)]\right\}\right\}=0,
\end{split}
\label{Eq:CF NHEigfreq}
\end{align}
which gives the parametric dependence of CF frequencies on $\omega$. Then, the CF modes $\tilde{\varphi}_n(x,\omega)$ are calculated as
\begin{align}
\begin{small}
\mathcal{N}_n
\begin{cases}
e^{-i \omega_n(\omega) (x-x_0+1)} \left[e^{2 i \omega_n(\omega) x}+(1-2 i \omega_n(\omega) \chi _L)\right] \left[e^{2 i \omega_n(\omega)(1-x_0)}+\left(1-2 i \omega_n(\omega) \chi_R\right)\right],&0<x<x_0\\
e^{-i \omega_n(\omega) (x_0-x+1)} \left[e^{2 i \omega_n(\omega) x_0}+(1-2 i \omega_n(\omega) \chi _L)\right] \left[e^{2 i \omega_n(\omega)(1-x)}+\left(1-2 i \omega_n(\omega) \chi_R\right)\right],&x_0<x<1\\
-2i\chi_R\omega_n(\omega)e^{-i\omega_n(\omega) (1+x_0)} \left[e^{+2i\omega_n(\omega) x_0}+(1-2i \chi_L \omega_n(\omega))\right]e^{+i\omega x},&x>1 \\
-2i\chi_L\omega_n(\omega)e^{-i\omega_n(\omega)(1-x_0)} \left[e^{2i\omega_n(\omega) (1-x_0)}+(1-2i \chi_R\omega_n(\omega))\right]e^{-i\omega x}. &x<0 \\
\end{cases}
\end{small}
\label{Eq:General NHEigFun}
\end{align}
These modes satisfy the biorthonormality condition
\begin{align}
\begin{split}
\int_{0}^{1}dx\chi(x,x_0)\bar{\tilde{\varphi}}_m^*(x,\omega)\tilde{\varphi}_n(x,\omega)=\delta_{mn},
\label{Eq:Open Ortho Cond-unsimp}
\end{split}
\end{align}
where $\{\bar{\tilde{\varphi}}_m(x,\omega)\}$ satisfy the Hermitian adjoint of eigenvalue problem~(\ref{Eq:Helmholtz Eq for Phi_n(x)}). In other words, $\tilde{\varphi}_n(x,\omega)$ and $\bar{\tilde{\varphi}}_n(x,\omega)$ are the right and left eigenfunctions and obey $\bar{\tilde{\varphi}}_n(x,\omega)=\tilde{\varphi}_n^*(x,\omega)$. The normalization of Eq.~(\ref{Eq:General NHEigFun}) is then fixed by setting $m=n$.

In terms of the CF modes, the spectral representation of the GF can then be constructed
\begin{align}
\tilde{G}(x,x',\omega)=\sum\limits_{n}\frac{\tilde{\varphi}_n(x,\omega)\bar{\tilde{\varphi}}_n^*(x',\omega)}{\omega^2-\omega_n^2(\omega)}.
\label{Eq:Spectral rep of G-Open}
\end{align}
Examining Eq.~(\ref{Eq:Spectral rep of G-Open}), we realize that there are two sets of poles of $\tilde{G}(x,x',\omega)$ in the complex $\omega$ plane. First, from setting the denominator of Eq.~(\ref{Eq:Spectral rep of G-Open}) to zero which gives $\omega=\omega_n(\omega)$. These are the quasi-bound eigenfrequencies that satisfy the transcendental characteristic equation  
\begin{align}
\begin{split}
&\left[e^{2i\omega_n}-(1-2i\chi_L\omega_n)(1-2i\chi_R\omega_n)\right]\\
&+\frac{i}{2}\chi_s\omega_n[e^{2i\omega_n x_0}+(1-2i\chi_L\omega_n)]\\
&\times[e^{2i\omega_n (1-x_0)}+(1-2i\chi_R\omega_n)]=0.
\end{split}
\label{Eq:Generic NHEigfreq}
\end{align}
The quasi bound solutions $\omega_n$ to Eq.~(\ref{Eq:Generic NHEigfreq}) reside in the lower half of complex $\omega$-plane and  come in symmetric pairs with respect to the $\Im\{\omega\}$ axis, i.e. both $\omega_{n}$  and $-\omega_{n}^*$ satisfy the transcendental Eq.~(\ref{Eq:Generic NHEigfreq}). Therefore, we can label the eigenfrequencies as
\begin{align}
\omega_n=\begin{cases}
-i\kappa_0, \quad & n=0\\
+\nu_n-i\kappa_n, \quad &n\in+\mathbb{N}\\
-\nu_n-i\kappa_n, \quad &n\in-\mathbb{N}
\end{cases}
\end{align}
where $\nu_n$ and $\kappa_n$ are positive quantities representing the oscillation frequency and decay rate of each quasi-bound mode. Second, there is an extra pole at $\omega=0$ which comes from the $\omega$-dependence of CF states $\tilde{\varphi}_n(x,\omega)$. We confirmed these poles by solving for the explicit solution $\tilde{G}(x,x',\omega)$ that obeys Eq.~(\ref{Eq:Helmholtz Eq for G(x,x0,W)}) with BCs~(\ref{Eq:Cont of G(x,x0,W) at x0}-\ref{Eq:Outgoing BC for G(x,x0,W))}) with Mathematica.
\section{Effective dynamics of transmon qubit}
\label{SubApp:Eff Dyn of transmon}

Note that Eqs.~(\ref{Eq:Res Dyn}-\ref{Eq:Side Res Dyn}) are linear in terms of $\hat{\varphi}(x,t)$ and $\hat{\varphi}_{R,L}(x,t)$ . Therefore, it is possible to eliminate these linear degrees of freedom and express the formal solution for $\hat{\varphi}(x,t)$ in terms of $\hat{\varphi}_j(t)$ and $G(x,t|x',t')$. At last, by plugging the result into the RHS of Eq.~(\ref{Eq:Res Dyn}) we find a closed equation for $\hat{\varphi}_j(t)$.

Let us denote the source term that appears on the RHS of Eq.~(\ref{Eq:Res Dyn}) as 
\begin{align}
S\left[\hat{\varphi}_j(t)\right]\equiv \chi_s\omega_j^2\sin{[\hat{\varphi}_j(t)]}.
\end{align}
Then, we write two equations for $\hat{\varphi}(x,t)$ and $G(x,t|x',t')$ \cite{Morse_Methods_1953} (See Sec. $7.3$) as
\begin{subequations}
\begin{align}
&\left[\partial_{x'}^2-\chi(x',x_0)\partial_{t'}^2\right]\hat{\varphi}(x',t')=S\left[\hat{\varphi}_j(t')\right]\delta(x'-x_0), 
\label{Eq:Eff Dyn-wave Eq for varphi_j}\\
&\left[\partial_{x'}^2 -\chi(x,x')\partial_{t'}^2\right]G(x,t|x',t')=\delta(x-x') \delta(t-t').
\label{Eq:Eff Dyn-wave Eq for GF}
\end{align}
\end{subequations}
In Eq.~(\ref{Eq:Eff Dyn-wave Eq for GF}) we have employed the reciprocity property of the GF 
\begin{align}
G(x,t|x',t')=G(x',-t'|x,-t),
\label{Eq:Reciprocity property of Green's function}
\end{align}
which holds since Eq.~(\ref{Eq:Eff Dyn-wave Eq for GF}) is invariant under
\begin{align}
x\leftrightarrow x', \quad t\leftrightarrow-t'.
\label{Eq:Sym of Wave Eq}
\end{align}

Multiplying Eq.~(\ref{Eq:Eff Dyn-wave Eq for varphi_j}) by $G(x,t|x',t')$ and Eq.~(\ref{Eq:Eff Dyn-wave Eq for GF}) by $\hat{\varphi}(x',t')$ and integrating over the dummy variable $x'$ in the interval $[0^-,1^+]$ and over $t'$ in the interval $[0,t^+]$ and finally taking the difference gives
\begin{align}
\begin{split}
&\int_{0}^{t^+}dt' \int_{0^-}^{1^+}dx'\left\{\underbrace{\left(G\partial_{x'}^2\hat{\varphi}-\hat{\varphi}\partial_{x'}^2G\right)}_{\bf(a)}\right.\\
&+\underbrace{\left[\chi(x,x')\hat{\varphi}\partial_{t'}^2G-\chi(x',x_0)G\partial_{t'}^2\hat{\varphi}\right]}_{\bf (b)}\\
&-\left.\underbrace{G S(\hat{\varphi}_j)\delta(x'-x_0)}_{\bf (c)}+\underbrace{\hat{\varphi}\delta(t-t')\delta(x-x')}_{\bf (d)}\right\}=0,
\end{split}
\label{Eq:Intermediate Step Derivation of Green's Foralism}
\end{align}
where we have used the shorthand notation $G\equiv G(x,t|x',t')$ and $\hat{\varphi}\equiv \hat{\varphi}(x',t')$. 

The term labeled as $(a)$ can be simplified further through integration by parts in $x'$ as
\begin{align}
\int_{0}^{t^+}dt' \left.\left(G\partial_{x'}\hat{\varphi}-\hat{\varphi}\partial_{x'}G\right)\right|_{x'=0^-}^{x'=1^+}
\end{align}
There are two contributions from term $(b)$. One comes from the constant capacitance per length in $\chi(x,x')$ and $\chi(x,x_0)$ that simplifies to
\begin{align}
\int_{0^-}^{1^+}\,dx' \left.\left(\hat{\varphi}\partial_{t'}G-G\partial_{t'}\hat{\varphi}\right)\right|_{t'=0},
\end{align}
where due to working with the retarded GF 
\begin{align}
G(x,t|x',t^+)=0,
\end{align}
hence the upper limit $t'=t^+$ vanishes. The second contribution comes from the Dirac $\delta$-functions in $\chi(x,x')$ and $\chi(x,x_0)$ which gives
\begin{align}
\chi_s\int_{0}^{t^+}dt'&\left[\hat{\varphi}(x,t')\partial_{t'}^2G(x,t|x,t')-G(x,t|x_0,t')\partial_{t'}^2\hat{\varphi}(x_0,t')\right]
\end{align}
Terms $(c)$ and $(d)$ get simplified due to Dirac $\delta$-functions as
\begin{align}
\int_{0}^{t^+}\,dt'G(x,t|x_0,t')S[\hat{\varphi}_j(t')],
\end{align}
and $\hat{\varphi}(x,t)$, respectively. 

At the end, we find a generic solution for the flux field $\hat{\varphi}(x,t)$ in the domain $[0^-,1^+]$ as
\begin{align}
\begin{split}
&\hat{\varphi}(x,t)=\underbrace{\int_{0}^{t^+}\,dt'G(x,t|x_0,t')S[\hat{\varphi}_j(t')]}_{Source \ Contribution}\\
&+\underbrace{\int_{0}^{t^+}\,dt' \left.\left[\hat{\varphi}(x',t')\partial_{x'}G(x,t|x',t')-G(x,t|x',t')\partial_{x'}\hat{\varphi}(x',t')\right]\right|_{x'=0^-}^{x'=1^+}}_{Boundary \ Contribution}\\
&+\underbrace{\int_{0^-}^{1^+}\,dx' \left.\left[\hat{\varphi}(x',t')\partial_{t'}G(x,t|x',t')-G(x,t|x',t')\partial_{t'}\hat{\varphi}(x',t')\right]\right|_{t'=0}}_{Initial \ Condition \ Contribution}\\
&+\underbrace{\chi_s\int_{0}^{t^+}dt'\left[\hat{\varphi}(x,t')\partial_{t'}^2G(x,t|x,t')-G(x,t|x_0,t')\partial_{t'}^2\hat{\varphi}(x_0,t')\right]}_{Feedback \ induced \ by \ transmon}.
\end{split}
\label{Eq:Generic Sol of varphi(x,t)}
\end{align}

According to Eq.~(\ref{Eq:Transmon Dyn}), the transmon is forced by the resonator flux field evaluated at $x=x_0$, i.e. $\hat{\varphi}(x_0,t)$. In the following, we rewrite the GF in terms of its Fourier representation for each term in Eq.~(\ref{Eq:Generic Sol of varphi(x,t)}) at $x=x_0$. The Fourier representation simplifies the boundary contribution further, while also allowing us to employ the spectral representation of GF discussed in Sec.~\ref{SubApp:Spec Rep of G-open}.

The source contribution can be written as
\begin{align}
\chi_s\int_{0}^{t}\,dt'\int_{-\infty}^{+\infty} \frac{d\omega}{2\pi} \tilde{G}(x_0,x_0,\omega)\omega_j^2 \sin{[\hat{\varphi}_j(t')]}e^{-i\omega(t-t')}.
\label{Eq:Freq rep of Source}
\end{align}

The boundary terms consist of two separate contributions at each end. Assuming that there is no radiation in the waveguides for $t<0$ we can write

\begin{subequations}
\begin{align}
&\hat{\varphi}_{R,L}(x,t)=\hat{\varphi}_{R,L}(x,t)\Theta(t),
\label{Eq:Causaltiy of Phi_R,L}\\
&\partial_x\hat{\varphi}_{R,L}(x,t)=\partial_x\hat{\varphi}_{R,L}(x,t)\Theta(t).
\label{Eq:Causaltiy of d_x Phi_R,L}
\end{align}
\end{subequations}

Using Eqs.~(\ref{Eq:Causaltiy of Phi_R,L}-\ref{Eq:Causaltiy of d_x Phi_R,L}) and causality of the GF, i.e. $G(x,t|x',t')\propto \Theta(t-t')$, we can extend the integration domain in $t'$ from $[0,t^+]$ to $[-\infty,\infty]$ without changing the value of integral since for an arbitrary integrable function $F(t,t')$, we have 
\begin{align}
\int_{0}^{t^+}dt' F(t,t')&\theta(t')\theta(t-t')=\int_{-\infty}^{+\infty}dt' F(t,t')\theta(t')\theta(t-t').
\end{align} 
This extension of integration limits becomes handy when we write both $\hat{\varphi}_R(x',t')$ and $G(x_0,t|x',t')$ in terms of their Fourier transforms in time. Focusing on the right boundary contribution at $x'=1^+$ we get
\begin{align}
\begin{split}
&\int _{-\infty}^{+\infty}dt' \int_{-\infty}^{+\infty}\frac{d\omega_1}{2\pi} \int_{-\infty}^{+\infty}\frac{d\omega_2}{2\pi} \left[\hat{\tilde{\varphi}}_R(x',\omega_1)\partial_{x'}\tilde{G}(x_0,x',\omega_2)\right.
\\
&\left.\left.-\tilde{G}(x_0,x',\omega_2)\partial_{x'}\hat{\tilde{\varphi}}_R(x',\omega_1)\right]\right|_{x'=1^+}e^{-i\omega_1 t'}e^{-i\omega_2 (t-t')}.
\end{split}
\label{Eq:Halfway simplified Boundary contribution}
\end{align}

Next, we write $\hat{\tilde{\varphi}}_R(x',\omega)$ as the sum of ``incoming'' and ``outgoing'' parts
\begin{align}
\hat{\tilde{\varphi}}_R(1^+,\omega_1)=\hat{\tilde{\varphi}}_R^{inc}(1^+,\omega_1)+\hat{\tilde{\varphi}}_R^{out}(1^+,\omega_1),
\label{Eq:Splitting varphi_R(x,t) into out/inc}
\end{align}
defined as
\begin{subequations}
\begin{align}
&\partial_{x'}\hat{\tilde{\varphi}}_R^{out}(x'=1^+,\omega_1)=+i\omega_1\hat{\tilde{\varphi}}_R^{out}(x'=1^+,\omega_1),
\label{Eq:outgoing part of varphi_R(x,t)}\\
&\partial_{x'}\hat{\tilde{\varphi}}_R^{inc}(x'=1^+,\omega_1)=-i\omega_1\hat{\tilde{\varphi}}_R^{inc}(x'=1^+,\omega_1).
\label{Eq:incoming part of varphi_R(x,t)}
\end{align}
\end{subequations}
On the other hand, since we are using a retarded GF with outgoing BC we have
\begin{align}
\partial_{x'}\tilde{G}(x_0,x'=1^+,\omega_2)=+i\omega_2\tilde{G}(x_0,x'=1^+,\omega_2).
\label{Eq:Outgoing BC for G of varphi_R(x,t)}
\end{align}
By substituting Eqs.~(\ref{Eq:outgoing part of varphi_R(x,t)}, (\ref{Eq:incoming part of varphi_R(x,t)}) and (\ref{Eq:Outgoing BC for G of varphi_R(x,t)}) into Eq.~(\ref{Eq:Halfway simplified Boundary contribution}), the integrand becomes
\begin{align}
i(\omega_1+\omega_2)\tilde{G}(x_0,1^+,\omega_2)\hat{\tilde{\varphi}}_R^{inc}(1^+,\omega_1)+i(\omega_2-\omega_1)\tilde{G}(x_0,1^+,\omega_2)\hat{\tilde{\varphi}}_R^{out}(1^+,\omega_1) 
\end{align}
By taking the integral in $t'$ as $\int_{-\infty}^{\infty} dt' e^{i(\omega_2-\omega_1)t'}=2\pi\delta(\omega_1-\omega_2)$, Eq.~(\ref{Eq:Halfway simplified Boundary contribution}) can be simplified as
\begin{align}
\int_{-\infty}^{+\infty}\frac{d\omega}{2\pi} \left[2i\omega\tilde{G}(x_0,x'=1^+,\omega)\hat{\tilde{\varphi}}_R^{inc}(0^-,\omega) \right]e^{-i\omega t},
\label{Eq:Freq rep of right drive}
\end{align}
which indicates that only the incoming part of the field leads to a non-zero contribution to the field inside the resonator. A similiar expression holds for the left boundary with the difference that the incoming wave at the left waveguide is ``right-going" in contrast to the right waveguide
\begin{align}
\int_{-\infty}^{+\infty}\frac{d\omega}{2\pi} \left[2i\omega\tilde{G}(x_0,x'=0^-,\omega)\hat{\tilde{\varphi}}_L^{inc}(0^-,\omega) \right]e^{-i\omega t}.
\label{Eq:Freq rep of left drive} 
\end{align}
The initial condition (IC) terms can be expressed in a compact form as
\begin{align}
\int_{x_1}^{x_2} dx'\int _{-\infty}^{\infty}\frac{d\omega}{2\pi} \left\{\chi(x',x_0)\tilde{G}(x_0,x',\omega)\left[\hat{\dot{\varphi}}(x',0)-i\omega\hat{\varphi}(x',0)\right]\right\}e^{-i\omega t}.
\end{align}
Gathering all the contributions, plugging it in the RHS of Eq.~(\ref{Eq:Transmon Dyn}) and defining a family of memory kernels
\begin{subequations}
\begin{align}
\mathcal{K}_n(\tau)\equiv\gamma\chi_s\int_{-\infty}^{+\infty} \frac{d\omega}{2\pi} \omega^n\tilde{G}(x_0,x_0,\omega)e^{-i\omega\tau},
\label{Eq:Def of K_n(tau)}
\end{align}
and transfer functions
\begin{align}
&\mathcal{D}_R(\omega)\equiv -2i\gamma\omega^3\tilde{G}(x_0,1^+,\omega),
\label{Eq:Def of D_R(om)}\\
&\mathcal{D}_L(\omega)\equiv -2i\gamma\omega^3\tilde{G}(x_0,0^-,\omega),
\label{Eq:Def of D_L(om)}\\
&\mathcal{I}(x',\omega)\equiv \gamma\omega^2\chi(x',x_0)\tilde{G}(x_0,x',\omega),
\label{Eq:Def of D_L(om)}
\end{align}
\end{subequations}
the effective dynamics of the transmon is found to be
\begin{align}
\begin{split}
&\hat{\ddot{\varphi}}_j(t)+(1-\gamma)\omega_j^2\sin{\left[\hat{\varphi}_j(t)\right]}=\\
+&\frac{d^2}{dt^2}\int_{0}^{t}dt'\mathcal{K}_0(t-t')\omega_j^2\sin{\left[\hat{\varphi}_j(t')\right]}\\
+&\int_{-\infty}^{+\infty}\frac{d\omega}{2\pi}\mathcal{D}_R(\omega)\hat{\tilde{\varphi}}_R^{inc}(1^+,\omega)e^{-i\omega t}\\
+&\int_{-\infty}^{+\infty}\frac{d\omega}{2\pi}\mathcal{D}_L(\omega)\hat{\tilde{\varphi}}_L^{inc}(0^-,\omega)e^{-i\omega t}\\
+&\int_{0^-}^{1^+}dx'\int_{-\infty}^{+\infty}\frac{d\omega}{2\pi}\mathcal{I}(x',\omega)\left[i\omega\hat{\varphi}(x',0)-\hat{\dot{\varphi}}(x',0)\right]e^{-i\omega t}.
\end{split}
\label{Eq:Red Dyn before trace}
\end{align}
This is Eq.~(\ref{eqn:Eff Dyn before trace}) in Sec.~\ref{Sec:Eff Dyn Of Transmon}.
\section{Effective dynamics for spontaneous emission}
\label{SubApp:SE Eff Dyn}
Equation~(\ref{Eq:Red Dyn before trace}) is the most generic effective dynamics of a transmon coupled to an open multimode resonator. In this section, we find the effective dynamics for the problem of spontaneous emission where the system starts from the IC
\begin{align}
\hat{\rho}(0)=\hat{\rho}_j(0)\otimes \ket{0}_{ph}\bra{0}_{ph}.
\label{Eq:SE-IC}
\end{align}
In the absence of external drive and due to the interaction with the leaky modes of the resonator, the system reaches its ground state $\hat{\rho}_g\equiv\ket{0}_{j}\bra{0}_{j}\otimes \ket{0}_{ph}\bra{0}_{ph}$ in steady state.

Note that due the specific IC~(\ref{Eq:SE-IC}), there is no contribution from IC of the resonator in Eq.~(\ref{Eq:Red Dyn before trace}). To show this explicitly, recall that at $t=0$ the interaction has not turned on and we can represent $\hat{\varphi}(x,0)$ and $\hat{\dot{\varphi}}(x,0)$ in terms of a set of Hermitian modes of the resonator as \cite{Malekakhlagh_Origin_2016}    
\begin{subequations}
\begin{align}
&\hat{\varphi}(x,0)=\hat{\mathbf{1}}_j\otimes\sum\limits_n\left(\frac{\hbar}{2\omega_n^{(H)} cL}\right)^{1/2}\left[\hat{a}_n(0)+\hat{a}_n^{\dag}(0)\right]\tilde{\varphi}_n^{(H)}(x),\\
&\hat{\dot{\varphi}}(x,0)=\hat{\mathbf{1}}_j\otimes\sum\limits_n -i\left(\frac{\hbar \omega_n^{(H)}}{2 cL}\right)^{1/2}\left[\hat{a}_n(0)-\hat{a}_n^{\dag}(0)\right]\tilde{\varphi}_n^{(H)}(x),
\end{align}
\end{subequations}
where we have used superscript notation $(H)$ to distinguish Hermitian from non-Hermitian modes. By taking the partial trace over the photonic sector we find 
\begin{align}
\Tr_{ph}\left\{\hat{\rho}_{ph}\left[\hat{a}_n(0)\pm\hat{a}_n^{\dag}(0)\right]\right\}=\bra{0}_{ph}\left[\hat{a}_n(0)\pm\hat{a}_n^{\dag}(0)\right]\ket{0}_{ph}=0.
\label{Eq:Why IC terms vanish}
\end{align}
With no external drive, $\hat{\tilde{\varphi}}_{R,L}^{inc}$ do not have a coherent part and their expectation value vanish due to the same reasoning as Eq.~(\ref{Eq:Why IC terms vanish}). Therefore, the effective dynamics for the spontaneous emission problem reduces to
\begin{align}
\begin{split}
&\hat{\ddot{\phi}}_j(t)+(1-\gamma)\omega_j^2 \Tr_{ph} \left\{\hat{\rho}_{ph}(0)\sin{\left[\hat{\varphi}_j(t)\right]}\right\}\\
&=\frac{d^2}{dt^2}\int_0^{t}dt' \mathcal{K}_0(t-t') \omega_j^2 \Tr_{ph}\left\{\hat{\rho}_{ph}(0)\sin{\left[\hat{\varphi}_j(t')\right]}\right\}.
\end{split}
\end{align}

Taking the second derivative of the RHS using Leibniz integral rule, and bringing the terms evaluated at the integral limits to the LHS gives
\begin{align}
\begin{split}
&\hat{\ddot{\phi}}_j(t)-\omega_j^2\mathcal{K}_0(0)\Tr_{ph}\left\{\hat{\rho}_{ph}(0)\cos{\left[\hat{\varphi}_j(t)\right]}\hat{\dot{\varphi}}_j(t)\right\}\\
&+\omega_j^2\left[1-\gamma+i\mathcal{K}_1(0)\right]\Tr_{ph}\left\{\hat{\rho}_{ph}(0)\sin{\left[\hat{\varphi}_j(t)\right]}\right\}\\
&=-\int_0^{t}dt'\mathcal{K}_2(t-t')\omega_j^2\Tr_{ph}\left\{\hat{\rho}_{ph}(0)\sin{\left[\hat{\varphi}_j(t')\right]}\right\},
\end{split}
\label{Eq:SE Nonlinear Problem}
\end{align}
where we have used Eq.~(\ref{Eq:Def of K_n(tau)}) to rewrite time-derivatives of $\mathcal{K}_0(\tau)$ in terms of $\mathcal{K}_n(\tau)$.
\section{Spectral representation of $\mathcal{K}_0$, $\mathcal{K}_1$ and $\mathcal{K}_2$}
\label{SubApp:Spec Rep of K}
In this section, we express the contributions from the kernels $\mathcal{K}_0(0)$, $\mathcal{K}_1(0)$ and $\mathcal{K}_2(\tau)$ appearing in Eq.~(\ref{Eq:SE Nonlinear Problem}) in terms of the spectral representation of the GF. For this purpose, we use the partial fraction expansion of the GF in agreement with \cite{Leung_Time-independent_1994, Leung_Two-component_1997, 
Servini_Second_2004, Muljarov_Brillouin_2010, 
Doost_Resonant_2013, Kristensen_Normalization_2015} in terms of its simple poles discussed in Sec.~\ref{SubApp:Spec Rep of G-open} as
\begin{align}
\tilde{G}(x,x',\omega)=\sum\limits_{n\in\mathbb{Z}}\frac{1}{2\omega}\frac{\tilde{\varphi}_n(x)\tilde{\varphi}_n(x')}{\omega-\omega_n},
\label{Eq:QB rep of G-Open}
\end{align}
where $\tilde{\varphi}_n(x)\propto \tilde{\varphi}_n(x,\omega=\omega_n)$ is the quasi-bound eigenfunction.

\begin{figure}
\centering
\subfloat[\label{subfig:IntegContour1}]{%
\includegraphics[scale=0.55]{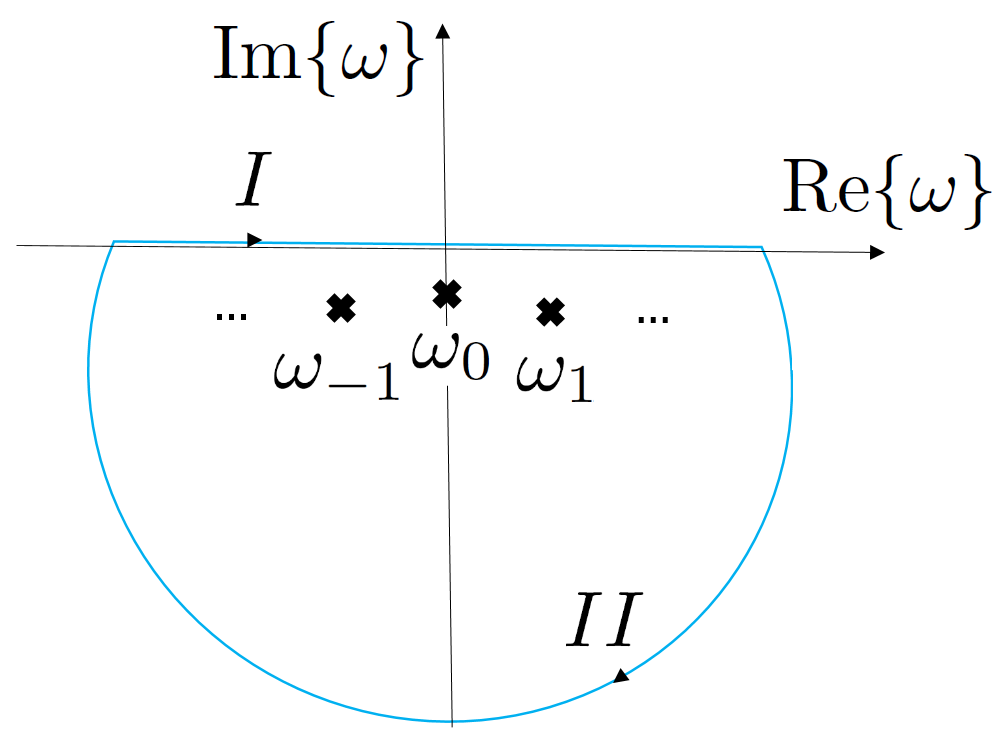}%
}
\subfloat[\label{subfig:IntegContour2}]{%
\includegraphics[scale=0.55]{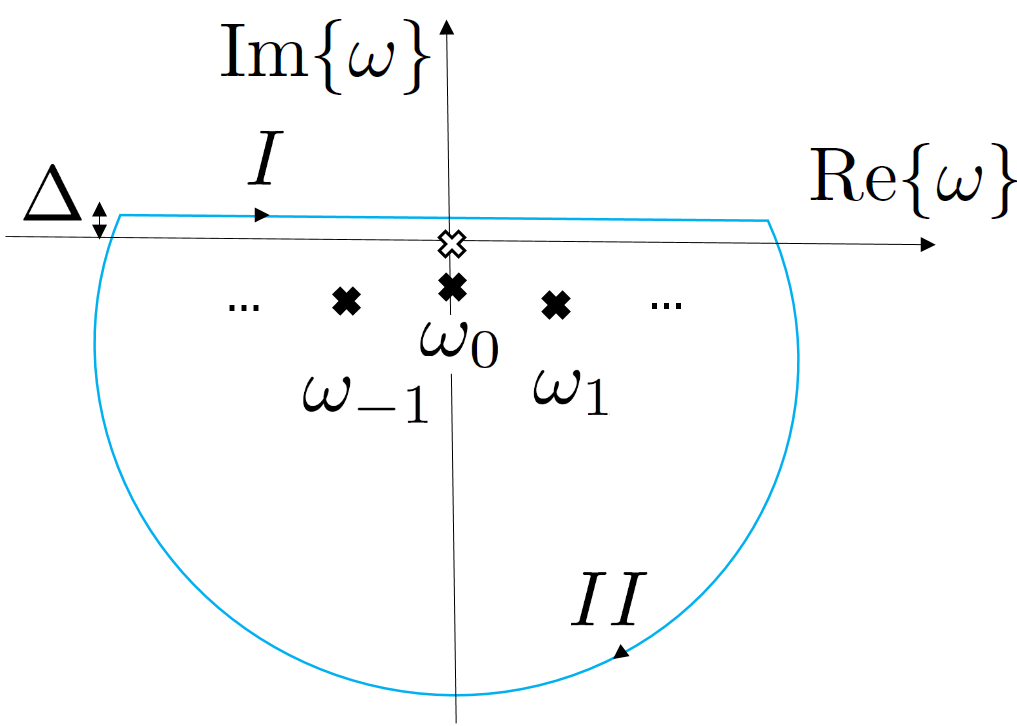}%
}
\caption{Integration contours: a) Integration contour that encloses the poles of $\omega^2\tilde{G}(x_0,x_0,\omega)$ and $\omega\tilde{G}(x_0,x_0,\omega)$; b) integration contour for $\tilde{G}(x_0,x_0,\omega)$, which has an extra pole at $\omega=0$.} 
\label{Fig:Inegration Contour for the Kernel}
\end{figure}

Let us first calculate $\mathcal{K}_2(\tau)$. By choosing an integration contour in the complex $\omega$-plane shown in Fig. \ref{subfig:IntegContour1} and applying Cauchy's residue theorem \cite{Mitrinovic_Cauchy_1984, Hassani_Mathematical_2013} we find
\begin{align}
\begin{split}
\oint_C d\omega \omega^2 \tilde{G}(x_0,x_0,\omega)e^{-i\omega \tau}&=\int_{I} d\omega \omega^2 \tilde{G}(x_0,x_0,\omega)e^{-i\omega \tau}+\int_{II} d\omega \omega^2 \tilde{G}(x_0,x_0,\omega)e^{-i\omega \tau}\\
&=-2\pi i\sum\limits_{n=0}^{\infty}\frac{1}{2}\left[\omega_n [\tilde{\varphi}_n(x_0)]^2 e^{-i\omega_n\tau}-\omega_n^*[\tilde{\varphi}_n^*(x_0)]^2e^{+i\omega_n^*\tau}\right]\\
&=-2\pi\sum\limits_{n=0}^{\infty}|\omega_n||\tilde{\varphi}_n(x_0)|^2\sin{[\nu_n\tau+\theta_n-2\delta_n(x_0)]}e^{-\kappa_n\tau},
\end{split}
\end{align} 
where due to nonzero opening of the resonator, both $\omega_n$ and $\tilde{\varphi}_n(x)$ are in general complex valued. Therefore, we have defined
\begin{align}
&\theta_n\equiv \arctan{\left(\frac{\kappa_n}{\nu_n}\right)},
\label{Eq:Def. of theta_n}\\
&\delta_n(x)\equiv \arctan{\left(\frac{\Im[{\tilde{\varphi}_n(x)}]}{\Re[{\tilde{\varphi}_n(x)}]}\right)}.
\label{Eq:Def. of delta_n(x)}
\end{align}
As the radius of the half-circle in Fig.~\ref{subfig:IntegContour1} is taken to infinity, $\int_{II}d\omega \omega^2 G(x_0,x_0,\omega)$ approaches zero. This can be checked by a change of variables  
\begin{align}
\omega=R_{II}e^{-i\psi}, \quad \psi \in [0,\pi]\rightarrow d\omega=-iR_{II}e^{-i\psi}d\psi
\end{align}
Substituting this into $\int_{II}$ and taking the limit $R_{II}\to \infty$ gives
\begin{align}
\begin{split}
&\lim\limits_{R_{II}\to\infty}\int_{II}d\omega\omega^2\tilde{G}(x_0,x_0,\omega)e^{-i\omega\tau}\\
&=\sum\limits_{n=0}^{\infty}\lim\limits_{R_{II}\to\infty} \int_{II} d\omega \frac{\omega(\omega+i\kappa_n)[\tilde{\varphi}_n(x_0)]^2}{(\omega-\omega_n)(\omega+\omega_n^*)}e^{-i\omega\tau}\\
&\propto\int_{0}^{\pi}d\psi\lim\limits_{R_{II}\to\infty}e^{-iR_{II}\tau\cos{(\psi)}}R_{II}e^{-R_{II}\tau\sin{(\psi)}}=0, \ \tau>0.
\end{split}
\end{align}
On the other hand, $\int_{I}$ in this limit reads 
\begin{align}
\lim\limits_{R_{II}\to\infty}\int_{I}d\omega\omega^2\tilde{G}(x_0,x_0,\omega)e^{-i\omega\tau}=\int_{-\infty}^{\infty}d\omega\omega^2\tilde{G}(x_0,x_0,\omega)e^{-i\omega\tau},
\end{align}
which is the quantity of interest. Therefore, we find
\begin{align}
\begin{split}
\int_{-\infty}^{\infty}d\omega\omega^2\tilde{G}(x_0,x_0,\omega)e^{-i\omega\tau}=-2\pi\sum\limits_{n=0}^{\infty}|\omega_n||\tilde{\varphi}_n(x_0)|^2\sin{\left[\nu_n\tau+\theta_n-2\delta_n(x_0)\right]}e^{-\kappa_n\tau}.
\end{split}
\end{align}
From this, we obtain the spectral representation of $\mathcal{K}_2(\tau)$ as
\begin{align}
&\mathcal{K}_2(\tau)=-\sum\limits_{n=0}^{\infty} A_n \sin{\left[\nu_n\tau+\theta_n-2\delta_n(x_0)\right]}e^{-\kappa_n \tau},
\label{Eq:K^(2)(tau)}
\end{align}
with $A_n\equiv \gamma\chi_s \sqrt{\nu_n^2+\kappa_n^2} \left|\tilde{\varphi}_n(x_0)\right|^2$.
%%%%%%%% Fig:Integration Contours %%%%%%%%%%
%%%%%%%%%%%%%%%%%%%%%%%%%%%%%%%%%%%%%%%%%%%

$\mathcal{K}_1(0)$ can be found through similar complex integration  
\begin{align}
\begin{split}
\oint_C d\omega \omega \tilde{G}(x_0,x_0,\omega)&=\int_{I} d\omega \omega \tilde{G}(x_0,x_0,\omega)+\int_{II} d\omega \omega \tilde{G}(x_0,x_0,\omega)\\
&=-2\pi i\sum\limits_{n=0}^{\infty}\left[\frac{[\tilde{\varphi}_n(x_0)]^2}{2}+\frac{[\tilde{\varphi}_{n}^*(x_0)]^2}{2}\right]\\
&=-2\pi i \sum\limits_{n=0}^{\infty}|\tilde{\varphi}_n(x_0)|^2\cos{[2\delta_{n}(x_0)]}
\end{split}
\end{align}
It can be shown again that $\int_{II}\to 0$ as $R_{II}\to\infty$ from which we find that
\begin{align}
i\mathcal{K}_1(0)=\gamma\chi_s\sum\limits_{n=0}^{\infty}|\tilde{\varphi}_n(x_0)|^2\cos{[2\delta_{n}(x_0)]}=\sum\limits_{n=0}^{\infty}\frac{A_n}{\sqrt{\nu_n^2+\kappa_n^2}}\cos{[2\delta_{n}(x_0)]}.
\label{Eq:iK^(1)(0)}
\end{align}

$\mathcal{K}_0(0)$ has an extra pole at $\omega=0$, so the previous contour is not well defined. Therefore, we shift the integration contour as shown in Fig. \ref{Fig:Inegration Contour for the Kernel}. Then, we have
\begin{align}
\begin{split}
\oint_C d\omega \tilde{G}(x_0,x_0,\omega)&=\int_{I}d\omega\tilde{G}(x_0,x_0,\omega)+\int_{II}d\omega\tilde{G}(x_0,x_0,\omega)\\
&=-2\pi i\sum\limits_{n=0}^{\infty}\frac{1}{2}\left[\frac{[\tilde{\varphi}_n(x_0)]^2}{\omega_n}-\frac{[\tilde{\varphi}_n^*(x_0)]^2}{\omega_n^*}\right]\\
&-2\pi i\sum\limits_{n=0}^{\infty}\frac{1}{2}\left[\frac{[\tilde{\varphi}_n(x_0)]^2}{-\omega_n}+\frac{[\tilde{\varphi}_n^*(x_0)]^2}{\omega_n^*}\right]=0,\\
\end{split}
\end{align}
where the first sum comes from the residues at $\omega=\omega_n$ and $\omega=-\omega_n^*$, while the last sum is the residue at $\omega=0$ and they completely cancel each other and we get
\begin{align}
\mathcal{K}_0(0)=0.
\label{Eq:K^(0)(0)}
\end{align}

From Eq.~(\ref{Eq:K^(0)(0)}) we find that the effective dynamics for the spontaneous emission problem simplifies to
\begin{align}
\begin{split}
&\hat{\ddot{\phi}}_j(t)+\omega_j^2\left[1-\gamma+i\mathcal{K}_1(0)\right]\Tr_{ph}\left\{\hat{\rho}_{ph}(0)\sin{\left[\hat{\varphi}_j(t)\right]}\right\}\\
&=-\int_0^{t}dt'\mathcal{K}_2(t-t')\omega_j^2\Tr_{ph}\left\{\hat{\rho}_{ph}(0)\sin{\left[\hat{\varphi}_j(t')\right]}\right\}.
\label{Eq:NL simplified SE Problem}
\end{split}
\end{align}
\chapter{Characteristic function $D_j(s)$ for the linear equations of motion}
\label{App:Char func D(s)}

Up to linear order, transmon acts as a simple harmonic oscillator and we find
we find 
\begin{align}
\hat{\ddot{X}}_j(t)&+\omega_j^2\left[1-\gamma+i\mathcal{K}_1(0)\right]\hat{X}_j(t)=-\int_0^{t}dt'\mathcal{K}_2(t-t')\omega_j^2\hat{X}_j(t').
\label{Eq:Lin SE Problem}
\end{align}
Equation~(\ref{Eq:Lin SE Problem}) is a linear integro-differential equation with a memory integral on the RHS, appearing as the convolution of the memory kernel $\mathcal{K}_2$ with earlier values of $\hat{X}_j$. It can be solved by means of unilateral Laplace transform \cite{Abramowitz_Handbook_1964, Korn_Mathematical_2000, Hassani_Mathematical_2013} defined as
\begin{align}
\tilde{f}(s)\equiv \int_{0}^{\infty}dt e^{-st}f(t).
\end{align} 
Employing the following properties of Laplace transform:
\begin{itemize}
\item[1)] Convolution
\begin{align}
\begin{split}
&\mathfrak{L}\left\{\int_0^t dt' f(t')g(t-t')\right\}=\mathfrak{L}\left\{\int_0^t dt' f(t-t')g(t')\right\}\\
&=\mathfrak{L}\left\{f(t)\right\}\cdot \mathfrak{L}\left\{g(t)\right\}=\tilde{f}(s)\tilde{g}(s),
\end{split}
\end{align}
\item[2)] General derivative
\begin{align}
\mathfrak{L}\left\{\frac{d^N}{dt^N}f(t)\right\}=s^N\tilde{f}(s)-\sum\limits_{n=1}^{N}s^{N-n}\left.\frac{d^{n-1}}{dt^{n-1}}f(t)\right|_{t=0},
\end{align}
\end{itemize} 
we can transform the integro-differential Eq.~(\ref{Eq:Lin SE Problem}) into a closed algebraic form in terms of $\hat{\tilde{X}}_j(s)$ as
\begin{align}
\hat{\tilde{X}}_j(s)=\frac{s\hat{X}_j(0)+\hat{\dot{X}}_j(0)}{D_j(s)}=\frac{s\hat{X}_j(0)+\omega_j\hat{Y}_j(0)}{D_j(s)},
\label{Eq:Sol of X_j(s)}
\end{align}
where we have defined
\begin{subequations}
\begin{align}
&D_j(s)\equiv s^2+\Omega^2(s),
\label{Eq:Def of D(s)}\\
&\Omega^2(s)\equiv\omega_j^2\left[1-\gamma+i\mathcal{K}_1(0)+\tilde{\mathcal{K}}_2(s)\right].
\label{Eq:Def of Omega2(s)}
\end{align}
\end{subequations}
and $\hat{Y}_j$ is the normalized charge variable and is canonically conjugate to $\hat{X}_j$ such that $[\hat{X}_j(0),\hat{Y}_j(0)]=2i$.

Note that in order to solve for $\hat{X}_j(t)$ from Eq.~(\ref{Eq:Sol of X_j(s)}), one has to take the inverse Laplace transform of the resulting algebraic form in $s$. This requires studying the denominator first which determines the poles of the entire system up to linear order. Using the expressions for $\mathcal{K}_2(\tau_1)$ and $i\mathcal{K}_1(0)$ given in Eqs.~(\ref{Eq:K^(2)(tau)}) and (\ref{Eq:iK^(1)(0)}) we find
\begin{align}
\begin{split}
&i\mathcal{K}_1(0)+\tilde{\mathcal{K}}_2(s)=\sum\limits_{n\in\mathbb{N}}\frac{A_n}{\sqrt{\nu_n^2+\kappa_n^2}}\cos{[2\delta_{n}(x_0)]}\\
&-\sum\limits_{n \in \mathbb{N}} A_n\frac{\cos{[\theta_n-2\delta_{n}(x_0)]}\nu_n+\sin{[\theta_n-2\delta_{n}(x_0)]}(s+\kappa_n)}{(s+\kappa_n)^2+\nu_n^2}.
\end{split}
\label{Eq:iK1(0)+K2(s)}
\end{align}
Expanding the sine and cosine in the numerator of the second term in Eq.~(\ref{Eq:iK1(0)+K2(s)}) as
\begin{align}
\begin{split}
&\cos{[\theta_n-2\delta_{n}(x_0)]}\nu_n+\sin{[\theta_n-2\delta_{n}(x_0)]}(s+\kappa_n)\\
&=\left\{\cos{(\theta_n)}\cos{[2\delta_{n}(x_0)]}+\sin{(\theta_n)}\sin{[2\delta_{n}(x_0)]}\right\}\nu_n\\
&+\left\{\sin{(\theta_n)}\cos{[2\delta_{n}(x_0)]}-\cos{(\theta_n)}\sin{[2\delta_n(x_0)]}\right\}(s+\kappa_n)\\
&=\frac{\left\{\kappa_n\cos{[2\delta_{n}(x_0)]}-\nu_n\sin{[2\delta_{n}(x_0)]}\right\}s}{\sqrt{\nu_n^2+\kappa_n^2}}\\
&+\frac{(\nu_n^2+\kappa_n^2)\cos{[2\delta_{n}(x_0)]}}{\sqrt{\nu_n^2+\kappa_n^2}},
\end{split}
\end{align} 
Eq.~(\ref{Eq:iK1(0)+K2(s)}) simplifies to
\begin{align}
\begin{split}
&\sum\limits_{n=0}^{\infty}\frac{A_n}{\sqrt{\nu_n^2+\kappa_n^2}}\left\{\cos{[2\delta_{n}(x_0)]}-\frac{(\nu_n^2+\kappa_n^2)\cos{[2\delta_{n}(x_0)]}}{(s+\kappa_n)^2+\nu_n^2}\right.\\
&-\left. \frac{\left\{\kappa_n\cos{[2\delta_{n}(x_0)]}-\nu_n\sin{[2\delta_{n}(x_0)]}\right\}s}{(s+\kappa_n)^2+\nu_n^2}\right\}\\
&=\sum\limits_{n=0}^{\infty}M_n\frac{s\{\cos{[2\delta_{n}(x_0)]}s+\sin{[2\delta_{n}(x_0)]}\nu_n\}}{(s+\kappa_n)^2+\nu_n^2},
\end{split}
\end{align}
where we have defined
\begin{align}
M_n\equiv\frac{A_n}{\sqrt{\nu_n^2+\kappa_n^2}}=\gamma\chi_s|\tilde{\varphi}_n(x_0)|^2.
\label{Eq:Def of Mn}
\end{align}
Therefore, $D_j(s)$ simplifies to
\begin{align}
D_j(s)=s^2+\omega_j^2+\underbrace{\omega_j^2\left\{-\gamma+\sum\limits_{n=0}^{\infty}M_n\frac{s\{\cos{[2\delta_{n}(x_0)]}s+\sin{[2\delta_{n}(x_0)]}\nu_n\}}{(s+\kappa_n)^2+\nu_n^2}\right\}}_{Modification \ due \ to \ memory}.
\label{Eq:simplified D(s)}
\end{align}
\chapter{Multi-Scale Analysis}
\label{App:MSPT}
In order to understand the application of MSPT on the problem of spontaneous emission, we have broken down its complexity into simpler toy problems, discussing each in a separate subsection. In Sec.~\ref{SubApp:ClDuffingDiss}, we revisit the classical Duffing oscillator problem \cite{Bender_Advanced_1999} in the presence of dissipation, to study the interplay of nonlinearity and dissipation. In Sec.~\ref{SubApp:QuDuffingNoMem}, we discuss the free quantum Duffing oscillator to show how the non-commuting algebra of quantum mechanics alters the classical solution. Finally, in Sec.~\ref{SubApp:QuDuffQuHarm}, we study the full problem and provide the derivation for the MSPT solution~(\ref{eqn:PertCorr-X^(0)(t) MSPT Sol}).
\section{Classical Duffing oscillator with dissipation}
\label{SubApp:ClDuffingDiss}
Consider a classical Duffing oscillator 
\begin{align}
\ddot{X}(t)+\delta\,\omega\dot{X}(t)+\omega^2\left[X(t)-\varepsilon X^3(t)\right]=0,
\label{Eq:ClDuffing Osc}
\end{align}
with initial condition $X(0)=X_0$, $\dot{X}(0)=\omega Y_0$. In order to have a bound solution, it is sufficient that the initial energy of the system be less than the potential energy evaluated at its local maxima, $X_{max}\equiv\pm\sqrt{1/3\varepsilon}$ , i.e. $E_0 < U(X_{max})$ which in terms of the initial conditions $X_0$ and $Y_0$ reads
\begin{align}
\frac{1}{2}Y_0^2+\frac{1}{2}\left(X_0^2-\varepsilon X_0^4\right)<\frac{5}{36\varepsilon}.
\end{align}

Note that a naive use of conventional perturbation theory decomposes the solution into a series $X(t)=X^{(0)}(t)+\varepsilon X^{(1)}(t)+\ldots$, which leads to unbounded (secular) solutions in time. In order to illustrate this, consider the simple case where $\delta=0$, $X_0=1$ and $Y_0=0$. Then, we find
\begin{subequations}
\begin{align}
&\mathcal{O}(1):\ddot{X}^{(0)}(t)+\omega^2 X^{(0)}(t)=0,\\
&\mathcal{O}(\varepsilon):\ddot{X}^{(1)}(t)+\omega^2 X^{(1)}(t)=\omega^2[X^{(0)}(t)]^3,
\end{align} 
\end{subequations}
which leads to $X^{(0)}(t)=\cos(\omega t)$ and $X^{(1)}(t)=\frac{1}{32}\cos(\omega t)-\frac{1}{32}\cos(3\omega t)+\frac{3}{8}\omega t \sin(\omega t)$. The latter has a secular contribution that grows unbounded in time.

The secular terms can be canceled order by order by introducing multiple time scales, which amounts to a resummation of the conventional perturbation series \cite{Bender_Advanced_1999}. We assume small dissipation and nonlinearity, i.e. $\delta,\varepsilon \ll 1$. This allows us to define additional slow time scales $\tau\equiv\varepsilon t$ and $\eta\equiv\delta t$ in terms of which we can perform a multi-scale expansion for $X(t)$ as
\begin{subequations}
\begin{align}
X(t)&=x^{(0)}(t,\tau,\eta)+\varepsilon x^{(1)}(t,\tau,\eta)+\delta y^{(1)}(t,\tau,\eta)+\mathcal{O}(\varepsilon^2,\delta^2,\varepsilon\delta).
\label{Eq:ClDuffing Expansion of X}
\end{align}
Using the chain rule, the total derivative $d/dt$ is also expanded as
\begin{align}
d_t=\partial_t+\varepsilon \partial_{\tau}+\delta\partial_{\eta}+\mathcal{O}(\varepsilon^2,\delta^2,\varepsilon\delta).
\label{Eq:ClDuffing Expansion of d/dt}
\end{align}
\end{subequations}
Plugging Eqs.~(\ref{Eq:ClDuffing Expansion of X}-\ref{Eq:ClDuffing Expansion of d/dt}) into Eq.~(\ref{Eq:ClDuffing Osc}) and collecting equal powers of $\delta$ and $\epsilon$ we find
\begin{subequations}
\begin{align}
&\mathcal{O}(1):\partial_t^2 x^{(0)}+\omega^2 x^{(0)}=0,
\label{Eq:ClDuffing-O(1)}\\
&\mathcal{O}(\delta):\partial_t^2 y^{(1)}+\omega^2 y^{(1)}=-\omega\partial_tx^{(0)}-2\partial_t\partial_{\eta}x^{(0)},
\label{Eq:ClDuffing-O(del)}\\
&\mathcal{O}(\varepsilon):\partial_t^2 x^{(1)}+\omega^2 x^{(1)}=\omega^2 \left[x^{(0)}\right]^3-2\partial_t\partial_{\tau}x^{(0)}.
\label{Eq:ClDuffing-O(eps)}
\end{align}
\end{subequations}

The general solution to $O(1)$ Eq.~(\ref{Eq:ClDuffing-O(1)}) reads
\begin{align}
x^{(0)}(t,\tau,\eta)=a(\tau,\eta)e^{-i\omega t}+a^*(\tau,\eta)e^{+i\omega t}.
\label{Eq:ClDuffing-O(1) Sol}
\end{align}
Plugging Eq.~(\ref{Eq:ClDuffing-O(1) Sol}) into Eq.~(\ref{Eq:ClDuffing-O(del)}) we find that in order to remove secular terms $a(\tau,\eta)$ satisfies 
\begin{align}
(2\partial_{\eta}+\omega)a(\tau,\eta)=0,
\label{ClDuffing-eta Sec Cond}
\end{align} 
which gives the $\eta$-dependence of $a(\tau,\eta)$ as
\begin{align}
a(\tau,\eta)=\alpha(\tau)e^{-\frac{\omega}{2}\eta}.
\label{Eq:ClDuffing-eta dep of a}
\end{align}

The condition that removes the secular term on the RHS of $\mathcal{O}(\varepsilon)$ Eq.~(\ref{Eq:ClDuffing-O(eps)}) reads
\begin{align}
2i\omega\partial_{\tau}a(\tau,\eta)+3\omega^2|a(\tau,\eta)|^2a(\tau,\eta)=0.
\label{ClDuffing-tau Sec Cond}
\end{align} 
Multipliying Eq.~(\ref{ClDuffing-tau Sec Cond}) by $a^*(\tau,\eta)$ and its complex conjugate by $a(\tau,\eta)$ and taking the difference gives
\begin{align}
\partial_{\tau}|a(\tau,\eta)|^2=0,
\end{align}
which together with Eq.~(\ref{Eq:ClDuffing-eta dep of a}) implies that
\begin{align}
|a(\tau,\eta)|^2=|\alpha(0)|^2e^{-\omega \eta}.
\end{align}
Then, $a(\tau,\eta)$ is found as 
\begin{align}
a(\tau,\eta)=\alpha(0)e^{-\frac{\omega}{2}\eta}e^{i\frac{3}{2}\omega|\alpha(0)|^2e^{-\omega \eta}\tau}.
\label{Eq:ClDuffing-sol of a(tau,eta)}
\end{align}

Replacing $\tau=\varepsilon t$ and $\eta=\delta t$, and, the general solution up to $\mathcal{O}(\varepsilon^2,\delta^2,\varepsilon\delta)$ reads
\begin{align}
\begin{split}
X^{(0)}(t)=x^{(0)}(t,\varepsilon t,\delta t)=e^{-\frac{\kappa}{2}t}\left[\alpha(0)e^{-i\bar{\omega}(t)t}+c.c.\right],
\end{split}
\label{Eq:ClDuffing-Sol of X^(0)(t)}
\end{align}
where we have defined the decay rate $\kappa\equiv\delta. \omega$ and a normalized frequency $\bar{\omega}(t)$ as
\begin{align}
\bar{\omega}(t)\equiv\left[1-\frac{3\varepsilon}{2}|\alpha(0)|^2e^{-\kappa t}\right]\omega.
\label{Eq:ClDuffing-Def of bar(om)(t)}
\end{align}
Furthermore, $\alpha(0)$ is determined based on initial conditions as $\alpha(0)=(X_0+iY_0)/2$. 
%%%%%%%%%%%%% Fig of ClDuffing %%%%%%%%%%%%%%
\begin{figure}
\centering
\subfloat[\label{subfig:XtClDuffingK001eps01}]{%
\includegraphics[scale=0.45]{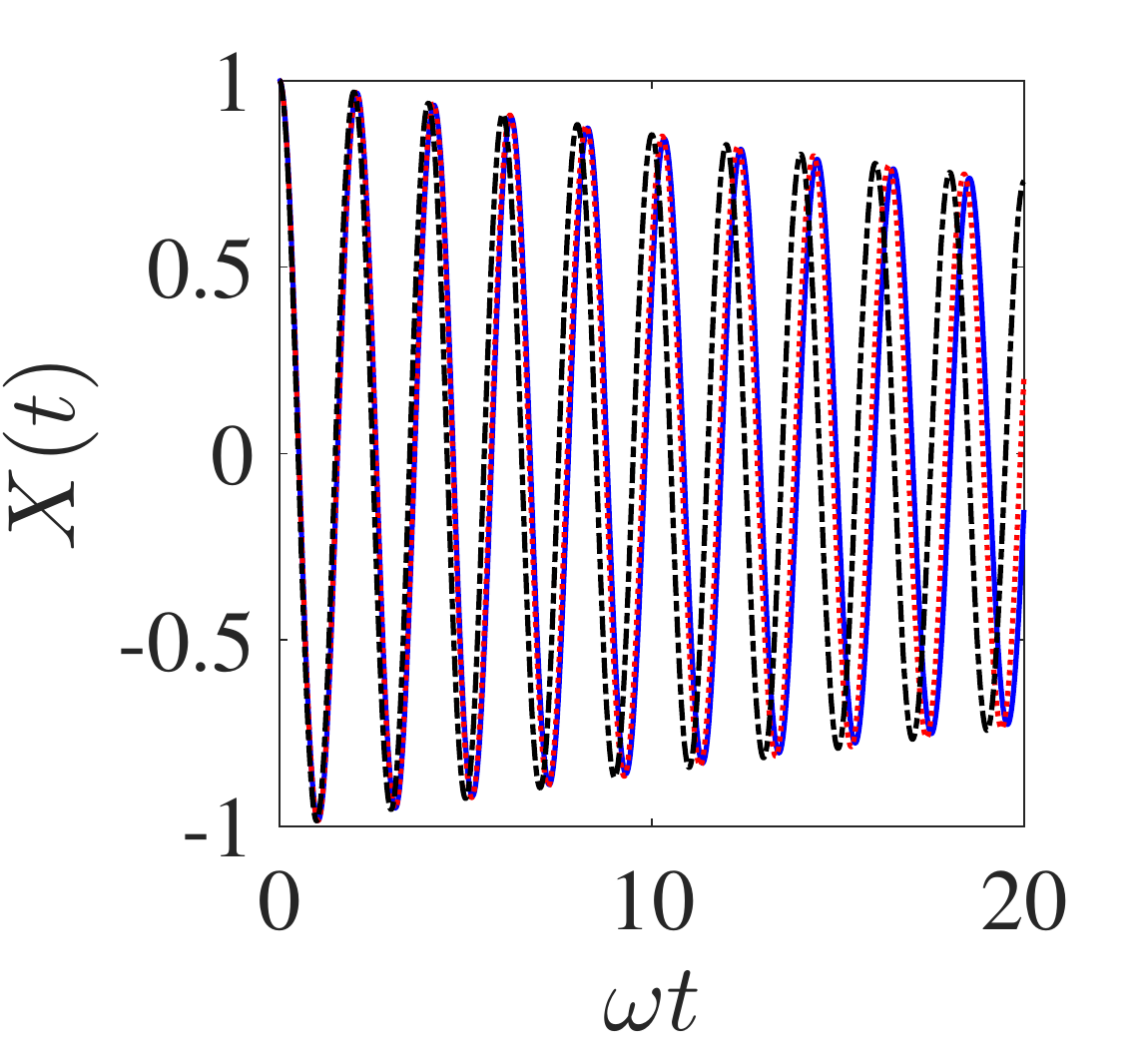}%
}
\subfloat[\label{subfig:XtClDuffingK001eps02}]{%
\includegraphics[scale=0.45]{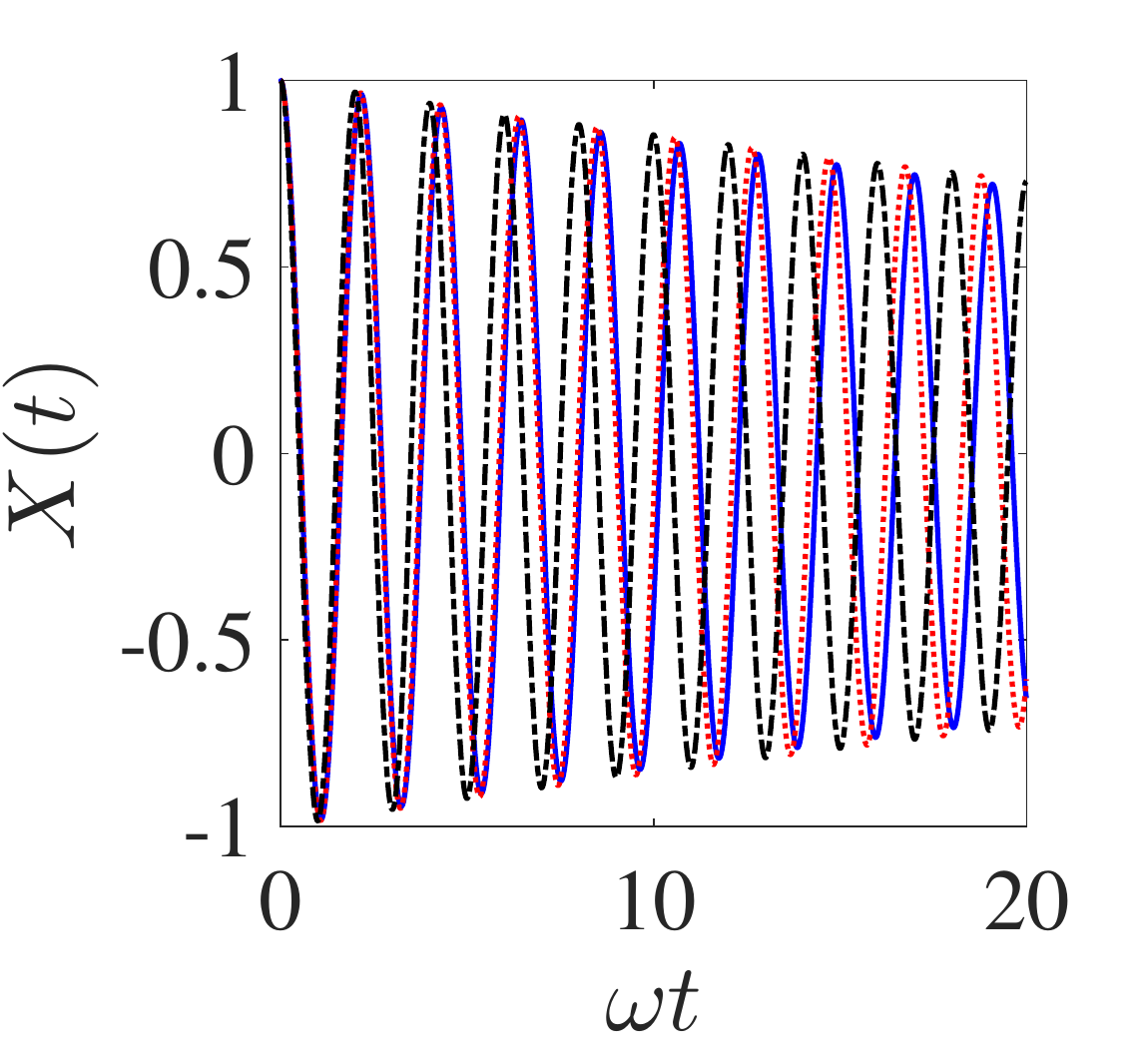}%
}
\caption{(Color online) Comparison of numerical solution (blue solid) with MSPT solution ~(\ref{Eq:ClDuffing-Sol of X^(0)(t)}) (red dotted) and linear solution, i.e. $\varepsilon=0$, (black dash-dot) of Eq.~(\ref{Eq:ClDuffing Osc}) for $\delta=0.01$ and ICs $X_0=1$, $Y_0=0$. a) $\varepsilon=0.1$ , b) $\varepsilon=0.2$.} 
\label{Fig:ClDuffing}
\end{figure}
%%%%%%%%%%%%%%%%%%%%%%%%%%%%%%%%%%%%%%%%%%%%%%% 

A comparison between the numerical solution (blue), $\mathcal{O}(1)$ MSPT solution~(\ref{Eq:ClDuffing-Sol of X^(0)(t)}) (red) and linear solution (black) is made in Fig.~\ref{Fig:ClDuffing} for the first ten oscillation periods. The MSPT solution captures the true oscillation frequency better than the linear solution. However, it is only valid for $\omega t\ll \varepsilon^{-2}, \delta^{-2}, \varepsilon^{-1}\delta^{-1}$ up to this order in perturbation theory.   
\section{A free quantum Duffing oscillator}
\label{SubApp:QuDuffingNoMem}
Consider a free quantum Duffing oscillator that obeys 
\begin{align}
\hat{\ddot{X}}(t)+\omega^2\left[\hat{X}(t)-\varepsilon \hat{X}^3(t)\right]=0,
\label{Eq:QuDuffingNoMem Osc}
\end{align}
with operator initial conditions
\begin{align}
\hat{X}(0),\quad 
\hat{\dot{X}}(0)=\omega \hat{Y}(0)
\label{Eq:Eq:QuDuffingNoMem ICs}
\end{align}
such that $\hat{X}(0)$ and $\hat{Y}(0)$ are canonically conjugate variables and obey $[\hat{X}(0),\hat{Y}(0)]=2i\hat{\mathbf{1}}$. 

Next, we expand $\hat{X}(t)$ and $d/dt$ up to $\mathcal{O}(\varepsilon^2)$ as
\begin{subequations}
\begin{align}
&\hat{X}(t)=\hat{x}^{(0)}(t,\tau)+\varepsilon \hat{x}^{(1)}(t,\tau)+\mathcal{O}(\varepsilon^2),
\label{Eq:QuDuffingNoMem Expansion of X}\\
&d_t=\partial_t+\varepsilon \partial_{\tau}+\mathcal{O}(\varepsilon^2).
\label{Eq:QuDuffingNoMem Expansion of d/dt}
\end{align}
\end{subequations}
Plugging this into Eq.~(\ref{Eq:QuDuffingNoMem Osc}) and collecting equal powers of $\varepsilon$ gives
\begin{subequations}
\begin{align}
&\mathcal{O}(1):\partial_t^2 \hat{x}^{(0)}+\omega^2 \hat{x}^{(0)}=0,
\label{Eq:QuDuffingNoMem-O(1)}\\
&\mathcal{O}(\varepsilon):\partial_t^2 \hat{x}^{(1)}+\omega^2 \hat{x}^{(1)}=\omega^2 \left[\hat{x}^{(0)}\right]^3-2\partial_t\partial_{\tau}\hat{x}^{(0)}.
\label{Eq:QuDuffingNoMem-O(eps)}
\end{align}
\end{subequations}
Up to $\mathcal{O}(1)$, the general solution reads
\begin{align}
&\hat{x}^{(0)}(t,\tau)=\hat{a}(\tau)e^{-i\omega t}+\hat{a}^{\dag}(\tau)e^{+i\omega t}
\label{Eq:QuDuffingNoMem-O(1) Ansatz}
\end{align}
Furthermore, from the commutation relation $[\hat{x}(t,\tau),\hat{y}(t,\tau)]=2i\hat{\mathbf{1}}$ we find that
$[\hat{a}(\tau),\hat{a}^{\dag}(\tau)]=\hat{\mathbf{1}}$. Substituting Eq.~(\ref{Eq:QuDuffingNoMem-O(1) Ansatz}) into the RHS of Eq.~(\ref{Eq:QuDuffingNoMem-O(eps)}) and setting the secular term oscillating at $\omega$ to zero we obtain
\begin{align}
2i\omega\frac{d \hat{a}(\tau)}{d \tau}+\omega^2\left[\hat{a}(\tau)\hat{a}(\tau)\hat{a}^{\dag}(\tau)+\hat{a}(\tau)\hat{a}^{\dag}(\tau)\hat{a}(\tau)+\hat{a}^{\dag}(\tau)\hat{a}(\tau)\hat{a}(\tau)\right]=0,
\label{Eq:QuDuffingNoMem-Secular Cond 1}
\end{align}
The condition that removes secular term at $-\omega$, appears as Hermitian conjugate of Eq.~(\ref{Eq:QuDuffingNoMem-Secular Cond 1}).

Using $[\hat{a}(\tau),\hat{a}^{\dag}(\tau)]=1$, Eq.~(\ref{Eq:QuDuffingNoMem-Secular Cond 1}) can be rewritten in a compact form
\begin{align}
\frac{d \hat{a}(\tau)}{d \tau}-i\frac{3\omega}{4}\left[\hat{\mathcal{H}}(\tau)\hat{a}(\tau)+\hat{a}(\tau)\hat{\mathcal{H}}(\tau)\right]=0,
\label{Eq:QuDuffingNoMem-Sec Cond  Simpl}
\end{align}
where 
\begin{align}
\hat{\mathcal{H}}(\tau)\equiv \frac{1}{2}\left[\hat{a}^{\dag}(\tau)\hat{a}(\tau)+\hat{a}(\tau)\hat{a}^{\dag}(\tau)\right].
\label{Eq:QuDuffingNoMem-Def of H}
\end{align}

Next, we show that $\hat{\mathcal{H}}(\tau)$ is a conserved quantity. Pre- and post-multiplying Eq.~(\ref{Eq:QuDuffingNoMem-Secular Cond 1}) by $\hat{a}^{\dag}(\tau)$, pre- and post-multiplying Hermitian conjugate of Eq.~(\ref{Eq:QuDuffingNoMem-Secular Cond 1}) by $\hat{a}(\tau)$ and adding all the terms gives
\begin{align}
\frac{d\hat{\mathcal{H}}(\tau)}{d\tau}=0,
\end{align}
which implies that $\hat{\mathcal{H}}(\tau)=\hat{\mathcal{H}}(0)$. Therefore, we find the solution for $\hat{a}(\tau)$ as
\begin{align}
\hat{a}(\tau)=\mathcal{W}\left\{\hat{a}(0)\exp\left[+i\frac{3\omega}{2}\hat{\mathcal{H}}(0)\tau\right]\right\},
\label{Eq:QuDuffingNoMem-a_j(tau_1) sol}
\end{align}
where $\mathcal{W}\{\bullet\}$ represents Weyl-ordering of operators \cite{Schleich_Quantum_2011}. The operator ordering $\mathcal{W}\left\{\hat{a}(0)f\left(\hat{\mathcal{H}}(0)\tau\right)\right\}$ is defined as follows:
\begin{enumerate}
\item
Expand $f\left(\hat{\mathcal{H}}(0)\tau\right)$ as a Taylor series in powers of operator $\hat{\mathcal{H}}(0)\tau$, \\
\item
Weyl-order the series term-by-term as:
\begin{align}
\mathcal{W}\left\{\hat{a}(0)\left[\hat{\mathcal{H}}(0)\right]^n\right\}\equiv\frac{1}{2^n}\sum\limits_{m=0}^{n} {{n}\choose{m}} \left[\hat{\mathcal{H}}(0)\right]^m \hat{a}(0)\left[\hat{\mathcal{H}}(0)\right]^{n-m}.
\end{align}
\end{enumerate}
The formal solution~(\ref{Eq:QuDuffingNoMem-a_j(tau_1) sol}) can be re-expressed in a closed form \cite{Bender_Resolution_1986, Bender_Continuous_1987, Bender_Polynomials_1988, Bender_Multiple_1996} using the properties of Euler polynomials \cite{Abramowitz_Handbook_1964} as 
\begin{align}
\hat{a}(\tau)=\frac{\hat{a}(0)e^{i\frac{3\omega}{2}\hat{\mathcal{H}}(0)\tau}+e^{i\frac{3\omega}{2}\hat{\mathcal{H}}(0)\tau}\hat{a}(0)}{2\cos\left(\frac{3\omega\tau}{4}\right)}.
\label{Eq:QuDuffingNoMem-Weyl Ord Simplified}
\end{align}
Plugging Eq.~(\ref{Eq:QuDuffingNoMem-Weyl Ord Simplified}) into Eq.~(\ref{Eq:QuDuffingNoMem-O(1) Ansatz}) and substituting $\tau=\varepsilon t$, we find the solution for $\hat{X}(t)$ up to $\mathcal{O}(\varepsilon)$ as
\begin{align}
\hat{X}^{(0)}(t)=\hat{x}^{(0)}(t,\varepsilon t)=\frac{\hat{a}(0)e^{-i\hat{\bar{\omega}} t}+e^{-i\hat{\bar{\omega}} t}\hat{a}(0)}{2\cos\left(\frac{3\omega}{4}\varepsilon t\right)}+\frac{\hat{a}^{\dag}(0)e^{+i\hat{\bar{\omega}} t}+e^{+i\hat{\bar{\omega}} t}\hat{a}^{\dag}(0)}{2\cos\left(\frac{3\omega}{4}\varepsilon t\right)},
\label{Eq:QuDuffingNoMem-X^(0)(t) sol}
\end{align}
where $\hat{\bar{\omega}}\equiv \omega[1-\frac{3\varepsilon}{2}\hat{\mathcal{H}}(0)]$ appears as a renormalized frequency operator. 

The physical quantity of interest is the expectation value of $\hat{X}^{(0)}(t)$ with respect to the initial density matrix $\hat{\rho}(0)$. The number basis of the simple harmonic oscillator is a complete basis for the Hilbert space of the Duffing oscillator such that
\begin{align}
\hat{\rho}(0)=\sum\limits_{mn}c_{mn}\ket{m}\bra{n}.
\end{align}
Therefore, calculation of $\braket{\hat{X}^{(0)}(t)}$ reduces to calculating the matrix element $\bra{m}\hat{a}(\varepsilon t)\ket{n}$. From Eq.~(\ref{Eq:QuDuffingNoMem-Weyl Ord Simplified}) we find that the only nonzero matrix element read
\begin{align}
\begin{split}
\bra{n-1}\hat{a}(\varepsilon t)\ket{n}&=\frac{\bra{n-1}\hat{a}(0)\ket{n}e^{i\frac{3\varepsilon\omega}{2}\bra{n}\hat{\mathcal{H}}(0)\ket{n}}}{2\cos\left(\frac{3\varepsilon\omega}{4}t\right)}\\
&+\frac{e^{i\frac{3\varepsilon\omega}{2}\bra{n-1}\hat{\mathcal{H}}(0)\ket{n-1}}\bra{n-1}\hat{a}(0)\ket{n}}{2\cos\left(\frac{3\varepsilon\omega}{4}t\right)}\\
&=\bra{n-1}\hat{a}(0)\ket{n}e^{i\frac{3n\varepsilon\omega}{2}t},
\end{split}
\end{align}
where we used that $\bra{n}\hat{\mathcal{H}}(0)\ket{n}=n+1/2$ is diagonal in the number basis.
\section{Quantum Duffing oscillator coupled to a set of quantum harmonic oscillators}
\label{SubApp:QuDuffQuHarm}
Quantum MSPT can also be applied to the problem of a quantum Duffing oscillator coupled to multiple harmonic oscillators. For simplicity, consider the toy Hamiltonian
\begin{align}
\hat{\mathcal{H}}\equiv \frac{\omega_j}{4}\left(\hat{\mathcal{X}}_j^2+\hat{\mathcal{Y}}_j^2-\frac{\varepsilon}{2}\hat{\mathcal{X}}_j^4\right)+\frac{\omega_c}{4}\left(\hat{\mathcal{X}}_c^2+\hat{\mathcal{Y}}_c^2\right)+g\hat{\mathcal{Y}}_j\hat{\mathcal{Y}}_c,
\label{Eq:QuDuffQuHarm-ToyH}
\end{align}
where the nonlinearity only exists in the Duffing sector of the Hilbert space labeled as $j$. Due to linear coupling there will be a hybridization of modes up to linear order. Therefore, Hamiltonian (\ref{Eq:QuDuffQuHarm-ToyH}) can always be rewritten in terms of the normal modes of its quadratic part as
\begin{align}
\hat{\mathcal{H}} \equiv \frac{\beta_j}{4}\left(\hat{\bar{\mathcal{X}}}_j^2+\hat{\bar{\mathcal{Y}}}_j^2\right)+\frac{\beta_c}{4}\left(\hat{\bar{\mathcal{X}}}_c^2+\hat{\bar{\mathcal{Y}}}_c^2\right)-\frac{\varepsilon\omega_j}{8}\left(u_j\hat{\bar{\mathcal{X}}}_j+u_c\hat{\bar{\mathcal{X}}}_c\right)^4,
\label{Eq:QuDuffQuHarm-ToyH NormModes}
\end{align}
where $u_{j,c}$ are real hybridization coefficients and $\hat{\bar{\mathcal{X}}}_{j,c}$ and $\hat{\bar{\mathcal{Y}}}_{j,c}$ represent $j$-like and $c$-like canonical operators. For $g=0$, $u_j\rightarrow 1$, $u_c\rightarrow 0$, $\hat{\bar{\mathcal{X}}}_{j,c}\rightarrow \hat{\mathcal{X}}_{j,c}$ and $\hat{\bar{\mathcal{Y}}}_{j,c}\rightarrow \hat{\mathcal{Y}}_{j,c}$. To find $u_{j,c}$ consider the Heisenberg equations of motion
\begin{subequations}
\begin{align}
&\hat{\ddot{\mathcal{X}}}_j(t)+\omega_j^2\hat{\mathcal{X}}_j(t)=-2g\omega_c\hat{\mathcal{X}}_c(t),
\label{Eq:QuDuffQuHarm-ddot(X)_j}\\
&\hat{\ddot{\mathcal{X}}}_c(t)+\omega_c^2\hat{\mathcal{X}}_c(t)=-2g\omega_j\hat{\mathcal{X}}_j(t).
\label{Eq:QuDuffQuHarm-ddot(X)_c}
\end{align}
\end{subequations}
Expressing $\vec{\mathcal{X}}\equiv(\hat{\mathcal{X}}_j \ \hat{\mathcal{X}}_c)^T$, the system above can be written as $\vec{\ddot{\mathcal{X}}}+V\vec{\mathcal{X}}=0$, where $V$ is a $2\times 2$ matrix. Plugging an Ansatz $\vec{\mathcal{X}}=\vec{\mathcal{X}}_0e^{i\lambda t}$ leads to an eigensystem whose eigenvalues are $\beta_{j,c}$ and whose eigenvectors give the hybridization coefficients $u_{j,c}$. 

The Heisenberg equations of motion for the hybdridized modes $\hat{\bar{\mathcal{X}}}_{l}(t)$, $l\equiv j,c$, reads
\begin{align}
\begin{split}
\hat{\ddot{\bar{\mathcal{X}}}}_{l}(t)+\beta_{l}^2\left\{\hat{\bar{\mathcal{X}}}_{l}(t)-\varepsilon_{l} \left[u_j\hat{\bar{\mathcal{X}}}_j(t)+u_c\hat{\bar{\mathcal{X}}}_c(t)\right]^3\right\}=0,
\end{split}
\label{Eq:QuDuffQuHarm Osc}
\end{align}
where due to hybridization, each oscillator experiences a distinct effective nonlinearity as $\varepsilon_{l}\equiv\frac{\omega_j}{\beta_{l}}u_{l}\varepsilon$. Therefore, we define two new time scales $\tau_{l}\equiv\varepsilon_{l}t$ in terms of which we can expand
\begin{subequations}
\begin{align}
&\hat{\bar{\mathcal{X}}}_{l}(t)=\hat{\bar{x}}_{l}^{(0)}(t,\tau_j,\tau_c)+\varepsilon_{l}\hat{\bar{x}}_{l}^{(1)}(t,\tau_j,\tau_c)+\varepsilon_{l'}\hat{\bar{y}}_{l}^{(1)}(t,\tau_j,\tau_c)+\mathcal{O}(\varepsilon_j^2,\varepsilon_c^2,\varepsilon_j\varepsilon_c),
\label{Eq:QuDuffQuHarm-Expansion of X}\\
&d_t=\partial_t+\varepsilon _j\partial_{\tau_j}+\varepsilon_c\partial_{\tau_c}+\mathcal{O}(\varepsilon_j^2,\varepsilon_c^2,\varepsilon_j\varepsilon_c).
\label{Eq:QuDuffQuHarm Expansion of d/dt}
\end{align}
\end{subequations}
where we have used the notation that if $l=j$, $l'=c$ and vice versa. Up to $\mathcal{O}(1)$ we find
\begin{align}
\mathcal{O}(1):\partial_t^2 \hat{\bar{x}}_{l}^{(0)}+\beta_{l}^2 \hat{\bar{x}}_{l}^{(0)}=0,
\label{Eq:QuDuffQuHarm-O(1)}
\end{align}
whose general solution reads
\begin{align}
\hat{\bar{x}}_{l}^{(0)}(t,\tau_j,\tau_c)=\hat{\bar{a}}_{l}(\tau_j,\tau_c)e^{-i\beta_{l}t}+\hat{\bar{a}}_{l}^{\dag}(\tau_j,\tau_c)e^{+i\beta_{l}t}.
\end{align}
where
\begin{align}
[\hat{\bar{a}}_{l_1},\hat{\bar{a}}_{l_2}^{\dag}]=\delta_{l_1l_2}\hat{\mathbf{1}}, \ [\hat{\bar{a}}_{l_1},\hat{\bar{a}}_{l_2}]=[\hat{\bar{a}}_{l_1}^{\dag},\hat{\bar{a}}_{l_2}^{\dag}]=0.
\label{Eq:QuDuffQuHarm-Commut Rels} 
\end{align}
There are $\mathcal{O}(\varepsilon_{l})$ and $\mathcal{O}(\varepsilon_{l'})$ equations of for each normal mode as
\begin{subequations}
\begin{align}
&\mathcal{O}(\varepsilon_{l}) \ of \ l: \partial_t^2 \hat{\bar{x}}_{l}^{(1)}+\beta_{l}^2 \hat{\bar{x}}_{l}^{(1)}=-2\partial_t\partial_{\tau_{l}}\hat{\bar{x}}_{l}^{(0)}-\beta_{l}^2\left[u_j\hat{\bar{x}}_j^{(0)}+u_c\hat{\bar{x}}_{c}^{(0)}\right]^3=0,
\label{Eq:QuDuffQuHarm-O(eps_pm) of pm}\\
&\mathcal{O}(\varepsilon_{l'}) \ of \ l:\partial_t^2 \hat{\bar{y}}_{l}^{(1)}+\beta_{l}^2 \hat{\bar{y}}_{l}^{(1)}=-2\partial_t\partial_{\tau_{l'}}\hat{\bar{x}}_{l}^{(0)}.
\label{Eq:QuDuffQuHarm-O(eps_mp) of pm}
\end{align}
\end{subequations}
By setting the secular terms on the RHS of Eq.~(\ref{Eq:QuDuffQuHarm-O(eps_mp) of pm}) we find that $\partial_{\tau_{l'}}\hat{b}_{l}=0$ which means that $q$ and $c$ sectors are only modified with their own time scale, i.e. $\hat{\bar{a}}_{l}=\hat{\bar{a}}_{l}(\tau_{l})$. Applying the same procedure on Eq.~(\ref{Eq:QuDuffQuHarm-O(eps_pm) of pm}) and using commutation relations (\ref{Eq:QuDuffQuHarm-Commut Rels}) we find
\begin{align}
\frac{d \hat{\bar{a}}_{l}}{d \tau_{l}}-i\frac{3\beta_{l}}{4}&\left\{u_{l}^3\left[\hat{\bar{\mathcal{H}}}_{l}\hat{\bar{a}}_{l}+\hat{\bar{a}}_{l}\hat{\bar{\mathcal{H}}}_{l}\right]2u_{l}u_{l'}^2\left[\hat{\bar{\mathcal{H}}}_{l'}\hat{\bar{a}}_{l}+\hat{\bar{a}}_{l}\hat{\bar{\mathcal{H}}}_{l'}\right]\right\}=0,
\label{Eq:QuDuffQuHarm-Sec Cond}
\end{align}
where
\begin{align}
\hat{\bar{\mathcal{H}}}_{l}(\tau_{l})\equiv \frac{1}{2}\left[\hat{\bar{a}}_{l}^{\dag}(\tau_{l})\hat{\bar{a}}_{l}(\tau_{l})+\hat{\bar{a}}_{l}(\tau_{l})\hat{\bar{a}}_{l}^{\dag}(\tau_{l})\right].
\label{Eq:QuDuffQuHarm-Def of H_pm}
\end{align}
By pre- and post-multiplying Eq.~(\ref{Eq:QuDuffQuHarm-Def of H_pm}) by $\hat{\bar{a}}_{l}^{\dag}(\tau_{l})$ and its Hermitian conjugate by $\hat{\bar{a}}_{l}(\tau_{l})$ and adding them we find that
\begin{align}
\frac{d\hat{\bar{\mathcal{H}}}_{l}(\tau_{l})}{d\tau_{l}}=0,
\end{align}
which means that the sub-Hamiltonians of each normal mode remain a constant of motion up to this order in perturbation. Therefore, in terms of effective Hamiltonians
\begin{align}
\hat{\bar{h}}_{l}(0)\equiv u_{l}^3\hat{\bar{\mathcal{H}}}_{l}(0)+2 u_{l}u_{l'}^2\hat{\bar{\mathcal{H}}}_{l'}(0),
\label{Eq:QuDuffQuHarm-Eff h}
\end{align}
Eq.~(\ref{Eq:QuDuffQuHarm-Sec Cond Simpl}) simplifies to
\begin{align}
\frac{d \hat{\bar{a}}_{l}}{d \tau_{l}}-i\frac{3\beta_{l}}{4}\left[\hat{\bar{h}}_{l}(0)\hat{\bar{a}}_{l}+\hat{\bar{a}}_{l}\hat{\bar{h}}_{l}(0)\right]=0.
\label{Eq:QuDuffQuHarm-Sec Cond Simpl}
\end{align}
Equation~(\ref{Eq:QuDuffQuHarm-Sec Cond Simpl}) has the same form as Eq.~(\ref{Eq:QuDuffingNoMem-Sec Cond  Simpl}) while the Hamiltonian $\mathcal{H}(0)$ is replaced by an effective Hamiltonian $\hat{\bar{h}}_{l}(0)$. Therefore, the formal solution is found as the Weyl ordering 
\begin{align}
\hat{\bar{a}}_{l}(\tau)=\mathcal{W}\left\{\hat{\bar{a}}_{l}(0)\exp\left[+i\frac{3\beta_{l}}{2}\hat{\bar{h}}_{l}(0)\tau_{l}\right]\right\}.
\label{Eq:QuDuffQuHarm-b_(eta) Sol}
\end{align}
Note that since $[\hat{\bar{a}}_{l},\hat{\bar{\mathcal{H}}}_{l'}(0)]=0$, the Weyl ordering only acts partially on the Hilbert space of interest which results in a closed form solution
\begin{align}
\hat{\bar{a}}_{l}(\tau_{l})=\frac{\hat{\bar{a}}_{l}(0)e^{i\frac{3\beta_{l}}{2}\hat{\bar{h}}_{l}(0)\tau_{l}}+e^{i\frac{3\beta_{l}}{2}\hat{\bar{h}}_{l}(0)\tau_{l}}\hat{\bar{a}}_{l}(0)}{2\cos\left(\frac{3 u_{l}^3\beta_{l}\tau_{l}}{4}\right)}.
\label{Eq:QuDuffQuHarm-Weyl Ord Simplified}
\end{align}
At last, $\hat{\bar{\mathcal{X}}}_{l}^{(0)}(t)$ is found by replacing $\tau_{l}=\varepsilon_{l} t$ as
\begin{align}
\hat{\bar{\mathcal{X}}}_{l}^{(0)}(t)=\hat{\bar{x}}_{l}^{(0)}(t,\varepsilon_{l}t)=\frac{\hat{\bar{a}}_{l}(0)e^{-i\hat{\bar{\beta}}_{l} t}+e^{-i\hat{\bar{\beta}}_{l} t}\hat{\bar{a}}_{l}(0)}{2\cos\left(\frac{3 u_{l}^3\beta_{l}\varepsilon_{l}}{4}t\right)}+\frac{\hat{\bar{a}}_{l}^{\dag}(0)e^{+i\hat{\bar{\beta}}_{l} t}+e^{+i\hat{\bar{\beta}}_{l} t}\hat{\bar{a}}^{\dag}(0)}{2\cos\left(\frac{3 u_{l}^3\beta_{l}\varepsilon_{l}}{4}t\right)},
\label{Eq:QuDuffQuHarm-Q_(eta)^(0)(t) sol}
\end{align}
where $\hat{\bar{\beta}}_{l}\equiv \beta_{l}\left[1-\frac{3\varepsilon_{l}}{2}\hat{\bar{h}}_{l}(0)\right]$. Plugging the expressions for $\varepsilon_{l}$ and $\hat{\bar{h}}_{l}(0)$, we find the explicit operator renormalization of each sector as
\begin{subequations}
\begin{align}
&\hat{\bar{\beta}}_j=\beta_j-\frac{3\varepsilon}{2}\omega_j\left[u_j^4\hat{\bar{\mathcal{H}}}_j(0)+2u_j^2u_c^2\hat{\bar{\mathcal{H}}}_c(0)\right],
\label{Eq:QuDuffQuHarm-bar(nu)_j}\\
&\hat{\bar{\beta}}_{c}=\beta_{c}-\frac{3\varepsilon}{2}\omega_j\left[u_c^4\hat{\bar{\mathcal{H}}}_c(0)+2u_c^2u_j^2\hat{\bar{\mathcal{H}}}_j(0)\right].
\label{Eq:QuDuffQuHarm-bar(nu)_c}
\end{align}
\end{subequations}
Equations.~(\ref{Eq:QuDuffQuHarm-bar(nu)_j}-\ref{Eq:QuDuffQuHarm-bar(nu)_c}) are symmetric under $j\leftrightarrow c$, implying that in the normal mode picture all modes are renormalized in the same manner. The terms proportional to $u_{j,c}^4$ and $u_{j,c}^2u_{c,j}^2$ are the self-Kerr and cross-Kerr contributions, respectively.

This analysis can be extended to the case of a Duffing oscillator coupled to multiple harmonic oscillators without further complexity, since the Hilbert spaces of the distinct normal modes do not mix to lowest order in MSPT. Consider the full Hamiltonian of our cQED system as
\begin{align}
\hat{\mathcal{H}} \equiv \frac{\omega_j}{4}\left(\hat{\mathcal{X}}_j^2+\hat{\mathcal{Y}}_j^2-\frac{\varepsilon}{2}\hat{\mathcal{X}}_j^4\right)+\sum\limits_{n}\frac{\omega_n}{4}\left(\hat{\mathcal{X}}_n^2+\hat{\mathcal{Y}}_n^2\right)+\sum\limits_{n}g_n\hat{\mathcal{Y}}_j\hat{\mathcal{Y}}_n,
\label{Eq:QuDuffQuHarm-H}
\end{align}
where here we label transmon operators with $j$ and all modes of the cavity by $n$. The coupling $g_n$ between transmon and the modes is given as \cite{Malekakhlagh_Origin_2016}
\begin{align}
g_n=\frac{1}{2}\gamma\sqrt{\chi_j}\sqrt{\omega_j\omega_n}\tilde{\Phi}_n(x_0).
\end{align}
Then, the Hamiltonian can be rewritten in a new basis that diagonalizes the quadratic part as
\begin{align}
\hat{\mathcal{H}} \equiv \frac{\beta_j}{4}\left(\hat{\bar{\mathcal{X}}}_j^2+\hat{\bar{\mathcal{Y}}}_j^2\right)+\sum\limits_n\frac{\beta_n}{4}\left(\hat{\bar{\mathcal{X}}}_n^2+\hat{\bar{\mathcal{Y}}}_n^2\right)-\frac{\varepsilon\omega_j}{8}\left(u_j \hat{\bar{\mathcal{X}}}_j+\sum\limits_nu_n \hat{\bar{\mathcal{X}}}_n\right)^4.
\label{Eq:QuDuffQuHarm-H NormModes}
\end{align}
The procedure to arrive at $u_{j,c}$ and $\beta_{j,c}$ is a generalization of the one presented under Eqs.~(\ref{Eq:QuDuffQuHarm-ddot(X)_j}-\ref{Eq:QuDuffQuHarm-ddot(X)_c}).

The Heisenberg dynamics of each normal mode is then obtained as
\begin{align}
\begin{split}
\hat{\ddot{\bar{\mathcal{X}}}}_{l}(t)+\beta_{l}^2\left\{\hat{\bar{\mathcal{X}}}_{l}(t)-\varepsilon_{l} \left[u_j\hat{\bar{\mathcal{X}}}_j(t)+\sum\limits_n u_n\hat{\bar{\mathcal{X}}}_n(t)\right]^3\right\}=0,
\end{split}
\label{Eq:QuDuffQuHarm Osc}
\end{align}
where $\varepsilon_{l}\equiv\frac{\omega_j}{\beta_{l}}u_{l}\varepsilon$ for $l\equiv j,n$. Up to lowest order in perturbation, the solution for $\hat{\bar{\mathcal{X}}}_{l}^{(0)}(t)$ has the exact same form as Eq.~(\ref{Eq:QuDuffQuHarm-Q_(eta)^(0)(t) sol}) with operator renormalization $\hat{\bar{\beta}}_{j}$ and $\hat{\bar{\beta}}_{n}$ as
\begin{subequations}
\begin{align}
&\hat{\bar{\beta}}_{j}=\beta_{j}-\frac{3\varepsilon}{2}\omega_j\left[u_j^4\hat{\bar{\mathcal{H}}}_j(0)+\sum\limits_{n}2u_j^2u_n^2\hat{\bar{\mathcal{H}}}_n(0)\right],
\label{Eq:QuDuffQuHarm-bar(beta)_j}\\
&\hat{\bar{\beta}}_{n}=\beta_{n}-\frac{3\varepsilon}{2}\omega_j\left[u_n^4\hat{\bar{\mathcal{H}}}_n(0)+2u_n^2u_j^2\hat{\bar{\mathcal{H}}}_j(0)+\sum\limits_{m\neq n}2u_n^2u_m^2\hat{\bar{\mathcal{H}}}_m(0)\right].
\label{Eq:QuDuffQuHarm-bar(beta)_n}
\end{align}
\end{subequations}

In App.~\ref{SubApp:ClDuffingDiss}, we showed that adding another time scale for the decay rate and doing MSPT up to leading order resulted in the trivial solution ~(\ref{Eq:ClDuffing-Sol of X^(0)(t)}) where the dissipation only appears as a decaying envelope. Therefore, we can immediately generalize the MSPT solutions~(\ref{Eq:QuDuffQuHarm-bar(beta)_j}-\ref{Eq:QuDuffQuHarm-bar(beta)_n}) to the dissipative case where the complex pole $p_{j}=-\alpha_j-i\beta_j$ of the transmon-like mode is corrected as
\begin{subequations}
\begin{align}
\hat{\bar{p}}_{j}=p_{j}+i\frac{3\varepsilon}{2}\omega_j\left[u_j^4\hat{\bar{\mathcal{H}}}_j(0)e^{-2\alpha_j t}+\sum\limits_{n}2u_j^2u_n^2\hat{\bar{\mathcal{H}}}_n(0)e^{-2\alpha_n t}\right],
\label{Eq:QuDuffQuHarm-bar(p)_j}
\end{align}
and resonator-like mode $p_{n}=-\alpha_n-i\beta_n$ as
\begin{align}
\hat{\bar{p}}_{n}=p_{n}+i\frac{3\varepsilon}{2}\omega_j\left[u_n^4\hat{\bar{\mathcal{H}}}_n(0)e^{-2\alpha_n t}+2u_n^2u_j^2\hat{\bar{\mathcal{H}}}_j(0)e^{-2\alpha_j t}+\sum\limits_{m\neq n}2u_n^2u_m^2\hat{\bar{\mathcal{H}}}_m(0)e^{-2\alpha_m t}\right].
\label{Eq:QuDuffQuHarm-bar(p)_n}
\end{align}
\end{subequations}
Then, the MSPT solution for $\hat{\mathcal{X}}_j^{(0)}(t)$ is obtained as
\begin{align}
\hat{\mathcal{X}}_j^{(0)}(t)=u_j\frac{\hat{\bar{a}}_j(0)e^{\hat{\bar{p}}_j t}+e^{\hat{\bar{p}}_j t}\hat{\bar{a}}_j(0)}{2\cos\left(\frac{3\omega_j}{4}u_j^4\varepsilon t e^{-2\alpha_j t}\right)}+H.c.+\sum\limits_n\left[u_n\frac{\hat{\bar{a}}_n(0)e^{\hat{\bar{p}}_n t}+e^{\hat{\bar{p}}_n t}\hat{\bar{a}}_n(0)}{2\cos\left(\frac{3\omega_j}{4}u_n^4\varepsilon t e^{-2\alpha_n t}\right)}+H.c.\right].
\label{Eq:QuDuffQuHarm-X_j^(0)(t) MSPT Sol}
\end{align}
Note that if there is no coupling, $u_j=1$ and $u_n=0$ and we retrieve the MSPT solution of a free Duffing oscillator given in Eq.~(\ref{Eq:QuDuffingNoMem-X^(0)(t) sol}).
\chapter{Reduced equation for the numerical solution}
\label{App:RedNumEq}
In this appendix, we provide the derivation for Eq.~(\ref{eqn:NumSim-QuDuffingOscMemReduced}) based on which we did the numerical solution for the spontaneous emission problem. Substituting Eq.~(\ref{eqn:Expansion of Sine}) into Eq.~(\ref{eqn:NL SE Problem}) we obtain the effective dynamics up to $\mathcal{O}(\varepsilon^2)$ as
\begin{align}
\begin{split}
&\hat{\ddot{X}}_j(t)+\omega_j^2\left[1-\gamma+i\mathcal{K}_1(0)\right]\left[\hat{X}_j(t)-\varepsilon \Tr_{ph}{\{\hat{\rho}_{ph}(0)\hat{\mathcal{X}}_j^3(t)\}}\right]\\
&=-\int_{0}^{t}dt'\omega_j^2 \mathcal{K}_2(t-t')[\hat{X}_j(t')-\varepsilon \Tr_{ph}{\{\hat{\rho}_{ph}(0)\hat{\mathcal{X}}_j^3(t')\}}].
\label{Eq:RedNumEq-QuDuffingOscMem}
\end{split}
\end{align}
If we are only interested in the numerical results up to linear order in $\varepsilon$ then we can write
\begin{align}
\hat{\mathcal{X}}_j(t)=\hat{\mathcal{X}}_j^{(0)}(t)+\varepsilon\hat{\mathcal{X}}_j^{(1)}(t)+\mathcal{O}(\varepsilon^2),
\end{align}
and we find that
\begin{align}
\varepsilon\Tr_{ph}\left\{\hat{\rho}_{ph}(0)\hat{\mathcal{X}}_j^3(t)\right\}=\varepsilon\Tr_{ph}\left\{\hat{\rho}_{ph}(0)\left[\hat{\mathcal{X}}_j^{(0)}(t)\right]^3\right\}+\mathcal{O}\left(\varepsilon^2\right).
\end{align}
Note that in this appendix $\hat{\mathcal{X}}_j^{(0)}(t)$ differs the MSPT notation in the main body and represents the linear solution. We know the exact solution for $\hat{\mathcal{X}}_j^{(0)}(t)$ via Laplace transform as
\begin{align}
\begin{split}
\hat{\mathcal{X}}_j^{(0)}(t)&=\mathfrak{L}^{-1}\left\{\frac{s\hat{\mathcal{X}}_j^{(0)}(0)+\omega_j\hat{\mathcal{Y}}_j^{(0)}(0)}{D_j(s)}\right\}\\
&+\mathfrak{L}^{-1}\left\{\frac{\sum\limits_n \left[a_n(s)\hat{\mathcal{X}}_n^{(0)}(0)+b_n(s)\hat{\mathcal{Y}}_n^{(0)}(0)\right]}{D_j(s)}\right\}\\
&=\mathfrak{L}^{-1}\left\{ \frac{s\hat{X}_j^{(0)}(0)+\omega_j\hat{Y}_j^{(0)}(0)}{D_j(s)}\right\}\otimes\hat{\mathbf{1}}_{ph}\\
&+\hat{\mathbf{1}}_j\otimes\mathfrak{L}^{-1}\left\{\frac{\sum\limits_n \left[a_n(s)\hat{X}_n^{(0)}(0)+b_n(s)\hat{Y}_n^{(0)}(0)\right]}{D_j(s)}\right\},
\end{split}
\label{Eq:RedNumEq-FormalSol Of Mathcal(X)_j}
\end{align}
where we have employed the fact that at $t=0$, the Heisenberg and Schr\"odinger operators are the same and have the following product form
\begin{subequations}
\begin{align}
&\hat{\mathcal{X}}_j^{(0)}(0)=\hat{X}_j^{(0)}(0)\otimes \hat{\mathbf{1}}_{ph},\\
&\hat{\mathcal{Y}}_j^{(0)}(0)=\hat{Y}_j^{(0)}(0)\otimes \hat{\mathbf{1}}_{ph},\\
&\hat{\mathcal{Y}}_n^{(0)}(0)=\hat{\mathbf{1}}_{j}\otimes\hat{Y}_n^{(0)}(0),\\
&\hat{\mathcal{X}}_n^{(0)}(0)=\hat{\mathbf{1}}_{j}\otimes\hat{X}_n^{(0)}(0).
\end{align}
\end{subequations}
The coefficients $a_n(s)$ and $b_n(s)$ can be found from the circuit elements and are proportional to light-matter coupling $g_n$. However, for the argument that we are are trying to make, it is sufficient to keep them in general form. 

Note that equation~(\ref{Eq:RedNumEq-FormalSol Of Mathcal(X)_j}) can be written formally as
\begin{align}
\hat{\mathcal{X}}_j^{(0)}(t)=\hat{X}_j^{(0)}(t)\otimes \hat{\mathbf{1}}_{ph}+\hat{\mathbf{1}}_{j} \otimes \hat{X}_{j,ph}(t).
\end{align}
Therefore, $\left[\hat{\mathcal{X}}_j^{(0)}(t)\right]^3$ is found as
\begin{align}
\begin{split}
&\left[\hat{\mathcal{X}}_j^{(0)}(t)\right]^3=\left[\hat{X}_j^{(0)}(t)\right]^3\otimes \hat{\mathbf{1}}_{ph}+\hat{\mathbf{1}}_{j} \otimes \hat{X}_{j,ph}^3(t)\\
&+3\left\{\left[\hat{X}_j^{(0)}(t)\right]^2\otimes \hat{X}_{j,ph}(t)+\hat{X}_j^{(0)}(t) \otimes \hat{X}_{j,ph}^2(t)\right\}.
\end{split}
\end{align}
Finally, we have to take the partial trace with respect to the photonic sector. For the initial density matrix $\hat{\rho}_{ph}(0)=\ket{0}_{ph}\bra{0}_{ph}$
\begin{align}
\braket{\hat{X}_{j,ph}(t)}_{ph}=\braket{\hat{X}_{j,ph}^3(t)}_{ph}=0.
\end{align}
The only nonzero expectation values in $\braket{\hat{\mathcal{X}}_{j,ph}^2(t)}_{ph}$ are $\braket{\hat{X}_n^2(0)}_{ph}=\braket{\hat{Y}_n^2(0)}_{ph}=1$. Therefore, the partial trace over the cubic nonlinearity takes the form
\begin{align}
\begin{split}
&\Tr_{ph}\left\{\hat{\rho}_{ph}(0)\left[\hat{\mathcal{X}}_j^{(0)}(t)\right]^3\right\}=\\
&\left[\hat{X}_j^{(0)}(t)\right]^3
+3\mathfrak{L}^{-1}\left\{\frac{\sum\limits_n\left[a_n^2(s)+b_n^2(s)\right]}{D_j(s)}\right\}\hat{X}_j^{(0)}(t).
\end{split}
\end{align}
The first term is the reduced transmon operator cubed. The second term is the sum over vacuum fluctuations of the resonator modes. Neglecting these vacuum expectation values we can write
\begin{align}
\Tr_{ph}\left\{\hat{\rho}_{ph}(0)\left[\hat{\mathcal{X}}_j^{(0)}(t)\right]^3\right\}\approx\left[\hat{X}_j^{(0)}(t)\right]^3=\hat{X}_j^3(t)+\mathcal{O}(\varepsilon^2),
\label{Eq:RedNumEq-trace simplification}
\end{align}
Substituting Eq.~(\ref{Eq:RedNumEq-trace simplification}) into Eq.~(\ref{Eq:RedNumEq-QuDuffingOscMem}) gives
\begin{align}
\begin{split}
&\hat{\ddot{X}}_j(t)+\omega_j^2\left[1-\gamma+i\mathcal{K}_1(0)\right]\left[\hat{X}_j(t)-\varepsilon\left[\hat{X}_j(t)\right]^3\right]\\
&=-\int_{0}^{t}dt'\omega_j^2 \mathcal{K}_2(t-t')\left[\hat{X}_j(t')-\varepsilon\left[\hat{X}_j(t')\right]^3\right]+\mathcal{O}(\varepsilon^2).
\label{Eq:RedNumEq-QuDuffingOscMemReduced}
\end{split}
\end{align}
\chapter{Canonical quantization of a closed cQED system}
\label{App:CanonicalQuant}
In this appendix, we discuss the quantization of a closed cQED system in terms of the modified CC basis found in Sec.~\ref{Sec:A2-ClassHam&CCBasis}.

First, we extend the classical variables into quantum operators by introducing a set of creation and annihilation operators that obey the usual bosonic commutation relations
\begin{subequations}
\begin{align}
[\hat{a}_n,\hat{a}^{\dagger}_m]&=i\hbar \delta_{nm}, \\
[\hat{a}_n,\hat{a}_m]&=0, \\
[\hat{a}^{\dagger}_n,\hat{a}^{\dagger}_m]&=0.
\end{align}
\end{subequations}
Remembering that $\{\tilde{\Phi}_n(x) \}$ represent Hermitian modes and thus real functions, we find the quantum operators $\hat{\Phi}(x,t)$ and $\hat{\rho}(x,t)$ to be
\begin{subequations}
\begin{align}
&\hat{\Phi}(x,t)=\sum\limits_n \left(\frac{\hbar}{2\omega_n cL}\right)^{\frac{1}{2}}\left(\hat{a}_n + \hat{a}_n^{\dagger}\right)\tilde{\Phi}_n(x),\\
&\hat{\rho}(x,t)=-i\sum\limits_n \left(\frac{\hbar \omega_n}{2cL}\right)^{\frac{1}{2}}\left(\hat{a}_n - \hat{a}_n^{\dagger}\right)c(x,x_0)\tilde{\Phi}_n(x). 
\end{align}
\end{subequations}
Inserting these spectral representations into $\hat{\mathcal{H}}_C^{mod}$ and using the orthogonality relations \ref{Eq: Orthogonality of Phi} and \ref{Eq: Orthogonality of d/dx Phi}  will result in
\begin{align}
\hat{\mathcal{H}}_C^{mod}=\sum\limits_n \frac{\hbar \omega_n}{2}\left(a_n^{\dagger}a_n +a_n a_n^{\dagger}\right)=\sum\limits_n \hbar \omega_n a_n^{\dagger}a_n+ \text{const},
\end{align}
which is a sum over energy of each independent mode as we expected.

Having found the the resonator's Hamiltonian in second quantized form, we have to calculate now the spectrum of transmon whose Hamiltonian is given as
\begin{align}
\hat{\mathcal{H}}_{A}=\frac{\hat{Q}_j^2}{2C_j}-E_j\cos{\left(2\pi\frac{\hat{\Phi}_j}{\Phi_0}\right)}.
\end{align}
choosing to solve for the spectrum in the flux basis $\{\ket{\Phi_j}\}$ where $\hat{Q}_j\equiv\frac{h}{i}\frac{\partial}{\partial\Phi_j}$, we find
\begin{align}
\left[-\frac{\hbar^2}{2C_j}\frac{d^2}{d \Phi_j^2}-E_j\cos{\left(2\pi\frac{\Phi_j}{\Phi_0}\right)}\right]\Psi_n(\Phi_j)=\hbar\Omega_n\Psi_n(\Phi_j).
\end{align}
The solution to the above equation is a set of real eigenenergies and eigenmodes $\{\hbar\Omega_n,\Psi_n(\phi_j)| n\in\mathbb{N}^{0}\}$ where any operator in the transmon's space has a spectral representation over them as
\begin{align}
\begin{split}
\hat{O}_T(t)=&\hat{U}(t)\hat{O}_T(0)\hat{U}^{\dagger}(t)\\
=&\hat{U}(t)\left(\sum_{m,n}\bra{m}\hat{O}_T(0)\ket{n}\hat{P}_{mn}\right)\hat{U}^{\dagger}(t)\\
=&\sum_{m,n}\bra{m}\hat{O}_T(0)\ket{n}\hat{P}_{mn}(t),
\end{split}
\end{align}
where $\{\hat{P}_{mn}=\ket{m}\bra{n}\}$ is a set of projection operators between states m and n , and
\begin{align}
\bra{m}\hat{O}_T(0)\ket{n}\equiv \int d\phi_j \Psi_m(\phi_j)\hat{O}_T\left[\phi_j,\frac{h}{i}\frac{\partial}{\partial\phi_j}\right]\Psi_n(\phi_j).
\end{align}

Next, we express $\hat{\Phi}_j$ and $\hat{Q}_j$ in their spectral representation. Notice that since the potential is an even function of $\phi_j$, the eigenmodes are either even or odd functions of $\phi_j$, so the diagonal matrix elements are zero, since
\begin{subequations}
\begin{align} 
&\bra{n}\hat{\Phi}_j\ket{n}=\int d\Phi_j \underbrace{\Phi_j\Psi_n(\Phi_j)\Psi_n(\Phi_j)}_{Odd}=0,\\
&\bra{n}\hat{Q}_j\ket{n}=\frac{\hbar}{i}\int d\Phi_j \underbrace{\Psi_n(\Phi_j)\frac{\partial}{\partial \Phi_j} \Psi_n (\Phi_j)}_{Odd}=0.
\end{align}
\end{subequations}
Therefore, we can express $\hat{\Phi}_j$ and $\hat{Q}_j$ as
\begin{subequations}
\begin{align}
&\hat{\Phi}_j(t)=\sum\limits_{m\neq n}\bra{m}\hat{\Phi}_j(0)\ket{n}\hat{P}_{mn}(t)=\sum\limits_{m<n}\bra{m}\hat{\Phi}_j(0)\ket{n}\left(\hat{P}_{mn}(t)+\hat{P}_{nm}(t)\right),\\
&\hat{Q}_j(t)=\sum\limits_{m\neq n}\bra{m}\hat{Q}_j(0)\ket{n}\hat{P}_{mn}(t)=\sum\limits_{m<n}\bra{m}\hat{Q}_j(0)\ket{n}\left(\hat{P}_{mn}(t)-\hat{P}_{nm}(t)\right).
\end{align}
\end{subequations}
where the second lines are written based on the observation that by working in a flux basis, matrix elements of $\hat{Q}_j$ and $\hat{\Phi}_j$ are purely real and imaginary respectively. Now that we know the spectrum of both the resonator and the qubit, we can easily write the interaction term as
\begin{align}
\begin{split}
&\gamma \hat{Q}_j \int_{0}^{L} dx \frac{\hat{\rho}(x,t)}{c(x,x_0)} \delta(x-x_0) =\\
&-i\gamma \sum\limits_{m<n,l}Q_{J,mn} (\hat{P}_{mn} -\hat{P}_{nm})\left(\frac{\hbar \omega_l}{2cL}\right)^{\frac{1}{2}}\left(\hat{a}_l - \hat{a}_l^{\dagger}\right)\tilde{\Phi}_l(x_0).
\end{split}
\end{align}
Defining the coupling intensity $g_{mnl}$ as
\begin{align}
\hbar g_{mnl} \equiv \gamma \left(\frac{\hbar \omega_l}{2cL}\right)^{\frac{1}{2}}(iQ_{J,mn})\tilde{\Phi}_l(x_0), 
\label{Eq:Expression for g- closed case}
\end{align}
the interaction takes the form
\begin{align}
-\sum\limits_{m<n,l} \hbar g_{mnl}(\hat{P}_{mn}-\hat{P}_{nm}) (\hat{a}_l-\hat{a}_l^{\dagger}).
\end{align} 

Finally, up to a unitary transformation $\hat{a}_l\rightarrow i\hat{a}_l$ and $\hat{P}_{mn}\rightarrow i\hat{P}_{mn}$ for $m<n$,  the Hamiltonian reads
\begin{align}
\hat{\mathcal{H}}=\underbrace{\sum\limits_n \hbar\Omega_n \hat{P}_{nn}}_{\hat{\mathcal{H}}_A}+\underbrace{\sum\limits_n \hbar \omega_n \hat{a}_n^{\dagger} \hat{a}_n}_{\hat{\mathcal{H}}_C^{mod}}+\underbrace{\sum\limits_{m<n,l} \hbar g_{mnl} \left(\hat{P}_{mn}+\hat{P}_{nm}\right)\left(\hat{a}_l+\hat{a}^{\dagger}_l\right)}_{\hat{\mathcal{H}}_{int}}
\end{align}
Note that by truncating transmon's space into its first two levels, we are able to recover a multimode Rabi Hamiltonian
\begin{align}
\hat{\mathcal{H}}=\underbrace{\frac{1}{2} \hbar \omega_{01} \hat{\sigma}^z}_{\hat{\mathcal{H}}_A^{tru}}+\underbrace{\sum\limits_n \hbar \omega_n \hat{a}_n^{\dagger} \hat{a}_n}_{\hat{\mathcal{H}}_C^{mod}}+\underbrace{\sum\limits_n \hbar g_n (\hat{\sigma}^- + \hat{\sigma}^+)(\hat{a}_n+\hat{a}^{\dagger}_n)}_{\hat{\mathcal{H}}_{int}},
\end{align}
where we have used the shorthand notation $\hat{\sigma}^{-}=\hat{P}_{01}$, $\hat{\sigma}^{+}=\hat{P}_{10}$ and $\omega_{01}=\Omega_1-\Omega_0$. $g_{mnl}$ is also reduced to $g_{n}\equiv g_{01n}$ given as
\begin{align}
\hbar g_{n} \equiv \gamma \left(\frac{\hbar \omega_n}{2cL}\right)^{\frac{1}{2}}(iQ_{J,01})\tilde{\Phi}_n(x_0). 
\end{align}

\chapter{Second quantized Hamiltonian of a Transmon qubit coupled to an open resonator}
\label{App:2ndQuantOpen}
\section{Lagrangian and modified eigenmodes}
The results from App.~\ref{App:CanonicalQuant} make it very easy to find the Lagrangian and hence the dynamics for the open case where now the end capacitors $C_R$ and $C_L$ have finite values as shown in Fig.\ref{Fig:cQEDOpen-EqCircuit}. Here, we have a finite length resonator which is capacitively coupled to two other microwave resonators at each end. Assuming that the transmon qubit is connected to the resonator at some arbitrary point $x=x_0$, the Lagrangian for this system can be written as
\begin{align}
\begin{split}
\mathcal{L}&=\frac{1}{2}C_j\dot{\Phi}_{j}(t)^2-U(\Phi_j(t)) \\
&+\int_{0^+}^{L^-}\,dx\left\{\frac{1}{2} c[\partial_t \Phi(x,t)]^2-\frac{1}{2l}[\partial_x \Phi(x,t)]^2\right\}\\
&+\int_{L^+}^{+\infty}\,dx\left\{\frac{1}{2} c[\partial_t \Phi_R(x,t)]^2-\frac{1}{2l}[\partial_x \Phi_R(x,t)]^2\right\}\\
&+\int_{-\infty}^{0^-}\,dx\left[\frac{1}{2} c[\partial_t \Phi_L(x,t)]^2-\frac{1}{2l}[\partial_x \Phi_L(x,t)]^2\right]\\
&+\frac{1}{2}C_L\left[\dot{\Phi}_L(0^-,t)-\dot{\Phi}(0^+,t)\right]^2\\&
+\frac{1}{2}C_R\left[\dot{\Phi}_R(L^+,t)-\dot{\Phi}(L^-,t)\right]^2\\
&+\frac{1}{2}C_g\left[\dot{\Phi}_j(t)-\dot{\Phi}(x_0,t)\right]^2,
\end{split}
\end{align}

We have already learned how the coupling intensity depends on the Hermition eigenmodes and eigenfrequencies of the resonator as well as the dipole moment of the transmon. Here we have the same situation except that due to the opening introduced by the finite end capacitors $C_R$ and $C_L$, we need to find the modified Hermitian modes of the open system. Therefore, let us for the moment forget about Lagrangian of the transmon and its coupling to the resonator and focus on the modification introduced by one of the end capacitors, for instance $C_L$. The trick is that we can write this contribution as sum of three separate terms
\begin{align}
\begin{split}
\frac{1}{2}C_L \left[\dot{\Phi}_L(0^-,t)-\dot{\Phi}(0^+,t)\right]^2&=\frac{1}{2}C_L \dot{\Phi}_L^2(0^-,t)+\frac{1}{2}C_L \dot{\Phi}^2(0^+,t)-C_L \dot{\Phi}(0^+,t)\dot{\Phi}_L(0^-,t). 
\end{split}
\end{align}
Notice that only the term $-C_L \dot{\Phi}(0^+,t)\dot{\Phi}_L(0^-,t)$ is responsible for coupling of the resonator to the bath and the other two can be considered as a modification on top of the closed case. Applying the same method for the right capacitor, we can define the modified Lagrangian for the left and right baths
\begin{subequations}
\begin{align}
&\mathcal{L}_{R}^{op}=\int_{L^+}^{\infty}\,dx\left\{\frac{1}{2} c[\partial_t \Phi_R(x,t)]^2-\frac{1}{2l}[\partial_x \Phi_R(x,t)]^2\right\}+\frac{1}{2}C_R \dot{\Phi}_R^2(L^+,t),\\
&\mathcal{L}_{L}^{op}=\int_{-\infty}^{0^-}\,dx\left\{\frac{1}{2} c[\partial_t\Phi_L(x,t)]^2-\frac{1}{2l}[\partial_x \Phi_L(x,t)]^2\right\}+\frac{1}{2}C_L \dot{\Phi}_R^2(0^-,t).
\end{align}
\end{subequations}
Moreover, the interaction Lagrangian is found as
\begin{align}
\mathcal{L}_{C,LR}=-C_L \dot{\Phi}(0^+,t)\dot{\Phi}_L(0^-,t)-C_R \dot{\Phi}(L^-,t)\dot{\Phi}_R(L^+,t),
\end{align}
which is also equal to the interaction Hamiltonian since by going from Lagrangian to Hamiltonian capacitive contributions (kinetic contributions) do not change sign. The idea is to find the Hermition modes governed only by each of these uncoupled modified contributions and finally write the interaction in terms of Hermitian modes of each subsystem.

Up to here, we have not considered the effect of transmon on capacitance per length as we found in \ref{Eq:modified capacitance per length}. It is not necessary to go over the derivation again, since these two effects, modification due to opening and due to transmon, are completely independent. By considering the inhomogeneity introduced by the transmon we have
\begin{align}
\mathcal{L}_{C}^{op}=\int_{0^+}^{L^-}\,dx\left\{\frac{1}{2} c_{op}(x)[\partial_t \Phi(x,t)]^2-\frac{1}{2l}[\partial_x \Phi(x,t)]^2\right\},
\end{align}
where $c_{op}(x,x_0)$ is given as
\begin{align}
\begin{split}
c_{op}(x,x_0)&=c+C_s \delta(x-x_0)+ C_R\delta(x-L^-)+C_L \delta(x-0^+).
\end{split}
\end{align}

The new delta functions in $c_{op}(x)$ are only important at the boundaries, which can be found by integrating the equation along an infinitesimal interval that includes the delta functions. In order to find the modes, we need to solve
\begin{align}
\left(\partial_x^2+lc\omega_n^2\right)\tilde{\Phi}_n(x)=0 ,\qquad x\neq x_0, 
\end{align}
with boundary conditions given as
\begin{subequations}
\begin{align}
&\left.\partial_x\tilde{\Phi}_n(x)\right|_{x=L^-}=lC_R \omega_n^2\tilde{\Phi}_n(L^-),\\
&\left.\partial_x\tilde{\Phi}_n(x)\right|_{x=0^+}=-lC_L\omega_n^2\tilde{\Phi}_n(0^+),\\
&\left.\partial_x\tilde{\Phi}_n(x)\right]_{x_0^-}^{x_0^+} + lC_s\omega_n^2 \tilde{\Phi}_n(x_0)=0, \\
&\tilde{\Phi}_n(x_0^+)=\tilde{\Phi}_n(x_0^-).
\end{align}
\end{subequations}

Defining unitless parameters $\chi_{R,L}\equiv \frac{C_{R,L}}{cL}$ as we did for $\chi_s$, we find eigenfrequencies satisfy a transcendental equation as
\begin{align}
\begin{split}
&+\left[1-\chi_R\chi_L (k_nL)^2\right]\sin{(k_n L)}\\
&+\left(\chi_R+\chi_L\right)k_nL\cos{(k_n L)}\\
&+\chi_s k_n L\cos{(k_n x_0)}\cos{[k_n (L-x_0)]}\\
&-\chi_R\chi_s (k_n L)^2 \cos{(k_n x_0)}\sin{[k_n (L-x_0)]}\\
&-\chi_L\chi_s (k_n L)^2 \sin{(k_n x_0)}\cos{[k_n (L-x_0)]}\\
&+\chi_R\chi_L\chi_s (k_n L)^3 \sin{(k_n x_0)}\sin{[k_n (L-x_0)]}=0.
\label{Eq: Open CC Eigenfrequencies}
\end{split}
\end{align}
The real-space representation of these modes read
\begin{align}
\tilde{\Phi}_n(x)\propto
\begin{cases}
\tilde{\Phi}_n^{<}(x) \quad 0<x<x_0\\
\tilde{\Phi}_n^{>}(x) \quad x_0<x<L
\end{cases}
\end{align}
where $\tilde{\Phi}_n^{<}(x)$ and $\tilde{\Phi}_n^{>}(x)$ are found as
\begin{subequations}
\begin{align}
\begin{split}
\tilde{\Phi}_n^{<}(x)=\left\{\cos{[k_n(L-x_0)]}-\chi_Rk_nL\sin{[k_n(L-x_0)]}\right\}\left\{\cos{(k_n x)}-\chi_L k_nL\sin{(k_n x)}\right\},
\end{split}\\
\begin{split}
\tilde{\Phi}_n^{>}(x)=\left\{\cos{(k_n x_0)}-\chi_L k_nL\sin{(k_n x_0)}\right\}\left\{\cos{[k_n(L-x)]}-\chi_Rk_nL\sin{[k_n(L-x)]}\right\}.
\end{split}
\end{align}
\end{subequations}
The normalization constant will be set by the orthogonality conditions
\begin{align}
\int_{0}^{L} dx \frac{c_{op}(x,x_0)}{c}\tilde{\Phi}_m(x)\tilde{\Phi}_n(x)=L\delta_{mn}
\end{align}
\begin{align}
\begin{split}
&\int _{0}^{L}dx \partial_x\tilde{\Phi}_m(x)\partial_x\tilde{\Phi}_n(x)-\frac{1}{2}\left(k_m^2+k_n^2\right)L\left[\chi_R\tilde{\Phi}_m(L^-)\tilde{\Phi}_n(L^-)+\chi_L\tilde{\Phi}_m(0^+)\tilde{\Phi}_n(0^+)\right]\\
&=k_m k_n L \delta_{mn}
\end{split}
\end{align}
\section{Canonical quantization}
Following the same quantization procedure as for the closed case we can write the field operators in terms of the eigenmodes and eigenfrequencies of each part of the circuit as 
\begin{subequations}
\begin{align}
&\hat{\Phi}(x,t)=\sum\limits_n \left(\frac{\hbar}{2\omega_n cL}\right)^{\frac{1}{2}}\left[\hat{a}_n(t) + \hat{a}_n^{\dagger}(t)\right]\tilde{\Phi}_n(x),\\
&\hat{\Phi}_R(x,t)=\sum \limits_n \left(\frac{\hbar}{2\omega_{n,R} cL_R}\right)^{\frac{1}{2}} \left[\hat{b}_{n,R}(t) + \hat{b}_{n,R}^{\dagger}(t)\right]\tilde{\Phi}_{n,R}(x),\\
&\hat{\Phi}_L(x,t)=\sum \limits_n \left(\frac{\hbar}{2\omega_{n,L} cL_L}\right)^{\frac{1}{2}} \left[\hat{b}_{n,L}(t) + \hat{b}_{n,L}^{\dagger}(t)\right]\tilde{\Phi}_{n,L}(x).
\end{align}
\end{subequations}
where we have also considered some finite length for the left and right resonators as well to keep the normalization constants meaningful. Now, we can derive an expression for resonator-bath coupling in terms of modes of each part. Consider coupling to the left bath for the moment
\begin{subequations}
\begin{align}
&\hat{\mathcal{H}}_{CL}=-C_L \dot{\hat{\Phi}}(0^+,t)\dot{\hat{\Phi}}_L(0^-,t),\\
&\hat{\mathcal{H}}_{CR}=-C_R \dot{\hat{\Phi}}(L^+,t)\dot{\hat{\Phi}}_R(L^-,t).
\end{align}
\end{subequations}

Considering that the time-dynamics of annihiliation and creation operators up to here are only governed by the free Lagrangian of each part, we have
\begin{align}
\begin{split}
\dot{\hat{\Phi}}(x,t)&=\sum\limits_n \left(\frac{\hbar}{2\omega_n cL}\right)^{\frac{1}{2}}\left[-i\omega_n \hat{a}_n(t) +i\omega_n \hat{a}_n^{\dagger}(t)\right]\tilde{\Phi}_n(x)\\
&=-i\sum\limits_n \left[\frac{\hbar \omega_n}{2cL}\right]^{\frac{1}{2}}\left(\hat{a}_n(t) - \hat{a}_n^{\dagger}(t)\right)\tilde{\Phi}_n(x)
\end{split}
\end{align} 
And we have the same type of expression for the end resonators as well. The interaction Hamiltonian then reads
\begin{align}
\begin{split}
\hat{\mathcal{H}}_{C,LR}=&-\sum_{m,n}\hbar \beta_{mn,R}\left(\hat{a}_m-\hat{a}_m^{\dagger}\right)\left(\hat{b}_{n,R}-\hat{b}_{n,R}^{\dagger}\right)\\
&-\sum_{m,n}\hbar \beta_{mn,L}\left(\hat{a}_m-\hat{a}_m^{\dagger}\right)\left(\hat{b}_{n,L}-\hat{b}_{n,L}^{\dagger}\right),
\end{split}
\end{align}
where we find $\beta_{mn,R}$ and $\beta_{mn,L}$ as
\begin{subequations}
\begin{align}
&\beta_{mn,R}=\frac{C_{R}}{2c\sqrt{L}\sqrt{L_{R}}}\omega_m^{\frac{1}{2}}\omega_{n,R}^{\frac{1}{2}}\tilde{\Phi}_m(L^-)\tilde{\Phi}_{n,R}(L^+),\\
&\beta_{mn,L}=\frac{C_{L}}{2c\sqrt{L}\sqrt{L_{L}}}\omega_m^{\frac{1}{2}}\omega_{n,L}^{\frac{1}{2}}\tilde{\Phi}_m(0^+)\tilde{\Phi}_{n,L}(0^-).
\label{Eq:beta{L,R}}
\end{align}
\end{subequations}

The expression for $g_n$ is the same as in \ref{Eq:Expression for g- closed case} but with the new set of Hermitian modes satisfying the open-boundary conditions discussed before. The interaction Hamiltonian then is found as
\begin{align}
-\sum\limits_{m<n,l} \hbar g_{mnl}(\hat{P}_{mn}-\hat{P}_{nm}) (\hat{a}_l-\hat{a}_l^{\dagger})
\end{align}  

Gathering all different contributions together and moving to a new frame where $\hat{P}_{mn}\to i\hat{P}_{mn}$ for $m<n$, $ a_n \to i a_n $ and $b_{n,L/R} \to i b_{n,L/R} $, the Hamiltonian in its $2^{nd}$ quantized form reads 
\begin{align}
\begin{split}
\hat{\mathcal{H}}&=\underbrace{\sum\limits_n \hbar\Omega_n \hat{P}_{nn}}_{\hat{\mathcal{H}}_A}+\underbrace{\sum\limits_n \hbar\omega_n \hat{a}_n^{\dagger} \hat{a}_n}_{\hat{\mathcal{H}}_C} +\underbrace{\sum\limits_{n,S=\{L,R\}} \hbar\omega_{n,S} \hat{b}_{n,S}^{\dagger}\hat{b}_{n,S}}_{\hat{\mathcal{H}}_B}\\
&+\underbrace{\sum\limits_{m<n,l} \hbar g_{mnl} \left(\hat{P}_{mn}+\hat{P}_{nm}\right)\left(\hat{a}_l+\hat{a}^{\dagger}_l\right)}_{\hat{\mathcal{H}}_{int}}\\
&+\underbrace{\sum\limits_{m,n,S=\{L,R\}} \hbar\beta_{mn,S}\left(\hat{a}_m+\hat{a}_m^{\dagger}\right)\left(\hat{b}_{n,S}+ \hat{b}_{n,S}^{\dagger}\right)}_{\hat{\mathcal{H}}_{CB}}
\end{split}
\label{Eq:2nd Quantized Hamiltonian_Open Case_APP}
\end{align}

\chapter{TRK sum rules for a transmon qubit}
\label{App:TRKSumRules}
Here, we first find a general sum rule in quantum mechanics and then apply the results to calculate upper bounds for matrix elements of charge and flux operator i.e. $\bra{m}\hat{Q}_j\ket{n}$ and $\bra{m}\hat{\Phi}_j\ket{n}$ for the case of a transmon qubit. 

Assume a Hamiltonian $\hat{\mathcal{H}}$ where its eigenmodes and eigenenergies are known as $\left\{\ket{n},E_n | n\in \mathbb{N}^{0}\right\}$. Consider an arbitrary Hermitian operator $\hat{\mathcal{A}}=\hat{\mathcal{A}}^{\dagger}$ where we define successive commutation of $\hat{\mathcal{H}}$ and $\hat{\mathcal{A}}$ as
\begin{align}
\hat{\mathcal{C}}_{\hat{\mathcal{A}}}^{(k)}\equiv\left[
\hat{\mathcal{H}},\hat{\mathcal{C}}_{\hat{\mathcal{A}}}^{(k-1)}\right],\quad 
\hat{\mathcal{C}}_{\hat{\mathcal{A}}}^{(0)}\equiv \hat{\mathcal{A}}
\label{Eq:TRK-Def of C_A^(k)}
\end{align}
From the definition~(\ref{Eq:TRK-Def of C_A^(k)}) we find that for any two arbitrary eigenstates $\ket{m}$ and $\ket{n}$ we have 
\begin{align}
\begin{split}
\bra{m}\hat{\mathcal{C}}_{\hat{\mathcal{A}}}^{(k)}\ket{n}=&(E_m-E_n)\bra{m}\hat{\mathcal{C}}_{\hat{\mathcal{A}}}^{(k-1)}\ket{n}\\
&\vdots\\
=&\left(E_m-E_n\right)^k \bra{m}\hat{\mathcal{A}}\ket{n}.
\end{split}
\label{Eq:TRK-Identitiy}
\end{align}
Using the identity~(\ref{Eq:TRK-Identitiy}) we can write 
\begin{align}
\begin{split}
\bra{m}\left[\hat{\mathcal{A}},\hat{\mathcal{C}}_{\hat{\mathcal{A}}}^{(k)}\right]\ket{m}&=\bra{m}\hat{\mathcal{A}}\underbrace{\hat{\mathbf{1}}}_{\sum\limits_n \ket{n}\bra{n}}\hat{\mathcal{C}}_{\hat{\mathcal{A}}}^{(k)}\ket{m}\\
&-\bra{m}\hat{\mathcal{C}}_{\hat{\mathcal{A}}}^{(k)}\underbrace{\hat{\mathbf{1}}}_{\sum\limits_n \ket{n}\bra{n}}\hat{\mathcal{A}}\ket{m}\\
=\sum\limits_n \left(E_n-E_m \right)&\left[1-(-1)^k\right]\left|\bra{m}\hat{\mathcal{A}}\ket{n}\right|^2.
\end{split}
\label{Eq:General Sum Rule in QM}
\end{align}
In the following, we use the sum rule~(\ref{Eq:General Sum Rule in QM}) to obtain upper bounds for the transition matrix elements of a transmon qubit.

Consider the Hamiltonian for a transmon qubit 
\begin{align}
\hat{\mathcal{H}}=\frac{\hat{Q}_j^2}{2C_j}+U(\hat{\Phi}_j),
\end{align} 
where $U(\hat{\Phi}_j)=-E_j\cos{\left(\frac{2\pi}{\Phi_0}\hat{\Phi}_j\right)}$. Applying the result found in Eq.~(\ref{Eq:General Sum Rule in QM}), we can write
\begin{align}
\bra{0}[\hat{\Phi}_j,\underbrace{[\hat{\mathcal{H}},\hat{\Phi}_j]}_{\hat{\mathcal{C}}_{\hat{\Phi}}^{(1)}}]\ket{0}=\sum\limits_{n>0} 2\left(E_n-E_0 \right)\left|\bra{0}\hat{\Phi}_j\ket{n}\right|^2,
\end{align}
where $\ket{0}$ represents the ground state. The left hand side can be calculated explicitly as $\frac{\hbar^2}{C_j}$ which leads to the sum rule for $\hat{\Phi}_j$ as
\begin{align}
\sum\limits_{n>0} 2\left(E_n-E_0 \right)\left|\bra{0}\hat{\Phi}_j\ket{n}\right|^2=\frac{\hbar^2}{C_j}.
\label{Eq:Sum Rule for Phi_J}
\end{align}
Noticing that all terms on the left hand side of Eq.~(\ref{Eq:Sum Rule for Phi_J}) are positive, we can find an upper bound for $\Phi_{j,01}$ as
\begin{align}
|\Phi_{j,01}|^2<\frac{\hbar^2}{2C_j(E_1-E_0)}\approx \frac{E_c}{\sqrt{8E_jE_c}-E_c}\left(\frac{\hbar}{e}\right)^2,
\end{align}
where we have defined the charging energy $E_c\equiv\frac{e^2}{2C_j}$. In a similar manner, it is possible to obtain a sum rule for $\hat{Q}_j$ as
\begin{align}
\bra{0}[\hat{Q}_j,\underbrace{[\hat{\mathcal{H}},\hat{Q}_j]}_{\hat{\mathcal{C}}_{\hat{Q}}^{(1)}}]\ket{0}=\sum\limits_{n>0} 2\left(E_n-E_0 \right)\left|\bra{0}\hat{Q}_j\ket{n}\right|^2.
\end{align}
Again, the left hand side can be explicitly calculated as
\begin{align}
[\hat{Q}_j,[\hat{\mathcal{H}},\hat{Q}_j]]=\hbar^2\frac{\partial^2U(\hat{\Phi}_j)}{\partial\hat{\Phi}_j^2}=\left(\frac{2\pi\hbar}{\Phi_0}\right)^2 E_j\cos{\left(\frac{2\pi}{\Phi_0}\hat{\Phi}_j\right)}.
\end{align}
Since $\frac{2\pi\hbar}{\Phi_0}=2e$ brings us the sum rule for $\hat{Q}_j$ as
\begin{align}
\sum\limits_{n>0} 2\left(E_n-E_0 \right)\left|\bra{0}\hat{Q}_j\ket{n}\right|^2=(2e)^2E_j \bra{0}\cos{\left(\frac{2\pi}{\Phi_0}\hat{\Phi}_j\right)}\ket{0}<(2e)^2E_j.
\label{Eq:Sum Rule for Q_J _app}
\end{align}
Again, due to positivity of terms on the left hand side of Eq.~(\ref{Eq:Sum Rule for Q_J _app}) we obtain
\begin{align}
|Q_{j,01}|^2<\frac{2e^2 E_j}{E_1-E_0}\approx \frac{2E_j}{\sqrt{8E_jE_c}-E_c}e^2 .
\end{align}

\chapter{Finite size Transmon}
\label{App:FiniteSizeTrans}
In this appendix, we study how the results from change if we assume a finite length $d$ for the transmon. In such a case we assume that the coupling is not local and spreads over whole length of transmon with mutual capacitance per length $c_g$. Following the same discrete to continuous approach one finds the Lagrangian as
\begin{align}
\begin{split}
\mathcal{L}&=\frac{1}{2}(C_j+c_g d)\dot{\Phi}^2_j(t)-U_j(\Phi_j(t))\\
&+\int_{0}^{L}\,dx\ \left\{\frac{1}{2}[c+c_g\pi_d(x,x_0)]\left[\partial_t\Phi(x,t)\right]^2-\frac{1}{2l}\left[\partial_x \Phi(x,t)\right]^2\right\}\\
&-\int_{0}^{L}\,dx\ c_g\pi_d(x,x_0)\dot{\Phi}_j(t)\partial_t \Phi(x,t), 
\end{split}
\end{align}
where $\pi_d(x,x_0)$ is a unit window of width $d$ that is defined in terms of Heaviside function $\theta(x)$ as $\pi_d(x,x_0)\equiv \theta(x-x_0+d/2)-\theta(x-x_0-d/2)$. The Euler-Lagrange equations of motion then read
\begin{subequations}
\begin{align}
&(C_j+c_gd)\ddot{\Phi}_j(t)+\frac{\partial U_j(\Phi_j)}{\partial\Phi_j}=\int_{0}^{L} dx c_g \pi_d(x,x_0)\partial_t^2 \Phi(x,t),
\label{Eq:FiniteSizeTrans-ddot(Phij)}\\
&\left\{\partial_x^2-l\left[c+c_g\pi_d(x,x_0)\right]\partial_t^2\right\} \Phi(x,t)=-lc_g\pi_d(x,x_0)\ddot{\Phi}_j(t).
\label{Eq:FiniteSizeTrans-ddot(Phi(x,t))}
\end{align}
\end{subequations}
Comparing Eqs.~(\ref{Eq:FiniteSizeTrans-ddot(Phij)}-\ref{Eq:FiniteSizeTrans-ddot(Phi(x,t))}) to the Euler-Lagrange equations of motion~(\ref{Eq:Transmon's E.O.M}-\ref{Eq:Resonator's E.O.M}) drived earlier in App.~\ref{SubApp:A^2-ContLimit}, it is clear that the structure of the equations has remained the same while $C_g\delta(x-x_0)$ has been replaced with $c_g\pi(x,x_0)$ . The Hamiltonian can be found through the usual Legendre transformation as
\begin{align}
\begin{split}
\mathcal{H}&=\underbrace{\frac{Q_j^2}{2C_j^{mod}}-E_j\cos\left(2\pi\frac{\Phi_j}{\Phi_0}\right)}_{\mathcal{H}_A^{mod}}\\
&+\underbrace{\int_{0}^{L}dx \left\{\frac{\rho^2(x,t)}{2c_d(x,x_0)}+\frac{1}{2l}\left[\partial_x\Phi(x,t)\right]^2\right\}}_{\mathcal{H}_C^{mod}} \\
&+\underbrace{\frac{Q_j}{C_j^{mod}}\int_{0}^{L}dx c_g\pi_d(x,x_0)\frac{\rho(x,t)}{c_d(x,x_0)}}_{\mathcal{H}_{int}},
\end{split}
\end{align}
where $c_d(x,x_0)$ is the modified capacitance per length and reads
\begin{align}
c_d(x,x_0)=c+c_g\pi_d(x,x_0)\circledS \frac{C_j}{d}.
\end{align}
The $\circledS$-notation represents series combination of two capacitors. Importantly, we observe that when the dimension of transmon is taken into account, the modification is mutual and transmon's spectrum is also influenced by the coupling such that $C_j^{mod}$ reads
\begin{align}
\begin{split}
C_j^{mod}=C_j+\int_{0}^{L} dx \underbrace{\frac{cc_g\pi_d(x,x_0)}{c+c_g\pi_d(x,x_0)}}_{c_g\pi_d(x,x_0)\circledS c}=C_j+(c_g \circledS c)d.
\end{split}
\end{align}
The results above are general such that one can replace the rectangular window $c_g\pi_d(x,x_0)$ in capacitance per length by any smooth capacitance per length $c_g(x,x_0)$ and the form of $\mathcal{H}$, $c_d(x,x_0)$ and $C_j^{mod}$ remain the same. 

\chapter{Hamiltonian and modified eigenmodes of a closed cavity-QED system}
\label{App:DerivationOfHamCQED}
\begin{figure}
\centering
\centerline{\includegraphics[scale=0.40]{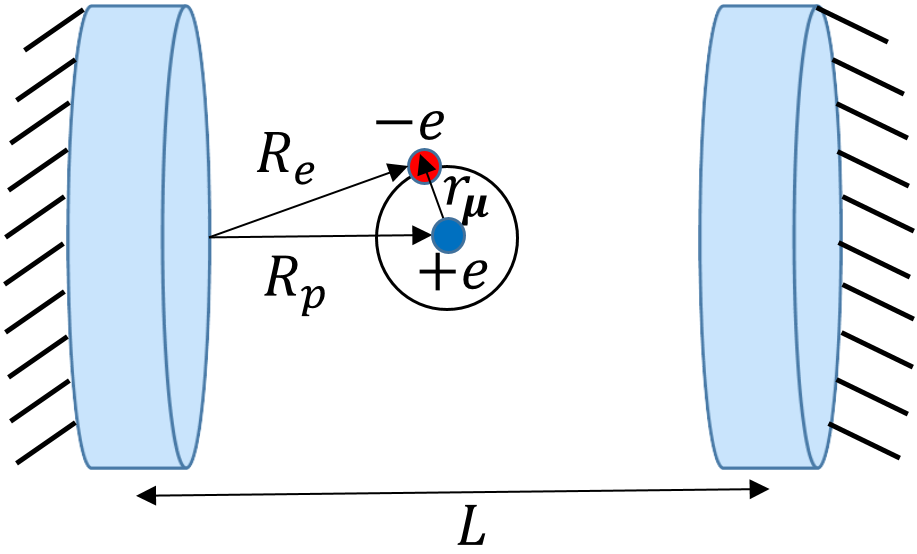}}
\caption{A single-electron atom interacting with the EM field inside a closed cavity of length $L$. } 
\label{Fig:DerOfCQEDHam-ClosedCQED}
\end{figure}

In this Appendix, first we derive the classical Hamiltonian for a general system containing finite number of point charges interacting with the EM field inside a closed cavity. This is achieved by expressing the Maxwell's and Newton's equations of motion in a Lagrangian formalism and then a Legendre transformation to find the Hamiltonian. This model is then reduced to describe a one-dimensional cavity shown in Fig. \ref{Fig:DerOfCQEDHam-ClosedCQED}. In order to emphasize on the resemblance to the cQED results we found earlier, it is assumed that there is only a single electron at  $\vec{R}_e(t)$, while all other electronic or nuclear degrees of freedom are frozen at $\vec{R}_{cm}$.

%%%%%%%%%%%%%% Fig: CQED Closed %%%%%%%%%%%%%
%%%%%%%%%%%%%%%%%%%%%%%%%%%%%%%%%%%%%%%%%%
\section{Classical Lagrangian}

Following the usual canonical quantization scheme, first we have to find the classical Lagranian. We already know the equations of motion for the EM fields to be the Maxwell's equations as
\begin{subequations}
\begin{align}
&\nabla.\vec{E}(\vec{r},t)=\frac{\rho(\vec{r},t)}{\epsilon_0},
\label{Eq:Maxwell Equations-Divergence of E}\\
&\nabla.\vec{B}(\vec{r},t)=0,
\label{Eq:Maxwell Equations-Divergence of B}\\
&\nabla\times\vec{E}(\vec{r},t)=-\partial_t\vec{B}(\vec{r},t),
\label{Eq:Maxwell Equations-Curl of E}\\
&\nabla\times\vec{B}(\vec{r},t)=\mu_0 \vec{J}(\vec{r},t)+\mu_0\epsilon_0\partial_t\vec{E}(\vec{r},t),
\label{Eq:Maxwell Equations-Curl of B}
\end{align}
\end{subequations}
where $\rho(\vec{r},t)$ and $\vec{J}(\vec{r},t)$ are scalar charge density and vector current density  and are given as 
\begin{subequations}
\begin{align}
&\rho(\vec{r},t)=\sum\limits_n q_n\delta^{(3)}\left(\vec{r}-\vec{r}_n(t)\right)
\label{Eq:Maxwell Eqautions-Charge density}\\
&\vec{J}(\vec{r},t)=\sum\limits_n q_n\dot{\vec{r}}_n(t)\delta^{(3)}\left(\vec{r}-\vec{r}_n(t)\right)
\label{Eq:Maxwell Eqautions-current density}
\end{align}
\end{subequations}
The remaining equation of motion is a Newton equation regarding the mechanical motion of the electron which reads
\begin{align}
m_n\ddot{\vec{r}}_n(t)&=\underbrace{q_n\left[\vec{E}(\vec{r}_n(t),t)+\dot{\vec{r}}_n(t)\times \vec{B}(\vec{r}_n(t),t)\right]}_{\rm Lorentz\quad Force}
\label{Eq:Newton's Law for Lorentz Force}
\end{align}

Based on Eqs.~(\ref{Eq:Maxwell Equations-Divergence of B}) and (\ref{Eq:Maxwell Equations-Curl of E}) we are able to express the physical fields $\vec{E}(\vec{r},t)$ and $\vec{B}(\vec{r},t)$ in terms of scalar potential $V(\vec{r},t)$ and vector potential $\vec{A}(\vec{r},t)$ up to a gauge degree of freedom as
\begin{subequations}
\begin{align}
&\vec{E}(\vec{r},t)=-\partial_t\vec{A}(\vec{r},t)-\nabla V(\vec{r},t) ),\\
&\vec{B}(\vec{r},t)=\nabla\times\vec{A}(\vec{r},t).
\end{align}
\end{subequations}

It is possible to write a Lagrangian that produces all previous equations of motion (\ref{Eq:Maxwell Equations-Divergence of E})-(\ref{Eq:Maxwell Equations-Curl of B}) and (\ref{Eq:Newton's Law for Lorentz Force}) as a result of the variational principle $\delta \mathcal{L}=0$. This Lagrangian reads 
\begin{align}
\begin{split}
\mathcal{L}&=\sum\limits_n\frac{1}{2}m_n\dot{\vec{r}}_n^2\\
&+\int d^3r\left[\frac{1}{2}\epsilon_0\left(\partial_t\vec{A}+\nabla V\right)^2-\frac{1}{2\mu_0}\left(\nabla \times \vec{A}\right)^2\right]\\
&+\int d^3r \left[\vec{J}.\vec{A}-\rho V\right].
\end{split}
\end{align}
In order to proceed further, we need to fix the gauge. Choosing to work in Coulomb gauge defined as $\nabla .\vec{A}=0$ and using Eq.~(\ref{Eq:Maxwell Equations-Divergence of E}) we find that the scalar potential $V(\vec{r},t)$ satisfies a Poisson equation as
\begin{align}
\nabla^2 V(\vec{r},t)=-\frac{\rho(\vec{r},t)}{\epsilon_0}.
\end{align}
Having the charge density in Eq.~(\ref{Eq:Maxwell Eqautions-Charge density}) we can solve the Poisson equation to obtain
\begin{align}
V(\vec{r},t)=\sum\limits_n \frac{q_n}{4\pi\epsilon_0 |\vec{r}-\vec{r}_n(t)|}.
\label{Eq:solution for V(r,t)}
\end{align}

Furthermore, this choice of gauge helps to simplify the Lagrangian since due to the divergence theorem
\begin{align}
\begin{split}
\int d^3r \partial_t \vec{A}.\nabla V&=\int d^3r \nabla.(V\partial_t \vec{A})-\int d^3r V\partial_t \underbrace{(\nabla. \vec{A})}_{0}=\oint d\vec{S}. \left(V\partial_t\vec{A}\right),
\end{split}
\end{align}
which means this term only contributes at the boundaries. Boundary terms do not affect equations of motion inside the cavity, however their existence are necessary to ensure the correct boundary conditions, i.e. continuity of parallel electric field and perpendicular magnetic field at the interface of cavity with the outside environment. As far as we fix these conditions properly, we can remove all surface terms form the Lagrangian. In a similar manner 
\begin{align}
\begin{split}
\int d^3 r \frac{1}{2}\epsilon_0(\nabla V)^2&=\frac{1}{2}\epsilon_0\oint d\vec{S}.\left(V\nabla V\right)-\int d^3 r\frac{1}{2}\epsilon_0 V\nabla^2 V\\
&=\frac{1}{2}\epsilon_0\oint d\vec{S}.\left(V\nabla V\right)+\int d^3r \frac{1}{2}\rho V.
\end{split}
\end{align}
Finally, by putting everything together and neglecting surface terms, we find the simplified Lagrangian as
\begin{align}
\begin{split}
\mathcal{L}&=\sum\limits_n\frac{1}{2}m_n\dot{\vec{r}}_n^2-\int d^3r \, \frac{1}{2}\rho V\\
&+\int d^3r\left[\frac{1}{2}\epsilon_0\left(\partial_t\vec{A}\right)^2-\frac{1}{2\mu_0}\left(\nabla \times \vec{A}\right)^2\right]\\
&+\int d^3r \, \vec{J}.\vec{A}.\\
\end{split}
\end{align}
\section{Classical Hamiltonian}

The first step is find the conjugate momenta as
\begin{align}
&\vec{p}_n \equiv \frac{\partial \mathcal{L}}{\partial{\dot{\vec{r}}_n}}=m_n\dot{\vec{r}}_n+q_n\vec{A}(\vec{r}_n(t),t),
\label{Eq: conjugate momentum of r(t)}\\
&\vec{\Pi}(\vec{r},t)\equiv \frac{\partial \mathcal{L}}{\partial{\dot{\vec{A}}}}= \epsilon_0 \partial_t \vec{A}(\vec{r},t).
\label{Eq: conjugate momentum of A(r,t)}
\end{align}
Then, the Hamiltonian is calculated via a Legendre transformation of the Lagrangian as
\begin{align}
\mathcal{H}=\sum\limits_n \vec{p}_n.\dot{\vec{r}}_n+\int d^3r \vec{\Pi}(\vec{r},t).\partial_t \vec{A}(\vec{r},t)-\mathcal{L}.
\end{align}
Substituting Eqs.~(\ref{Eq: conjugate momentum of r(t)}) and~(\ref{Eq: conjugate momentum of A(r,t)}) into the expression for Hamiltonian we find
\begin{align}
\begin{split}
\mathcal{H}&=\sum\limits_n \frac{1}{2}m\dot{\vec{r}}_n^2+\sum\limits_n \frac{1}{2}q_n V(\vec{r}_n)\\
&+\int d^3 r\left\{\frac{1}{2}\epsilon_0[\partial_t \vec{A}(r,t)]^2+\frac{1}{2 \mu_0}\left[\vec{\nabla}\times\vec{A}(\vec{r},t)\right]^2\right\}.
\end{split}
\end{align}
By replacing $\partial_t \vec{A}(\vec{r},t)$ and $\dot{\vec{r}}_n(t)$ in terms of conjugate momenta $\vec{\Pi}(\vec{r},t)$ and $\vec{p}_n(t)$ respectively, the Hamiltonian can be rewritten as
\begin{align}
\begin{split}
\mathcal{H}= \sum\limits_n\frac{\left[\vec{p}_n-q_n\vec{A}(\vec{r}_n,t)\right]^2}{2m_n}+\sum\limits_n \frac{1}{2}q_n V(\vec{r}_n)+\int d^3 r \left\{\frac{\vec{\Pi}^2(\vec{r},t)}{2\epsilon_0}+\frac{\left[\nabla \times \vec{A}(\vec{r},t)\right]^2}{2\mu_0}\right\},
\end{split}
\end{align}
which can be written in a more instructive way as
\begin{align}
\begin{split}
\mathcal{H}&=\underbrace{\sum\limits_n \frac{\vec{p}_n^2}{2m_n}+\sum\limits_n\frac{1}{2}q_n V(\vec{r}_n)}_{\mathcal{H}_A}\\
&+\underbrace{\int d^3 r \left\{\frac{\vec{\Pi}^2(\vec{r},t)}{2\epsilon_0}+\frac{\left[\nabla \times \vec{A}(\vec{r},t)\right]^2}{2\mu_0}\right\}}_{\mathcal{H}_C}\\
&+\underbrace{\int d^3r \sum\limits_n\frac{q_n^2}{2m_n}\vec{A}^2(\vec{r},t)\delta^{(3)}(\vec{r}-\vec{r}_n)}_{\mathcal{H}_C^{mod}}\\
&\underbrace{-\sum\limits_n \frac{q_n}{m_n}\vec{p}_n.\vec{A}(\vec{r}_n,t)}_{\mathcal{H}_{int}}.
\end{split}
\label{Eq:closed cavity-QED Hamiltonian}
\end{align}

Equation~(\ref{Eq:closed cavity-QED Hamiltonian}) is the most general form of the classical Hamiltonian of a finite number of charges interacting with the electromagnetic field inside a closed cavity. In what follows, we make a few assumptions to reduce this model for the system shown in Fig. \ref{Fig:DerOfCQEDHam-ClosedCQED}. First of all, we assume that the wavelength of EM field is much larger than atomic scale $\vec{r}_e$ such that we can apply zero-order dipole approximation $\vec{A}(\vec{R}_e(t),t)\approx \vec{A}(\vec{R}_0(t),t)\approx \vec{A}(\vec{R}_{cm},t)$ in both $\mathcal{H}_C^{mod}$ and $\mathcal{H}_{int}$ where in the last step $\vec{R}_{cm}$ is the center of mass of the electron and the nucleus. By rewriting the Hamiltonian in terms of new coordinates 
\begin{align}
\vec{r}_{\mu}\equiv \vec{R}_e-\vec{R}_p, \quad \vec{R}_{cm}\equiv\frac{m_e\vec{R}_e+m_p\vec{R}_p}{m_e+m_p}, 
\end{align}
and new momentum
\begin{align}
\vec{p}_{\mu} \equiv \frac{m_p\vec{P}_e-m_e\vec{P}_p}{m_e+m_p}, \quad \vec{P}_{cm}\equiv\vec{P}_e+\vec{P}_p 
\end{align}
and neglecting the center of mass kinetic energy we find (for a more detailed discussion see chapter 14 of \cite{Schleich_Quantum_2011}) 
\begin{align}
\begin{split}
\mathcal{H}&\approx\underbrace{\frac{\vec{p}_{\mu}^2}{2m_{\mu}}-eV(\vec{r}_{\mu})}_{\mathcal{H}_A}\\
&+\underbrace{\int d^3 r \left\{\frac{\vec{\Pi}^2(\vec{r},t)}{2\epsilon_0}+\frac{\left[\nabla \times \vec{A}(\vec{r},t)\right]^2}{2\mu_0}\right\}}_{\mathcal{H}_C}\\
&+\underbrace{\int d^3r \frac{e^2}{2m_{\mu}}\vec{A}^2(\vec{r},t)\delta^{(3)}(\vec{r}-\vec{R}_{cm})}_{\mathcal{H}_C^{mod}}\\
&\underbrace{-\frac{e}{m_{\mu}}\vec{p}_{\mu}.\vec{A}(\vec{R}_{cm},t)}_{\mathcal{H}_{int}},
\end{split}
\label{Eq:closed cavity-QED Hamiltonian}
\end{align}
where $m_{\mu}\equiv\frac{m_em_p}{m_e+m_P}$ is the reduced mass.
\section{Modified cavity eigenmodes and eigenfrequencies}

Having found the modification introduced by the $\vec{A}^2$ term in the previous section, we can now calculatae the effect it has on the structure of the modes. Specifically, we are after eigenmodes of the modified Hamiltonian for the cavity given as $\mathcal{H}_C^{mod}\equiv\mathcal{H}_C+\mathcal{H}^{mod}$. This can be done by finding the Hamilton equation of motion for the conjugate fields as
\begin{align}
&\partial_t \vec{A}(\vec{r},t)=\frac{1}{\epsilon_0}\vec{\Pi}(\vec{r},t),
\label{Eq: time-evolution of A(x,t)}\\
&\partial_t\vec{\Pi}(\vec{r},t)=-\frac{1}{\mu_0}\nabla \times [\nabla \times \vec{A}(\vec{r},t)]-\frac{e^2}{m_{\mu}}\vec{A}(\vec{r},t)\delta^{(3)}(\vec{r}-\vec{R}_{cm}).
\label{Eq: time-evolution of Pi(x,t)}
\end{align}
Combining these two equations and applying the gauge condition $\nabla .\vec{A}=0$ we find
\begin{align}
\left(\nabla^2 -\mu_0\epsilon_0 \partial_t^2\right)\vec{A}(\vec{r},t)=\frac{\mu_0 e^2}{m_{\mu}}\vec{A}(\vec{r},t)\delta^{(3)}(\vec{r}-\vec{R}_{cm}),
\end{align}
which is a wave equation with an extra term on the right hand side due to $\vec{A}^2$ modification. The coefficient $\frac{\mu_0 e^2}{m_{\mu}}$ can be expressed in terms of fine structure constant $\alpha$ and Bohr's radius $a_0$ as $4\pi\alpha^2 a_0$. Performing a Fourier transform 
\begin{align}
\tilde{\vec{A}}(\vec{r},t)=\frac{1}{2\pi}\int_{-\infty}^{+\infty} d\omega \tilde{\vec{A}}(\vec{r},\omega)e^{-i\omega t},
\end{align}
we can separate the time and spatial dependences to obtain
\begin{align}
&\left[\nabla^2+\left(\frac{\omega}{c}\right)^2-4\pi\alpha^2a_0\delta^{(3)}(\vec{r}-\vec{R}_{cm})\right]\tilde{\vec{A}}(\vec{r},\omega)=0
\label{Eq:seperation of variables A(r,omega)}
\end{align}

Assuming a closed cavity case and remembering that $\vec{E}_{\parallel}$ and $\vec{B}_{\bot}$ are continuous across the cavity, the boundary conditions read
\begin{subequations}
\begin{align}
&\vec{n}_{\parallel}\times \left.\left(- i\omega\tilde{\vec{A}}(\vec{r},\omega)+\nabla \tilde{V}\right)\right|_{B}=0,
\label{Eq:continuity of parallel Electric field}\\
&\vec{n}_{\bot}. \left.\left(\nabla \times \tilde{\vec{A}}(\vec{r},\omega)\right)\right|_{B}=0.
\label{Eq:continuity of perpendicular Electric field}
\end{align}
\end{subequations}
where $\vec{n}_{\bot}$ and $\vec{n}_{\parallel}$ represent perpendicular and parallel unit vectors on the boundaries of the cavity and $\tilde{V}$ is the time-Fourier transform of the scalar potential. Equation~(\ref{Eq:seperation of variables A(r,omega)}) with the boundary conditions above provide a discrete set of modes due to finite volume of the cavity. For notation simplicity, we label the eigenfrequencies as $\omega_{\lambda}$ and the modes as $\tilde{\vec{A}}_{\lambda}(\vec{r})\equiv\tilde{\vec{A}}(\vec{r},\omega_{\lambda})$ , while in reality $\lambda $ denotes multiple sets of discrete numbers each for a separate dimension of the cavity. These modes satisfy the general orthogonality relation
\begin{align}
&\int d^3 r \tilde{\vec{A}}_{\lambda}(\vec{r}).\tilde{\vec{A}}_{\lambda'}(\vec{r})=\mathcal{V}\delta_{\lambda\lambda'},
\end{align}
where we have set the normalization such that the modes are dimensionless. Another orthogonality relation can be found in terms of $\nabla\tilde{\vec{A}}_{\lambda}(\vec{r})$ as
\begin{align}
\begin{split}
&+\int d^3 r \nabla\tilde{\vec{A}}_{\lambda}(\vec{r}).\nabla\tilde{\vec{A}}_{\lambda'}(\vec{r})\\
-&\frac{1}{2}\oint d\vec{S}.\left[\tilde{\vec{A}}_{\lambda}(\vec{r}).\nabla\tilde{\vec{A}}_{\lambda'}(\vec{r})+\tilde{\vec{A}}_{\lambda'}(\vec{r}).\nabla\tilde{\vec{A}}_{\lambda}(\vec{r})\right]\\
&+4\pi\alpha^2 a_0\tilde{\vec{A}}_{\lambda}(\vec{R}_{cm}).\tilde{\vec{A}}_{\lambda'}(\vec{R}_{cm})=k_{\lambda}k_{\lambda'}\mathcal{V}\delta_{\lambda\lambda'}.
\end{split}
\end{align}

Up to this point, we have considered the mode structure for a general cavity with any arbitrary geometry. In order to demonstrate the connection to the results for a one dimensional cQED system, we have to make a few assumptions about the geometry of the cavity.  We assume that the cavity's length is much larger than the diameter of its cross section, i.e. $L\gg\sqrt{S}$. By considering variation of the eigenmodes only along this dimension we can write $\vec{A}(\vec{r},t)=\vec{u}_z A(x,t)$ and thus $\vec{B}(\vec{r},t)=-\vec{u}_y \partial_x A(x,t)$. The Hamiltonian is then reduced to
\begin{align}
\begin{split}
\mathcal{H}&=\underbrace{\frac{\vec{p}_{\mu}^2}{2m_{\mu}}-eV(\vec{r}_{\mu})}_{\mathcal{H}_A}\\
&+\underbrace{\int d^2s \int_{0}^{L} dx \left\{\frac{\Pi^2(x,t)}{2\epsilon_0}+\frac{\left[\partial_x A(x,t)\right]^2}{2\mu_0}\right\}}_{\mathcal{H}_C}\\
&+\underbrace{\int d^2s \int_{0}^{L} dx \frac{e^2}{2m_{\mu}}A^2(x_{cm},t)\delta^2(\vec{s}-\vec{s}_{cm})\delta(x-x_{cm})}_{\mathcal{H}^{mod}}\\
&\underbrace{-\frac{e}{m_{\mu}}p_{\mu}^z A(x_{cm},t)}_{\mathcal{H}_{int}}.
\end{split}
\end{align}

Following the same procedure, we can find a modified wave equation as a result of $\hat{\mathcal{H}}_C^{mod}$  
\begin{align}
\left[\frac{d^2}{dx^2}+\left(\frac{\omega}{c}\right)^2-\frac{4\pi\alpha^2a_0}{S}\delta(x-x_{cm})\right]\tilde{A}(x,\omega)=0
\label{Eq:seperation of variables A(x,omega)}
\end{align}
Assuming that the atom is fixed at point $x_{cm}$, $\nabla \tilde{V}$ only affects the bouundary condtion for the zero frequency mode which we are not interested in. Therefore, by applying the boundary conditions
\begin{align}
\left. \tilde{A}(x,\omega)\right|_{x=0,L}=0.
\end{align}
we find the normalized eigenfrequencies to satisfy a transcendental equation as
\begin{align}
\sin{(k_nL)}+\chi_c \frac{\sin{(k_nx_{cm})}\sin{[k_n(L-x_{cm})]}}{k_nL}=0,
\label{Eq: Closed Cavity Eigenfrequencies}
\end{align}
where we have defined the unitless parameter $\chi_c \equiv 4\pi\alpha^2 a_0 \frac{L}{S}$. 
The real-space representation of the eigenmodes read
\begin{align}
\tilde{A}_n(x)\propto
\begin{cases}
\sin{\left[k_n(L-x_{cm})\right]}\sin{(k_n x)}&0<x<x_{cm}\\
\sin{(k_n x_{cm})}\sin{[k_n(L-x)]} &x_{cm}<x<L
\end{cases}.
\label{Eq: Closed Cavity Eigenmodes}
\end{align}
Furthermore, it can be shown that these eigenmodes satisfy the orthogonality relations:
\begin{subequations}
\begin{align}
&\int_{0}^{L}dx \tilde{A}_m(x)\tilde{A}_n(x)=L\delta_{mn},
\label{Eq:Orthogonality condition for A _ app}\\
&\int_{0}^{L} dx \partial_x \tilde{A}_m\partial_x \tilde{A}_n+\frac{\chi_c}{L}\tilde{A}_m(x_{cm})\tilde{A}_n(x_{cm})=k_mk_nL\delta_{mn}.
\label{Eq:Orthogonality condition for d/dx A _app}
\end{align}
\end{subequations}

Note that only the ratio $\frac{L}{S}$ is determined by the geometry of the cavity, while the pre-factor $4\pi\alpha^2 a_0 \approx 3.54\times 10^{-14} m$ is a universal length scale. This implies that the modification is only visible when $\frac{S}{L}$ is around the same order as $4\pi\alpha^2 a_0$. 
%For instance, consider the optical regime where $400 nm \lessapprox \lambda_{opt}\lessapprox 700 nm$ . Roughly speaking, we expect the cavity $L$ to be of the same order as wavelength to produce a reasonably low mode number, which implies that the diameter of the cross section $d \propto \sqrt{S}=\sqrt{4\pi\alpha^2 a_0 L}$ should be approximately $10^{-10} m$! 
\section{Canonical quantization}

Now that we have the proper set of eigenmodes and eigenfrequencies that diagonalizes the classical Hamiltonian for a one-dimensional closed cavity-QED system, we can move forward and extend the classical variables into quantum operators by introducing the necessary commutation relation between conjugate pairs. Let's consider the conjugate fields for the cavity first. We can expand these fields in terms of the proper eigenmodes as 
\begin{subequations}
\begin{align}
\hat{A}(x,t)=\sum\limits_n \left(\frac{\hbar}{2\omega_n \epsilon_0 SL}\right)^{\frac{1}{2}}\left(\hat{a}_n+\hat{a}_n^{\dagger}\right)\tilde{A}_n(x),\\
\hat{\Pi}(x,t)=-i\sum\limits_n \left(\frac{\hbar \epsilon_0 \omega_n}{2SL}\right)^{\frac{1}{2}}\left(\hat{a}_n-\hat{a}_n^{\dagger}\right)\tilde{A}_n(x).
\end{align}
\end{subequations}
where $\hat{a}_n$  and $\hat{a}_n^{\dagger}$ are annihilation and creation operators for each mode. Inserting the above equations and using the orthogonality conditions~(\ref{Eq:Orthogonality condition for A _ app}) and~(\ref{Eq:Orthogonality condition for d/dx A _app}), $\hat{\mathcal{H}}_C^{mod}=\hat{\mathcal{H}}_C+\hat{\mathcal{H}}^{mod}$ becomes diagonal as
\begin{align}
\hat{\mathcal{H}}_C^{mod}=\sum\limits_n \frac{\hbar \omega_n}{2}\left(a_n^{\dagger}a_n +a_n a_n^{\dagger}\right)=\sum\limits_n \hbar \omega_n a_n^{\dagger}a_n+ \text{const}.
\end{align}

The next step is to obtain the spectrum of $\hat{\mathcal{H}}_A$ by solving a Schrodinger equation in real-space basis as
\begin{align}
\left[-\frac{\hbar^2}{2m_{\mu}}\nabla_{\mu}^2-e V(\vec{r}_{\mu})\right]\Psi_n(\vec{r}_{\mu})=\hbar\Omega_n \Psi_n(\vec{r}_{\mu}),
\end{align}
where we have denoted the eigenmodes and eigenenergies by $\{\Psi_n(\vec{r}_{\mu}),E_n=\hbar \Omega_n | n\in \mathbb{N}^{0}\}$. $\hat{\mathcal{H}}_A$ can be decomposed as
\begin{align}
\hat{\mathcal{H}}_A=\sum\limits_n \hbar\Omega_n \hat{P}_{nn}. 
\end{align} 
$\vec{p}_{\mu}$ also has a spectral decomposition over this basis. Since the Coulomb potential $V(\vec{r}_{\mu})$ is an even function of $\vec{r}_{\mu}$, as we explained in the case of charge qubit only diagonal matrix elements of $\vec{p}_{\mu}$ are nonzero and we can write
\begin{align}
\begin{split}
\hat{\vec{p}}_{\mu} =\sum\limits_{m\neq n} \bra{m}\vec{p}_{\mu}\ket{n}\hat{P}_{mn},
\end{split}
\end{align}
where matrix elements $\vec{p}_{\mu,mn}$ can be calculated as
\begin{align}
\bra{m}\vec{p}_{\mu}\ket{n}=\int d^3 \vec{r}_{\mu} \Psi_m(\vec{r}_{\mu})\frac{\hbar}{i} \vec{\nabla} \Psi_n(\vec{r}_{\mu}).
\end{align}
Working in a basis where $\Psi_{n}(\vec{r}_{\mu})$ are real functions, the dipole matrix elements are purely imaginary which allows us to write 
\begin{align}
\begin{split}
\hat{\vec{p}}_{\mu} =\sum\limits_{m>n} \bra{m}\vec{p}_{\mu}\ket{n}(\hat{P}_{mn}-\hat{P}_{nm}).
\end{split}
\end{align}

At last, for sake of resemblance to the cQED results, we move to a new frame $\hat{P}_{mn} \to i \hat{P}_{mn}$ for $m<n$ to obtain the Hamiltonian as
\begin{align}
\hat{\mathcal{H}}=\underbrace{\sum\limits_n \hbar\Omega_n \hat{P}_{nn}}_{\hat{\mathcal{H}}_A}+\underbrace{\sum\limits_n \hbar \omega_n \hat{a}_n^{\dagger} \hat{a}_n}_{\hat{\mathcal{H}}_C^{mod}}+\underbrace{\sum\limits_{m>n,l} \hbar g_{mnl} \left(\hat{P}_{mn}+\hat{P}_{nm}\right)\left(\hat{a}_l+\hat{a}^{\dagger}_l\right)}_{\hat{\mathcal{H}}_{int}},
\end{align}
where the coupling strength $g_{mnl}$ reads
\begin{align}
\hbar g_{mnl}=\frac{e}{m_{\mu}}(i\vec{p}_{e,mn}.\vec{u}_z)\left(\frac{\hbar}{2\epsilon_0\omega_l SL}\right)^{\frac{1}{2}}\tilde{A}_l(x_{cm})
\end{align}

\chapter{Asymptotic behavior of light-matter coupling $g_n$}
\label{App:AsymptoticOfgn}
In this section we find the asymptotic behavior of the eigenfrequencies $\omega_n$ and eigenmodes $\tilde{\varphi}_n(x)$ of the resonator discussed in the main text. This provides an analytical understanding of the high frequency suppression in the light-matter coupling $g_n\propto \sqrt{\omega_n}\tilde{\varphi}_n(x_0)$. 

To point out the origin of the suppression that arise from a nonzero $\chi_s$, let us consider the closed resonator ($\chi_{R,L}=0$) case. Consider the special case of $x_0=0^+$ first. This is of experimental interest in order to achieve the maximum coupling to all modes of a resonator. Then, the transcendental Eq.~(\ref{Eq:Hermitian Eigenfrequencies}) simplifies to
\begin{align}
\sin(\omega_n)+\chi_s\omega_n\cos(\omega_n)=0,
\label{Eq:HermEigFreq-x0=0}
\end{align}
which can be rewritten as 
\begin{align}
\tan(\omega_n)=-\chi_s\omega_n.
\label{Eq:HermEigFreq-x0=0-Simplified}
\end{align}
The large $\omega_n$ solution for $\chi_s\neq0$ is then obtained 
\begin{align}
\lim_{n\to\infty}\omega_n=n\pi-\frac{\pi}{2},
\end{align} 
which is independent of the value for $\chi_s$. This implies that the effect of a nonzero $\chi_s$ on $\omega_n$ is a total shift $\pi/2$ (half of the free spectral range) in comparison with the case $\chi_s=0$. Substituting $x_0=0^+$ in Eq.~(\ref{Eq:Sol of Phi_n(x)-closed}), the normalization factor $\mathcal{N}_n$ is found via Eq.~(\ref{Eq:Closed Orthogonality Condition}) as
\begin{align}
\int_0^{1}dx\cos^2[\omega_n(1-x)]+\chi_s\cos^2(\omega_n)=\frac{1}{\mathcal{N}_n^2},
\end{align}
which gives
\begin{align}
\mathcal{N}_n=\frac{\sqrt{2}}{\sqrt{1+\chi_s\cos^2(\omega_n)}}.
\label{Eq:Normalization_x0=0}
\end{align}
Therefore the eigenmode is found as
\begin{align}
\tilde{\varphi}_n(x_0=0^+)=\frac{\sqrt{2}\cos(\omega_n)}{\sqrt{1+\chi_s\cos^2(\omega_n)}}.
\label{Eq:HermEigMode@x0-x0=0}
\end{align}
Using the trigonometric identity 
\begin{align}
\cos^2(\omega_n)=\frac{1}{1+\tan^2(\omega_n)}
\end{align}
and Eq.~(\ref{Eq:HermEigFreq-x0=0-Simplified}) we can rewrite Eq.~(\ref{Eq:HermEigMode@x0-x0=0}) as
\begin{align}
\tilde{\varphi}_n(x_0=0^+)=\frac{\sqrt{2}}{\sqrt{1+\chi_s+\chi_s^2\omega_n^2}},
\label{Eq:HermEigMode@x0-x0=0-Simplified}
\end{align}
which now provides the algebraic dependence of $\tilde{\varphi}_n(x_0)$ on $\omega_n$. According to Eq.~(\ref{Eq:HermEigMode@x0-x0=0-Simplified}), for large enough $\omega_n$ ($\chi_s\omega_n \gg 1+\chi_s$), we find
\begin{align}
\tilde{\varphi}_n(x_0)\thicksim\frac{1}{\omega_n},
\label{Eq:AsymptOfPhi_n}
\end{align}
where the symbol $\thicksim$ represents asymptotic equivalence. This imposes a natural cut-off on the light matter coupling for $n\to\infty$, since
\begin{align}
g_n\propto \sqrt{\omega_n}\tilde{\varphi}_n(x_0)\thicksim \frac{1}{\sqrt{\omega_n}}.
\label{Eq:AsymptOfg_n}
\end{align}

Next, we would like to find the asymptotic behavior of $\omega_n$ and $\tilde{\varphi}_n(x_0)$ for a general $x_0$. In order to bring Eq.~(\ref{Eq:Hermitian Eigenfrequencies}) into a similar form to Eq.~(\ref{Eq:HermEigFreq-x0=0-Simplified}), we first replace $\sin(\omega_n)=\sin[\omega_nx_0+\omega_n(1-x_0)]$ and then divide by $\cos(\omega_n x_0)\cos[\omega_n (1-x_0)]$ to obtain
\begin{align}
\tan(\omega_n x_0)+\tan[\omega_n (1-x_0)]=-\chi_s\omega_n.
\label{Eq:HermEigFreq-Simplified}
\end{align}
Next, the normalization factor $\mathcal{N}_n$ is found from Eq.~(\ref{Eq:Closed Orthogonality Condition}) as
\begin{align}
\mathcal{N}_n=\frac{\sqrt{2}}{\sqrt{x_0\cos^2[\omega_n(1-x_0)]+(1-x_0)\cos^2(\omega_n x_0)+\chi_s\cos^2[\omega_n(1-x_0)]\cos^2(\omega_nx_0)}},
\label{Eq:Normalization-Simplified}
\end{align}
Plugging this into Eq.~(\ref{Eq:Sol of Phi_n(x)-closed}) we find
\begin{align}
\tilde{\varphi}_n(x_0)=\frac{\sqrt{2}}{\sqrt{1+\chi_s+x_0\tan^2(\omega_nx_0)+(1-x_0)\tan^2[\omega_n(1-x_0)]}}
\label{Eq:HermEigMode@x0-Simplified}
\end{align}
Equations~(\ref{Eq:HermEigFreq-Simplified}) and (\ref{Eq:HermEigMode@x0-Simplified}) provide the asymptotic behavior of $\omega_n$, $\tilde{\varphi}_n(x_0)$ and $g_n$ for a general $x_0$. 
\chapter{A proof for convergence of characteristic function $D_j(s)$}
\label{App:ConvOfDj}
Here, we derive the expression for the characteristic function $D_j(s)$ and compare its convergence in number of resonator modes with and without the modification we found for $g_n$. 

The unitless classical Hamiltonian for the cQED system is found as
\begin{align}
\begin{split}
\mathcal{H}_{sys}&=4\mathcal{E}_c n_j^2(t)-\mathcal{E}_j\cos{[\varphi_j(t)]}\\
&+\int_{0}^{1}dx \left\{\frac{n^2(x,t)}{2\chi(x,x_0)}+\frac{1}{2}\left[\partial_x \varphi(x,t)\right]^2\right\}\\
&+2\pi\gamma z n_j(t)\int_{0}^{1}dx\frac{n(x,t)}{\chi(x,x_0)}\delta(x-x_0),
\end{split}
\label{eqn:model-closed cQED H}
\end{align}
where $z\equiv Z/R_Q$ where $Z\equiv\sqrt{l/c}$ is the characteristic impedance of the resonator and $R_Q\equiv h/(2e)^2$ is the superconducting resistance quantum. The modification in capacitance per length originates from the system Lagrangian that contains the gauge-invariant qubit-resonator coupling $\chi_g[\dot{\varphi}_j(t)-\dot{\varphi}(x_0,t)]^2/2$. In contrast, a phenomenological product coupling $\chi_g\dot{\varphi}_j(t)\dot{\varphi}(x_0,t)$ would yield a $\mathcal{H}_{sys}$ with $\chi_s=0$ which results in bare resonator modes.

For the purpose of quantizing $\mathcal{H}_{sys}$, we find the spectrum of the resonator by solving the corresponding Helmholtz eigenvalue problem that has been discussed in Sec.~(\ref{SubApp:Spec Rep of G-closed}). We find the second quantized Hamiltonian as

\begin{align}
\hat{\mathcal{H}}_{\text{sys}} \equiv \frac{\omega_j}{4}\left\{\hat{\mathcal{Y}}_j^2-\frac{\sqrt{2}}{\epsilon}\cos\left[(2\epsilon^2)^{1/4}\hat{\mathcal{X}}_j\right]\right\}+\sum\limits_{n}\left\{\frac{\omega_n}{4}\left[\hat{\mathcal{X}}_n^2+\hat{\mathcal{Y}}_n^2\right]+g_n\hat{\mathcal{Y}}_j\hat{\mathcal{Y}}_n\right\},
\label{Eq:2nd quantized closed H}
\end{align}
where have defined the canonically conjugate variables $\hat{\mathcal{X}}_l\equiv(\hat{a}_l+\hat{a}_l^{\dag})$ and $\hat{\mathcal{Y}}_l\equiv -i(\hat{a}_l-\hat{a}_l^{\dag})$, where $\hat{a}_{l}$ represent the boson annihilation operator of sector $l\equiv j,c$. Moreover, $\omega_j\equiv \sqrt{8\mathcal{E}_j\mathcal{E}_c}$ and $\epsilon\equiv\sqrt{\mathcal{E}_c/\mathcal{E}_j}$ is a measure for the strength of transmon nonlinearity. For $\epsilon=0$, we recover $\omega_j(\hat{\mathcal{X}}_j^2+\hat{\mathcal{Y}}_j^2)/4$, the Hamiltonian of a simple harmonic oscillator. In the transmon regime where $\epsilon\ll 1$, the leading contribution is $-\sqrt{2}\epsilon\omega_j\hat{\mathcal{X}}_j^4/48$. The coupling between qubit and the $n$th CC mode of the resonator is
\begin{align} 
g_n=\frac{1}{2}\gamma\sqrt{\chi_j}\sqrt{\omega_j\omega_n}\tilde{\varphi}_n(x_0).
\label{eqn:Light-Matter g_n} 
\end{align}

The Heisenberg-Langevin equations of motion corresponding to Hamiltonian~(\ref{Eq:2nd quantized closed H}) in the linear regime ($\epsilon=0$) for $\hat{\mathcal{X}}_{j,n}(t)$ as
\begin{subequations}
\begin{align}
&\left(d_t^2+\omega_j^2\right)\hat{\mathcal{X}}_j(t)=-\sum\limits_{n}2g_n\omega_n\hat{\mathcal{X}}_n(t),
\label{Eq:Heis Eq of Xj}\\
&\left(d_t^2+2\kappa_n d_t+\omega_n^2\right)\hat{\mathcal{X}}_n(t)=-2g_n\omega_j\hat{\mathcal{X}}_j(t)-\hat{f}_n(t),
\label{Eq:Heis Eq of Xn}
\end{align}
\end{subequations}
where $\kappa_n$ and $\hat{f}_n$ are the decay rate and noise operator coming from coupling to the waveguide degrees of freedom \cite{Senitzky_Dissipation_1960}.

Equations~(\ref{Eq:Heis Eq of Xj}-\ref{Eq:Heis Eq of Xn}) are linear constant coefficient ODEs and can be solved exactly via the unilateral Laplace transform 
\begin{align}
\tilde{h}(s)=\int_{0}^{\infty}dt h(t)e^{-st}.
\label{Eq:Def of Lap transform}
\end{align}

Taking the Laplace transform of Eqs.~(\ref{Eq:Heis Eq of Xj}-\ref{Eq:Heis Eq of Xn}) we obtain
\begin{subequations}
\begin{align}
&\left(s^2+\omega_j^2\right)\hat{\tilde{\mathcal{X}}}_j(s)+\sum\limits_n2g_n\omega_n \hat{\tilde{\mathcal{X}}}_n(s)=
s\hat{\mathcal{X}}_j(0)+\hat{\dot{\mathcal{X}}}_j(0),
\label{Eq:Laplace Eq of Xj}\\
&\left(s^2+2\kappa_ns+\omega_n^2\right)\hat{\tilde{\mathcal{X}}}_n(s)+2g_n\omega_j\hat{\tilde{\mathcal{X}}}_j(s)=(s+2\kappa_n)\hat{\mathcal{X}}_n(0)+\hat{\dot{\mathcal{X}}}_n(0)+\hat{\tilde{f}}(s).
\label{Eq:Laplace Eq of Xn}
\end{align}
\end{subequations}
The solution for $\hat{\tilde{\mathcal{X}}}_j(s)$ then reads
\begin{align}
\hat{\tilde{\mathcal{X}}}_j(s)=\frac{\hat{N}_j(s)}{D_j(s)},
\label{Eq:Laplace Sol of Xj}
\end{align}
where the numerator
\begin{align}
\hat{N}_j(s)=s\hat{\mathcal{X}}_j(0)+\hat{\dot{\mathcal{X}}}_j(0)-\sum\limits_n\frac{2g_n\omega_n\left[(s+2\kappa_n)\hat{\mathcal{X}}_n(0)+\hat{\dot{\mathcal{X}}}_n(0)-\hat{\tilde{f}}_n(s)\right]}{s^2+2\kappa_n s+\omega_n^2},
\label{eqn:Def of Nj(s)}
\end{align}
contains the operator initial conditions and the denominator
\begin{align}
D_j(s)\equiv s^2+\omega_j^2-\sum\limits_n\frac{4g_n^2\omega_j\omega_n}{s^2+2\kappa_n s+\omega_n^2}.
\label{Eq:Def of Dj(s)}
\end{align}
is the characteristic function whose roots give the hybridized poles of the full system. Therefore, we can represent $D_j(s)$ as
\begin{align}
D_j(s)=(s-p_j)(s-p_j^*)\prod\limits_n\frac{(s-p_n)(s-p_n^*)}{(s-z_n)(s-z_n^*)},
\label{Eq:Formal rep of Dj(s)}
\end{align}
where $p_{j,n}\equiv -\alpha_{j,n}-i\beta_{j,n}$ stand for the transmon-like and the $n$th resonator-like poles, respectively. Furthermore, $z_n \equiv-\kappa_n-i\sqrt{\omega_n^2-\kappa_n^2}$ is the $nth$ \textit{bare} non-Hermitian resonator mode. The notation ($p$ for poles and $z$ for zeros) is chosen based on $1/D_j(s)$ that appears in the Laplace solution~(\ref{Eq:Laplace Sol of Xj}). 

In order to compute the hybridized poles in practice, we need to truncate the number of resonator modes in $D_j(s)$. This truncation is only justified if the function $D_j(s)$ converges as we include more and more modes. First, note that without the correction give by $\chi_s$ this sum is divergent, since $g_n\thicksim \sqrt{\omega}_n\thicksim \sqrt{n}$ and for a fixed s we obtain
\begin{align}
\frac{4g_n^2\omega_j\omega_n}{s^2+2\kappa_n s+\omega_n^2}\thicksim \frac{\omega_n^2}{\omega_n^2}\thicksim 1.
\end{align}
Hence, the series in divergent. On the other hand, we found that for a non-zero $\chi_s$, $g_n\thicksim 1/\sqrt{\omega_n}\thicksim 1/\sqrt{n}$. Therefore we find
\begin{align}
\frac{4g_n^2\omega_j\omega_n}{s^2+2\kappa_n s+\omega_n^2}\thicksim \frac{1}{\omega_n^2}\thicksim \frac{1}{n^2},
\label{Eq:Convergence of Dj(s)}
\end{align}
and the series becomes convergent. In writing Eq.~(\ref{Eq:Convergence of Dj(s)}), we used the fact that $\omega_n\thicksim n$ and $\kappa_n$ has a sublinear asymptotic behavior found numerically.
\chapter{Divergence in the Wigner-Weisskopf theory of spontaneous emission}
\label{App:WignerWeisskopf}
Divergence of the Purcell decay rate appears in other frameworks besides the dispersive limit Jaynes-Cummings model as well. In this appendix, we show that the spontaneous decay rate of a qubit coupled to continuum of modes is also \textit{divergent}, unless the gauge invariance of the interaction is incorporated as presented in this manuscript. The impression of an (erroneous) finite decay rate in free space goes back to Wigner and Weisskopf's original work on spontaneous atomic decay, which implicitly makes a Markov approximation (See Sec.~$6.3$ of \cite{Scully_Quantum_1997}). We emphasize that employing the Markov approximation always yields a finite value for the decay rate regardless of the form of spectral function for electromagnetic background. 

To see this explicitly, we go over the Wigner-Weisskopf theory of spontaneous emission for a two-level system coupled to a continuum of modes inside an infinitely long 1D medium.  In interaction picture, the Hamiltonian reads 
\begin{align}
\hat{\mathcal{H}}_I=\sum\limits_{k}\hbar\left[g_{k}^*(x_0)\hat{\sigma}^+\hat{a}_{k}e^{i(\omega_j-\omega_{k})t}+H.c.\right],
\label{Eq:WW-Def of H_I}
\end{align}
which conserves the total number of excitations 
\begin{align}
\hat{N}\equiv \hat{\sigma}^+\hat{\sigma}^-+\sum\limits_{\vec{k}}\hat{a}_{\vec{k}}^{\dag}\hat{a}_{\vec{k}}.
\label{Eq:WW-Ansatz for Psi}
\end{align} 
As a result, a number conserving Ansatz for the wavefunction can be written as
\begin{align}
\ket{\Psi(t)}=c_e(t)\ket{e,0}+\sum\limits_{k}c_{g,k}(t)\ket{g,1_{k}},
\end{align}
where there is either no photon in the cavity and the qubit is in excited state $\ket{e}$, or there is a photon at frequency $\omega_k$ with qubit in the ground state $\ket{g}$. By solving the Schrodinger equation we obtain the time evolution of the unknown probability amplitudes $c_e(t)$ and $c_{g,k}(t)$. Combining these equations yields an effective equation for $c_e(t)$ as
\begin{align}
\dot{c}_e(t)=-\int_{0}^{t}dt'\mathcal{K}(t-t')c_e(t'),
\label{Eq:WW-Voltera Eq for Ce(t) 1}
\end{align}
where the memory Kernel $\mathcal{K}(\tau)$ is given by
\begin{align}
\mathcal{K}(\tau)\equiv\sum\limits_{k}|g_{k}(x_0)|^2e^{i(\omega_j-\omega_{k})t}.
\label{Eq:WW-Def of K(tau) 1}
\end{align}
Next, we replace the expression for $g_k(x_0)$ as
\begin{align}
|g_{k}(x_0)|^2=\frac{\gamma\chi_s}{4}\omega_j\omega_k|\tilde{\varphi}_k(x_0)|^2.
\label{Eq:WW-Exp for g_k}
\end{align}
Note that without respecting the gauge symmetry of interaction $|\tilde{\varphi}_k(x_0)|=\mathcal{N}(x_0)$ is $k$-independent. Moreover, the sum over $k$ can be replaced as
\begin{align}
\sum\limits_{k}\rightarrow \frac{L}{2\pi}\int_{0}^{\infty}dk=\frac{L}{2\pi v_p}\int_{0}^{\infty}d\omega_{k},
\label{Eq:WW-Disc To Cont sum}
\end{align}
for a continuum of modes, where $v_p$ is the phase velocity of the medium. Inserting Eqs.~(\ref{Eq:WW-Exp for g_k}) and~(\ref{Eq:WW-Disc To Cont sum}) into the effective Eq.~(\ref{Eq:WW-Voltera Eq for Ce(t) 1}) we obtain
\begin{align}
\begin{split}
\dot{c}_e(t)&=-\frac{1}{2\pi}\frac{\gamma\chi_s\omega_j \mathcal{N}^2(x_0) L}{4v_p}\\
&\times\int_{0}^{\infty}d\omega_k \omega_k \int_{0}^{t}dt' e^{i(\omega_j-\omega_k)(t-t')}c_e(t')
\end{split}
\label{Eq:WW-Voltera Eq for Ce(t) 2}
\end{align}
Importantly, the integral over $\omega_k$ in Eq.~(\ref{Eq:WW-Voltera Eq for Ce(t) 2}) does not converge since the integrand grows unbounded as $\omega_k\to\infty$. To resolve this, Wigner and Weisskopf assumed that the dominant contribution comes from those modes of continuum whose frequency are close to the qubit frequency. Therefore, the factor $\omega_k$ can be replaced by $\omega_j$ and by extending the lower limit of integral over $\omega_k$ to $-\infty$ we can use the identity
\begin{align}
\int_{-\infty}^{+\infty}d\omega_{k}e^{i(\omega_j-\omega_{k})(t-t')}=2\pi\delta(t-t'),
\end{align}
to arrive at a \textit{finite} value for the spontaneous decay as
\begin{subequations}
\begin{align}
&\dot{c}_e(t)\approx-\frac{\Gamma_{sp}}{2}c_e(t),\\
&\Gamma_{sp}\equiv \frac{\gamma\chi_s\omega_j^2 \mathcal{N}^2(x_0) L}{2v_p}.
\end{align}
\end{subequations}
It is worth mentioning that using Markov approximation, one always obtains a finite expression for the spontaneous decay rate regardless of the form for the spectral function. This happens because instead of integrating over the entire frequency span, the Markov approximation picks a small window around qubit frequency.

Next, we show how our natural high frequency cut-off for light-matter coupling resolves the divergence of Wigner-Weisskopf theory. First, note that applying Markov approximation is indeed unnecessary, since the Volterra Eq.~(\ref{Eq:WW-Voltera Eq for Ce(t) 1}) with the memory kernel 
\begin{align}
\mathcal{K}(\tau)=\frac{1}{2\pi}\frac{\gamma\chi_s\omega_jL}{4v_p}\int_0^{\infty}d\omega_k \omega_k|\tilde{\varphi}_k(x_0)|^2e^{i(\omega_j-\omega_k)\tau},
\label{Eq:WW-Def of K(tau) 2}
\end{align}
has an exact solution in Laplace domain as
\begin{align}
\tilde{c}_e(s)=\frac{c_e(0)}{s+\tilde{\mathcal{K}}(s)},
\end{align}
where $\tilde{\mathcal{K}}(s)\equiv \int_{0}^{\infty}d\tau\mathcal{K}(\tau)e^{-s\tau}$ is the Laplace transform and is found as
\begin{align}
\tilde{\mathcal{K}}(s)=\frac{1}{2\pi}\frac{\gamma\chi_s\omega_jL}{4v_p}\int_0^{\infty}d\omega_k \frac{\omega_k|\tilde{\varphi}_k(x_0)|^2}{s+i(\omega_k-\omega_j)}.
\label{Eq:WW-tilde(K)(s)}
\end{align}
Second, when the gauge-invariance of the interaction is incorporated, the mode amplitude is frequency dependent that experiences a high frequency suppression as
\begin{align}
|\tilde{\varphi}_k(x_0)|\sim\frac{1}{\omega_k}.
\label{Eq:WW-Def of tilde(phi)_k(x_0)}
\end{align}
Replacing Eq.~(\ref{Eq:WW-Def of tilde(phi)_k(x_0)}) into expression~(\ref{Eq:WW-tilde(K)(s)}) for $\tilde{\mathcal{K}}(s)$ we obtain
\begin{align}
\tilde{\mathcal{K}}(s)\propto\int d\omega_k\frac{1}{\omega_k[s+i(\omega_k-\omega_j)]}.
\end{align}
Interestingly, with the corrected expression for the eigenmodes, the integrand behaves like $1/\omega_k^2$ at $\omega_k\to \infty$, and as a result the integral converges. Otherwise, the integrand behaves like a constant at $\omega_k\to \infty$ and the result is divergent.
% Make the bibliography single spaced
\singlespacing
\bibliographystyle{plain}
% add the Bibliography to the Table of Contents
\cleardoublepage
\ifdefined\phantomsection
  \phantomsection  % makes hyperref recognize this section properly for pdf link
\else
\fi
\addcontentsline{toc}{chapter}{Bibliography}
% include your .bib file
%%%%%%%%%%%%%%% Publications %%%%%%%%%%%%%%
\chapter*{Copyright Permissions}
\label{App:Copyright}
\begin{itemize}
%%%%%%%%%%%%%%%%%%%%%%%%%%%%%%%%%%%%%%%%%%%%%
\item
Figures~\ref{Fig:ChargeNoise} and~\ref{Fig:Background-TransCoupledRes} are used with permissions from: \\
Jens Koch, Terri M. Yu, Jay Gambetta, A. A. Houck, D. I. Schuster, J. Majer, Alexandre Blais, M. H. Devoret, S. M. Girvin, and R. J. Schoelkopf, \href{https://journals.aps.org/pra/abstract/10.1103/PhysRevA.76.042319}{Phys. Rev. A 76, 042319} \\
Copyright (2007) by the American Physical Society\\
%%%%%%%%%%%%%%%%%%%%%%%%%%%%%%%%%%%%%%%%%%%%%
\item
Figures~\ref{Fig:Background-Rabi-Gfunction} and~\ref{Fig:Background-JC Spectrum} are used with permissions from: \\
D. Braak, \href{https://journals.aps.org/prl/abstract/10.1103/PhysRevLett.107.100401}{Phys. Rev. Lett. 107, 100401} \\
Copyright (2011) by the American Physical Society
\end{itemize}
%%%%%%%%%%%%%%%%%%%%%%%%%%%%%%%%%%%%%%%%%%%
\bibliographystyle{unsrt}
\bibliography{Dissertation-Malekakhlagh-2017}
\end{document}